\documentclass[sigconf,10pt]{acmart}

\usepackage{placeins}
\usepackage[english]{babel}
\usepackage{blindtext}
\usepackage{subfigure}
\usepackage{booktabs} %
\usepackage{subcaption}
\usepackage{tcolorbox}
\usepackage{wrapfig}

\newcommand{\parab}[1]{\vspace{0.05in}\noindent\textbf{#1}}

\usepackage{enumitem}
\usepackage{pbox}
\usepackage{wrapfig}

\renewcommand\footnotetextcopyrightpermission[1]{} %
\setcopyright{none}

\acmDOI{}

\acmISBN{}

\acmPrice{}

\begin{document}
\title{An In-Depth Investigation of \\ LEO Satellite Topology Design Parameters}

\author{Wenyi Zhang}
\authornote{Wenyi and Zihan contributed equally to this work.}
\affiliation{%
  \institution{\textit{University of California, San Diego}}
}

\author{Zihan Xu}
\authornotemark[1]
\affiliation{%
  \institution{\textit{Carnegie Mellon University}}
}

\author{Sangeetha Abdu Jyothi}
\affiliation{%
  \institution{\textit{University of California, Irvine \& VMware Research}}
}

\renewcommand{\shortauthors}{X.et al.}

\begin{abstract}
Low Earth Orbit (LEO) satellite networks are rapidly gaining traction today. Although several real-world deployments exist, our preliminary analysis of LEO topology performance with the soon-to-be operational Inter-Satellite Links (ISLs) reveals several interesting characteristics that are difficult to explain based on our current understanding of topologies. For example, a real-world satellite shell with a low density of satellites offers better latency performance than another shell with nearly double the number of satellites. In this work, we conduct an in-depth investigation of LEO satellite topology design parameters and their impact on network performance while using the ISLs. In particular, we focus on three design parameters: the number of orbits in a shell, the inclination of orbits, and the number of satellites per orbit. Through an extensive analysis of real-world and synthetic satellite configurations, we uncover several interesting properties of satellite topologies. 
Notably, there exist thresholds for the number of satellites per orbit and the number of orbits below which the latency performance degrades significantly. Moreover, network delay between a pair of traffic endpoints depends on the alignment of the satellite's orbit (Inclination) with the geographic locations of endpoints.

\end{abstract}

\maketitle

\section{Introduction}

Low Earth Orbit (LEO) satellite networks are rapidly becoming a significant component of the backbone Internet infrastructure. They offer low latency and connectivity to remote areas not served by terrestrial fiber/wireless infrastructure. Today, several commercial providers are deploying massive LEO satellite constellations, including Starlink~\cite{Starlink}, Kuiper~\cite{Kuiper}, OneWeb~\cite{OneWeb}, and Telesat~\cite{Telesat}. Unlike Geosynchronous Earth Orbit (GEO) and Medium Earth Orbit (MEO) satellite networks, LEO networks require thousands of satellites to provide good coverage globally. The most prominent LEO provider today, Starlink, has already launched over 3,000 satellites and expects to complete its first phase of 4408 satellites before 2024~\cite{sesnic2023starlink}. Eventually, Starlink is slated to have over 12,000 satellites in their LEO constellation.

A distinguishing feature of LEO satellites, compared with GEO and MEO satellites, is the presence of Inter-Satellite Links (ISLs). These links allow LEO satellites to communicate with each other directly at the speed of light without relying on an intermediate ground station. Each Starlink satellite, for example, is equipped with four optical space laser generators and four laser signal receiver surfaces, which can support optical laser communication between neighboring satellites up to 80 Gbps capacity~\cite{Starlink,nextbigfuture,defence2018towards,handley2018delay,hecht2020laser,bhattacherjee2020orbit}. 

ISL-based communication is currently tested only in limited areas, and its present use is limited to areas where ground stations are missing~\cite{smutny20095,muncheberg2019development,LaserLink1}. However, it is expected that a network backbone entirely based on ISLs for end-to-end communication will be realized in the near future. In this context, the most studied topology is the +Grid topology~\cite{wood2001internetworking,del2019technical,de2004staged, bhattacherjee2019network,sidibeh2008adaption} in which each satellite has links to the two nearest satellites in its own orbit and to the two nearest satellites on adjacent orbits on either side. Communication using the +Grid inter-satellite topology is shown to reduce temporal variations in latency, offer greater resilience to weather variations typically encountered by ground stations, and achieve $3\times$ the throughput compared to comparable LEO network without ISLs~\cite{gavish1997impact,hauri2020internet,bhattacherjee2019network,kassing2020exploring}.

Although there are several ongoing constellations deployments, our understanding of the impact of various design parameters on the performance of LEO topologies is limited, particularly in the presence of soon-to-be operational ISLs. Our preliminary analysis reveals several performance characteristics that cannot be explained based on the current understanding of LEO topology design. We analyze the latency characteristics of the five shells of the Starlink constellation with +Grid topology and observe that a shell with fewer satellites offers better latency performance than another shell with nearly double the number of satellites. More specifically, Starlink shell 1 contains 1584 satellites, and shell 2 contains 720 satellites. However, both the mean and tail latency of shell 2 is significantly lower than shell 1 (detailed in \S~\ref{sec:motivation}), achieving nearly 13.5\% lower latency with fewer than half the number of satellites of shell 1. While past work has explored the topology design space in terms of various configurations of ISL links~\cite{bhattacherjee2019network,zhang2022enabling,yang2022constellation,chen2021analysis,pan2019opspf,werner2001topological,chang1995topological}, the peculiarities discovered in our preliminary experiments underscore the need for systematic investigation of LEO topology characteristics.

In this work, we conduct an in-depth investigation of LEO satellite topology design parameters to help guide the design of future LEO topologies as well as routing and traffic engineering schemes. In particular, we analyze the impact of three design parameters---the number of orbits in a shell, the inclination of orbits, and the number of satellites per orbit---on network performance. We also analyze how the relationship between the geographical distribution of traffic sources and the inclination of satellite orbits affects the performance of the constellation. While past work has analyzed the performance characteristics of a limited set of proposed commercial topologies~\cite{kassing2020exploring,perdices2022satellite,michel2022first,kassem2022browser}, our work presents the first comprehensive study across topology design parameters in a broad design space.

We analyze topologies of real-world LEO constellations: Starlink, Kuiper, Telesat, and a broad set of synthetic configurations. Since ISL usage is currently limited in real-world constellations, we rely on simulations to understand the performance of future ISL-enabled LEO backbone networks. We generate the synthetic topologies by varying the design parameters (number of orbits per shell, inclination of orbit, and number of satellites per orbit) over a wide range. We evaluate each configuration under a synthetic traffic matrix commonly used in prior work, a traffic matrix between the top 100 GDP city pairs~\cite{perrig2017scion,nordhaus2016global} spanning a significant fraction of Internet users. Our findings reveal several insights that are expected to hold across a broad range of traffic matrices.

Based on extensive experiments, we identify several important characteristics of LEO topology design. We highlight two key observations here. First, there exists a threshold for the number of satellites per orbit below which the performance degrades significantly. For example, when the number of satellites per orbit is below 28, there is a significant performance drop even when the total number of orbits is increased to a very high value of 59. Second, we show that the latency performance is better when the angle between the traffic endpoints is closely aligned with the satellite orbit inclination. In this case, most hops in the path follow intra-orbit links instead of inter-orbit links, thereby significantly improving the latency performance.

In summary, we make the following contributions:
\begin{itemize}[leftmargin=*,nolistsep]
\item We conduct an in-depth investigation of the impact of three previously unexplored topology design parameters in LEO constellations: the number of orbits per shell, the number of satellites per orbit, and the inclination of orbits. 

\item We investigate three real-world constellations and a broad range of synthetic constellations to understand the impact of design parameters on several network performance metrics: RTT, number of path changes, hop count, etc.

\item We identify the threshold for various design parameters below which the constellation performance degrades significantly. Using these observations, we are able to explain the better network performance of real-world constellations with fewer satellites.

\item We examine the impact of the relationship between traffic distribution and the inclination of satellite orbits on network performance.

\item We will open-source the code and data in this paper.
\end{itemize}

This work does not raise any ethical concerns.

\section{Background \& Related Work}
This section provides an overview of LEO satellite networks and the recent work on LEO topology design and LEO satellite network measurement.

\parab{LEO Constellation Structure.}
The Low Earth Orbit (LEO) satellite network relies on a group of satellites orbiting the Earth at low altitudes, ranging from 500 to 1500 km, much lower than traditional geostationary satellites at an altitude of 36,000 km. A large LEO satellite constellation may contain thousands of satellites that are distributed across a number of orbits, where each orbit is a circular path around the globe. 

\begin{wrapfigure}{r}{0.25\textwidth} %
  \centering %
  \includegraphics[width=0.25\textwidth]{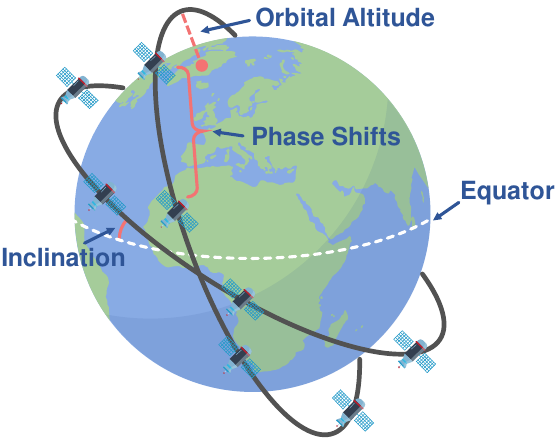}  %
  \vspace{-0.25cm}
  \caption{Parameters of LEO constellation.}    
  \label{constellation parameters}                             
\end{wrapfigure}

In order to fully describe the LEO mega-constellation, several parameters are required, as shown in Fig.~\ref{constellation parameters}: (a) The \textit{orbital altitude} of a satellite is its height above sea level, which, in turn, determines its velocity and coverage area; (b) The \textit{number of orbits} in a shell determines the coverage. (c) The \textit{number of satellites per orbit} in a constellation determines the density of the entire constellation. The density of the constellation affects the availability and redundancy of the network service; (d) The \textit{inclination} of an orbital plane is the angle between the Equator and a satellite in that orbit traveling northward. The inclination of orbits determines the coverage region of the constellation. A lower inclination provides better coverage in equatorial regions while a higher inclination provides better coverage in polar regions; (e) \textit{Phase shifts} of a satellite constellation determines the difference in position between one orbital plane and another. Existing constellations such as Starlink~\cite{Starlink}, Telesat~\cite{Telesat}, and Kuiper~\cite{Kuiper} apply uniform phase shift deployment, which evenly places the orbital planes in a full circle.

\parab{Inter-Satellite Connectivity.}
Traditionally, satellite communication relied on Ground Station-Satellite Links (GSLs) using a bent-pipe space-ground architecture, where satellites only act as relays that transmit data to the next ground station~\cite{zhang2022enabling,angeletti2008bent}. However, this architecture's performance may be compromised by several factors. The signal transmission time between the satellite and the ground stations can result in significant latency. Also, GSLs are vulnerable to interference from other communication signals, weather conditions, and other sources~\cite{leyva2020leo,del2019technical, motzigemba2019optical}. In addition, the satellites in LEO constellations are highly dynamic, resulting in limited time windows to communicate with ground stations.

Today, satellites are equipped with lasers that enable Inter-Satellite Links (ISLs) to provide a promising improvement over GSLs. ISL-based communication architecture allows the satellites to communicate directly with each other and, in turn, reduces the reliance on ground stations. For example, each satellite in Starlink~\cite{Starlink} has the hardware capabilities for four ISL links. The most studied topology is the +Grid topology~\cite{bhattacherjee2019network,hu2020directed,wang2006benchmarkinng,mclaughlin2023grid}, where each satellite connects to two adjacent satellites in the same orbit and the two nearest satellites on each of its neighboring orbits. The topology of the three existing major constellations is shown in Appendix~\ref{apeendix: Visualization 1}.

\parab{LEO Constellation Performance Analysis.} 
Several past works measure the performance of existing satellite constellations. A recent study~\cite{perdices2022satellite} investigates the performance of GEO satellite networks, highlighting the importance of ground station selection. Michel et al.~\cite{michel2022first} compare the performance of Starlink with GEO networks and show that the LEO constellation can achieve higher throughput and lower data transmission latency. Kassem et al.~\cite{kassem2022browser} examine the connectivity of Starlink, suggesting that the current bent-pipe connection results in high performance variability during bad weather conditions.

Due to limited practical deployments, several works resorted to simulation-based performance analysis of LEO networks. 
Kassing et al.~\cite{kassing2020exploring} developed a satellite network simulator and compared three prominent LEO satellite constellations (Starlink, Kuiper, and Telesat). Izhikevich et al.~\cite{izhikevich2023democratizing} introduced LEO HitchHiking to democratize LEO satellite measurements. Bhosale et al.~\cite{bhosale2023astrolabe} developed a model that captures the RTT variability exhibited in LEO networks.

Some recent works analyze the impact of satellite constellation parameter settings on performance, albeit under limited scope. Bhosale et al.~\cite{bhosale2023characterization} characterized route variability in LEO satellite networks, focusing on route churn and RTT variability. Basak et al.~\cite{basak2023exploring} focused on designing constellations under a specified budget for target markets. All prior work on simulation-based LEO constellation analysis~\cite{peluso2022megaconstellations, bhosale2023characterization, basak2023exploring} explore the performance of constellations over a limited range of parameter settings and provide a high-level overview of a small subset of satellite network characteristics. Moreover, certain performance characteristics of the constellation are inexplicable based on our current understanding. Therefore, a comprehensive investigation of the impact of parameter settings on constellation performance is currently needed. This work aims to fill this gap.

\section{Motivation}
\label{sec:motivation}

\begin{figure}[tp] %
  \centering %
  \includegraphics[width=0.6\linewidth]{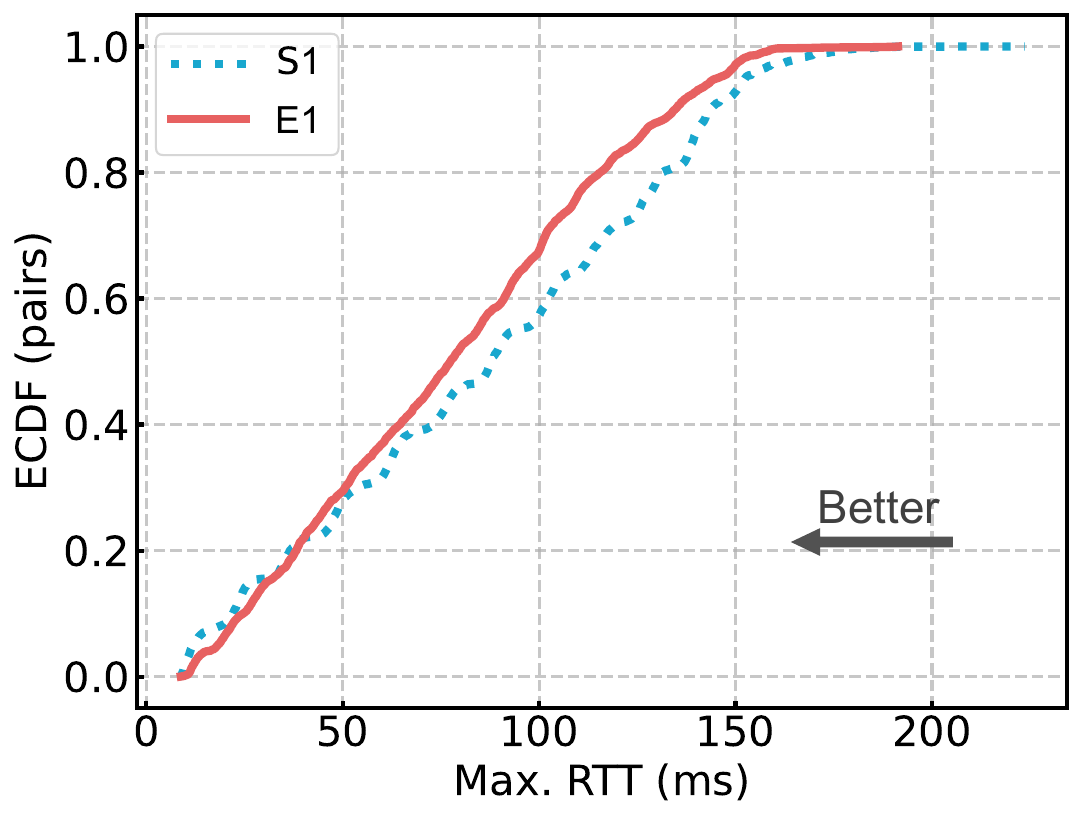}  %
  \vspace{-0.25cm}
  \caption{Distribution of maximum RTT (ms) across top 100 city pairs using Starlink S1 and Custom-designed topology, Example E1 .}    
  \label{s1 vs s2 rtt}                             
  \vspace{-0.25cm}
\end{figure}

To emphasize the importance of design parameter analysis, we compare the latency performance of two satellite constellations: Starlink shell 1 (S1) with 1584 satellites and our custom-designed example shell (E1) with 720 satellites. Both constellations share the same altitude but differ in orbit count (S1: 72, E1: 20), satellites per orbit (S1: 22, E1: 36), and inclination (S1: 53°, E1: 70°). Using these configurations in a +Grid topology, we evaluate latency between the top 100 most populous city pairs (details in~\S~\ref{sec:Design}).  In Fig.~\ref{s1 vs s2 rtt}, we present the ECDF of maximum RTT for each constellation, revealing that E1 exhibits lower mean and tail latencies. This unexpected result challenges the assumption that higher satellite counts inherently improve performance, as S1, despite having nearly double the satellites, shows inferior latency. Given the disparities in all three design parameters between S1 and E1, identifying the specific factor contributing to S1's latency issues proves cumbersome.

\section{Experiment Design}
\label{sec:Design}
In this section, we give an overview of the evaluation metrics and experiment configurations that we use for understanding the impact of various satellite topology design parameters.

\subsection{Evaluation Metrics}

The major factors that affect the user experience of data transmission over a satellite network are latency and network stability. In order to quantify these factors, we organize the evaluation metrics into three categories: \textit{(1) Round-trip time (RTT), (2) Path changes,} and \textit{(3) Hop count}.

\parab{Round-Trip Time (RTT).}
 The round-trip time denotes the propagation latency, i.e., the time taken for a packet of data to travel from a sender to a receiver and back to the sender. The RTT depends on the total distance between the sender and receiver, which includes 1) the line-of-sight distance from the sender to the first-hop satellite, 2) the sum of distances between satellites along the routing path, and 3) the distance from the last-hop satellite to the receiver. The RTT is computed as the total distance divided by the speed of light in vacuum (3*$10^8$ m/s). For satellite network evaluation, we consider the following characteristics of RTT: \\
 \textit{(a) Max. (Min.) RTT}, which denotes the maximum (minimum) RTT observed between the sender and the receiver during a specified period. RTT variations occur in satellite networks due to the changes in topology caused by satellite motion. \\
\textit{(b) Max. RTT - Min. RTT}, is the difference between the maximum RTT and minimum RTT observed between the sender and the receiver during a specified period. This metric reflects the fluctuations in RTT experienced by a connection.

Since RTT measurements are dependent on the distance between the sender and the receiver, we define two additional metrics that facilitate comparison of performance across different pairs of endpoints.

\noindent
\textit{(c) Max. to Min. Slowdown} captures the extent of variation in RTT observed between a pair of endpoints. It is defined as ${S}_{m} = \frac{max. RTT }{min. RTT }$, where $max.$ $(min.)$ $RTT$ is the maximum (minimum) RTT during a period of time. \\ 
\textit{(d) Geodesic Slowdown} is the slowdown with respect to the best possible path. It is defined as ${S}_{g} = \frac{max. RTT }{ideal. RTT }$, where $ideal.RTT$ is the RTT obtained when directly transmitting data between two endpoints through the shortest Euclidean distance on the earth's surface at the speed of light.

\parab{Path Changes.}
The dynamic nature of LEO networks with constantly moving satellites results in changes in the routing paths over time. Frequent changes in these paths can lead to jitter in data transmission, high computational overhead for updating routing tables, and loss of connectivity. \\
\textit{(e) The number of path changes} between two endpoints over a period of time denotes the extent of network stability.

\parab{Hop count.}
While using ISLs, data is transmitted between satellites over a series of hops. Each hop will result in additional transmission delay as well as energy consumption. If the number of hops increases significantly, it may indicate high latency in transmitting data. Hence, we measure the following characteristics of hops between two endpoints:  \\
\textit{(f) Average hop count} denotes the average number of hops between a pair of endpoints during a specified period of time. \\
\textit{(g) Max. hop count - Min. hop count} is the difference in the number of hops across paths during a specified period. \\
\textit{(h) Max. hop count / Min. hop count} denotes the relative change in the number of hops between a pair of endpoints during a specific period of time.

\subsection{Experiment Configurations}

The end-to-end network performance over LEO constellations can vary widely based on a variety of configuration parameters. Under a given budget of the number of satellites, it is unclear what set of configurations would lead to optimal performance. Therefore, it is essential to understand the impact of each configuration parameter on network performance.  What are the most important parameters affecting LEO mega-constellation network performance? In what ways and along what dimensions do they influence the performance of satellite networks? How do we design an optimal LEO mega-constellation? In order to scientifically and rigorously answer the above questions, we begin with the existing commercial LEO mega-constellations and expand to an in-depth analysis of synthetic LEO mega-constellations generated across a wide range of parameter settings. We design three key experiments to explore the impact of LEO mega-constellation parameter settings.

\parab{(i) Exploring different shells of the existing LEO mega-constellations.} We first simulate three leading existing LEO mega-constellations and analyze the network performance of their shells: five shells for Starlink Phase I, three shells for Kuiper, and two shells for Telesat. Detailed parameter settings are given in~\S~\ref{Commercial LEO}. We focus on four of the most representative metrics---Max. RTT, Geodesic Slowdown, (Max. RTT - Min. RTT), Number (\#) of Path Changes---to evaluate these shells, detailed experimental results are in~\S~\ref{Comparing Different Shells}.

\parab{(ii) Exploring different synthetic configurations of LEO mega-constellations.} In order to explore the impact of parameters on network performance, we analyze different settings for three parameters---Orbit Number, Satellite Number per Orbit, and Inclination. We vary each parameter independently while fixing others and generate 20 synthetic LEO mega-constellations (the settings are shown in~\S~\ref{Synthetic LEO Mega-constellations}). We select the three most representative metrics: Max. RTT, Geodesic Slowdown, and Number (\#) of Path Changes for evaluation, (detailed results in~\S~\ref{Impact of Different Configurations of LEO Mega-constellations}).

\parab{(iii) Understanding the Impact of User Endpoints Location.} Visualization of LEO paths reveals varying levels of zig-zag in paths between various user endpoint pairs (Appendix~\ref{apendix:visulization}). We conjecture that the degree of alignment between the Inclination of the satellite orbit and the Geographic Angle~\cite{bearangle} of the plane between source and destination endpoint locations could influence the latency performance. To verify this conjecture, we divide the traffic between user endpoints into nine categories based on the Geographic Angle of the plane connecting them (detailed in Appendix~\ref{Appendix:geographic angle}) and evaluate the network performance of these nine groups under different Inclination values of synthetic constellations. We choose the two most representative indicators, average hops and Geodesic Slowdown, for analysis. Max. RTT is not a suitable metric in this analysis since endpoint pairs in the nine groups have widely different distance distributions.

\section{Evaluation}
In this section, we provide an overview of the datasets used in our analysis, the implementation of our analysis framework, and our key observations while answering the following questions:

\begin{itemize}[leftmargin=*]
\item What are the key parameters that influence the performance of an LEO network?
\item Are there thresholds for each parameter above/below which the performance degrades significantly?
\item Why does a shell with a low number of satellites outperform another shell with nearly double the number of satellites (S1 vs. S2 of Starlink)?
\item How does the difference between satellite orbit inclination and the geographical angle of the plane containing the source and destination location affect performance?
\end{itemize}

\subsection{Datasets}

\begin{table}
\small
    \centering
\begin{tabular}{cccccc}
\toprule  %
Name & H (km) & Orb. & Sats/Orb & Total Sats & Incl.\\
\midrule  %
S1 & 550 & 72 & 22 & 1584 & 53.0°\\
S2 & 570 & 20 & 36 & 720 &70.0°\\
S3 & 560 & 6 & 58 & 348 &97.6°\\
S4 & 540 & 72 & 22 & 1584 &53.2°\\
S5 & 560 & 4 & 43 & 172 &97.6°\\
\midrule  %
K1 & 630 & 34 & 34 & 1156 & 51.9°\\
K2 & 610 & 36 & 36 & 1296 & 42°\\
K3 & 590 & 28 & 28 & 784 &33°\\
\midrule  %
T1 & 1015 & 27 & 13 & 351 & 98.98°\\
T2 & 1325 & 40 & 33 & 1320 & 50.88°\\
\bottomrule %
\end{tabular}
  \caption{Shell configurations for Starlink’s first phase (S1-S5),
Kuiper (K1-K3), and Telesat (T1-T2).}
\label{Shells 1}
\vspace{-0.95cm}
\end{table}

\subsubsection{Commercial LEO Mega-constellations}
\label{Commercial LEO}
In Table~\ref{Shells 1}, we detail the configurations of the three leading commercial LEO mega-constellations used in our evaluation: Starlink~\cite{Starlink}, Kuiper~\cite{Kuiper}, and Telesat~\cite{Telesat}.

\subsubsection{Synthetic LEO Mega-constellations}
\label{Synthetic LEO Mega-constellations}
We generate multiple synthetic constellations to understand the impact of parameters better. We fix each parameter to a constant value and vary others. First, we fix the inclination angle to 53°, covering more than 90\% of the cities on the earth and almost all the Internet user population, and synthesize satellite shells with different number of orbits (20, 33, 46, and 59) and number of satellites per orbit (20, 28, 36, and 44). To understand the impact of inclination, we fix the number of orbits to 33 and the number of satellites per orbit to 28 and vary the inclination across four values (45°, 55°, 65°, and 75°). Thus, we have a total of 20 synthetic shells.
We use a fixed orbital altitude of 570km for all synthetic shells. Note that we analyze the design parameter, the altitude of the satellite orbit, and find that variations in altitude had minimal impact on network performance in the altitude range of LEO satellites (Appendix~\ref{apeendix: Altitude}).

\subsubsection{Traffic Matrix}
To evaluate the performance of the LEO satellite-based networks, based on the known correlation between Internet penetration and GDP~\cite{amiri2013internet}, we choose the Top 100 Gross Domestic Product (GDP) cities as a representative to generate the traffic matrices for Internet use requests, following the methodology of past work on satellite network performance~\cite{bhattacherjee2019network}. The top 100 cities with the highest GDPs encompass a significant portion of the global Internet population~\cite{amiri2013internet}.  We collect the latitude and longitude of these Top 100 GDP cities and generate traffic matrices with one city as the source and another as the destination, with a total of 9,900 city pairs. We maintain constant traffic between the 9900 pairs of cities during the period of evaluation. For each time step in the simulation, we take into account the communication load between all pairs of cities and estimate their instantaneous network performance. (detailed in Appendix~\ref{appendix: Visualization of User Endpoints}). Note that several observations in this paper hold independent of the traffic matrix.

\subsection{Implementation}
\subsubsection{LEO Mega-constellations Generation}

Using an open-source LEO satellite network analysis platform, Hypatia~\cite{kassing2020exploring}, we generate the dynamic states of the LEO constellations under various configurations. Each experiment is conducted for a duration of 400 seconds. We snapshot the link pattern of LEO mega-constellations in 1-second increments and evaluate the network performance characteristics at each time step. Note that the topology patterns of constellations exhibit cyclical behavior with an approximate time period of 200 seconds. Hence, an evaluation period of 400 seconds is sufficient to cover the complete span of topology configurations and comprehensively reflect the performance patterns. 

We use the NetworkX~\cite{hagberg2008exploring} library to generate a network graph of the LEO satellite-based network for each time interval, considering the satellite positions and link lengths between satellites and ground stations. We calculate the forwarding state for each node based on a range of routing strategies. Our implementation employs shortest-path routing, computed using the Floyd-Warshall algorithm~\cite{hougardy2010floyd}. The latest forwarding state is read into static routing table entries during each change event in the network. Based on the forwarding state in each snapshot, we estimate the various evaluation metrics using the city pairs-based traffic matrix. In order to simulate the dynamic states and compute the evaluation metrics, we rely on three files: the TLE file, the ISL file, and the TM file.

\noindent \textbf{TLE File: } The TLE format is a two-line element (TLE) standard format for representing the trajectory of an Earth-orbiting object~\cite{TLE}. It is essential for satellite mobility modeling. Using the configuration parameters of real-world and synthetic constellations, we calculate the movement of satellites in space and save their positions in the TLE file.

\noindent \textbf{ISLs File: }
We connect each satellite using the +Grid topology and build the network for the LEO mega-constellations. We store these edges between satellites in an ISLs file, which contains the network graph.

\noindent \textbf{TM File: }
To balance the coverage of users and computational overhead, we selected 100 cities with the highest Gross Domestic Product (GDP) as traffic endpoints. We consider all pairs of cities in this set and store the traffic endpoints in the Traffic Matrix file (TM file). This file contains the latitude, longitude, and city ID of all pairs of traffic endpoints.

\subsubsection{Geographic Angle Between Traffic Endpoints Implementation}
\label{subsec:Different Geographic Angle Ground Station Implementation}
To investigate the impact of the alignment of the plane containing the source and destination of traffic with the satellite orbit inclination on network performance, we group the traffic endpoint pairs into nine categories based on their geographic angle (intervals of 10° ranging from 0°-10° to 80°-90°). The geographic angle is the angle between the plane connecting the endpoint pair and the equator (Appendix~\ref{appendix: Geographic Angle (Bearing Angle)}).

\subsection{Results}
\subsubsection{Real-World LEO Constellations' Performance}\ 
\label{Comparing Different Shells}

\begin{figure*}[htp]
	\centering
	\subfigure[Starlink shell 1-5]{
		\begin{minipage}[t]{0.24\linewidth}
			\centering
			\includegraphics[scale=.25]{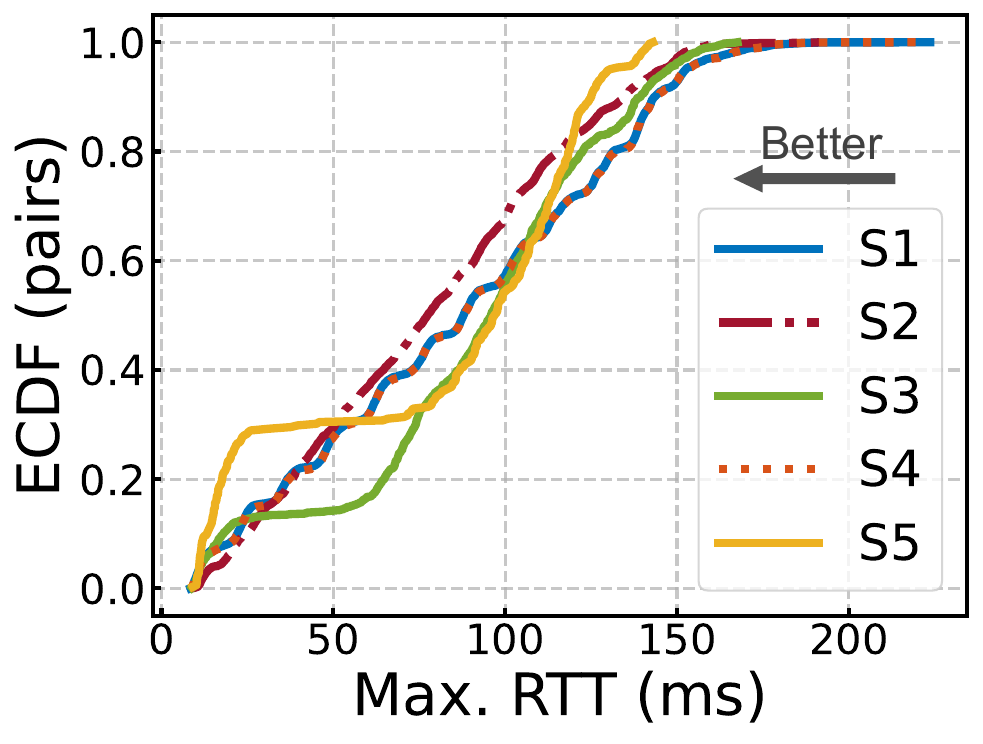}
			\vspace{-0.5cm}
   \label{starlink max rtt}
		\end{minipage}%
	}%
	\subfigure[Starlink shell 1-5]{
		\begin{minipage}[t]{0.24\linewidth}
			\centering
			\includegraphics[scale=.25]{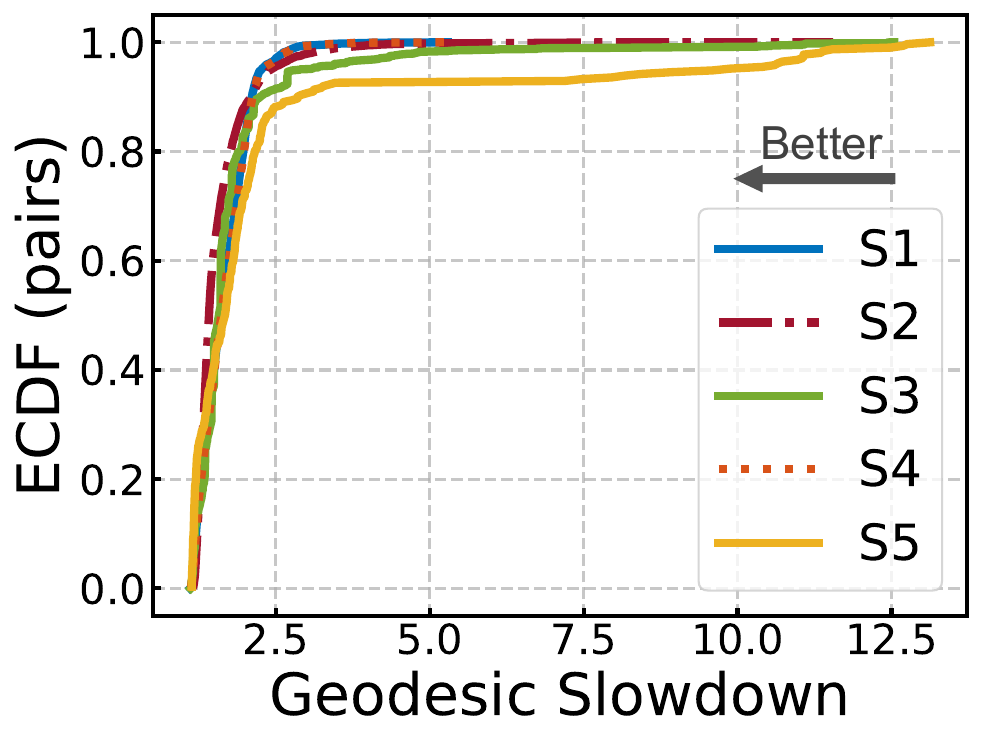}
			\vspace{-0.5cm}
   \label{starlink geodesic slowdown}
		\end{minipage}%
	}%
	\subfigure[Starlink shell 1-5]{
		\begin{minipage}[t]{0.24\linewidth}
			\centering
			\includegraphics[scale=.25]{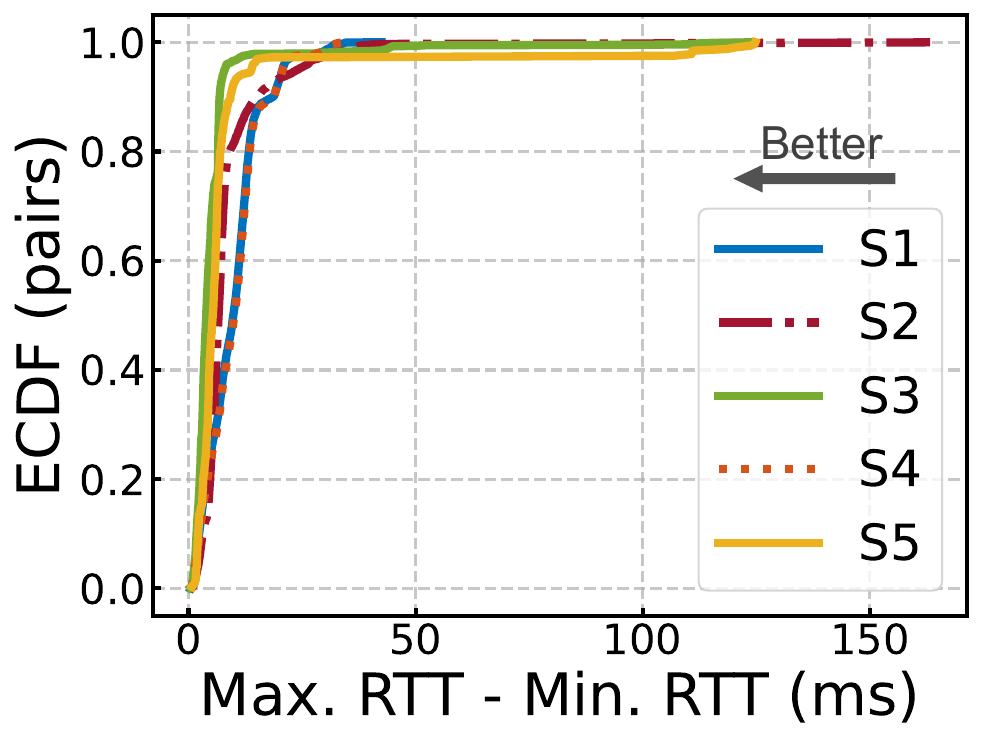}
			\vspace{-0.5cm}
   \label{starlink max minus min rtt}
		\end{minipage}%
	}%
	\subfigure[Starlink shell 1-5]{
		\begin{minipage}[t]{0.24\linewidth}
			\centering
			\includegraphics[scale=.25]{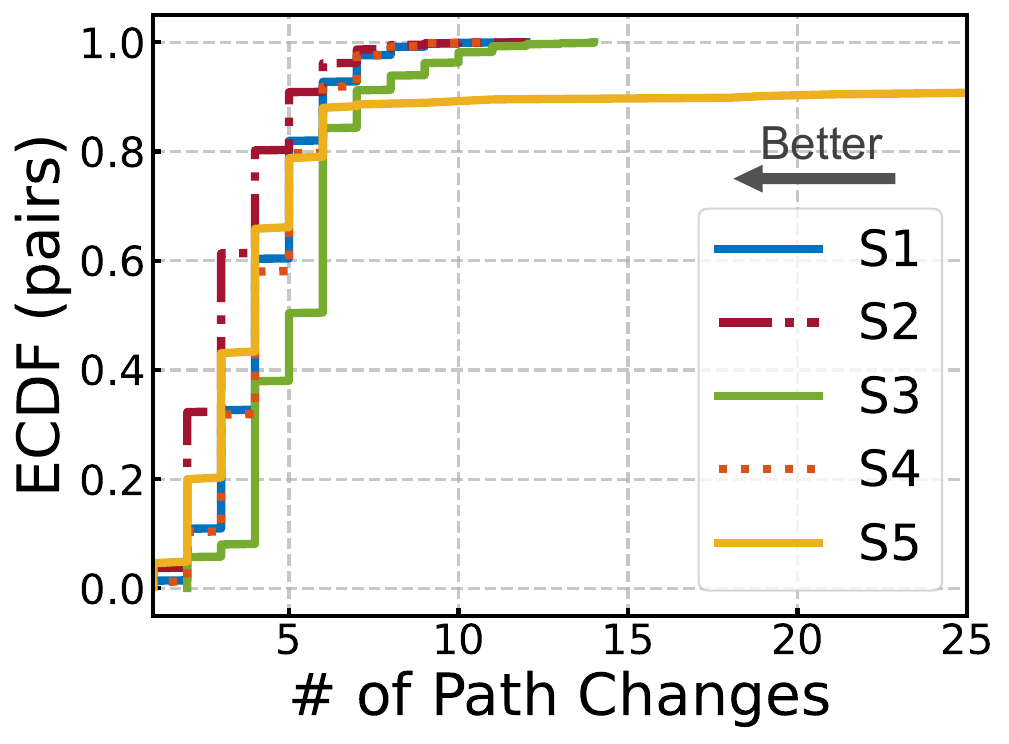}
			\vspace{-0.5cm}
   \label{starlink path changes}
		\end{minipage}%
	}%
    \vspace{-0.5cm}
	\caption{The network performance of Starlink shells 1-5 (S1-S5) of Phase I. For all curves, lower values indicate better performance. Long tails indicate outliers with poor performance. The tail of S5 is truncated in (d) due to a very high number of path changes.}
	\vspace{-0.25cm}
	\label{starlink shells}
\end{figure*}

\begin{figure*}[htp]
	\centering
	\subfigure[Kuiper and Telesat shells]{
	\vspace{-0.5cm}
		\begin{minipage}[t]{0.24\linewidth}
			\centering
			\includegraphics[scale=.25]{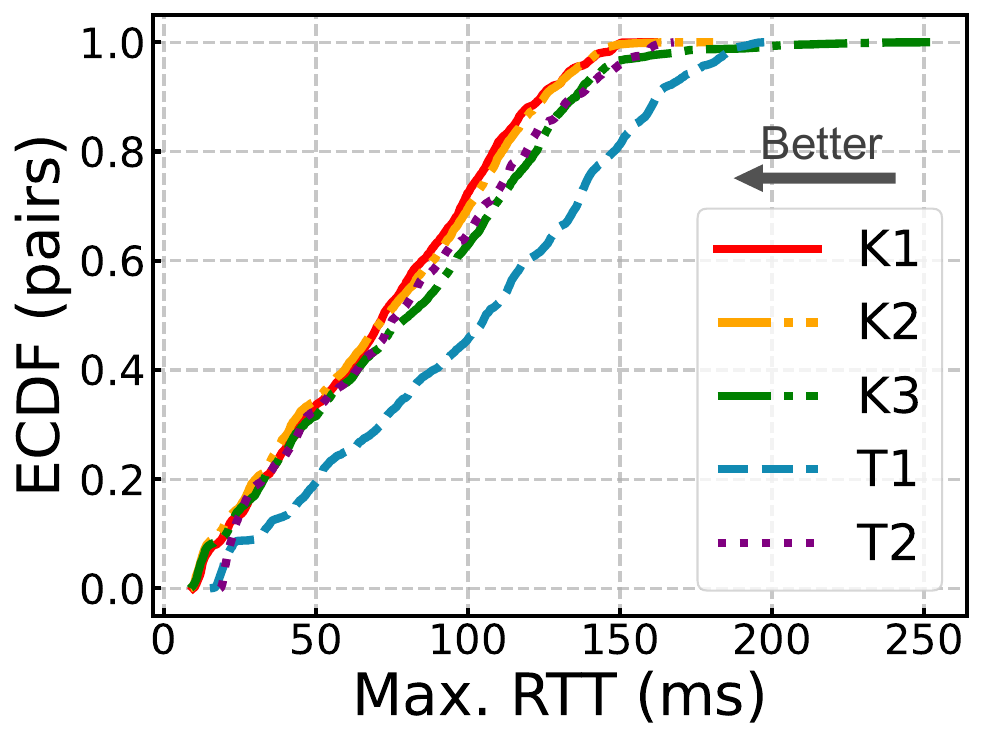}
			\vspace{-0.5cm}
   \label{kuiper & telesat max rtt}
		\end{minipage}%
	}%
	\subfigure[Kuiper and Telesat shells]{
	\vspace{-0.5cm}
		\begin{minipage}[t]{0.24\linewidth}
			\centering
			\includegraphics[scale=.25]{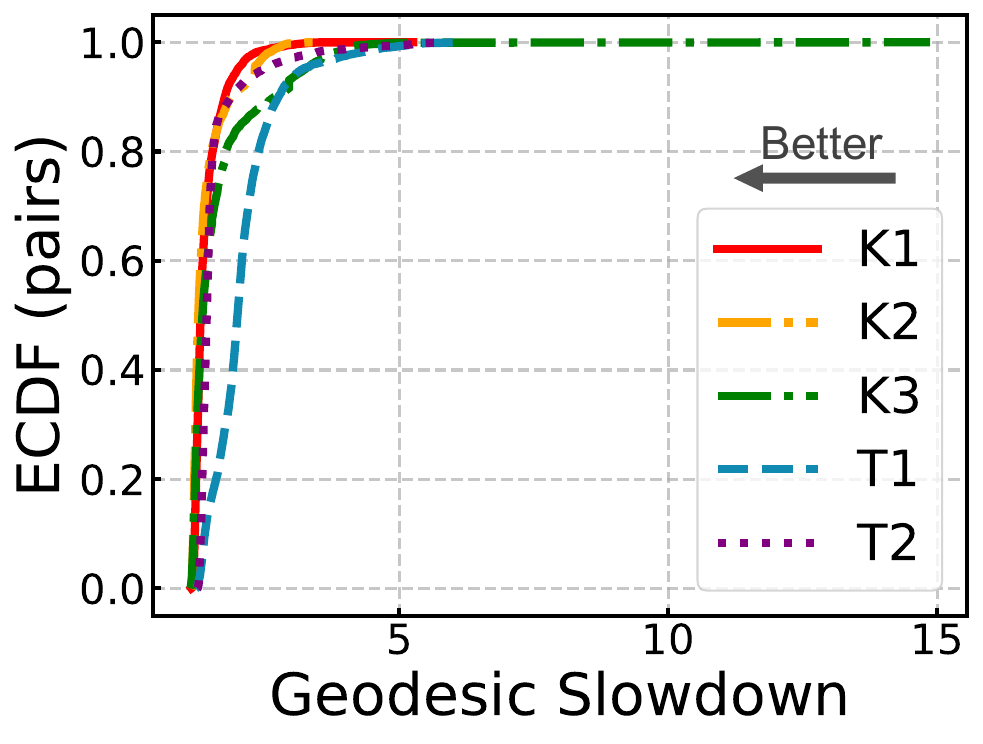}
			\vspace{-0.5cm}
   \label{kuiper & telesat geodesic slowdown}
		\end{minipage}%
	}%
	\subfigure[Kuiper and Telesat shells]{
	\vspace{-0.5cm}
		\begin{minipage}[t]{0.24\linewidth}
			\centering
			\includegraphics[scale=.25]{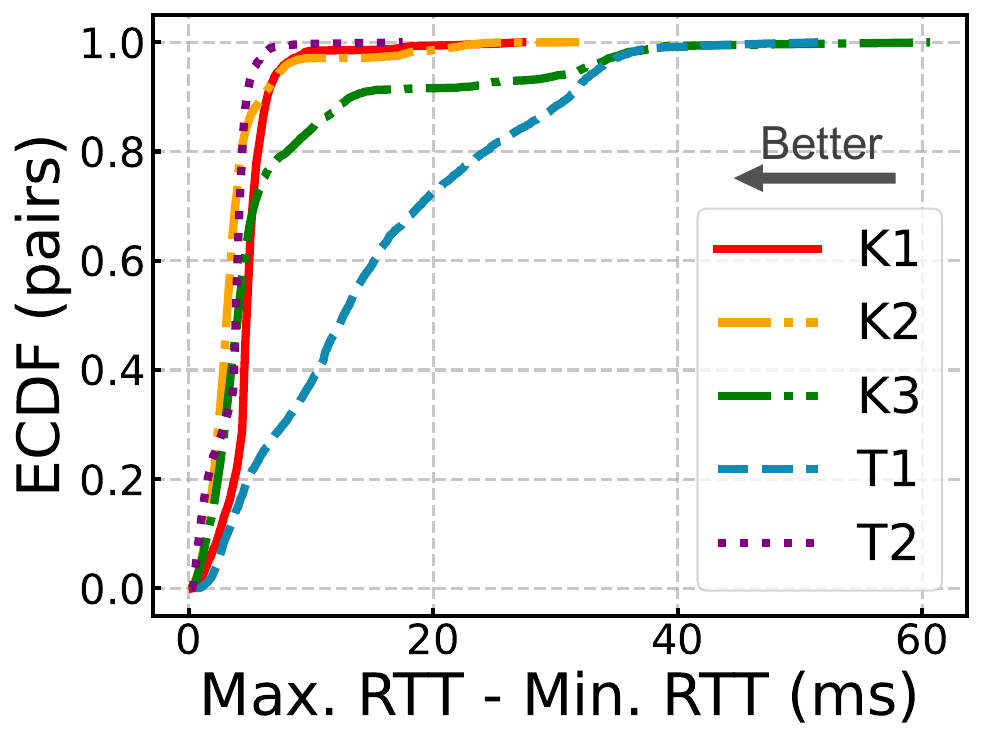}
			\vspace{-0.5cm}
   \label{kuiper & telesat max minus min rtt}
		\end{minipage}%
	}%
	\subfigure[Kuiper and Telesat shells]{
	\vspace{-0.5cm}
		\begin{minipage}[t]{0.24\linewidth}
			\centering
			\includegraphics[scale=.25]{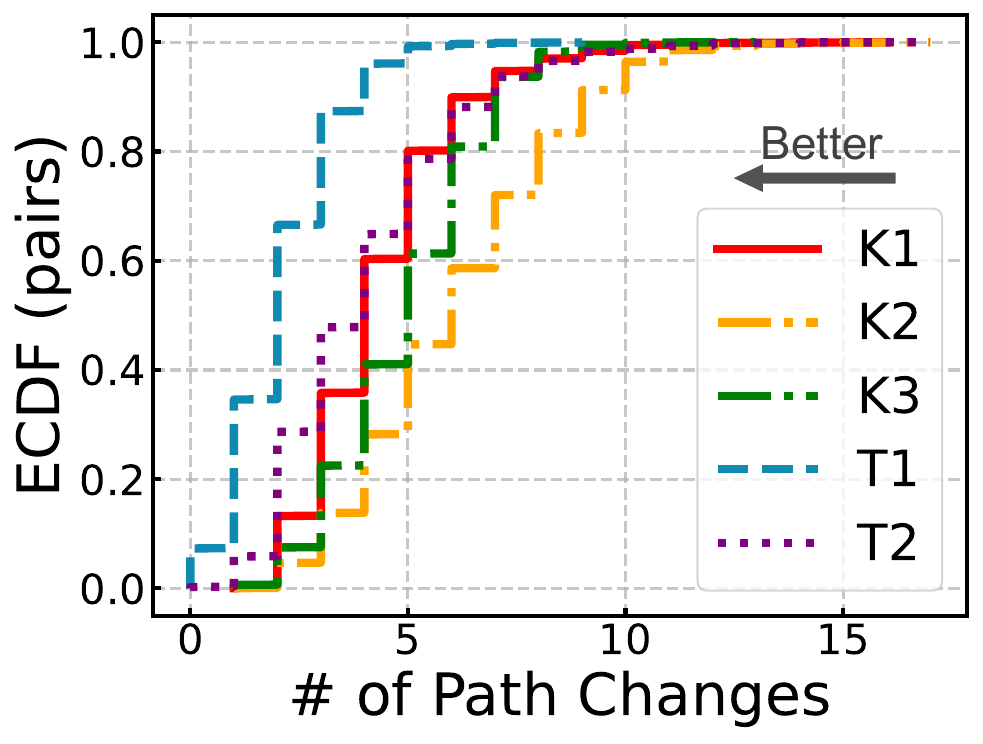}
			\vspace{-0.5cm}
   \label{kuiper & telesat path changes}
		\end{minipage}%
	}%
\vspace{-0.5cm}
	\caption{The network performance of shells 1-3 (K1-K3) of Kuiper and shell 1 and 2 (T1, T2) of Telesat. For all curves, lower values indicate better performance. Long tails indicate outliers with poor performance.}
	\vspace{-0.25cm}
	\label{kuiper & telesat shells}
\end{figure*}

\parab{Max. RTT: } We evaluate the maximum RTT observed between city pairs over a period of 400s. In Fig.~\ref{starlink max rtt}, we observe that different shells of Starlink have significant variations in maximum RTT. This behavior can be explained based on the shell configurations. Shells 1,2 and 4 have a large number of satellites and offer good coverage. However, shells 3 and 5 are sparse with many endpoints away from the poles unreachable during significant durations, causing large fluctuations in Max. RTT. In Fig.~\ref{kuiper & telesat max rtt}, we observe that both Kuiper and Telesat with shells that have large coverage also have an even distribution of Max. RTT.

\noindent\textbf{Geodesic Slowdown: }
We investigate the geodesic slowdown of the three constellations and find that over 80\% of the endpoint pairs have a maximum RTT less than 2.5$\times$ geodesic RTT, showing good latency reduction compared to today’s terrestrial Internet networks (terrestrial fiber paths are often long-winded, and the speed of light in fiber is roughly two-thirds of the speed of light in air leading to poor latency performance on land~\cite{internet-slow}). However, some shells have a long tail for the geodesic slowdown. In Fig.~\ref{starlink geodesic slowdown}, shells 2, 3, and 5 have longer tails than shells 1 and 4. Fig.~\ref{kuiper & telesat geodesic slowdown} also shows that when the total number of satellites decreases, the slowdown tends to have a longer tail. Prior work~\cite{kassing2020exploring} indicated that a drastic increase in the ratio might be caused by short-distance data transmission—the latency overhead of ground station-satellite link resulting in significant inflation of geodesic slowdown. Besides the overhead, the sparsity of the satellite constellation also causes an increase in geodesic slowdown. Routing across a limited number of satellites results in a longer path than a fully covered satellite shell.  

\noindent\textbf{Max. RTT - Min. RTT: }
To examine the latency variation, we further explore the difference between maximum and minimum RTT over time. Fig.~\ref{starlink max minus min rtt} shows that for Starlink, over 90\% of the endpoint pairs have latency variation less than 25ms. Fig.~\ref{kuiper & telesat max minus min rtt} shows that for Kuiper, over 90\% of the endpoint pairs have latency variation less than 20ms. We observe that shell 2 of Telesat has the lowest latency variation, with almost 100\% of the endpoint pairs having less than 10ms variation. Similar to the geodesic slowdown, the total number of satellites affects the tail of the latency variation.

\noindent\textbf{Number of path changes: }
We investigate the number of path changes to understand the stability of LEO mega-constellations. Fig.~\ref{kuiper & telesat path changes} shows that with fewer satellites, the shell has less number of path changes. However, in Fig.~\ref{starlink path changes}, we observe that with the smallest number of satellites, shells 3 and 5 have the greatest number of path changes. When the coverage of a satellite constellation is above a threshold, the number of path changes increases as the satellite density increases due to frequent shifts in the best available shortest path in a dense and dynamic network. However, below this threshold, the number of path changes is high due to poor coverage and loss of paths (detailed results in Appendix~\ref{appendix: Commercial LEO mega-constellations}).

\begin{figure*}[t]
	\centering
	\subfigure[20 Orbits]{
	\vspace{-0.5cm}
		\begin{minipage}[t]{0.24\linewidth}
			\centering
			\includegraphics[scale=.25]{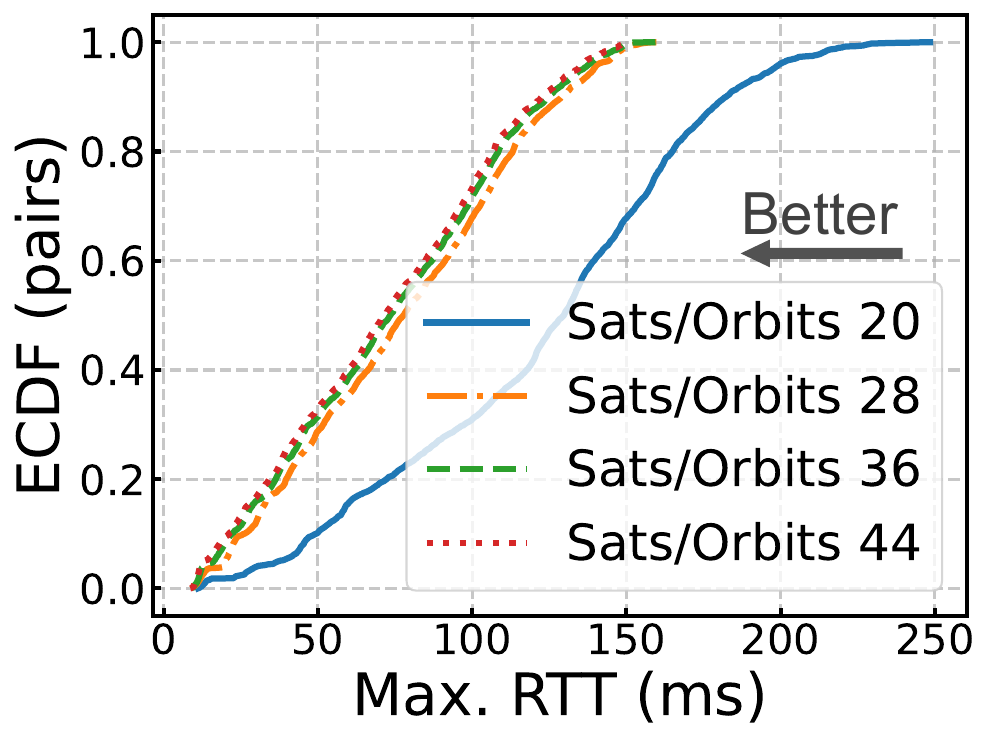}
			\vspace{-0.5cm}
      	\label{RTT Sats/Orbits 1}
		\end{minipage}%
	}%
	\subfigure[33 Orbits]{
	\vspace{-0.5cm}
		\begin{minipage}[t]{0.24\linewidth}
			\centering
			\includegraphics[scale=.25]{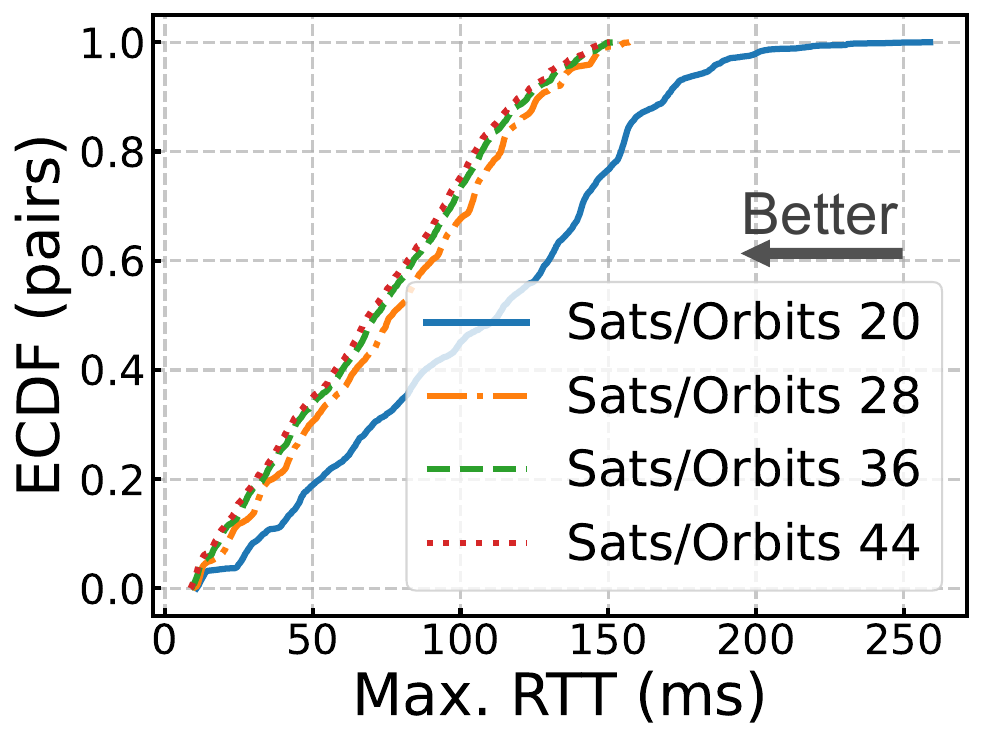}
			\vspace{-0.5cm}
      	\label{RTT Sats/Orbits 2}
		\end{minipage}%
	}%
	\subfigure[46 Orbits]{
	\vspace{-0.5cm}
		\begin{minipage}[t]{0.24\linewidth}
			\centering
			\includegraphics[scale=.25]{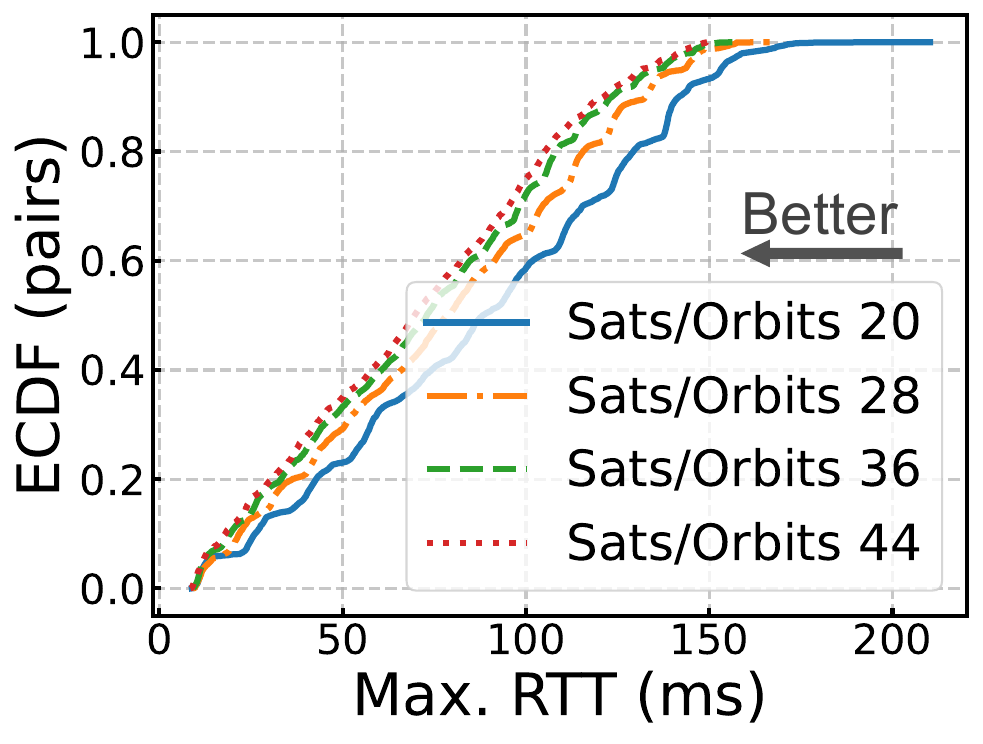}
			\vspace{-0.5cm}
      	\label{RTT Sats/Orbits 3}
		\end{minipage}%
	}%
	\subfigure[59 Orbits]{
	\vspace{-0.5cm}
		\begin{minipage}[t]{0.24\linewidth}
			\centering
			\includegraphics[scale=.25]{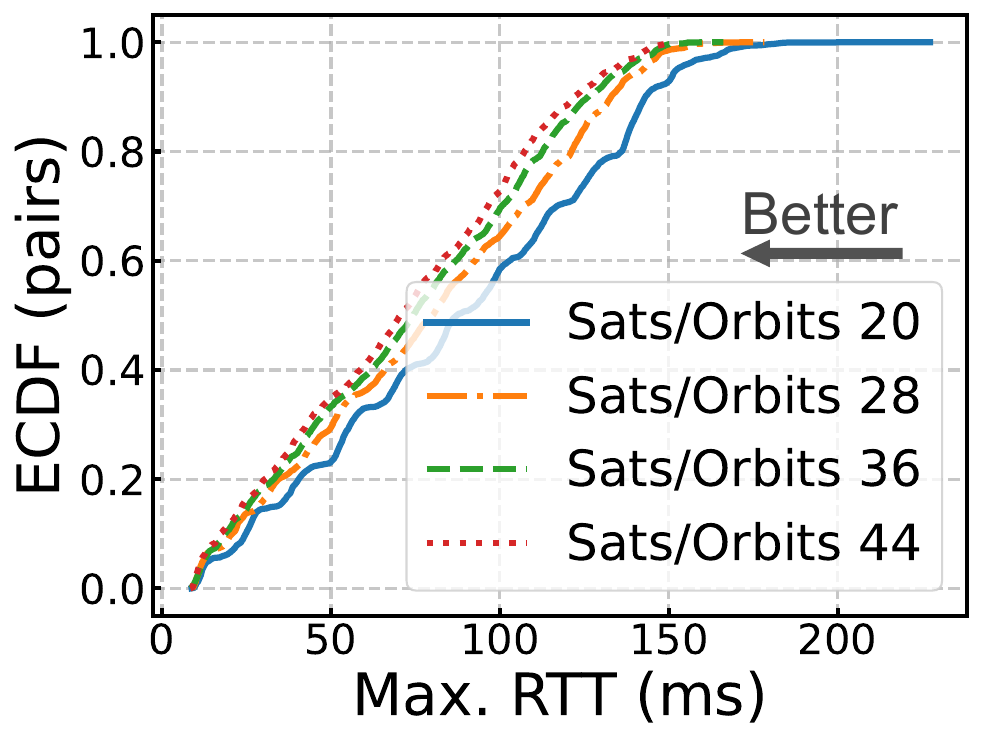}
			\vspace{-0.5cm}
   	\label{RTT Sats/Orbits 4}
		\end{minipage}%
	}%
    \vspace{-0.5cm}
	\caption{The distribution of Maximum RTT(ms) while varying the number of satellites per orbit (Sats/Orbit). For all curves, lower values indicate better performance. Long tails indicate outliers with poor performance.}
     \vspace{-0.25cm}
	\label{RTT Sats/Orbits}
\end{figure*}

\begin{figure*}[t]
	\centering
	\subfigure[20 Orbits]{
	\vspace{-0.5cm}
		\begin{minipage}[t]{0.24\linewidth}
			\centering
			\includegraphics[scale=.25]{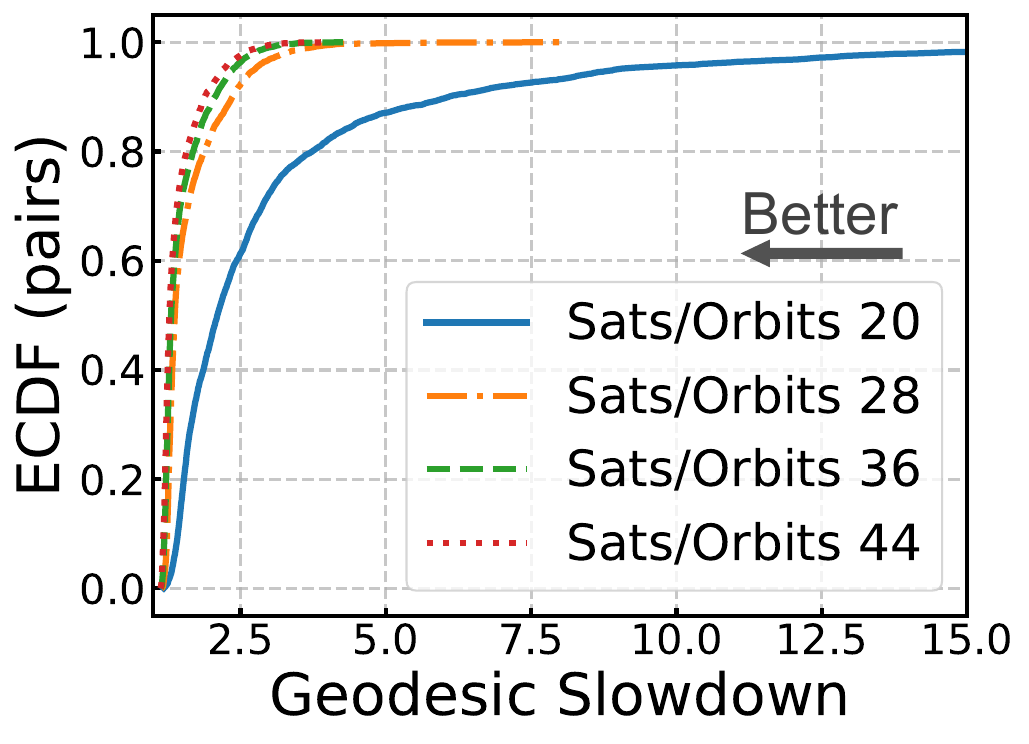}
			\vspace{-0.5cm}
   \label{Slowdown Sats/Orbits 1}
		\end{minipage}%
	}%
	\subfigure[33 Orbits]{
	\vspace{-0.5cm}
		\begin{minipage}[t]{0.24\linewidth}
			\centering
			\includegraphics[scale=.25]{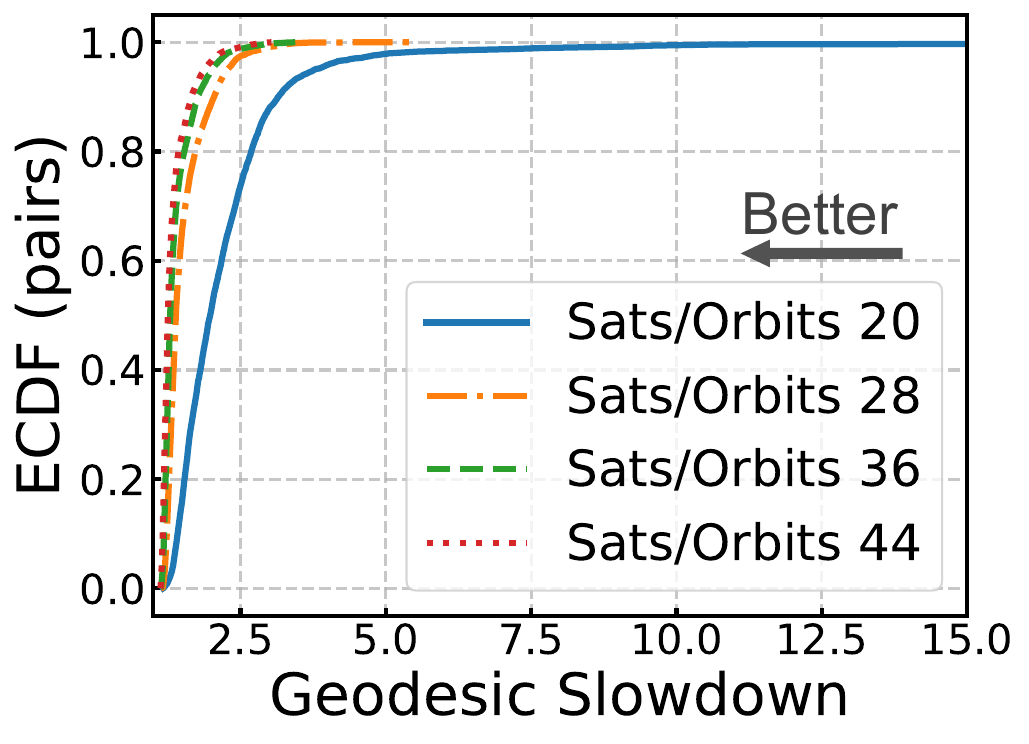}
			\vspace{-0.5cm}
   \label{Slowdown Sats/Orbits 2}
		\end{minipage}%
	}%
	\subfigure[46 Orbits]{
	\vspace{-0.5cm}
		\begin{minipage}[t]{0.24\linewidth}
			\centering
			\includegraphics[scale=.25]{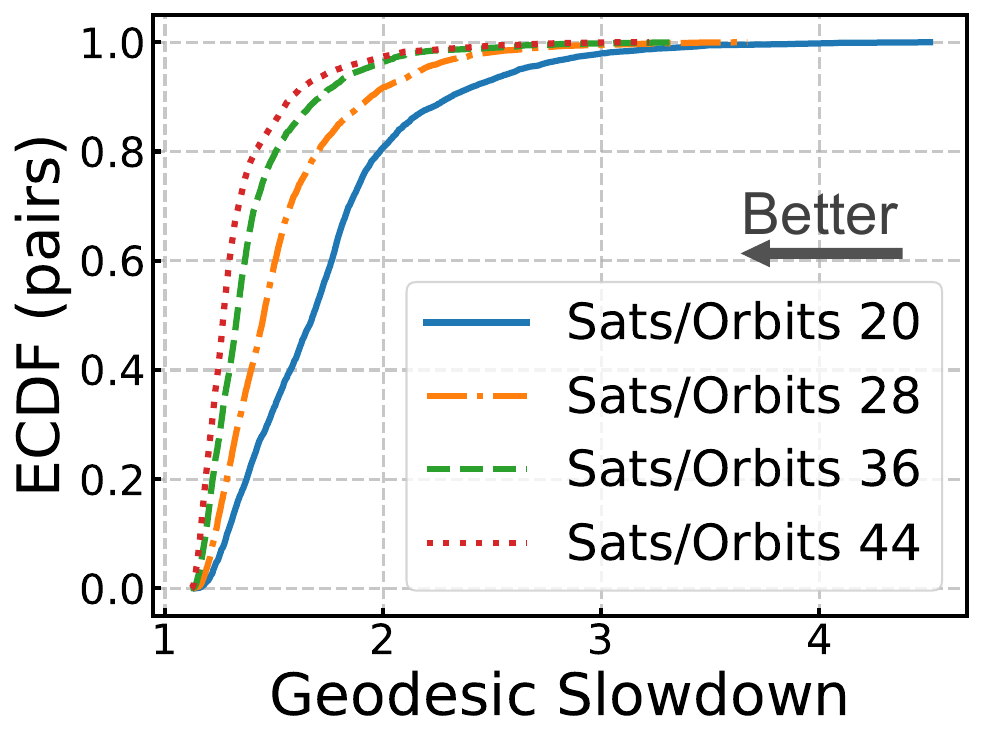}
			\vspace{-0.5cm}
	\label{Slowdown Sats/Orbits 3}
		\end{minipage}%
	}%
	\subfigure[59 Orbits]{
	\vspace{-0.5cm}
		\begin{minipage}[t]{0.24\linewidth}
			\centering
			\includegraphics[scale=.25]{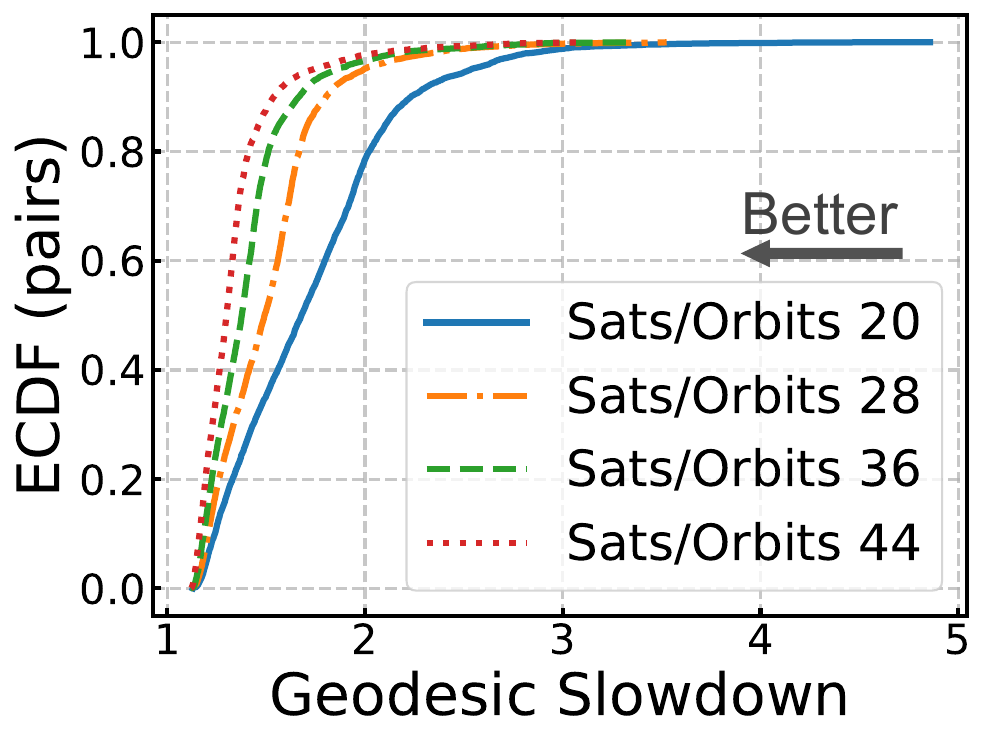}
			\vspace{-0.5cm}
	\label{Slowdown Sats/Orbits 4}
		\end{minipage}%
	}%
    \vspace{-0.5cm}
	\caption{The distribution of Geodesic Slowdown while varying the number of satellites per orbit (Sats/Orbit). For all curves, lower values indicate better performance. Long tails indicate outliers with poor performance. Sats/Orbit at 20 has a very long tail and is truncated in (a) and (b).}
	\label{Slowdown Sats/Orbits}
     \vspace{-0.25cm}
\end{figure*}

\begin{figure*}[t]
	\centering
	\subfigure[20 Orbits]{
	\vspace{-0.5cm}
		\begin{minipage}[t]{0.24\linewidth}
			\centering
			\includegraphics[scale=.25]{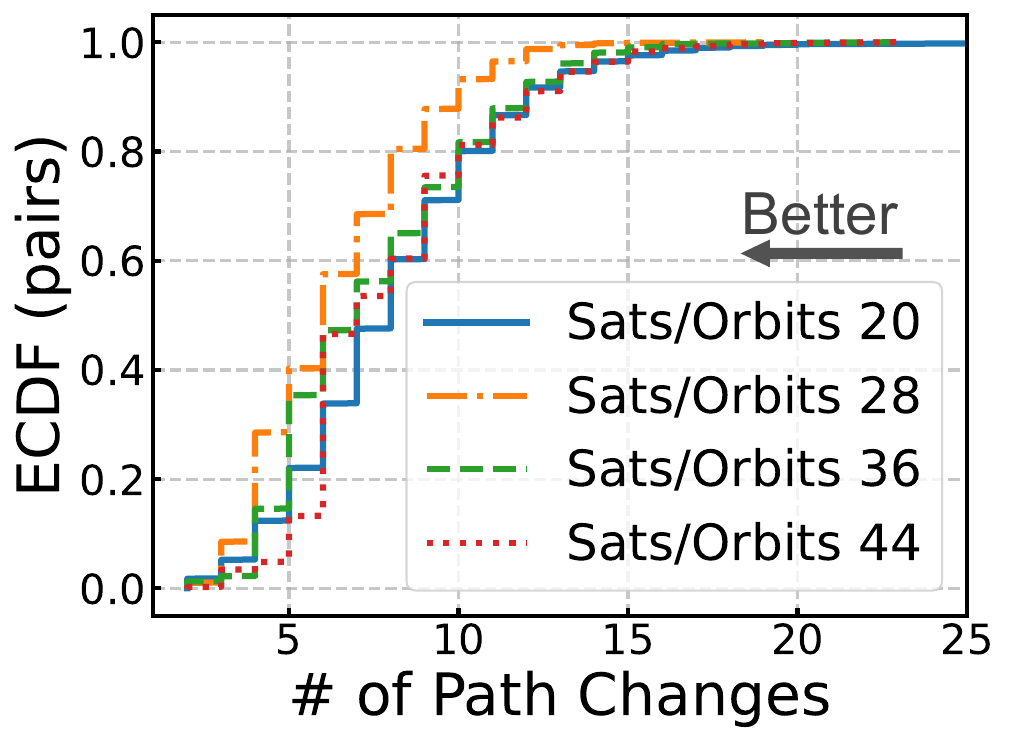}
			\vspace{-0.5cm}
   \label{Path Sats/Orbits 1}
		\end{minipage}%
	}%
	\subfigure[33 Orbits]{
	\vspace{-0.5cm}
		\begin{minipage}[t]{0.24\linewidth}
			\centering
			\includegraphics[scale=.25]{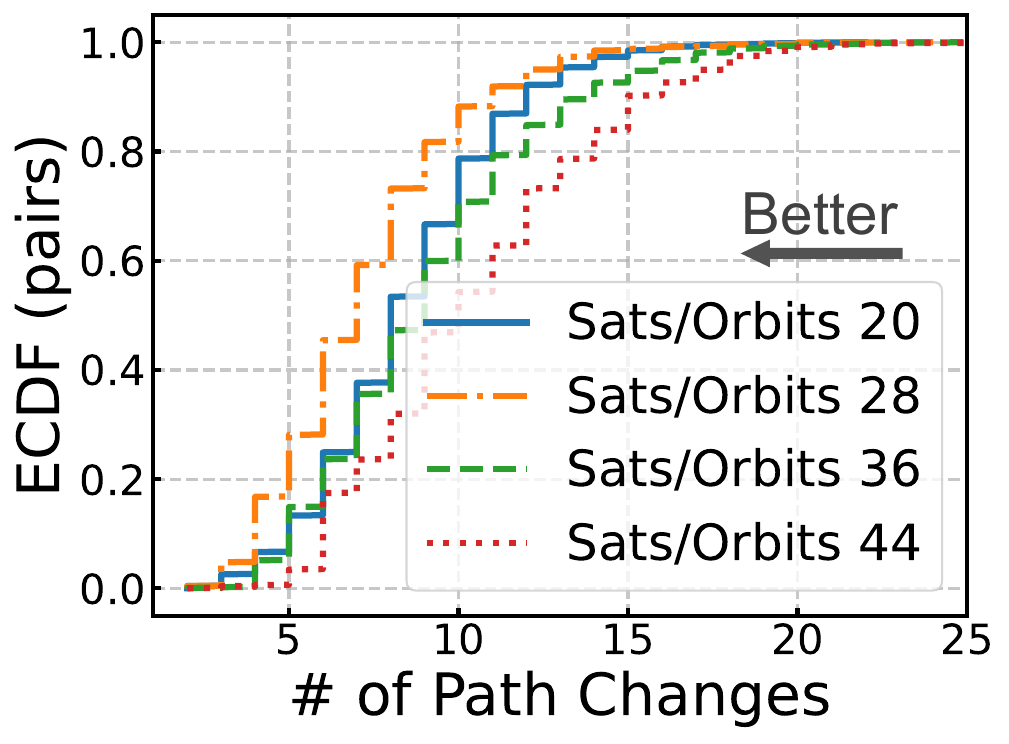}
			\vspace{-0.5cm}
   \label{Path Sats/Orbits 2}
		\end{minipage}%
	}%
	\subfigure[46 Orbits]{
	\vspace{-0.5cm}
		\begin{minipage}[t]{0.24\linewidth}
			\centering
			\includegraphics[scale=.25]{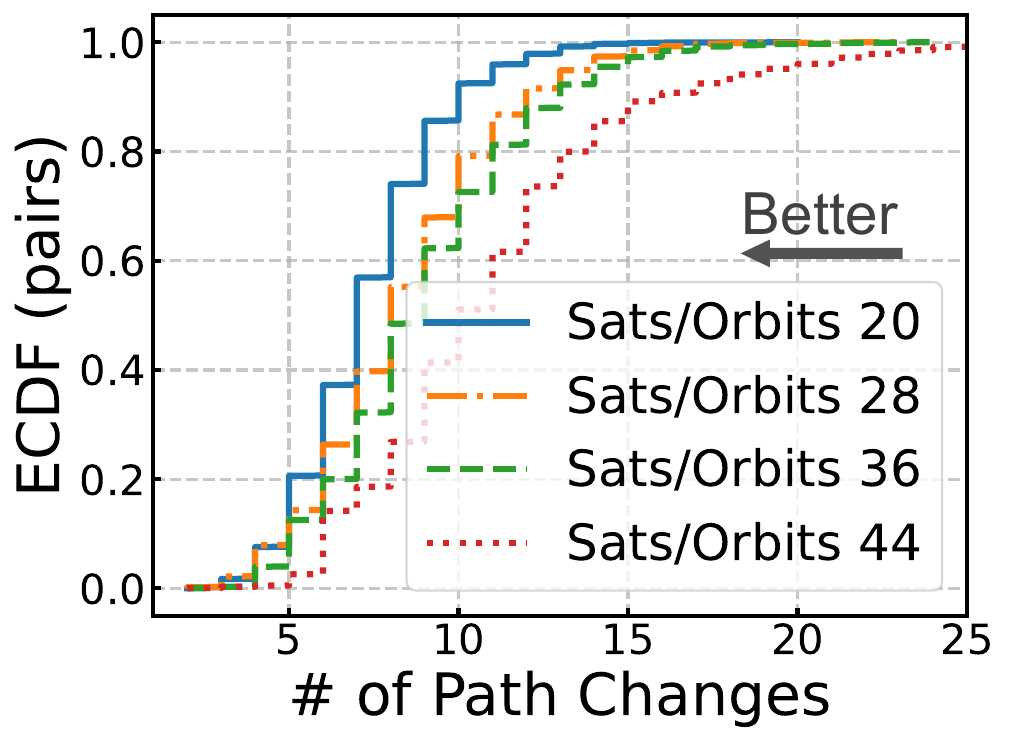}
			\vspace{-0.5cm}
   \label{Path Sats/Orbits 3}
		\end{minipage}%
	}%
	\subfigure[59 Orbits]{
	\vspace{-0.5cm}
		\begin{minipage}[t]{0.24\linewidth}
			\centering
			\includegraphics[scale=.25]{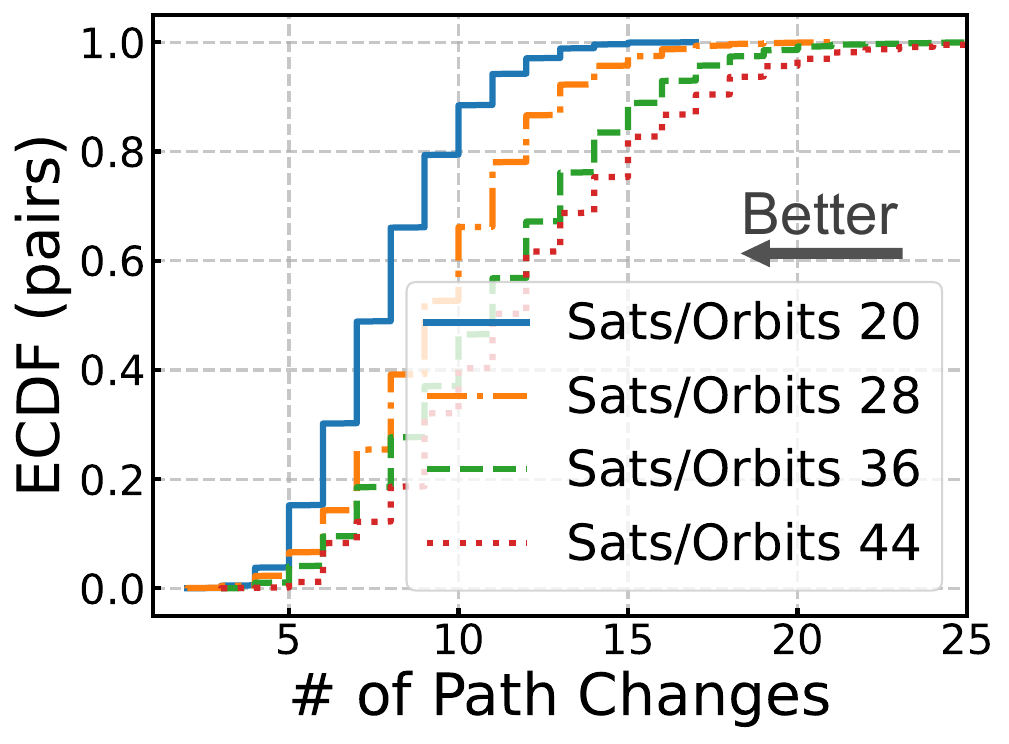}
			\vspace{-0.5cm}
   \label{Path Sats/Orbits 4}
		\end{minipage}%
	}%
    \vspace{-0.5cm}
	\caption{The distribution of the number of Path Changes while varying the number of satellites per orbit (Sats/Orbit). For all curves, lower values indicate better performance. Long tails indicate outliers with poor performance. Long tails are truncated in all plots.}
	\vspace{-0.25cm}
	\label{Path Sats/Orbits}
\end{figure*}
\vspace{2mm}
\noindent
\textbf{Real-world constellation analysis key takeaways: }
\begin{itemize}[leftmargin=*]    
    \item  Very sparse shells can lead to a high variance in RTT.
    \item  Path stability is affected by the number of satellites in the shell. The stability offered by a shell depends on its coverage. For shells with good coverage, the number of path changes increases as shell density increases.
    \item Typically, shells with a larger number of satellites and higher coverage offer a lower RTT, lower geodesic slowdown, and lower variance in RTT. However, there are some caveats. When the constellation parameters are in a certain range, the performance may be poor (for example, shell 1 vs. shell 2 in Starlink). This calls for a more thorough investigation of design parameters.  
\end{itemize}

\subsubsection{In-Depth Analysis of Synthetic Configurations}
\label{Impact of Different Configurations of LEO Mega-constellations}

\hfill

\noindent
\textbf{Number of Satellites Per Orbit: } At an inclination angle of 53°, we measure the network performance of synthetic LEO constellations by varying the number of satellites per orbit across (20, 28, 36, 44) and the number of orbits across (20, 33, 46, 59). The measurement results are shown in Fig.~\ref{RTT Sats/Orbits}, Fig.~\ref{Slowdown Sats/Orbits}, and Fig.~\ref{Path Sats/Orbits}. Note that the tails are very long in certain cases, and the ECDF is cropped in some plots for better visualization. More detailed experimental results are in Appendix~\ref{appendix: Satellites Per Orbit}.

\begin{figure*}[t]
	\centering
	\subfigure[20 Sats/Orbit]{
	\vspace{-0.75cm}
		\begin{minipage}[t]{0.24\linewidth}
			\centering
			\includegraphics[scale=.25]{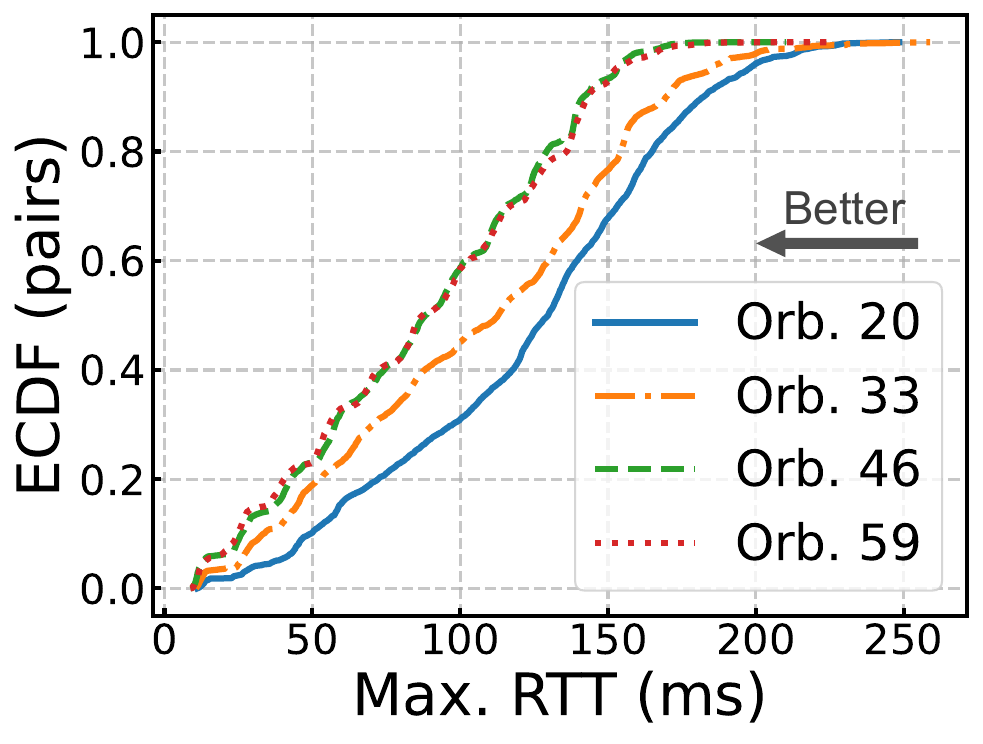}
			\vspace{-0.75cm}
      	\label{RTT Orbits 1}
		\end{minipage}%
	}%
	\subfigure[28 Sats/Orbit]{
	\vspace{-0.75cm}
		\begin{minipage}[t]{0.24\linewidth}
			\centering
			\includegraphics[scale=.25]{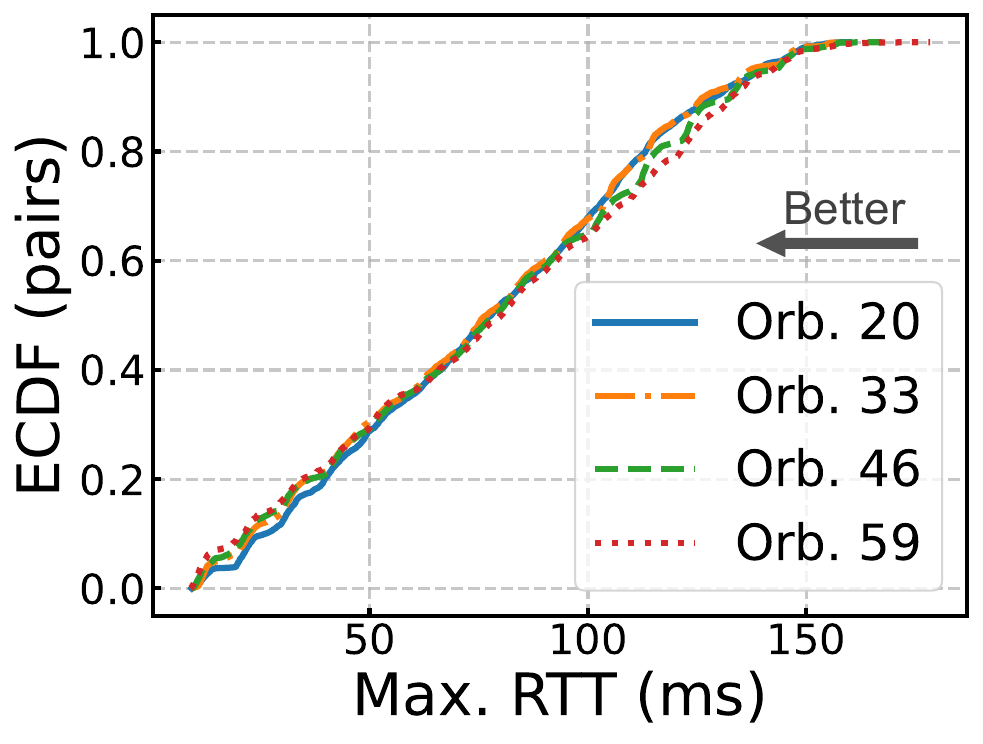}
			\vspace{-0.75cm}
      	\label{RTT Orbits 2}
		\end{minipage}%
	}%
	\subfigure[36 Sats/Orbit]{
	\vspace{-0.75cm}
		\begin{minipage}[t]{0.24\linewidth}
			\centering
			\includegraphics[scale=.25]{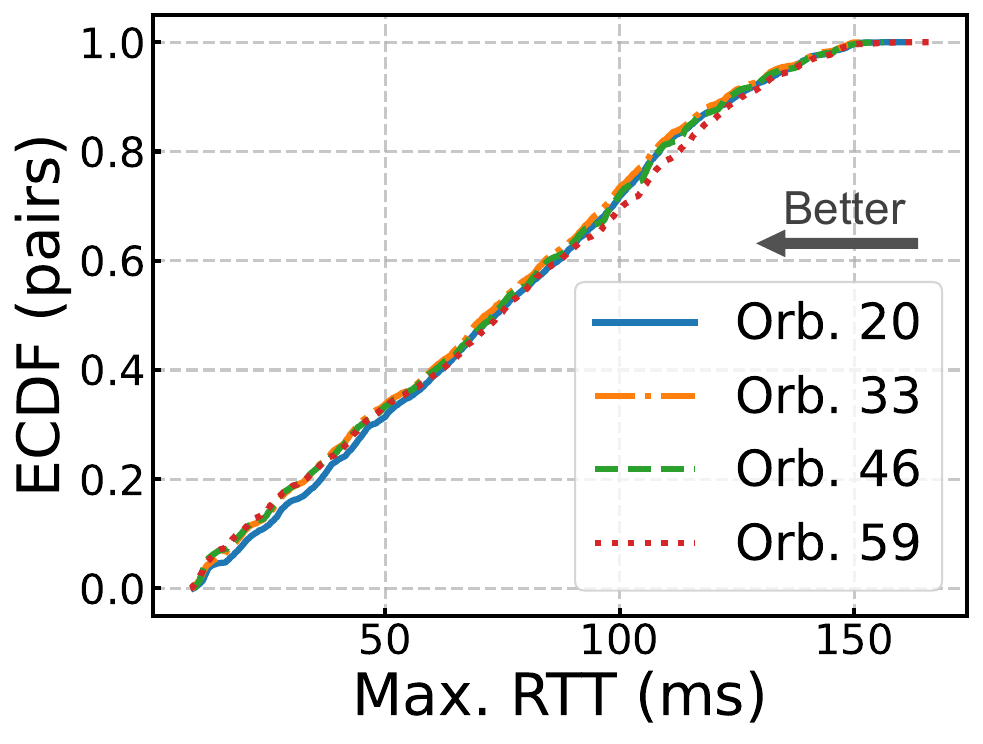}
			\vspace{-0.75cm}
      	\label{RTT Orbits 3}
		\end{minipage}%
	}%
	\subfigure[44 Sats/Orbit]{
	\vspace{-0.75cm}
		\begin{minipage}[t]{0.24\linewidth}
			\centering
			\includegraphics[scale=.25]{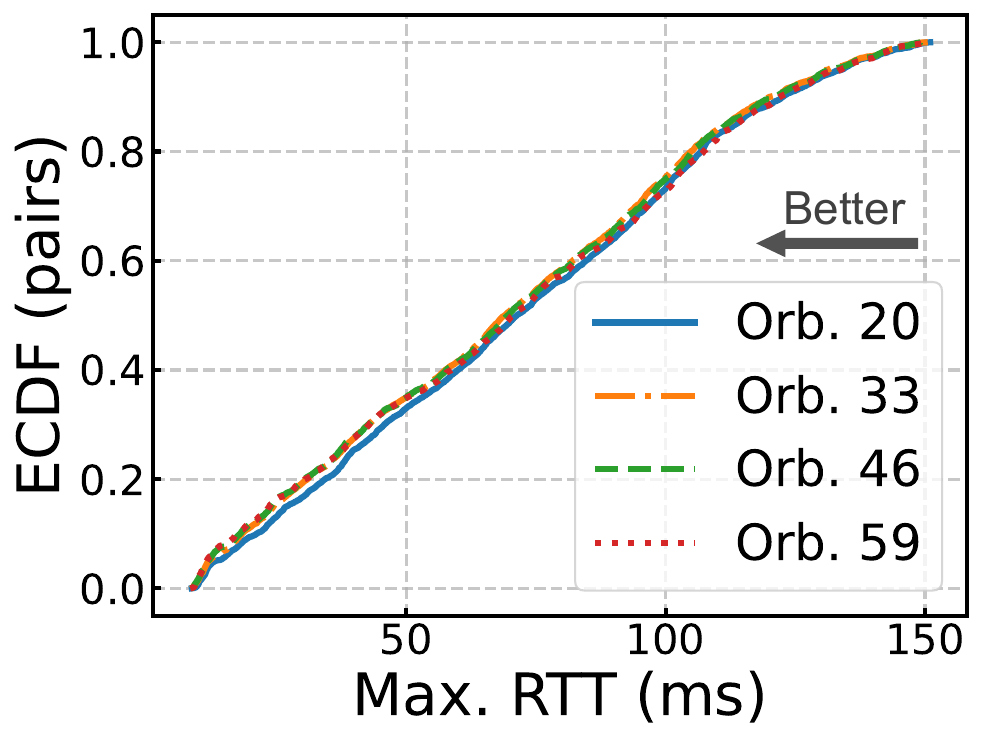}
			\vspace{-0.75cm}
      	\label{RTT Orbits 4}
		\end{minipage}%
	}%
    \vspace{-0.5cm}
	\caption{The distribution of Maximum RTT(ms) while varying the number of orbits. For all curves, lower values indicate better performance. Long tails indicate outliers with poor performance.}
	\vspace{-0.25cm}
   	\label{RTT Orbits}
\end{figure*}

\begin{figure*}[t]
	\centering
	\subfigure[20 Sats/Orbit]{
	\vspace{-0.75cm}
		\begin{minipage}[t]{0.24\linewidth}
			\centering
			\includegraphics[scale=.25]{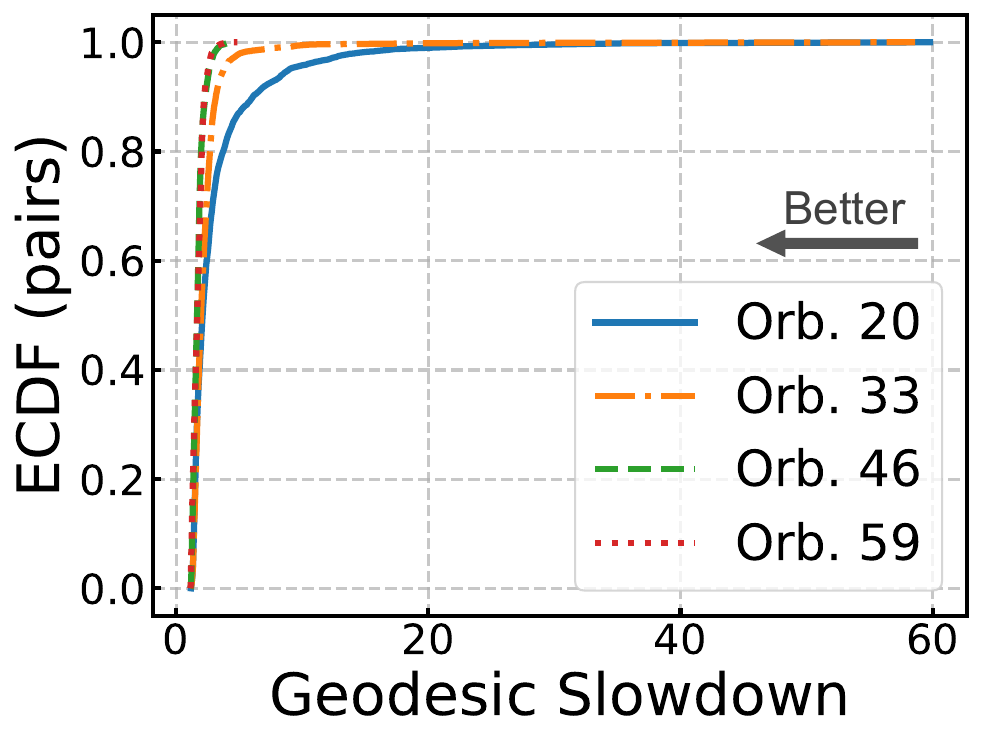}
			\vspace{-0.75cm}
      	\label{Slowdown Orbits 1}
		\end{minipage}%
	}%
	\subfigure[28 Sats/Orbit]{
	\vspace{-0.75cm}
		\begin{minipage}[t]{0.24\linewidth}
			\centering
			\includegraphics[scale=.25]{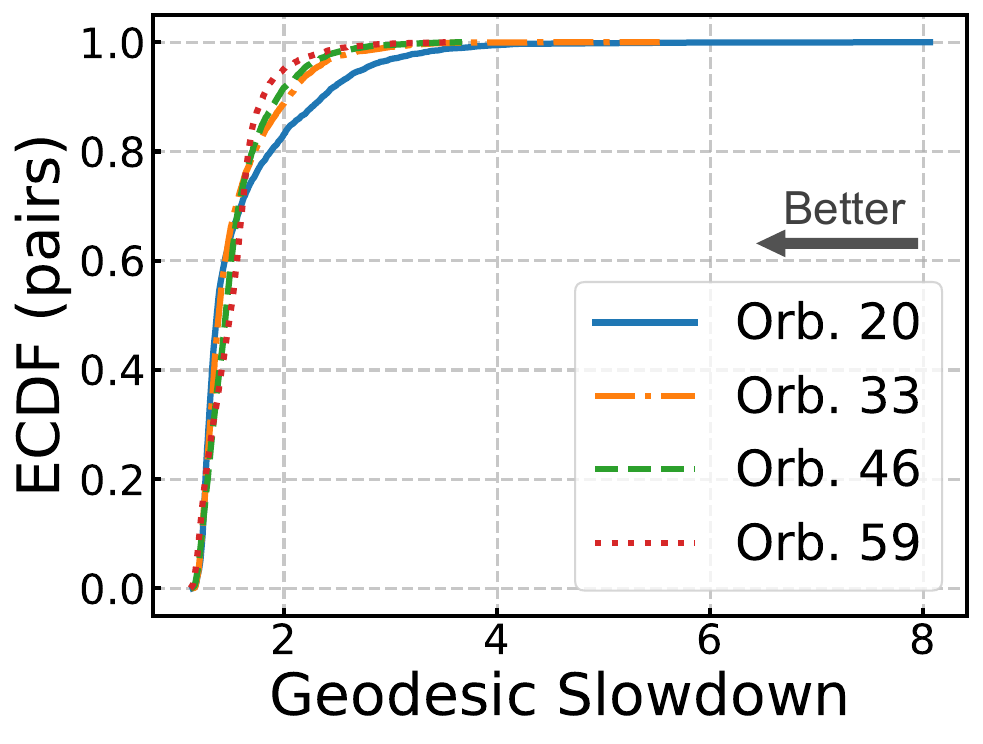}
			\vspace{-0.75cm}
      	\label{Slowdown Orbits 2}
		\end{minipage}%
	}%
	\subfigure[36 Sats/Orbit]{
	\vspace{-0.75cm}
		\begin{minipage}[t]{0.24\linewidth}
			\centering
			\includegraphics[scale=.25]{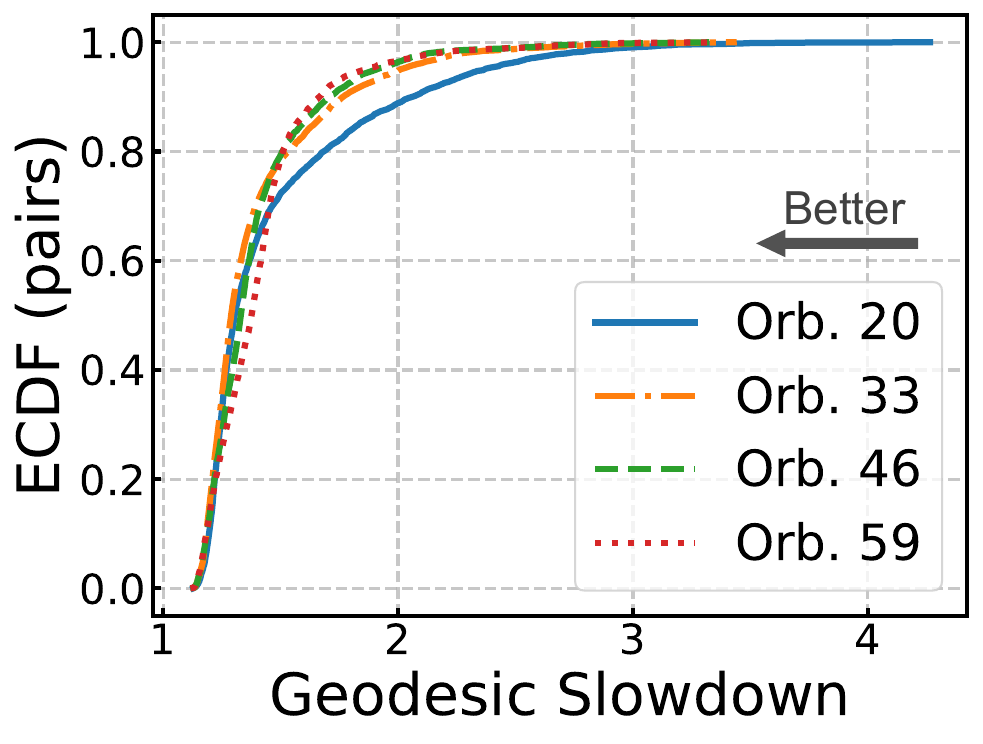}
			\vspace{-0.75cm}
      	\label{Slowdown Orbits 3}
		\end{minipage}%
	}%
	\subfigure[44 Sats/Orbit]{
	\vspace{-0.75cm}
		\begin{minipage}[t]{0.24\linewidth}
			\centering
			\includegraphics[scale=.25]{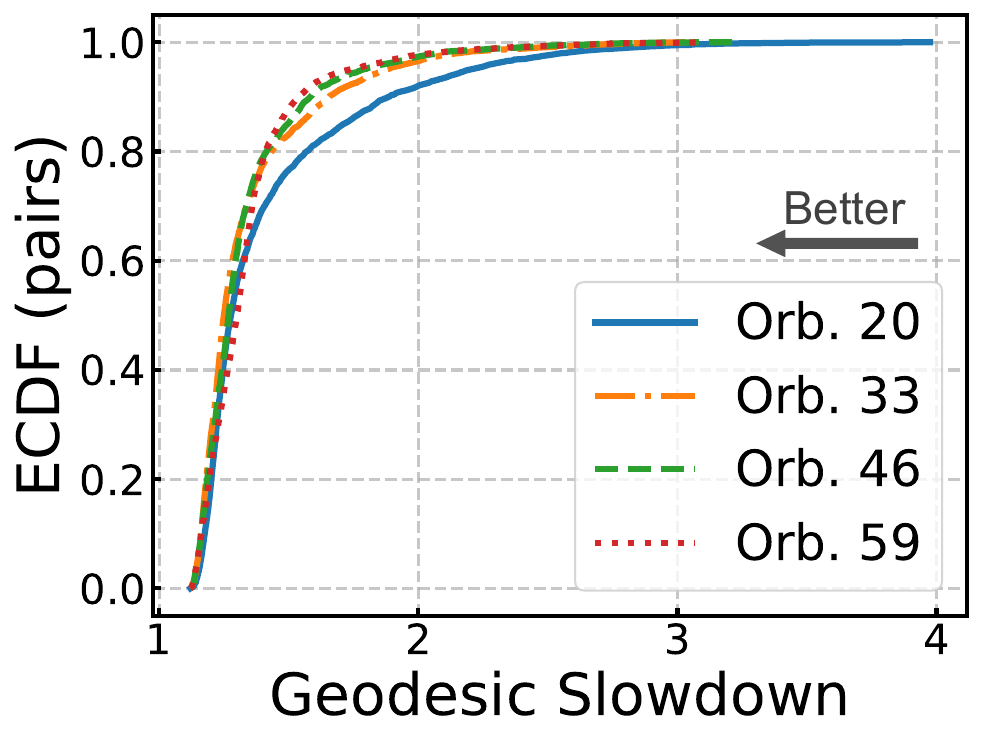}
			\vspace{-0.75cm}
   	\label{Slowdown Orbits 4}
		\end{minipage}%
	}%
    \vspace{-0.5cm}
	\caption{The distribution of Geodesic Slowdown while varying the number of orbits. For all curves, lower values indicate better performance. Long tails indicate outliers with poor performance. }
	\vspace{-0.25cm}
    \label{Slowdown Orbits}
\end{figure*}

\begin{figure*}[tp]
	\centering
	\subfigure[20 Sats/Orbit]{
	\vspace{-0.75cm}
		\begin{minipage}[t]{0.24\linewidth}
			\centering
			\includegraphics[scale=.25]{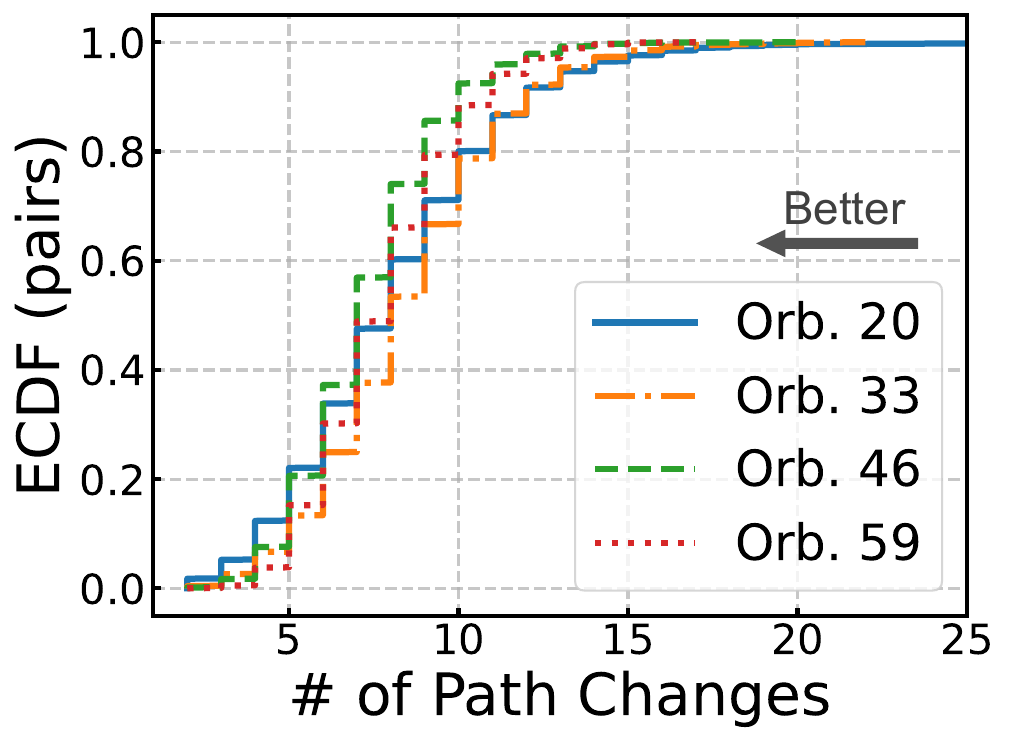}
			\vspace{-0.75cm}
   	\label{Path Orbits 1}
		\end{minipage}%
	}%
	\subfigure[28 Sats/Orbit]{
	\vspace{-0.75cm}
		\begin{minipage}[t]{0.24\linewidth}
			\centering
			\includegraphics[scale=.25]{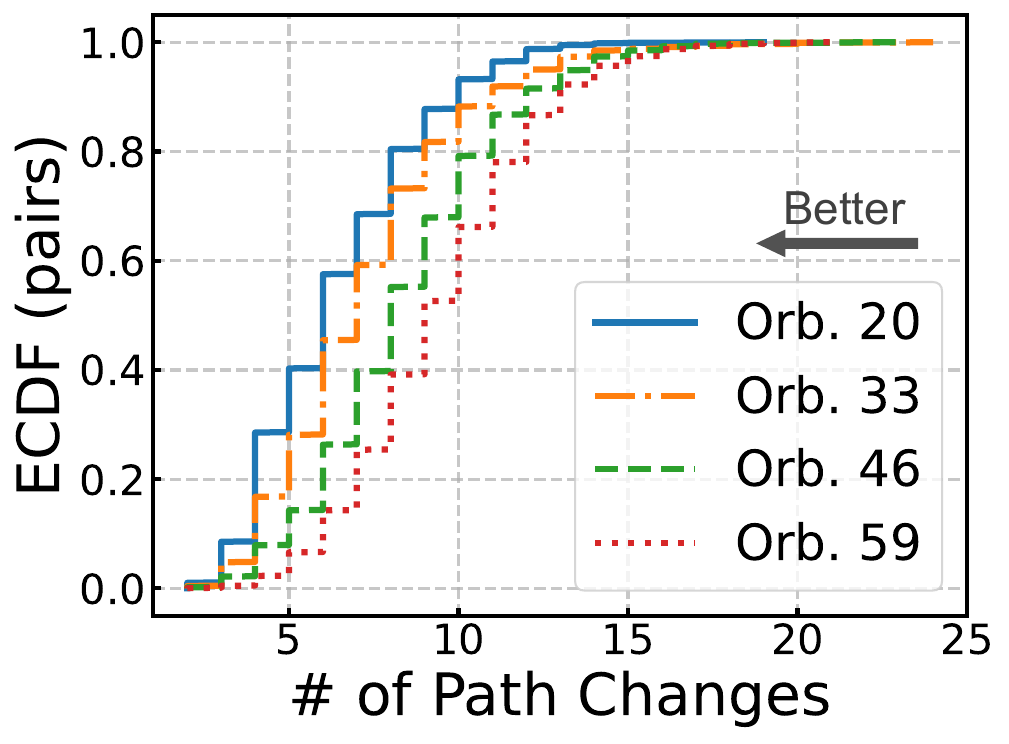}
			\vspace{-0.75cm}
   	\label{Path Orbits 2}
		\end{minipage}%
	}%
	\subfigure[36 Sats/Orbit]{
	\vspace{-0.75cm}
		\begin{minipage}[t]{0.24\linewidth}
			\centering
			\includegraphics[scale=.25]{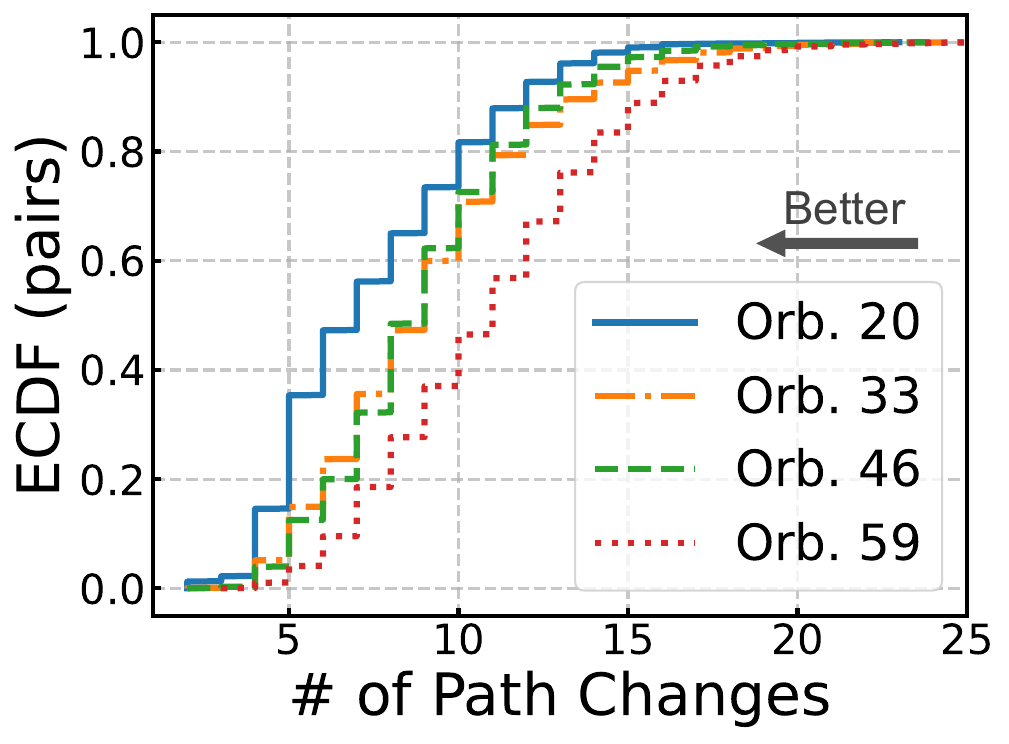}
			\vspace{-0.75cm}
   	\label{Path Orbits 3}
		\end{minipage}%
	}%
	\subfigure[44 Sats/Orbit]{
	\vspace{-0.75cm}
		\begin{minipage}[t]{0.24\linewidth}
			\centering
			\includegraphics[scale=.25]{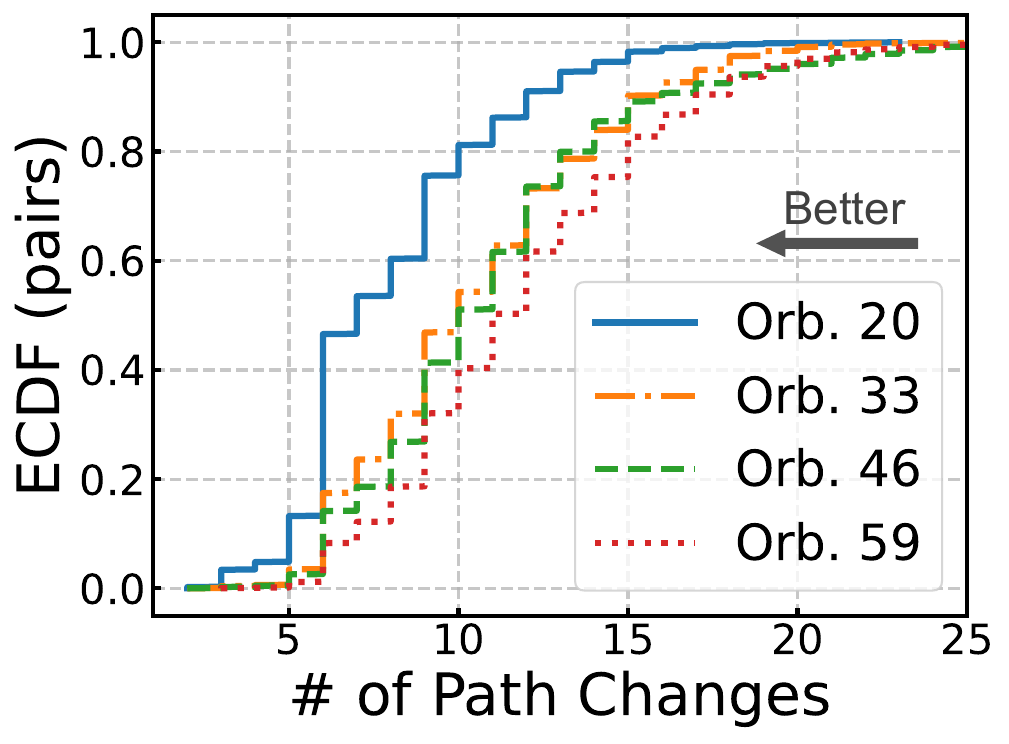}
			\vspace{-0.75cm}
   	\label{Path Orbits 4}
		\end{minipage}%
	}%
    \vspace{-0.5cm}
	\caption{The distribution of the number of Path Changes while varying the number of orbits. For all curves, lower values indicate better performance. Long tails indicate outliers with poor performance. Long tails are truncated in some plots.}
 	\vspace{-0.25cm}
	\label{Path Orbits}
\end{figure*}
In Fig.~\ref{RTT Sats/Orbits} and Fig.~\ref{Slowdown Sats/Orbits}, we observe that the distribution of Max. RTT and Geodesic Slowdown demonstrate similar behavior irrespective of the number of orbits. 
As the number of satellites per orbit (Sats/Orbit) increases, Max. RTT and Geodesic Slowdown decrease for a given orbit number. We also find that when the number of satellites per orbit is 20, the Max. RTT and Geodesic Slowdown are much higher than others (e.g., $2\times$ higher than 28 Sats/Orbit in Fig.~\ref{RTT Sats/Orbits 1} and Fig.~\ref{Slowdown Sats/Orbits 1}). This effect is especially pronounced when the Orbit number is less than 33.  The long tail effect is also more pronounced when Sats/Orbit = 20 (tail at 250ms+ vs. 150ms when Sats/Orbit = 28). When Sats/Orbit increases beyond 28, the performance gains are marginal. For example, a 33\% increase in satellite density from 36 to 48 results in a less than 2\% reduction in Max. RTT and less than 1\% reduction in Geodesic slowdown.

Note that while the differences between performance curves may be minimal in some experiments, we are interested in trends in performance changes as we vary each parameter. Hence, we highlight the performance trends by comparing plots across multiple parameter settings in our analysis.

The variations in path change reflect a different behavior. When the number of orbits increases, the path stability of constellations with a lower number of satellites per orbit improves. For example, in Fig.~\ref{Path Sats/Orbits 1},  Sats/Orbits = 20 leads to the highest number of path changes (across over 80\% of endpoint pairs); however, in Fig.~\ref{Path Sats/Orbits 4}, Sats/Orbits = 20 leads to the least number of path changes (across over 90\% of endpoint pairs). Also, in Fig.~\ref{Path Sats/Orbits 2}, for a fixed number of Orbit = 33, the median number of path changes increases from 6 to 10 as the number of satellites per orbit increases. 

\begin{figure*}[tp]
	\centering
	\subfigure[Maximum RTT]{
	\vspace{-0.5cm}
		\begin{minipage}[t]{0.24\linewidth}
			\centering
			\includegraphics[scale=.25]{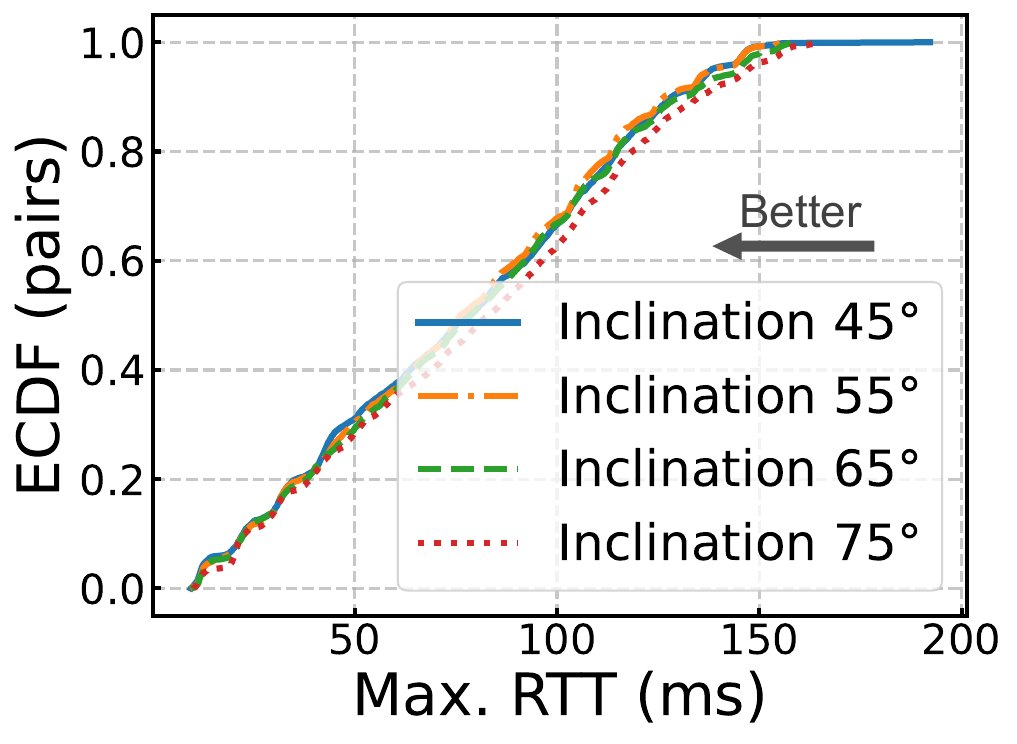}
			\vspace{-0.5cm}
   \label{Inclination 1}
		\end{minipage}%
	}%
	\subfigure[Geodesic Slowdown]{
	\vspace{-0.5cm}
		\begin{minipage}[t]{0.24\linewidth}
			\centering
			\includegraphics[scale=.25]{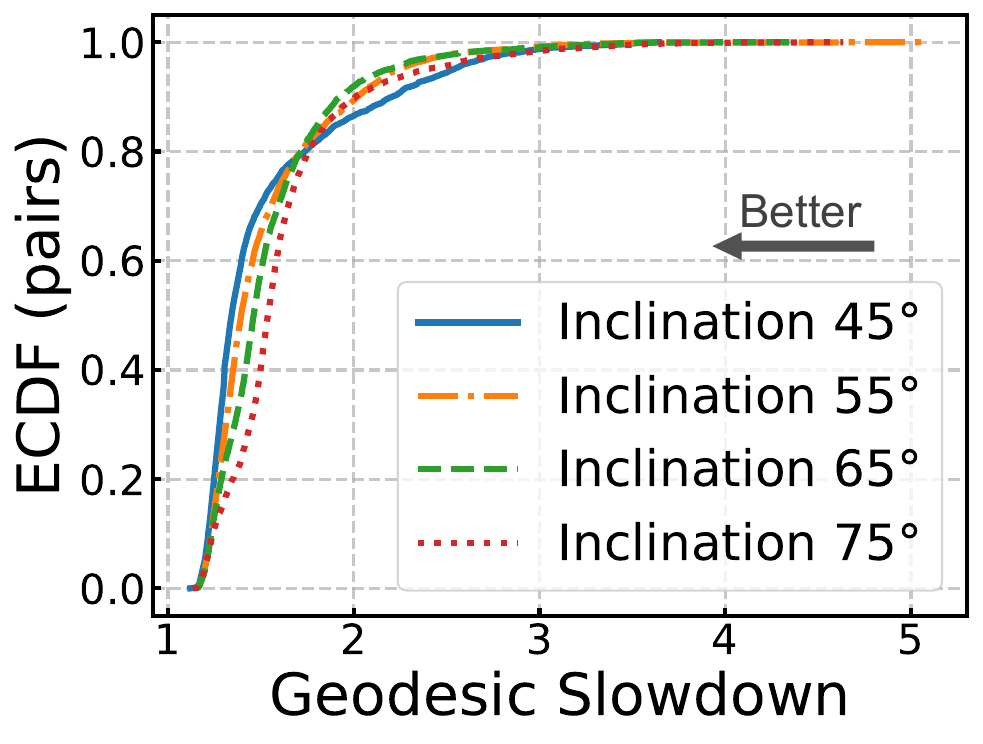}
			\vspace{-0.5cm}
   \label{Inclination 2}
		\end{minipage}%
	}%
	\subfigure[Max. RTT - Min. RTT (ms)]{
	\vspace{-0.5cm}
		\begin{minipage}[t]{0.24\linewidth}
			\centering
			\includegraphics[scale=.25]{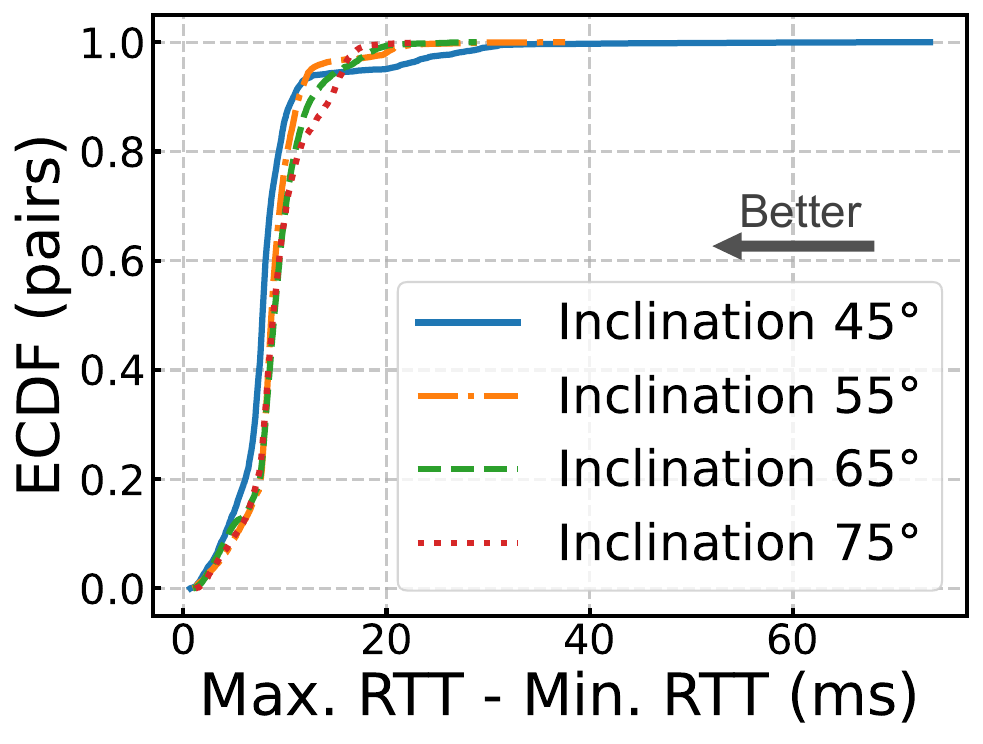}
			\vspace{-0.5cm}
   \label{Inclination 3}
		\end{minipage}%
	}%
	\subfigure[Number of path changes]{
	\vspace{-0.5cm}
		\begin{minipage}[t]{0.24\linewidth}
			\centering
			\includegraphics[scale=.25]{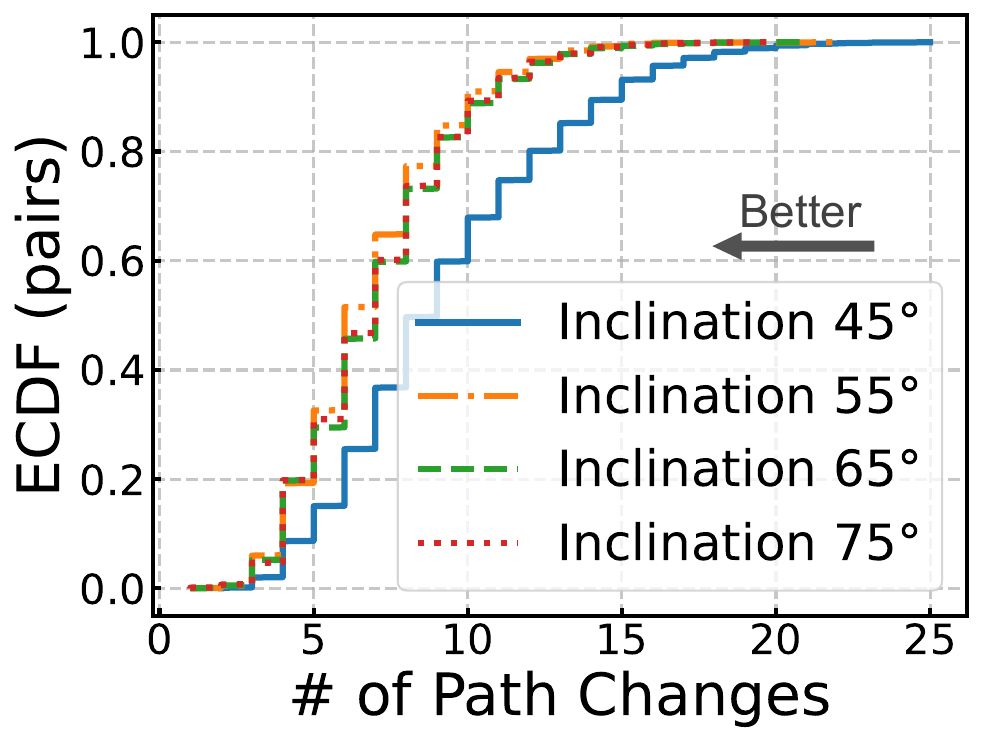}
			\vspace{-0.5cm}
   \label{Inclination 4}
		\end{minipage}%
	}%
\vspace{-0.5cm}
	\caption{The performance characteristics under varying values for Inclination. We use a constant number of orbits (33) and satellites per orbit (28) while varying the inclination.}
	\vspace{-0.25cm}
	\label{Inclination}
\end{figure*}
These experiments reveal a strong correlation between the number of satellites per orbit (Sats/Orbit) and the network performance of the constellations. The Sats/Orbit will determine the distance between satellites in each orbit, with a lower distance between satellites as Sats/Orbit increases. With the gridded satellite network, a higher value of Sats/Orbit can lead to the routing path in the +Grid LEO constellation being more closer to the optimal geodesic transmission path. However, with the high level of dynamism under a high density of satellites, the optimal shortest paths will change at a faster rate. Therefore, the trade-off between latency and path stability needs to be considered during constellation design. 20 Sats/Orbit offers poor network performance, and hence, 28 or more Sats/Orbit are needed to achieve the combination of relatively low delay and a lower number of path changes.

\noindent\textbf{Number of Orbits: }
With a constant inclination of 53° and constant satellites per orbit in the set (20, 28, 36, 44), we evaluate the variation in performance by varying the number of orbits across 20, 33, 46, and 59. More detailed experimental results are in Appendix~\ref{appendix: Orbit Number}. In Fig.~\ref{RTT Orbits 1}, we see that when the Sats/Orbit = 20 (i.e., a large distance between satellites), the performance improves significantly as the number of orbits is increased from 20 to 59. Compared to 20 Orbits, a 65\%, 120\%, and 195\% increase in orbits densities to 33, 44, and 59 Orbits leads to a 20\%, 50\%, and 50\% decrease in the median value of Max. RTT. Similarly, significant improvement is observed in Geodesic Slowdown with Sats/Orbit = 20 in Fig.~\ref{Slowdown Orbits 1}. Moreover, from Fig.~\ref{RTT Orbits} and Fig.~\ref{Slowdown Orbits}, we observe marginal benefits when Sats/Orbit density increases beyond a threshold. Beyond 28 Sats/Orbit, an increase in the number of orbits has a minimal effect on the maximum RTT, reducing the maximum RTT from 175ms to 150ms. 

In Fig.~\ref{Slowdown Orbits}, we see that when Sats/Orbit is greater than 28, we achieve minimal improvement by increasing the number of orbits. 
Interestingly, most endpoint pairs have a Geodesic Slowdown in the range 1.4-2.5 $\times$. With the number of path changes, we observe similar behavior as previously noted, i.e., when the number of satellites per orbit increases, the path stability is higher with a lower number of orbits. For Sats/Orbits = 28, the median number of path changes increases from 6 to 9 when the number of orbits increases.

The above results align with our predictions for the LEO mega-constellation network. At a given number of satellites per orbit, as the number of orbits increases, the density of the satellite network also increases. This causes the inter-satellite routing route to be closer to the optimal geographical distance but leads to a higher number of shortest path changes. Also shown in Fig.~\ref{RTT Orbits} and Fig.~\ref{Path Orbits}, when the number of satellites per orbit is low, i.e., the satellite network has a low density, 20 Orbits cannot offer the ideal satellite network performance, and at least 33 Orbits are required to achieve relatively low delay and the number of path changes.

\noindent\textbf{Inclination: }
With a fixed number of Sats/Orbits = 28 and a number of orbits of 33, we measure the variations in network performance with varying Inclinations (45°, 55°, 65°, and 75°) and show the results in Fig.~\ref{Inclination}. More detailed experimental results are in Appendix~\ref{appendix: Inclination}. In Fig.~\ref{Inclination 1}, the variations in maximum RTT across inclinations are minimal, although Inclination = 75° has the largest Max. RTT. Moreover, the difference between maximum and minimum RTT, shown in Fig.~\ref{Inclination 3}, is below 16 ms for nearly 90\% of endpoint pairs; however, at 45°, the tail can be as high as 60 ms. With Geodesic Slowdown in Fig.~\ref{Inclination 2}, the lower the inclination angle, the larger the fraction of endpoint pairs with a slowdown factor below 1.8$\times$. In addition, as shown in Fig.~\ref{Inclination 4}, an inclination of 55° has the lowest median value of the number of path changes. The median value for inclinations 75° and 65° are similar, while the number of path changes at an inclination of 45° is the largest, with a median of 9. Overall, it is difficult to identify clear trends in network performance associated with variations in inclination. We conclude that the influence of Inclination on network performance is multi-dimensional and defer a detailed study at the intersection of patterns in the traffic matrix and inclination to~\S~\ref{subsubsec:IURD}.

\vspace{1mm}
\noindent
\textbf{Synthetic constellation analysis key takeaways}:
\begin{itemize}[leftmargin=*]    
    \item  The number of satellites per orbit is a key determinant of network performance. As the number of satellites per orbit increases, the latency and geodesic slowdown in the network decreases (Fig.~\ref{RTT Sats/Orbits} \& \ref{Slowdown Sats/Orbits}).
    \item We identify a threshold for the number of satellites per orbit, 28, below which the network performance degrades significantly (Fig.~\ref{RTT Sats/Orbits} \& \ref{Slowdown Sats/Orbits}).
    \item Increasing the number of orbits is beneficial for performance only when the number of satellites per orbit is very low, i.e., below the identified threshold of 28 (Fig.~\ref{RTT Orbits} \& \ref{Slowdown Orbits}).
    \item Inclination has a complex relationship with network performance and needs to be evaluated in conjunction with the traffic matrix for a better understanding (Fig.~\ref{Inclination}).
\end{itemize}

\noindent
\textbf{Why does Starlink Shell 2 offer better performance compared to Shell 1?} 
We circle back to the observation that led to our analysis:  S2 offers better latency performance compared to S1 in spite of having only nearly half the number of satellites. Based on our synthetic constellation analysis, we identify that the poor performance of S1 is because its number of satellites per orbit (20) is lower than the minimum threshold required for good performance (28). When the number of satellites is below this threshold, it is difficult to compensate for the drop in performance, even with a much larger number of satellites. S2, on the other hand, has all its key design
parameters above the performance threshold. Hence, S2 has better performance than S1.

\subsubsection{Impact of Traffic Matrix}
\label{subsubsec:IURD}

\begin{figure}[tp]
	\centering
	\subfigure[40-50° Endpoints Pairs]{
	\vspace{-0.75cm}
		\begin{minipage}[t]{0.49\linewidth}
			\centering
			\includegraphics[scale=.14]{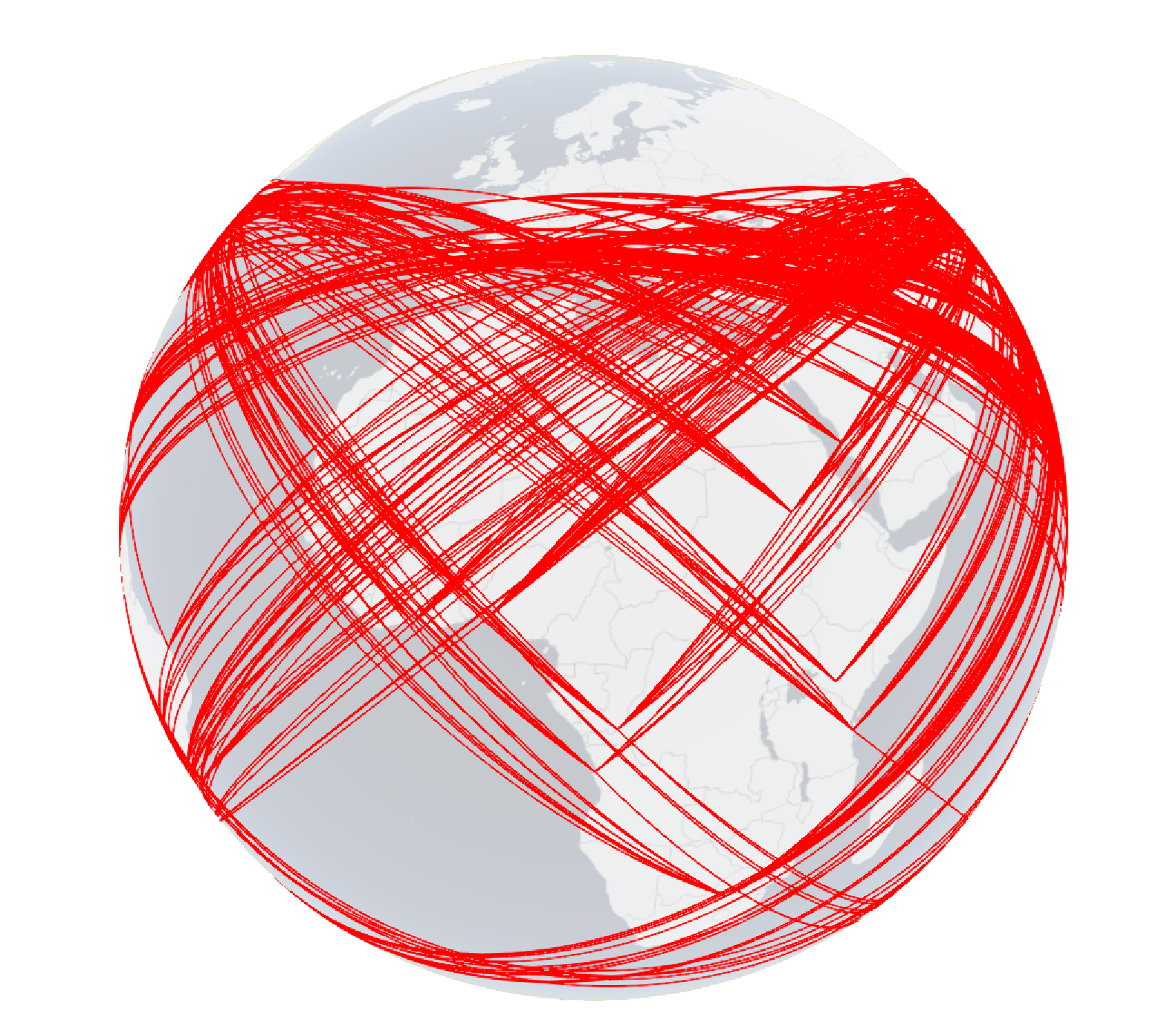}
			\vspace{-0.75cm}
   \label{gs pairs example}
		\end{minipage}%
	}%
	\subfigure[Geographic Angles Distribution]{
	\vspace{-0.75cm}
		\begin{minipage}[t]{0.49\linewidth}
			\centering
			\includegraphics[scale=.26]{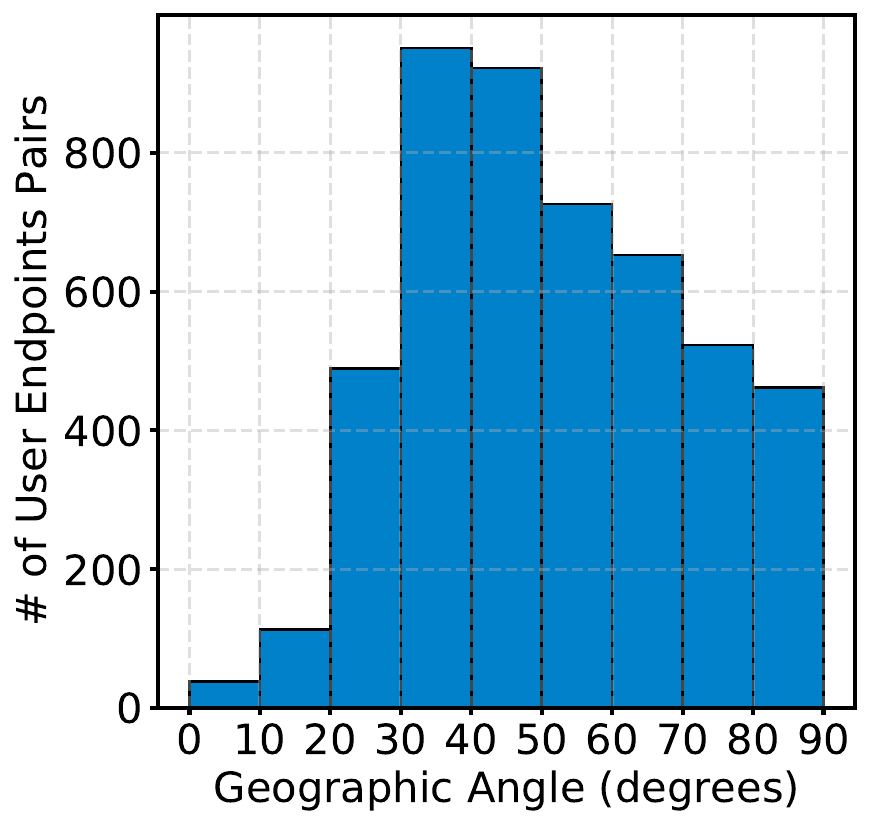}
			\vspace{-0.75cm}
   \label{gs pairs distribution}
		\end{minipage}%
	}%
    \vspace{-0.5cm}
	\caption{Geographic angle analysis: (a) Visualization of all city pairs with geographic angles in the range 40-50°. (b) The distribution of geographic angles of all 9900 city pairs used in our analysis.}
	\vspace{-0.45cm}
	\label{gs pairs hist}
\end{figure}

\hfill

\noindent
\textbf{Traffic Distribution: } We use the geographic angle of endpoint pairs to investigate the impact of traffic matrices (user distribution) on performance. The geographic angle is defined as the angle between the plane connecting the endpoint pair and the equator. We divide endpoint pairs into nine groups according to their geographic angle. For example, Fig.~\ref{gs pairs example} shows endpoint pairs with geographic angles in the range 40-50°(detailed information in Appendix~\ref{appendix: Visualization of User Endpoints}). Fig.~\ref{gs pairs distribution} shows the distribution of geographic angles of all city pairs. The majority of endpoint pairs are distributed within the range 20-90°, with the highest number of pairs in the range 30-40°.

\begin{figure*}[tp]
 	\centering

	\subfigure[40-50° Endpoints pairs.]{
	\vspace{-0.5cm}
		\begin{minipage}[t]{0.24\linewidth}
			\centering
			\includegraphics[scale=.26]{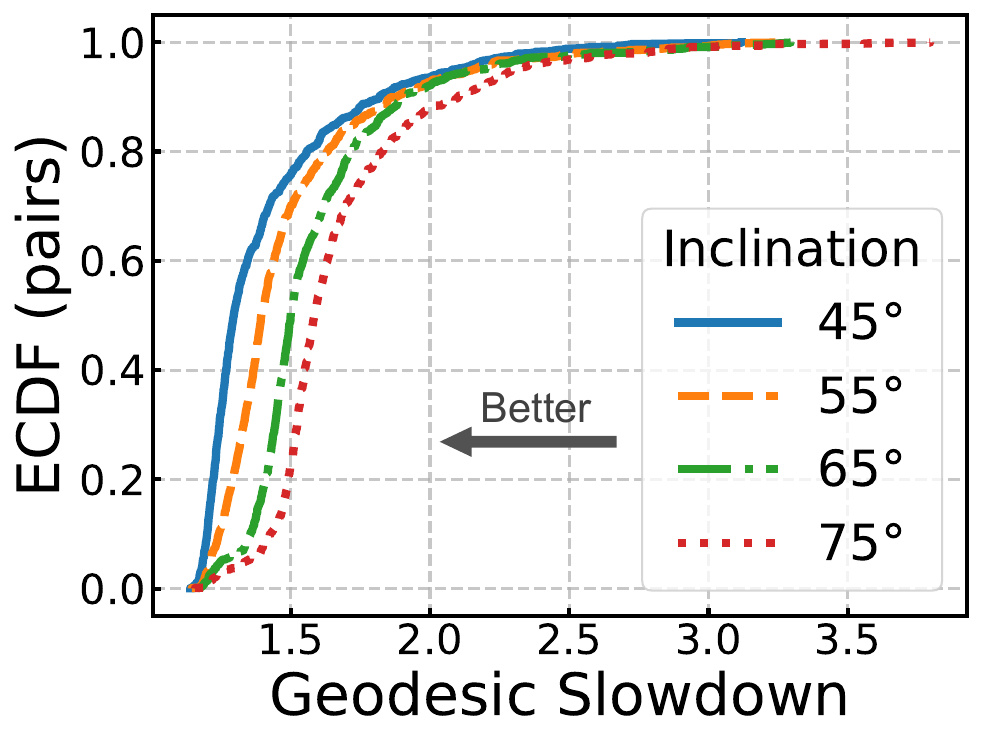}
			\vspace{-0.5cm}
                \label{appendix:inclination geodesic slowdown 5}
		\end{minipage}%
	}%
	\subfigure[50-60° Endpoints pairs.]{
	\vspace{-0.5cm}
		\begin{minipage}[t]{0.24\linewidth}
			\centering
			\includegraphics[scale=.26]{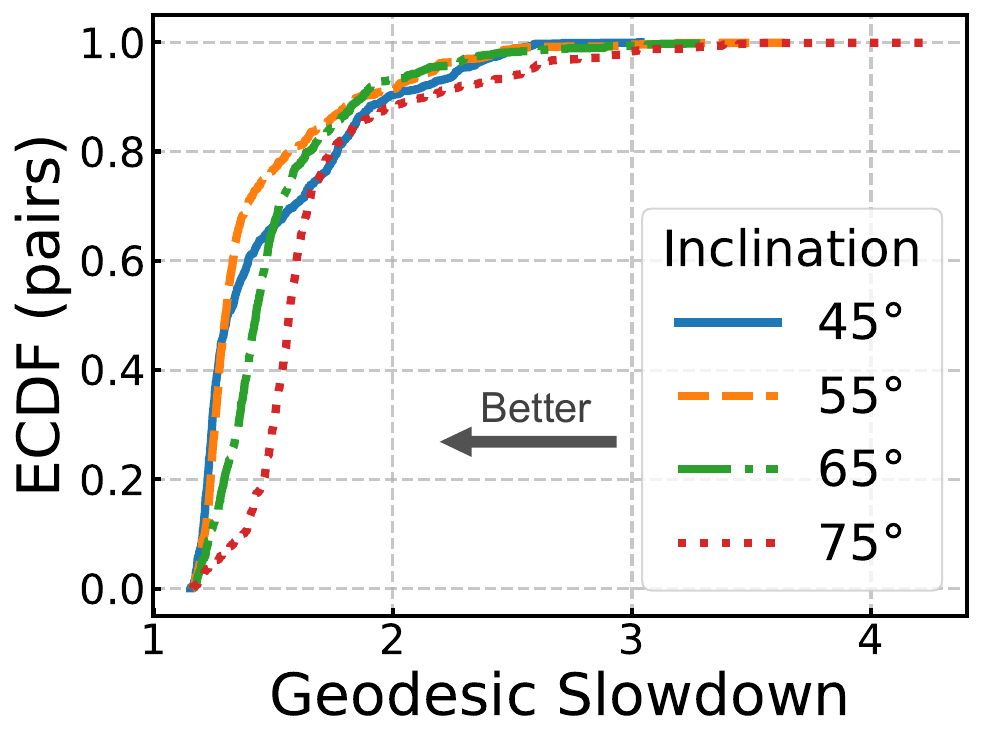}
			\vspace{-0.5cm}
                \label{appendix:inclination geodesic slowdown 6}
		\end{minipage}%
	}%
	\subfigure[60-70° Endpoints pairs.]{
	\vspace{-0.5cm}
		\begin{minipage}[t]{0.24\linewidth}
			\centering
			\includegraphics[scale=.26]{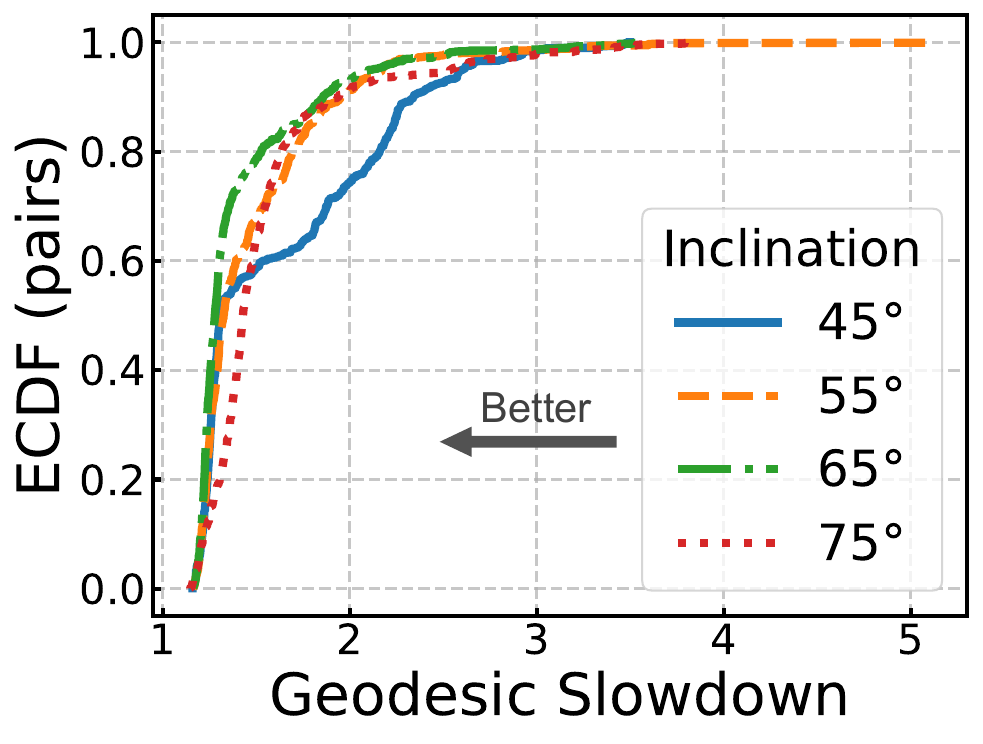}
			\vspace{-0.5cm}
                \label{appendix:inclination geodesic slowdown 7}
		\end{minipage}%
	}%
	\subfigure[70-80° Endpoints pairs.]{
	\vspace{-0.5cm}
		\begin{minipage}[t]{0.24\linewidth}
			\centering
			\includegraphics[scale=.26]{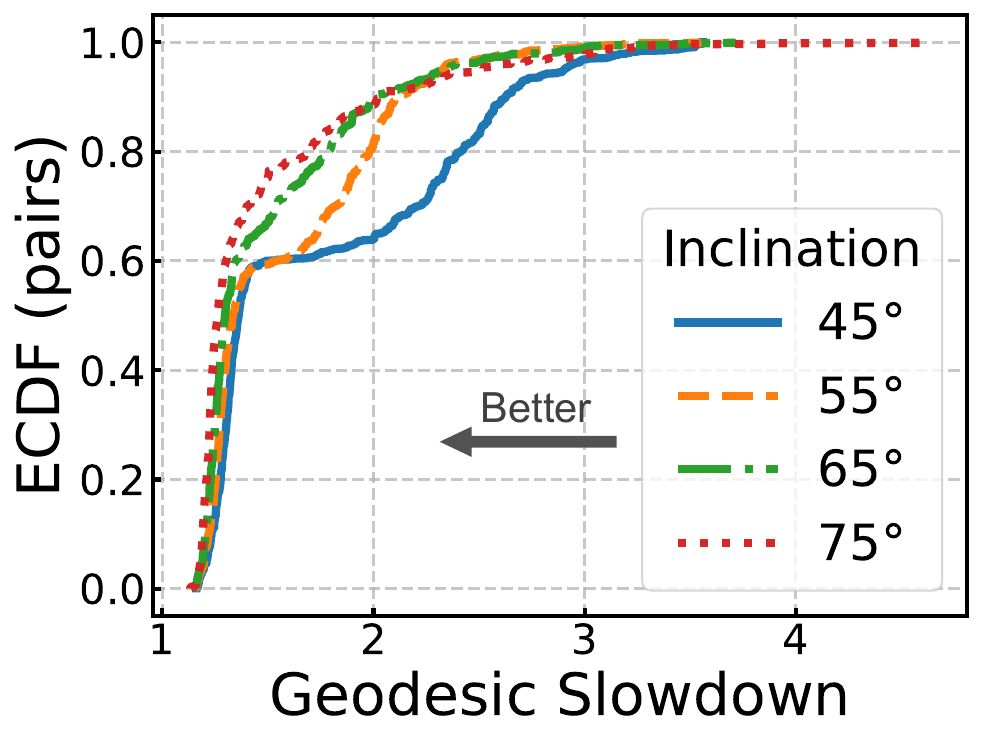}
			\vspace{-0.5cm}
                \label{appendix:inclination geodesic slowdown 8}
		\end{minipage}%
	}%
        	\vspace{-0.5cm}
	\caption{Evaluation of Geodesic Slowdown of nine groups of endpoint pairs with varying Inclination of orbits (1).  For all curves, lower values indicate better performance. Long tails indicate outliers with poor performance.}
	\vspace{-0.25cm}
	\label{inclination geodesic slowdown}
\end{figure*}

\noindent\textbf{Geodesic Slowdown across endpoint pairs at different Inclinations:}  Fig.~\ref{inclination geodesic slowdown} shows the geodesic slowdown of four groups of endpoint pairs from 40° to 80° in intervals of 10° under four values of orbit inclination (45°, 55°, 65°, and 75°). The performance of endpoint pairs varies based on the geographic angle. The plots for all nine bins (0°-90°) are given in Fig.~\ref{appendix:inclination geodesic slowdown} (Appendix~\ref{appendix: Geodesic Slowdown}). We observe that the latency performance is the best when the geographic angle of the endpoint pairs is aligned with the orbital Inclination. The shell with a 45° inclination offers the lowest slowdown for endpoint pairs with a geographic angle in the range 0-50°. The remaining shells also perform the best across endpoint pairs which have a geographic angle closest to the orbit's Inclination. We surmise that when the orbital inclination aligns with the geographic angle, the shortest path between endpoints follows more intra-orbital hops within the same orbit, leading to a better latency performance. We observe that this pattern holds across all ranges of geographic angles in Fig.~\ref{appendix:inclination all pairs}.

\begin{figure}[ht]
  \begin{center}
    \includegraphics[width=0.8\linewidth]{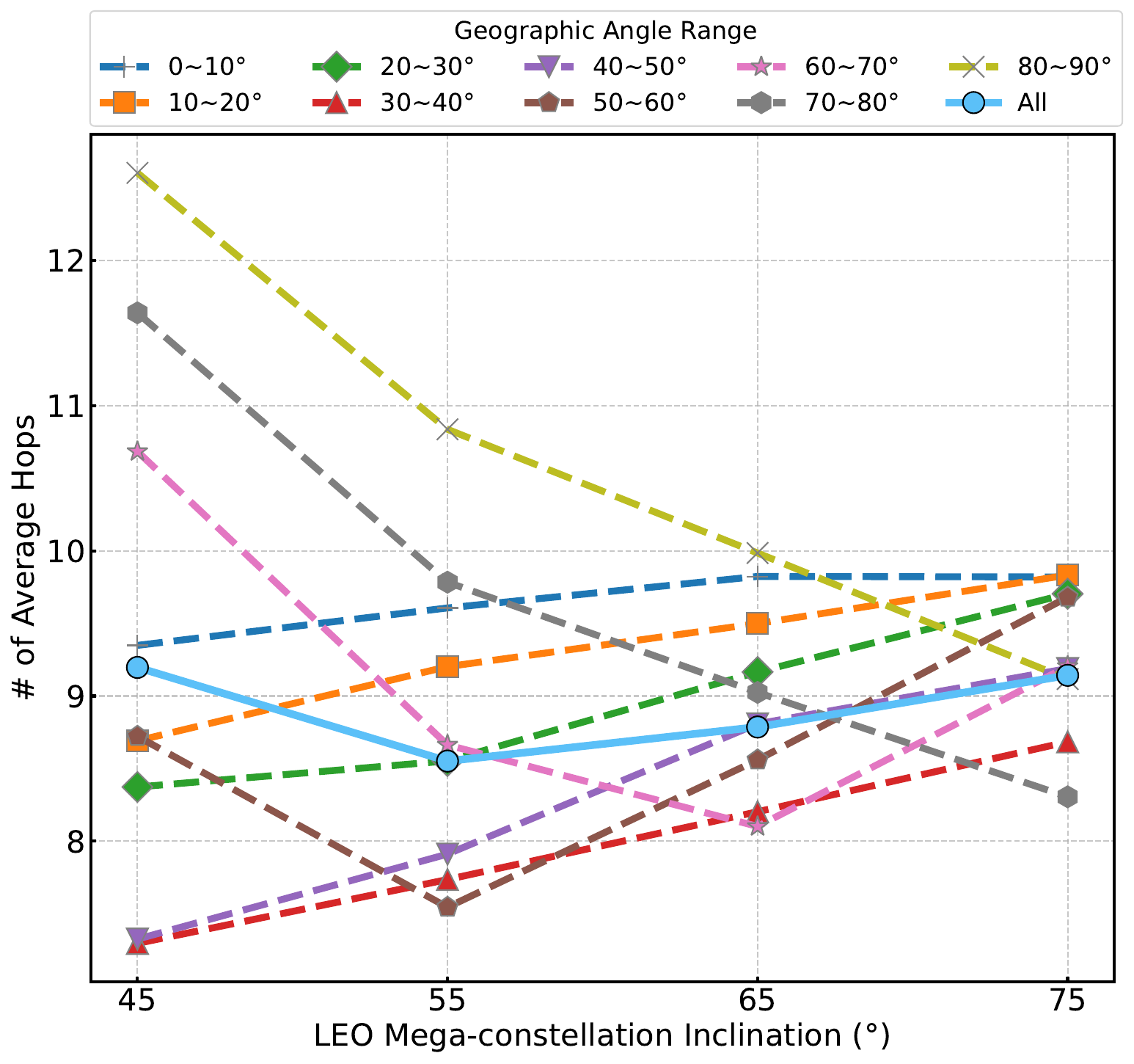}
  \end{center}
    \vspace{-0.25cm}
    \caption{Evaluation of Average Hop Count of nine groups of endpoint pairs under varying Inclination.}    
  \label{avrage hop 1}  
    \vspace{-0.5cm}
\end{figure}

\noindent\textbf{Average Hop Count for different
endpoint pairs under varying Inclination:} We evaluate the average hop count for all nine groups of endpoint pairs under different inclinations and present the results in Fig.~\ref{avrage hop 1}. Across all endpoint pairs, an inclination of 55° offers the best performance. Most commercial constellations today use an orbital angle very close to this value. Another interesting observation is that the inclination that offers the smallest average hop count is the one closest to the geographic angle of endpoint pairs. For example, for endpoint pairs with a geographic angle in the range of 40-50°, the constellation with an inclination of 45° offers the least average hop count. Similarly, 50-60° endpoint pairs fare best under an inclination of 55° and so on. When endpoint pairs have a geographic angle greater than 70°, the average hop count gradually decreases with an increase in inclination from 45° to 75°. Similarly, for endpoint pairs with a geographic angle of less than 40°, the average hop count gradually decreases as inclination decreases from 75° to 45°.

\begin{figure*}[ht]
	\centering
    \includegraphics[scale=.24]{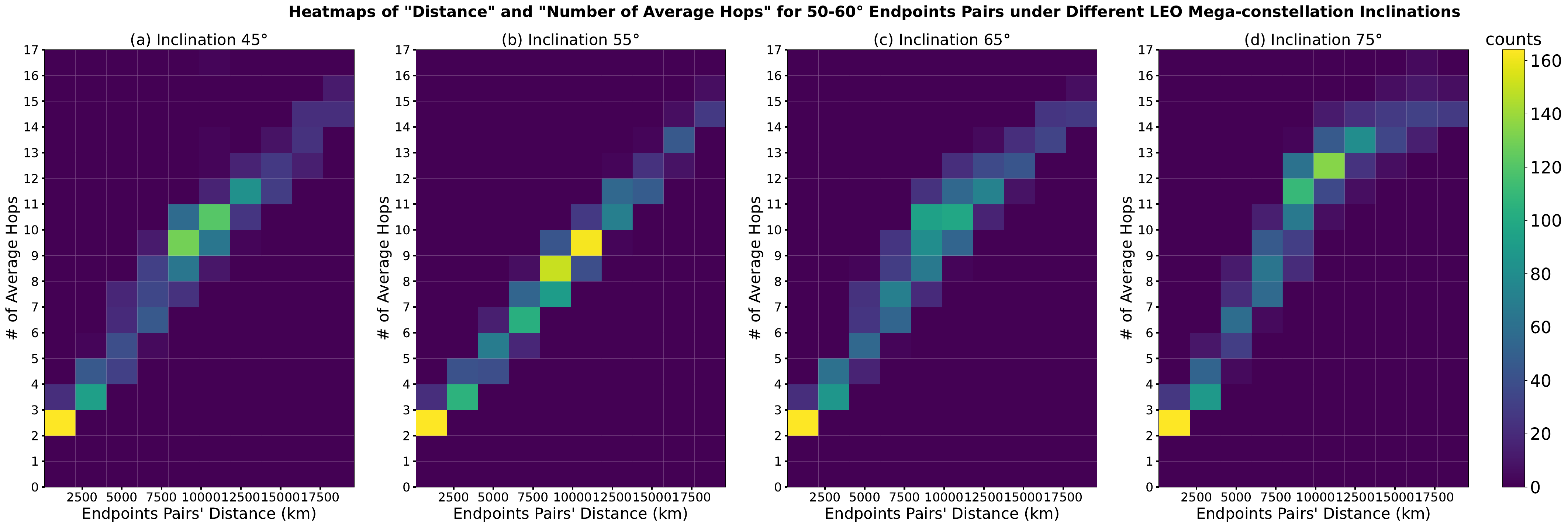}
	\vspace{-0.25cm}

	\caption{\small Heatmap of "Distance" vs. "Number of Average Hops" for 50-60° Endpoints Pairs under different LEO mega-constellation inclinations (a) 45°, (b) 55° (c) 65° and (d) 75°. The 50-60° city pairs are further divided into bins based on their geodesic distance (along the x-axis).  Comparing the same bin across (a) - (d) helps us understand the change in the number of average hops for the same city pairs under different inclinations. For example, 50°-60° city pairs have a lower average hop count in (b) when the inclination angle of the orbit is more closely aligned with their geographic angle.}
	\vspace{-0.25cm}
	\label{heatmap new}
\end{figure*}

We present a more fine-grained analysis of the variation of average hop count across endpoint pairs based on both geographic angle and geographic distance.(Full experimental results in Appendix~\ref{appendix: Average Hop}) We choose four representative values of 50-60° endpoints pairs under different orbital inclination, 45°, 55°, 65° and 75°, in Fig.~\ref{heatmap new}, respectively. Note that across both figures, the set of endpoint pairs that belong to each bin along the x-axis remains the same (since the distance between endpoint pairs is fixed). However, their average hop count (y-axis) can vary based on the inclination of the orbit. 

By comparing the same column in Fig.~\ref{heatmap new}(a)-(d), we observe that the number of average hops changes with the inclination. For example, consider endpoint pairs within the distance interval of 10000-12500km and a geographic angle of 50-60° (i.e., the sixth column in each subplot of Fig.~\ref{heatmap new}). Under the inclination of 45° (Fig.~\ref{heatmap new}(a)), more than 120 endpoint pairs have an average hop count in the range 10-11, with only a tiny fraction with an average hop count of 9-10 or 11-12. However, under the inclination of 55° (Fig.~\ref{heatmap new}(b)), most of the endpoint pairs have a smaller average hop count than in Fig.~\ref{heatmap new}(a). Less than 40 endpoint pairs have an average hop count in the range of 10-11 and the remaining have an average hop count in the range of 8-9, thereby decreasing the overall average hop count. In contrast, for the sixth column's color block, as we increase the orbital angle (Fig.~\ref{heatmap new}(b) vs. Fig.~\ref{heatmap new}(c) and  Fig.~\ref{heatmap new}(d)), the average hop count increases. This is because the orbital angle of 55° (Fig.~\ref{heatmap new}(b)) is closer to the geographic angle for 50-60° endpoint pairs. The same pattern holds for all other groups of endpoint pairs and their more aligned inclinations (Appendix~\ref{appendix: Average Hop}).

From these two experiments, we confirm that alignment between the inclination or orbit and the geographic angle of endpoint pairs is positively correlated with a reduced average hop count and improved performance. When an orbit is more aligned with the geographic angle of endpoint pairs, the satellite path uses more intra-orbit hops and reduces inter-orbit zig-zag hops, which in turn reduces the hop count. Therefore, if endpoint traffic distributions can be known a priori, designing orbital angles that align closely with endpoint traffic distributions can improve performance.

\vspace{2mm}
\noindent
\textbf{Traffic Matrix Analysis Key Takeaways: }
\vspace{-1mm}
\begin{itemize}[leftmargin=*, noitemsep]    
    \item The network delay is closest to optimal when the geographic angle of the traffic endpoints aligns closely with the orbital angle of the constellation.
    \item Average hop count is lower when the geographic angle of the traffic endpoints aligns closely with the orbital angle of the constellation.
\end{itemize}

\subsection{Summary of Results}

\begin{itemize}[leftmargin=*,nolistsep]
\item The design of most commercial constellations (except S3 and S5) is well-suited to support communication between the top 100 cities. Considering both delay and link stability, S2, K2, and T2 offer the best performance in each of the three constellations, respectively. 

\item The two design parameters---the number of satellites per orbit and the number of orbits---strongly influence network performance. For high performance, the number of satellites per orbit should be above a threshold, which we empirically estimate as 28. Similarly, the performance is high when the number of orbits is at least 33.

\item Network performance is not strongly correlated with the total number of satellites. S2 offers better latency performance than S1 in spite of much fewer satellites because both its key design parameters are above the performance threshold. The poor performance of S1 is due to its low number of satellites per orbit (20 vs. threshold of 28).

\item The inclination of the LEO mega-constellation is important. The alignment between the inclination and the geographic angle of endpoint pairs is positively correlated with network performance. When the endpoint pairs are more aligned with the orbital angle, they experience lower delay and lower average hop count. 
\end{itemize}

\section{Discussion}

In this section, we provide suggestions for the design of LEO mega-constellations and identify avenues for future work.

\noindent
\textbf{Suggestions for the design of LEO mega-constellations:}
We identify the thresholds for optimal performance as 33 for the number of orbits and 28 for the number of satellites per orbit. Moreover, network delay is minimized when the geographic angle 
between endpoint pairs is aligned with the orbital angle. Based on the known user distribution on Earth, the range between 45° to 55° is the ideal orbital inclination.

\noindent
\textbf{Impact of topology variations: }
The recommendations of this study are based on the assumption of LEO mega-constellations relying on ISL links and the +Grid topology. Although the +Grid topology is the most intuitive and popular structure, several recent papers investigate designs for better ISLs topologies~\cite{soret2019inter,beech1999study,akyildiz2002mlsr,bai2004distributed,song2014tlr, taleb2008explicit}. Our evaluation needs to be repeated under novel topologies. 

\noindent
\textbf{Routing and traffic engineering: }
In this work, we assume shortest path based routing scheme and ideal network conditions. In practice, the observed delay and hop count can vary based on the routing and traffic engineering schemes employed by the constellation operator. Moreover, the network load may not be uniformly distributed, with high congestion closer to heavily populated areas. We do not make assumptions about any intelligent routing or TE schemes.

\noindent
\textbf{Multi-Shell Analysis: } Our analysis focuses on the network performance of a single shell. Currently deployed satellites do not support inter-shell communication even within the same LEO mega-constellation; their ISLs are designed to communicate with satellites in the same shell. Understanding the aggregate network performance of multiple shells, each with different configuration settings, is an open problem.

\section{Conclusion}
We evaluate three key design parameters---the number of orbits per shell, the inclination of orbit, and the number of satellites per orbit---on real-world topologies of Starlink, Kuiper, Telesat, and a broad set of synthetic topologies. Our analysis reveals interesting insights and answers to previously inexplicable behavior patterns. We identify the parameter thresholds for optimal network performance, the causes for the poor performance of a high-density shell (S1) of Starlink compared to one with lower density (S2), and determine the impact of traffic endpoint locations on performance by analyzing their geographic angles. Our interesting observations underscore the need for a systematic analysis of the impact of design parameters in the context of the emerging domain of LEO mega-constellations. We highlight several avenues for future research in this space, including the impact of topology in relation to design parameters and multi-shell analysis.

\bibliographystyle{ACM-Reference-Format}
\bibliography{reference}

\clearpage
\appendix 

\section{Ethics}
This work does not raise any ethical issues. All the code and data used in this work were available in open source. This work does not rely on any data that compromises the privacy of end users or the security of enterprises.

\section{Altitude}
\label{apeendix: Altitude}

\begin{figure}[!htp]
	\centering
	\subfigure[33 Orbits, 28 Sats/Orbits, 53°]{
		\begin{minipage}[t]{0.5\linewidth}
			\centering
			\includegraphics[scale=.25]{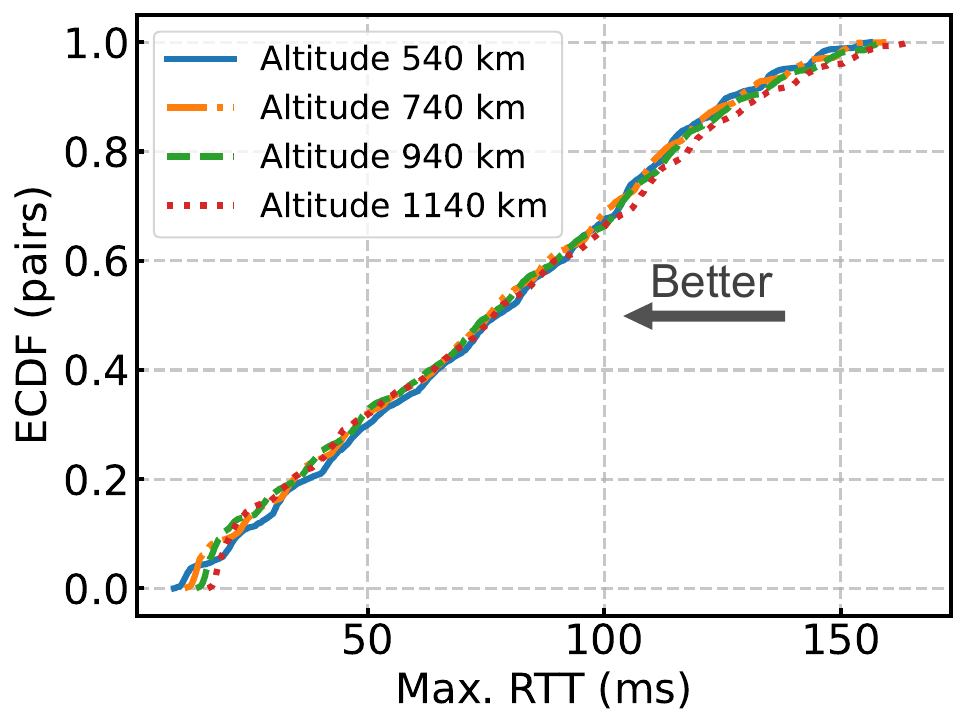}
   \label{Altitude 1}
		\end{minipage}%
	}%
	\subfigure[33 Orbits, 28 Sats/Orbits, 53°]{
		\begin{minipage}[t]{0.5\linewidth}
			\centering
			\includegraphics[scale=.25]{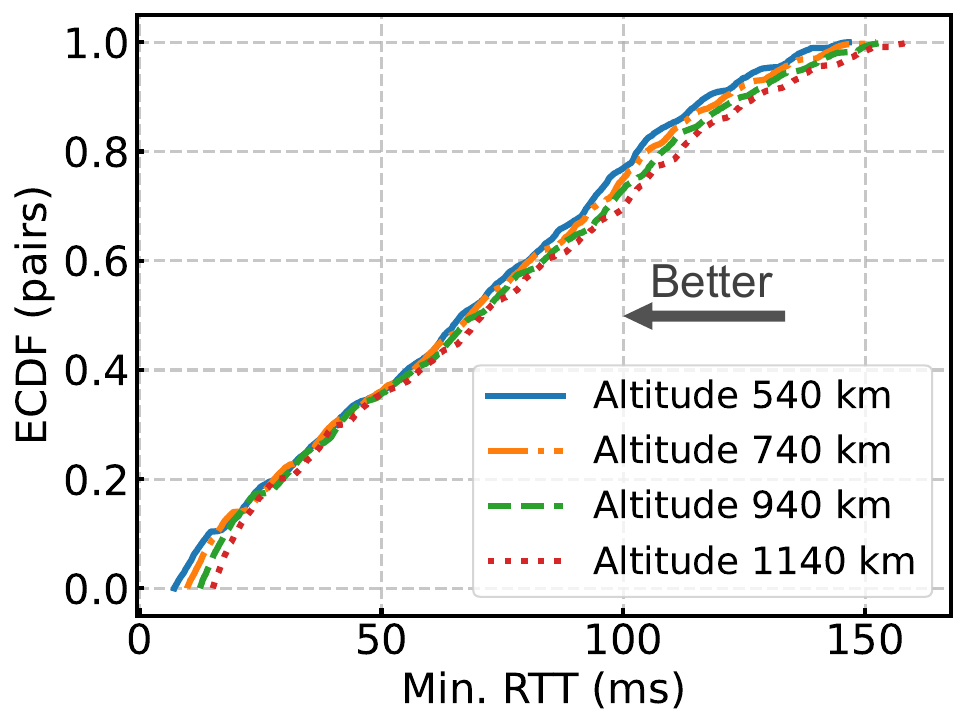}
   \label{Altitude 2}
		\end{minipage}%
	}%
 
	\subfigure[33 Orbits, 28 Sats/Orbits, 53°]{
		\begin{minipage}[t]{0.5\linewidth}
			\centering
			\includegraphics[scale=.25]{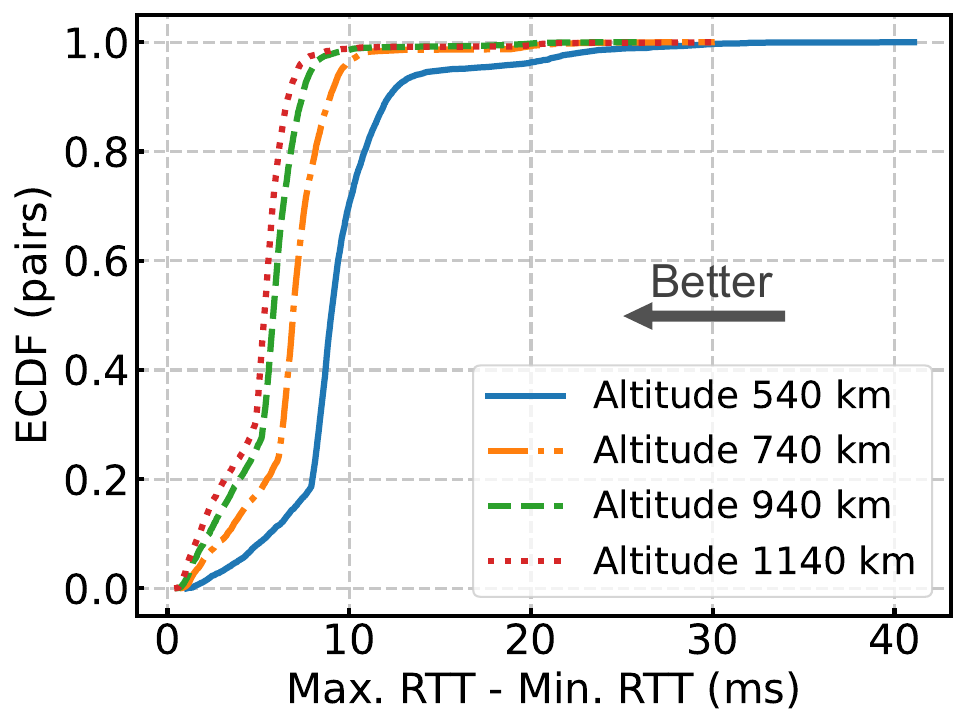}
   \label{Altitude 3}
		\end{minipage}%
	}%
	\subfigure[33 Orbits, 28 Sats/Orbits, 53°]{
		\begin{minipage}[t]{0.5\linewidth}
			\centering
			\includegraphics[scale=.25]{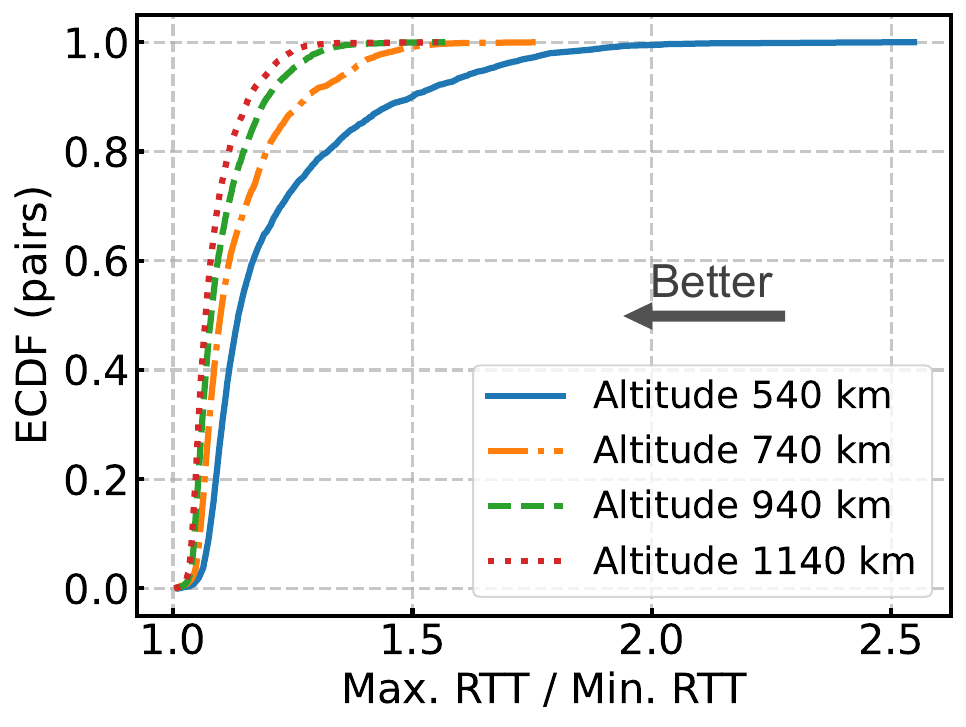}
   \label{Altitude 4}
		\end{minipage}%
	}%

\subfigure[33 Orbits, 28 Sats/Orbits, 53°]{
		\begin{minipage}[t]{0.5\linewidth}
			\centering
			\includegraphics[scale=.25]{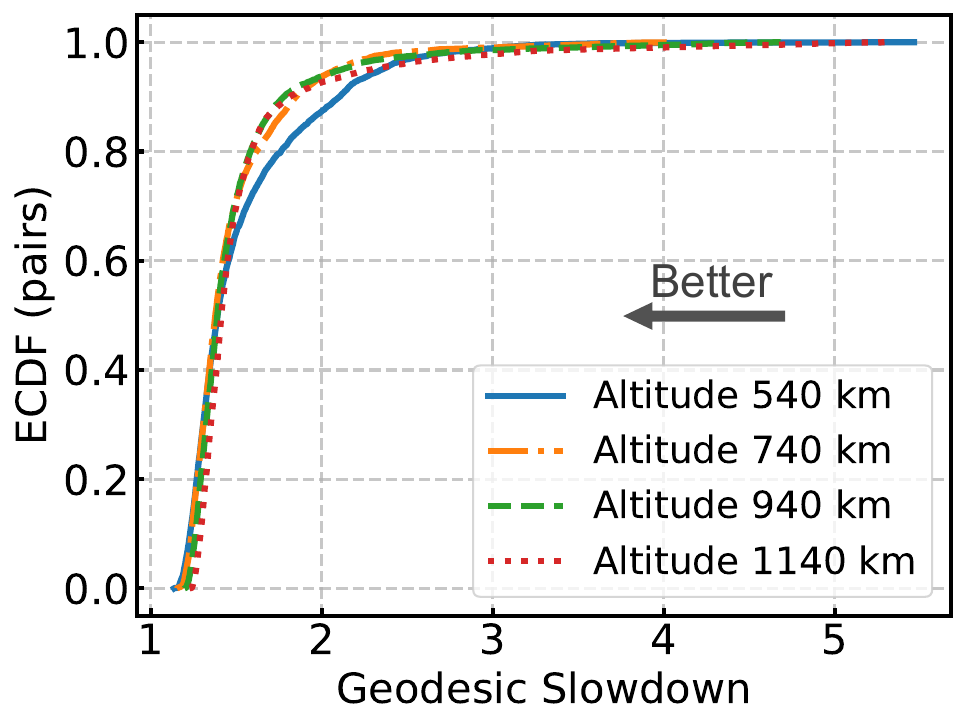}
   \label{Altitude 1}
		\end{minipage}%
	}%
	\subfigure[33 Orbits, 28 Sats/Orbits, 53°]{
		\begin{minipage}[t]{0.5\linewidth}
			\centering
			\includegraphics[scale=.25]{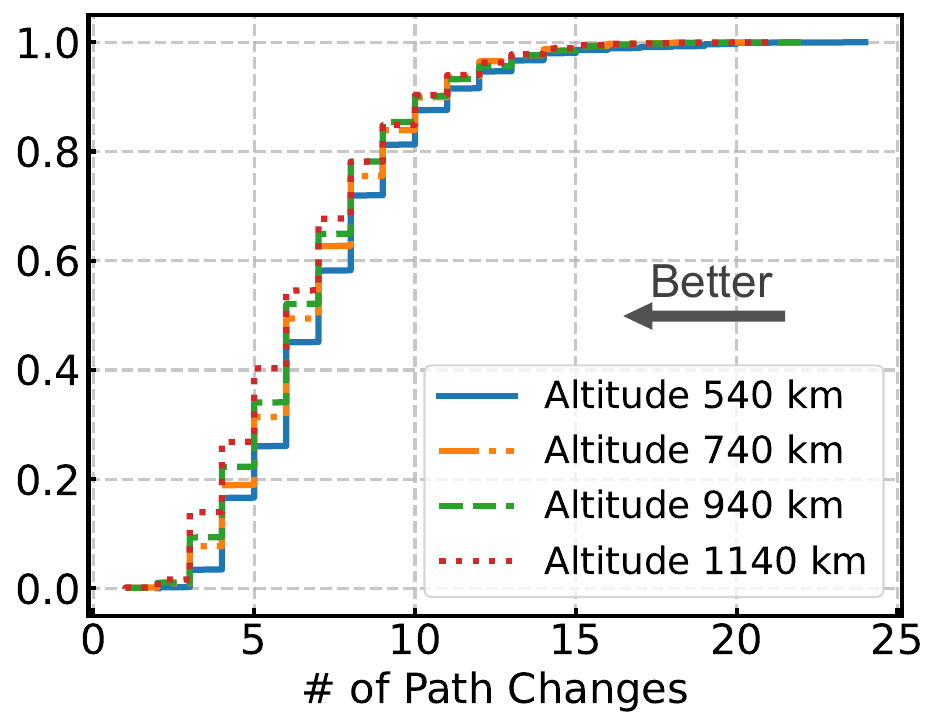}
   \label{Altitude 2}
		\end{minipage}%
	}%
 
	\subfigure[33 Orbits, 28 Sats/Orbits, 53°]{
		\begin{minipage}[t]{0.5\linewidth}
			\centering
			\includegraphics[scale=.25]{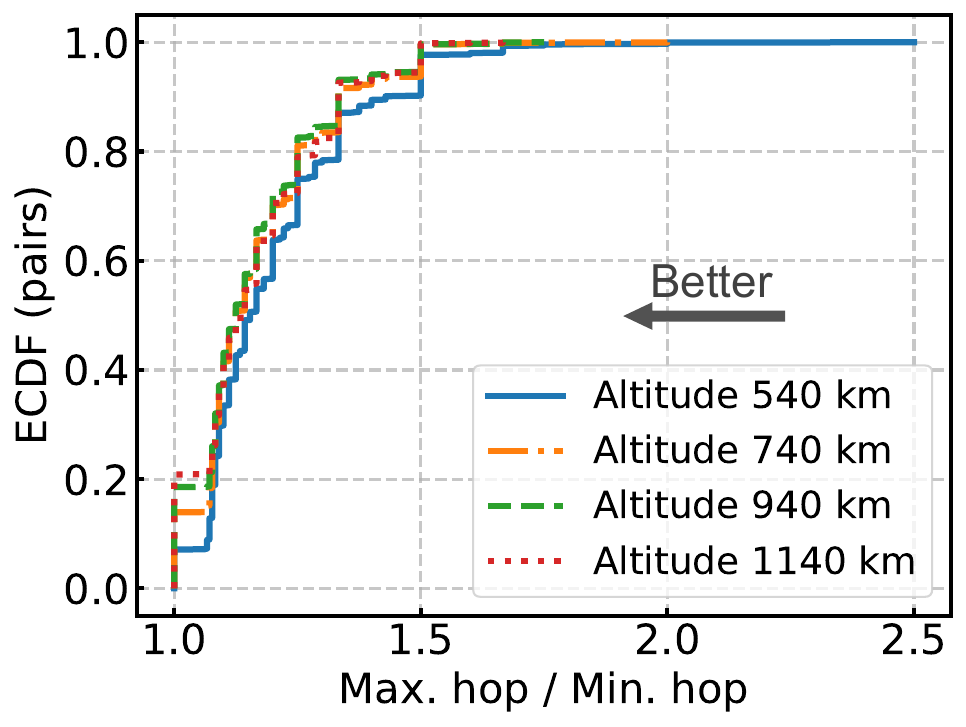}
   \label{Altitude 3}
		\end{minipage}%
	}%
	\subfigure[33 Orbits, 28 Sats/Orbits, 53°]{
		\begin{minipage}[t]{0.5\linewidth}
			\centering
			\includegraphics[scale=.25]{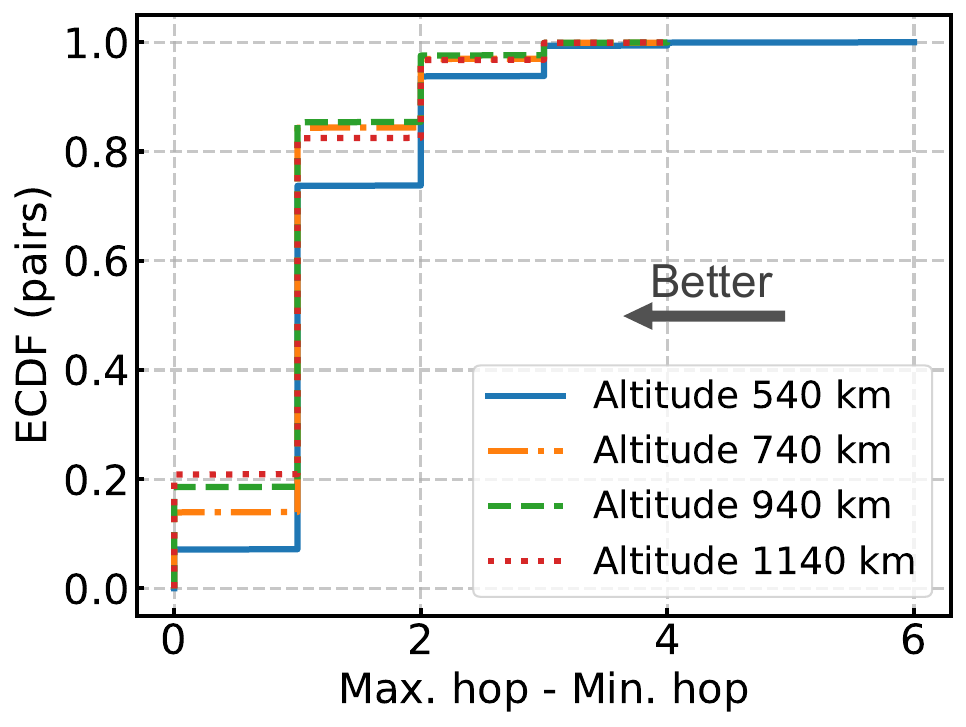}
   \label{Altitude 4}
		\end{minipage}%
	}
\vspace{-0.5cm}
	\caption{The performance of changing LEO mega-constellations' Altitude.}
	\label{Altitude}
\end{figure}

Altitude is one of the potential parameters that could affect the network performance of the LEO mega-constellation. However, altitude may not be a factor affecting the absoluteness of the LEO mega-constellation. From the parameter setting of the LEO mega constellation, we can find that although different commercial satellites have different altitude Settings, the most extensive range ranges from 540km for Starlink to 1325 km for Telesat. However, a change of a few hundred kilometers is only a hundredth of a change for an Earth over ten thousand kilometers across. It is reasonable to assume that the change in distance of several hundred kilometers may make altitude not a factor that can significantly affect the LEO mega-constellation. Our experiment confirms our hypothesis. We estimate the LEO mega-constellations with altitude = 540, 740, 940, and 1140 km, respectively. The other configurations are Orbits = 33, Sats/Orbits = 28, Inclination = 53°. 

\noindent\textbf{Altitude is not the critical parameter.}
The experiment result of a) Max. RTT, b) Min. RTT, c) Max. RTT - Min. RTT, d) Max. RTT / Min. RTT, e) Geodesic Slowdown, f) \# of Path Change, g) Max. Hop / Min. Hop, h) Max. Hop - Min. Hop are shown in Fig.~\ref{Altitude}. From the result, we can see the difference in Max. RTT and Min, RTT between different altitude constellations are negligible. The Geodesic Slowdown, \# of Path Change, Max. Hop / Min. Hop, Max. Hop - Min. Hop follows the same rules. Only the Max. RTT - Min. RTT, Max. RTT / Min. RTT shows little difference between different altitude constellations: when the altitude increases, the fluctuation of RTT becomes less. However, the RTT fluctuation of most endpoints pairs is still within 10ms. From what has been discussed above, we may conclude that altitude is not the critical parameter that could significantly determine the network performance of LEO mega-constellations.

\section{Visualization of Multishells}
\label{apeendix: Visualization 1}
\begin{figure*}[htp]
	\centering  %
	\subfigure{
		\includegraphics[width=0.16\textwidth]{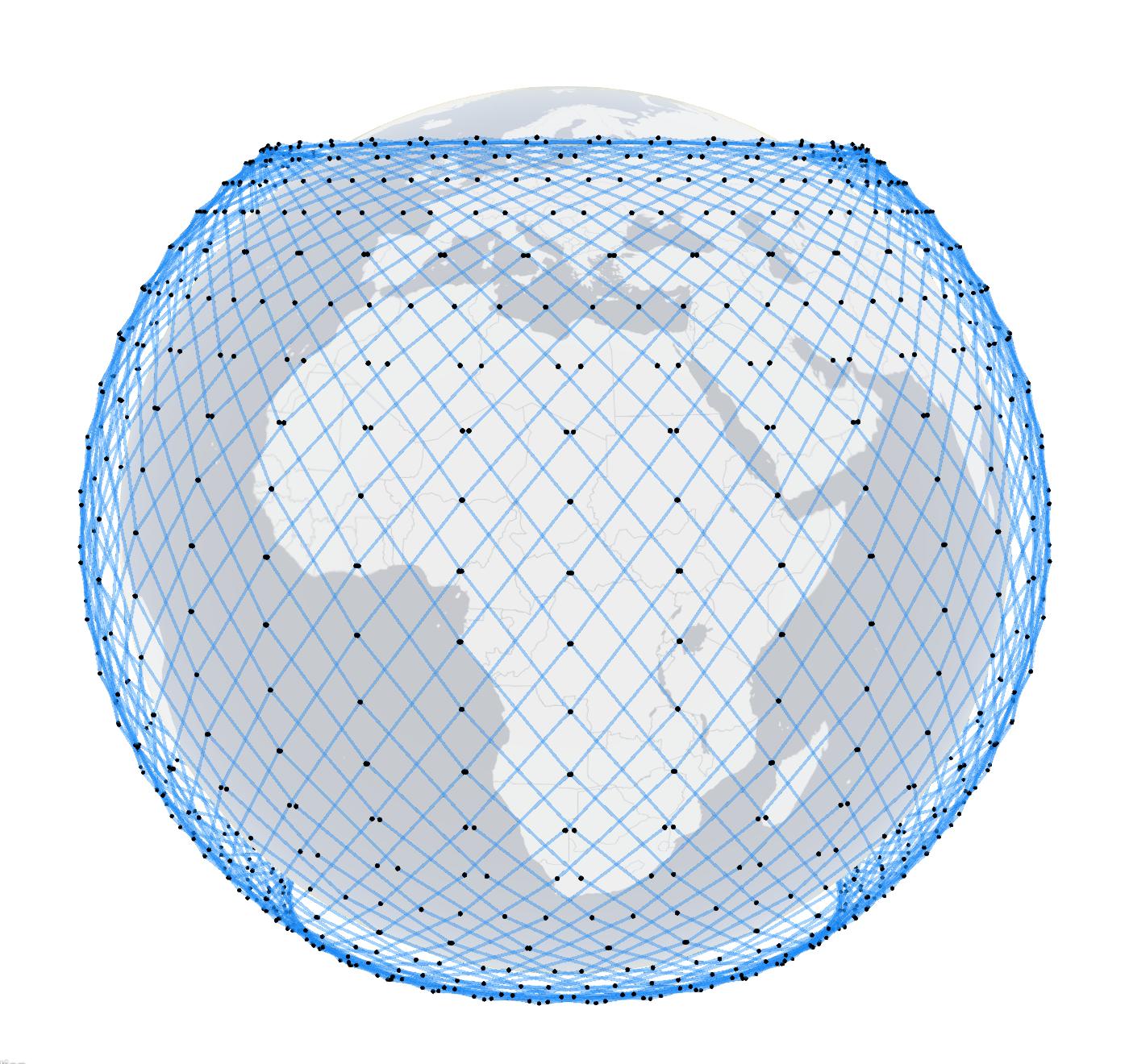}
    }
	\subfigure{
		\includegraphics[width=0.16\textwidth]{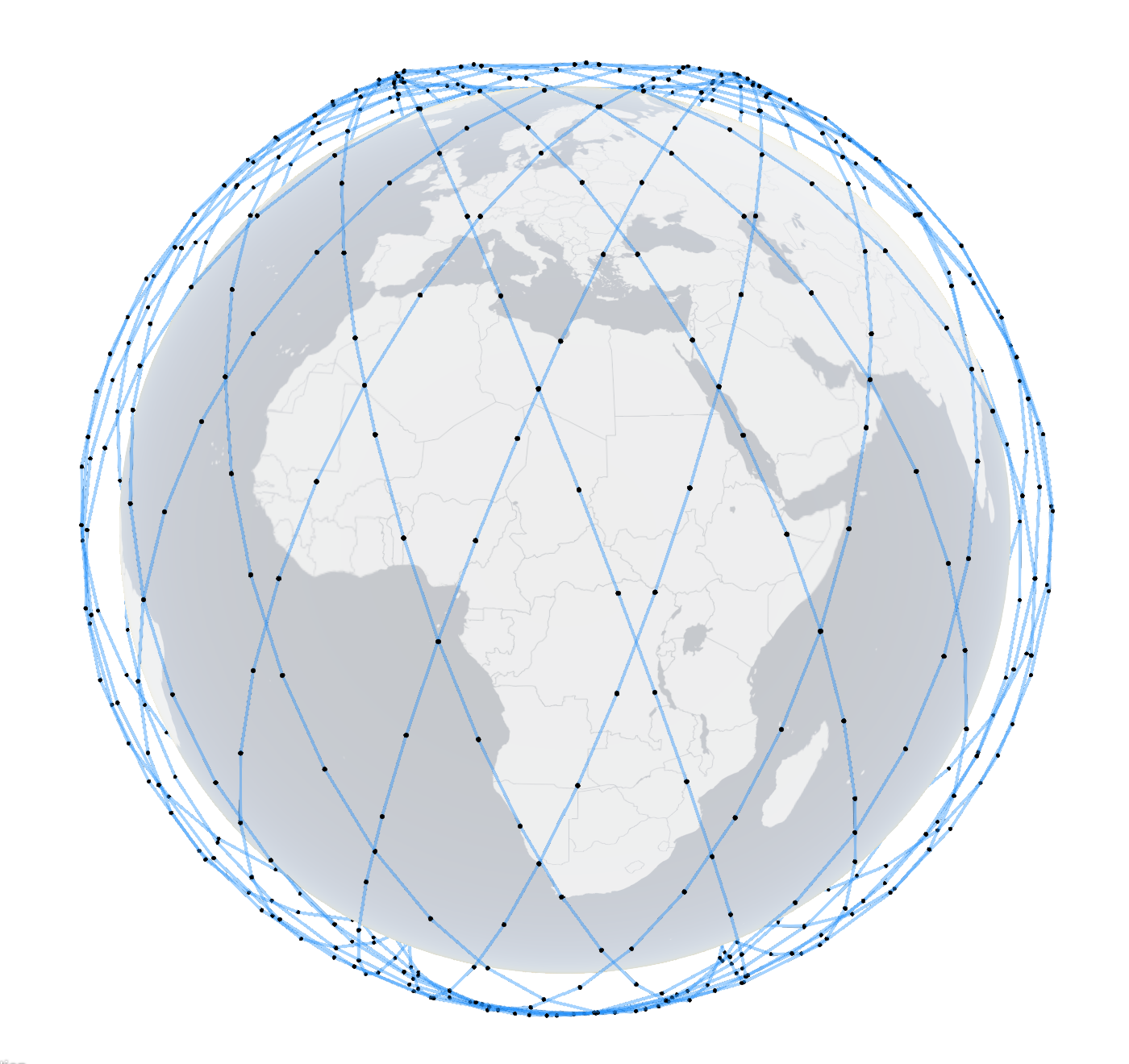}}
	\subfigure{
		\includegraphics[width=0.16\textwidth]{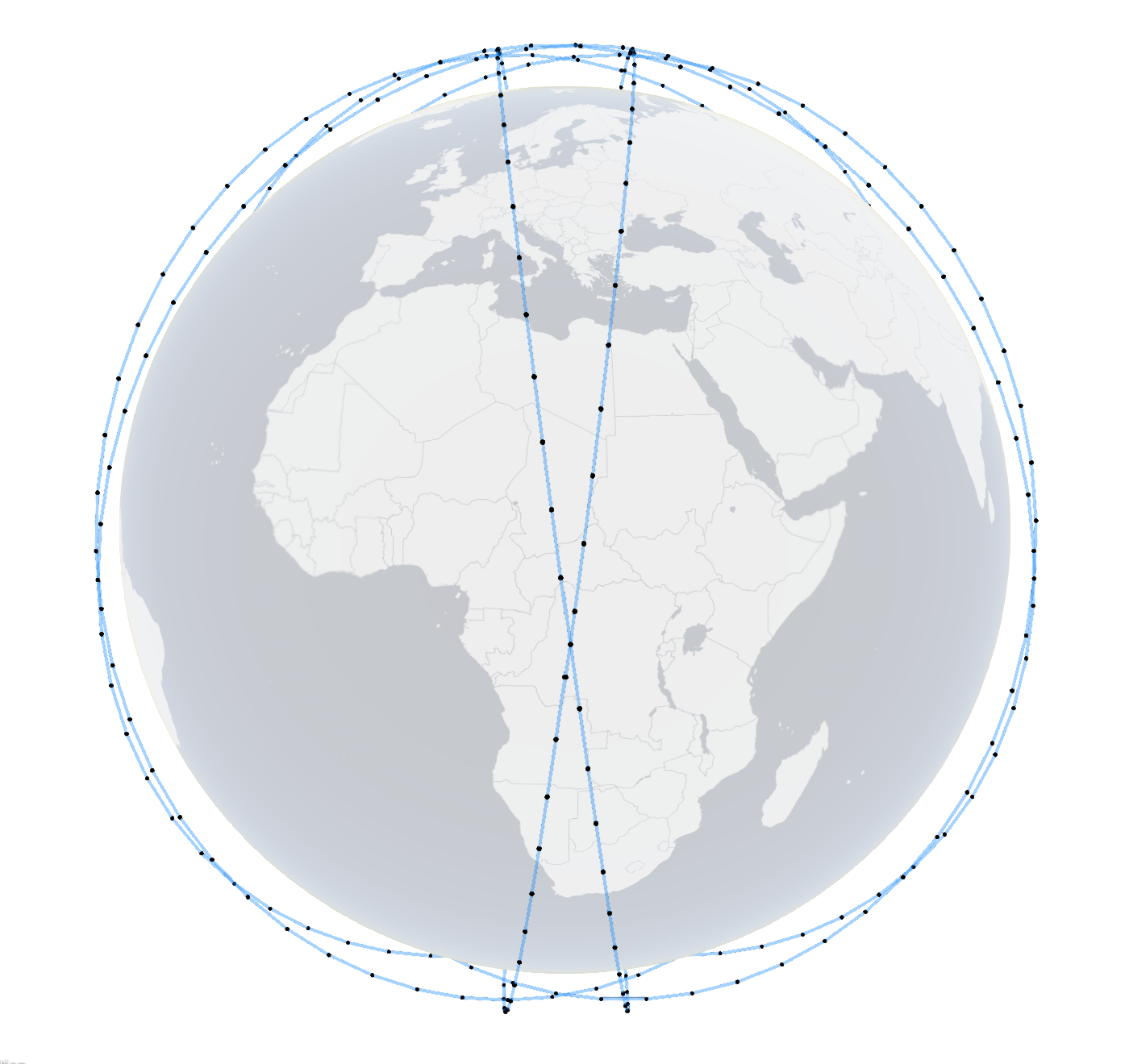}}
	\subfigure{
		\includegraphics[width=0.16\textwidth]{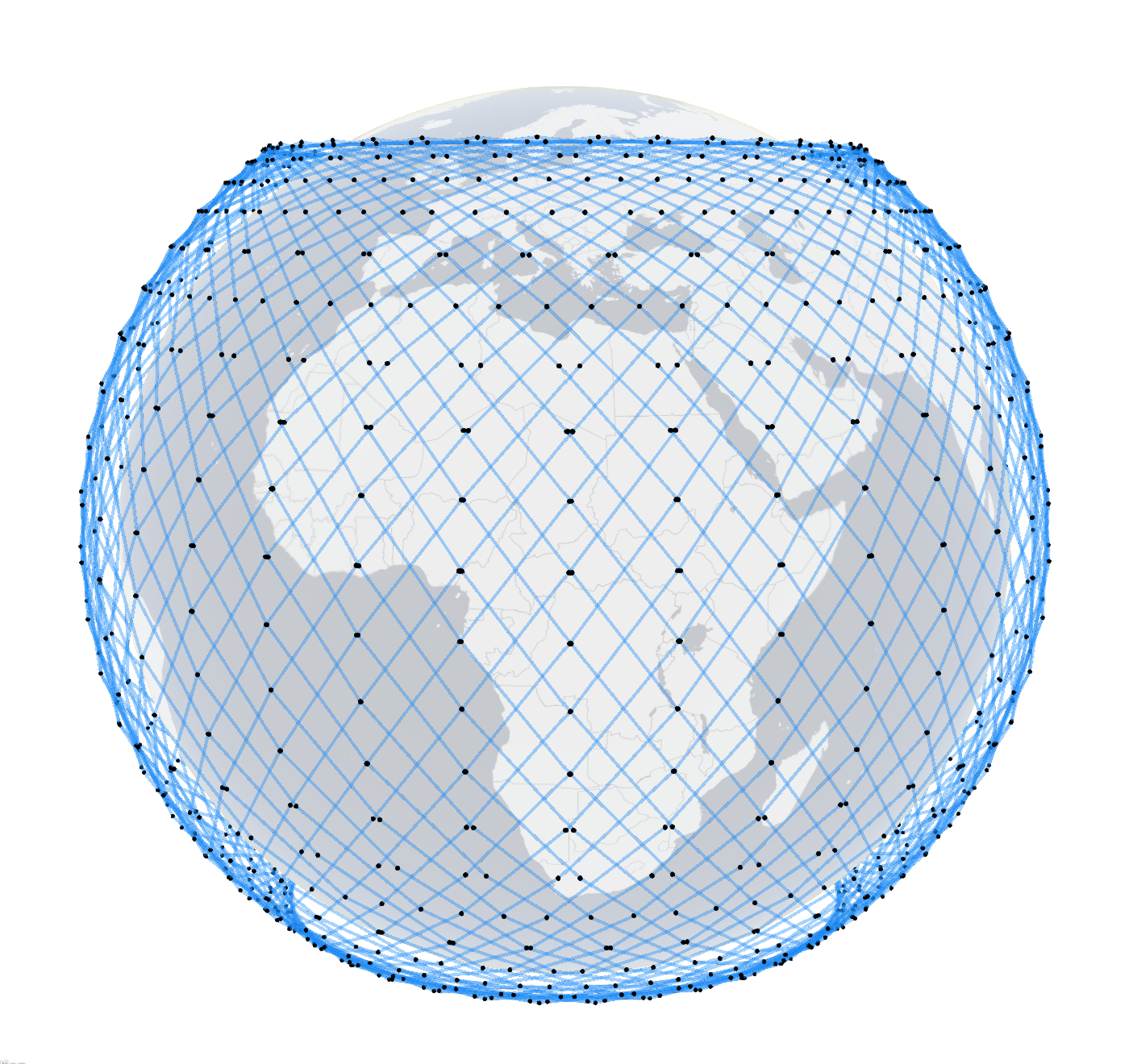}}
        \subfigure{
		\includegraphics[width=0.16\textwidth]{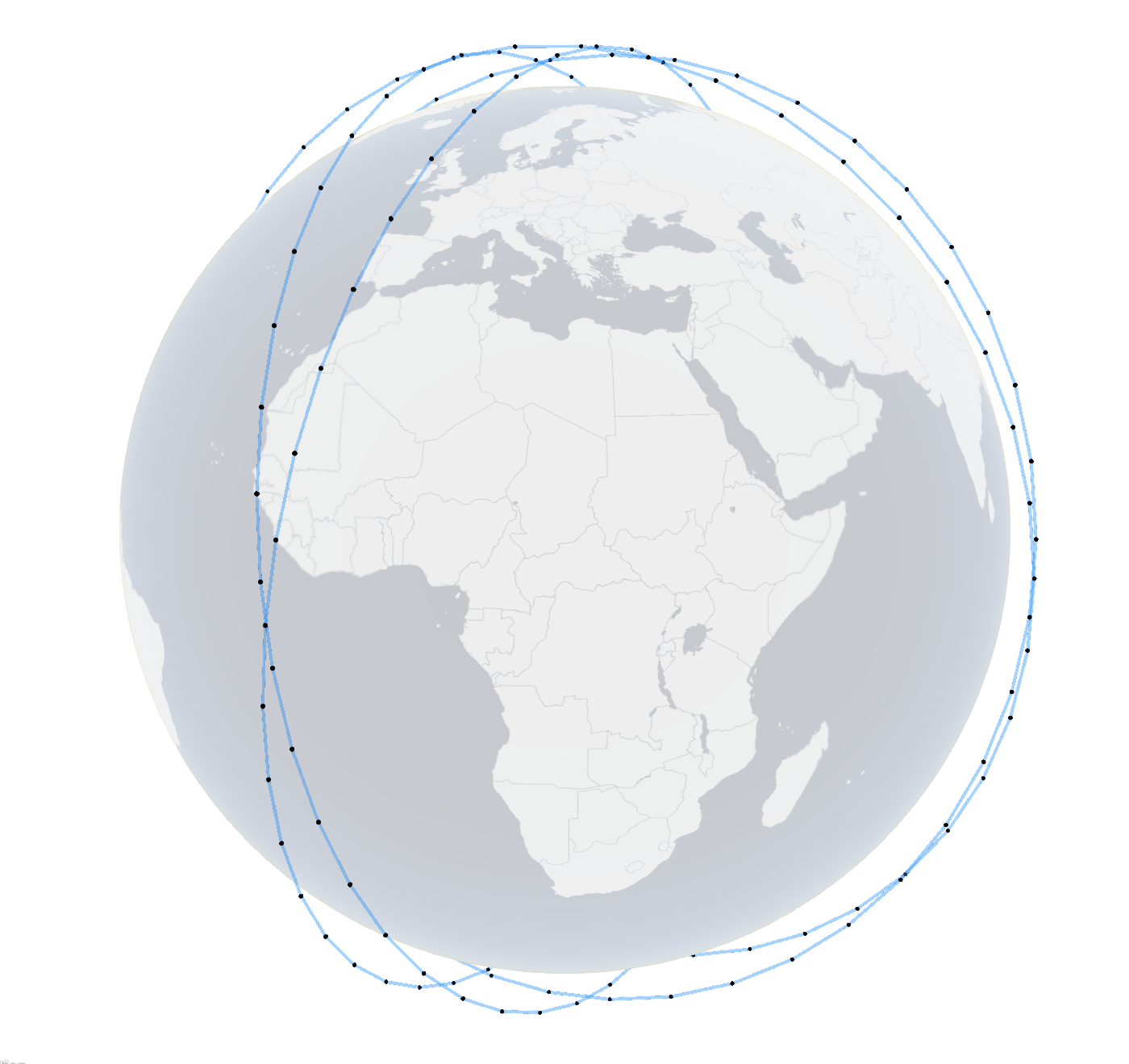}}
        \subfigure{
		\includegraphics[width=0.16\textwidth]{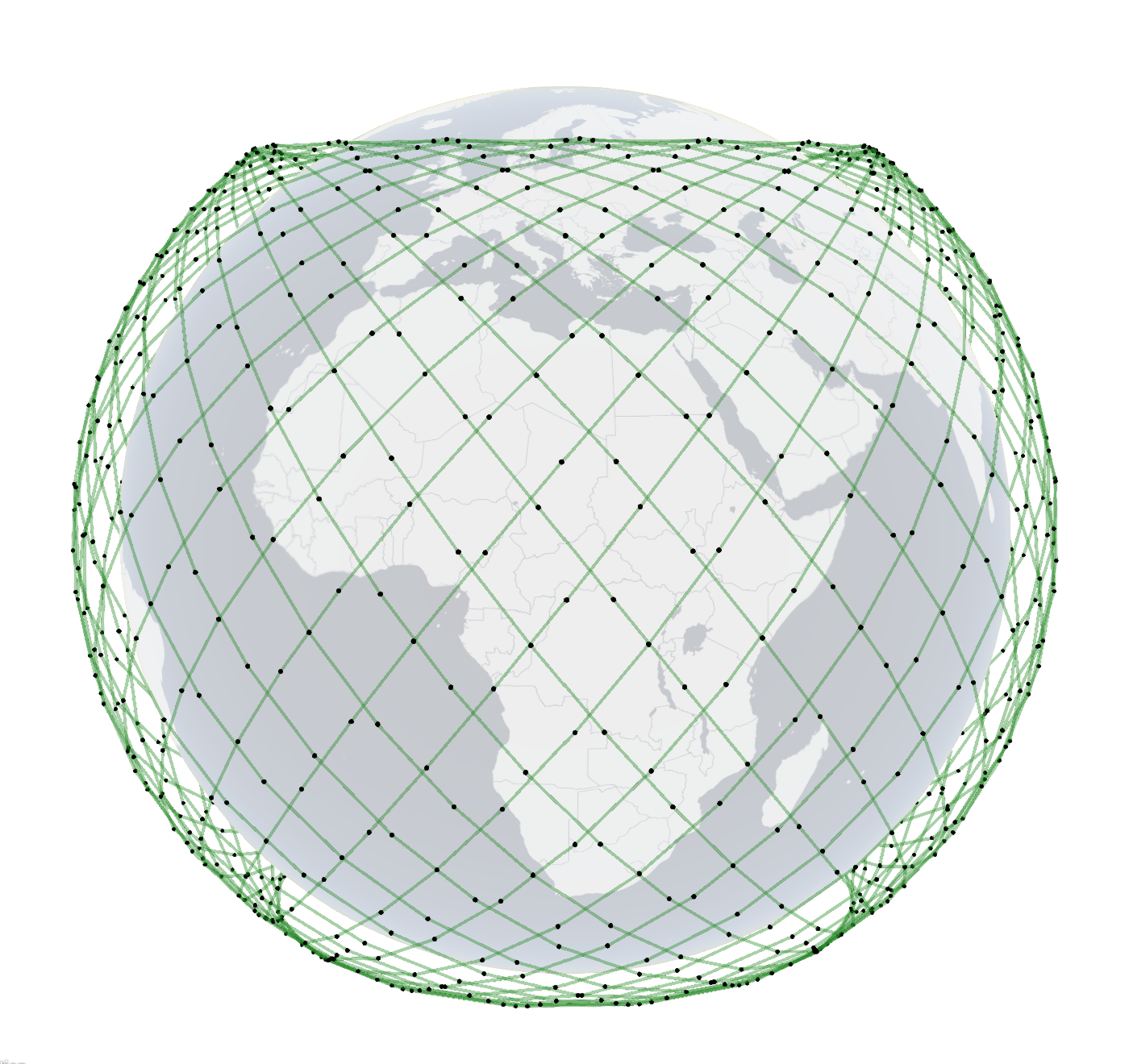}}
	\subfigure{
		\includegraphics[width=0.16\textwidth]{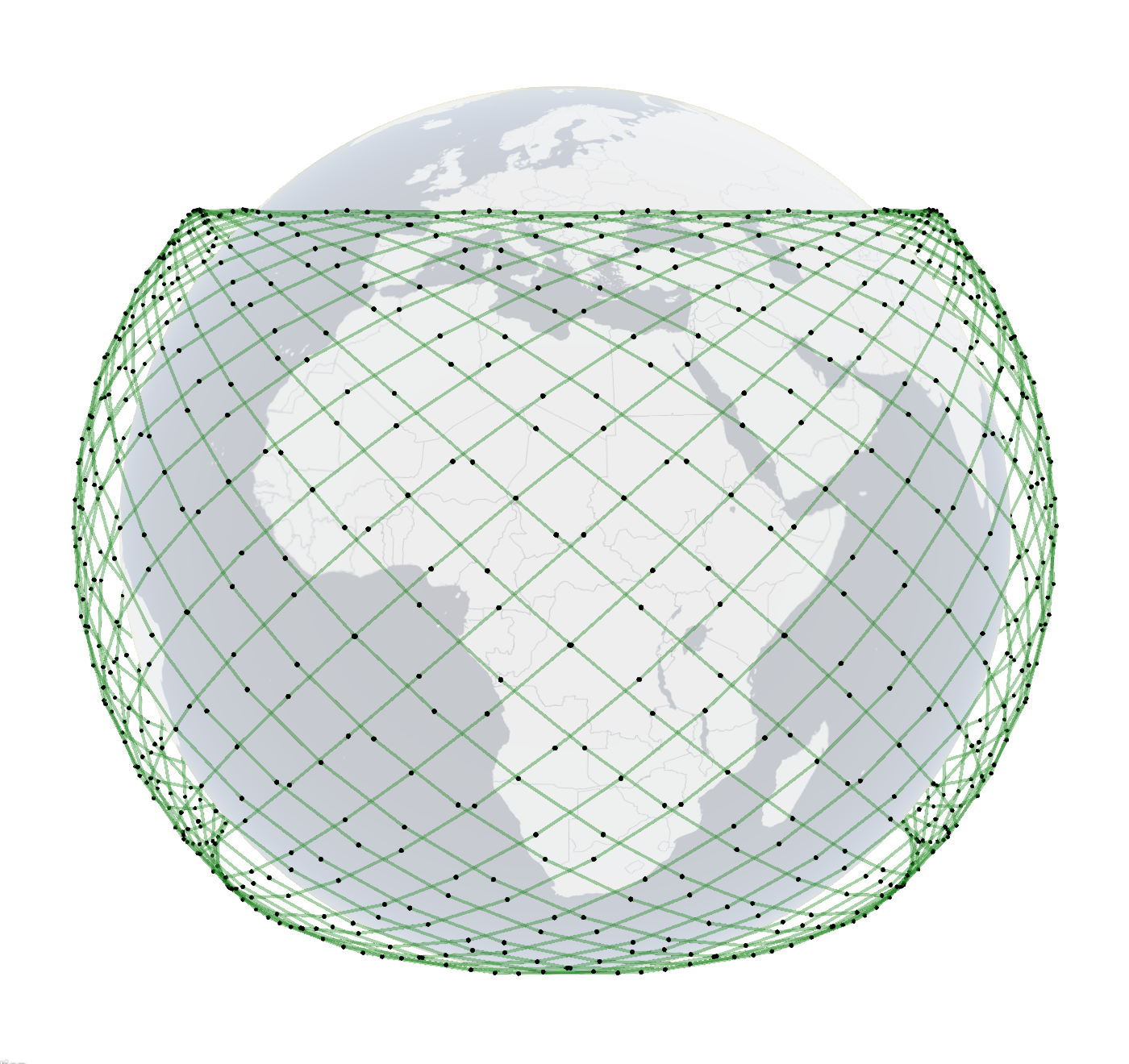}}
	\subfigure{
		\includegraphics[width=0.16\textwidth]{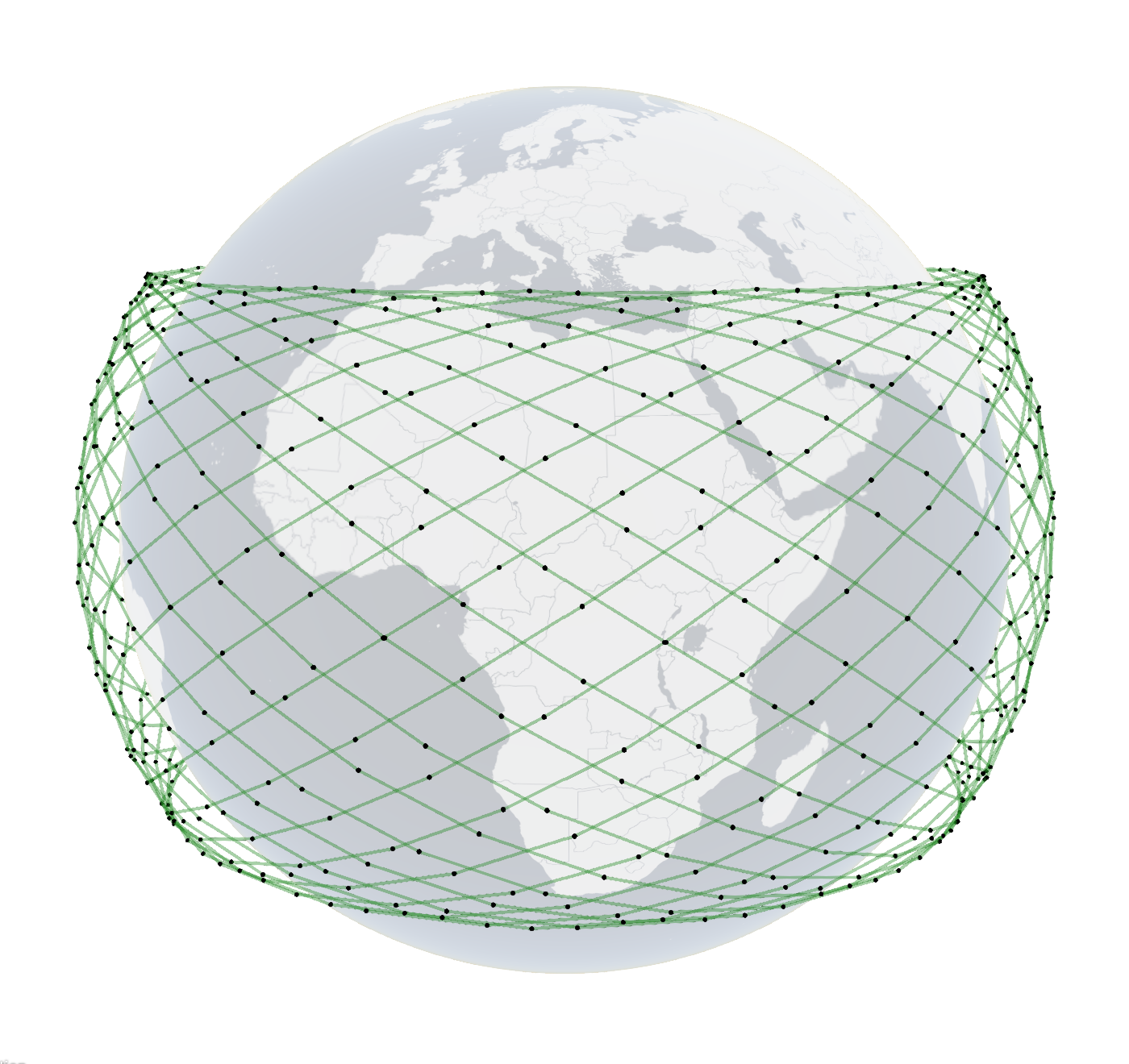}}
	\subfigure{
		\includegraphics[width=0.16\textwidth]{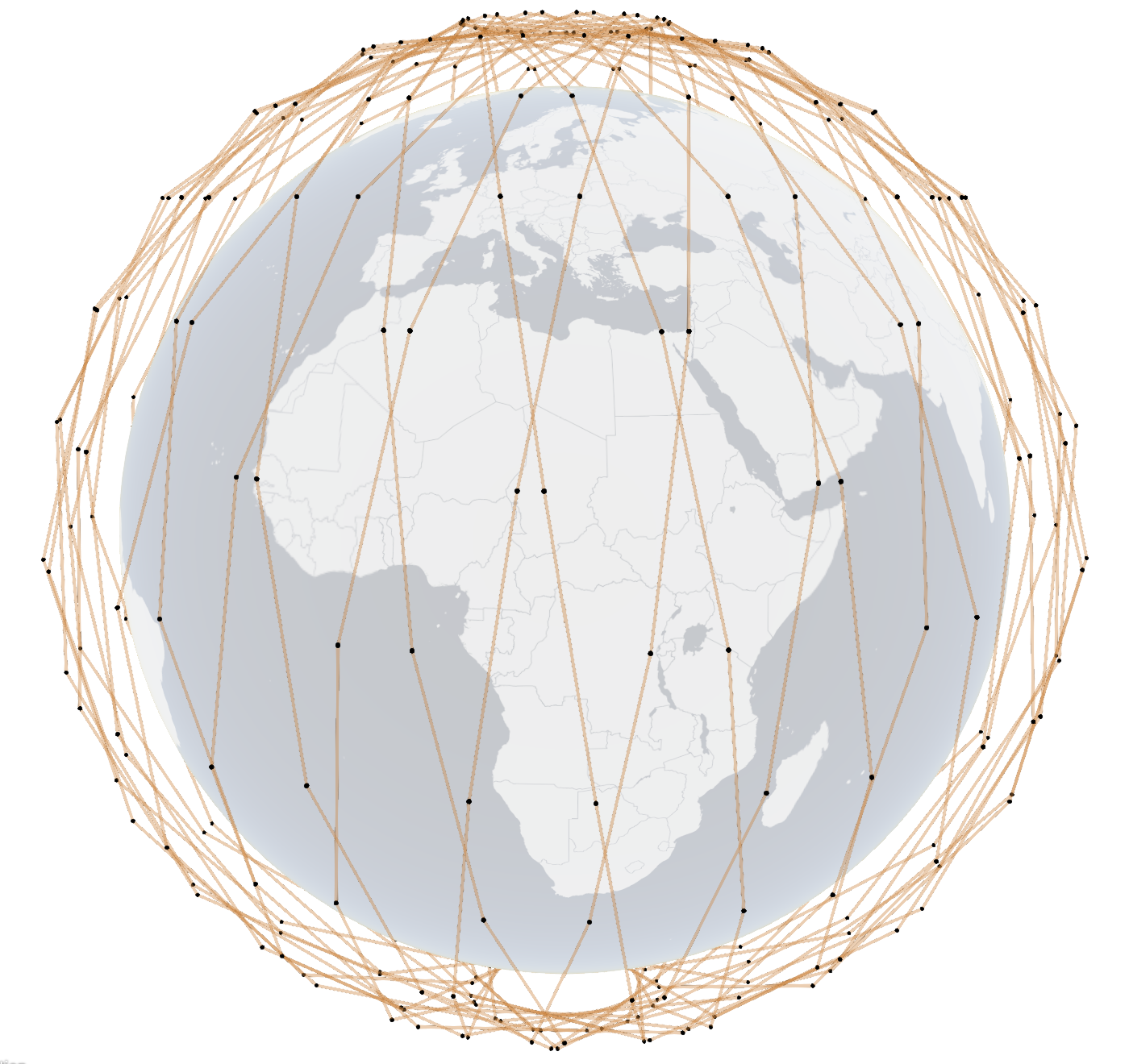}}
        \subfigure{
		\includegraphics[width=0.16\textwidth]{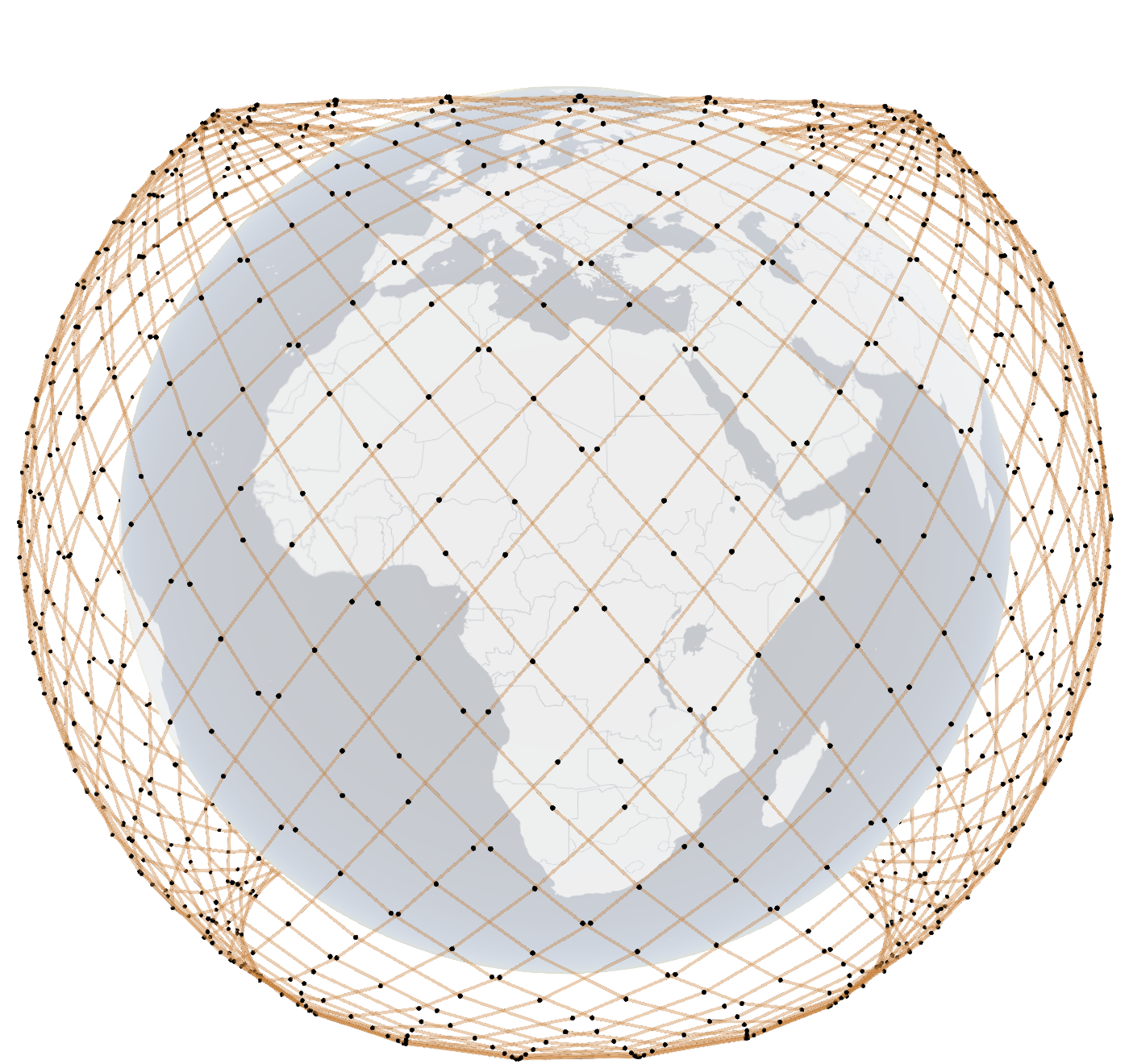}}
  \vspace{-0.5cm}
	\caption{Visualization of Starlink shells 1-5 (blue), Kuiper shells 1-3 (green), Telesat shells 1-2 (brown).}
  \vspace{-0.35cm}
        \label{VisShell}
\end{figure*}

Fig.~\ref{VisShell} demonstrates the topology of the three existing major constellations.

\section{Commercial LEO mega-constellations}
\label{appendix: Commercial LEO mega-constellations}
\begin{figure*}[htp]
	\subfigure[]{
	\vspace{-0.5cm}
		\begin{minipage}[t]{0.49\linewidth}
			\centering
			\includegraphics[scale=.28]{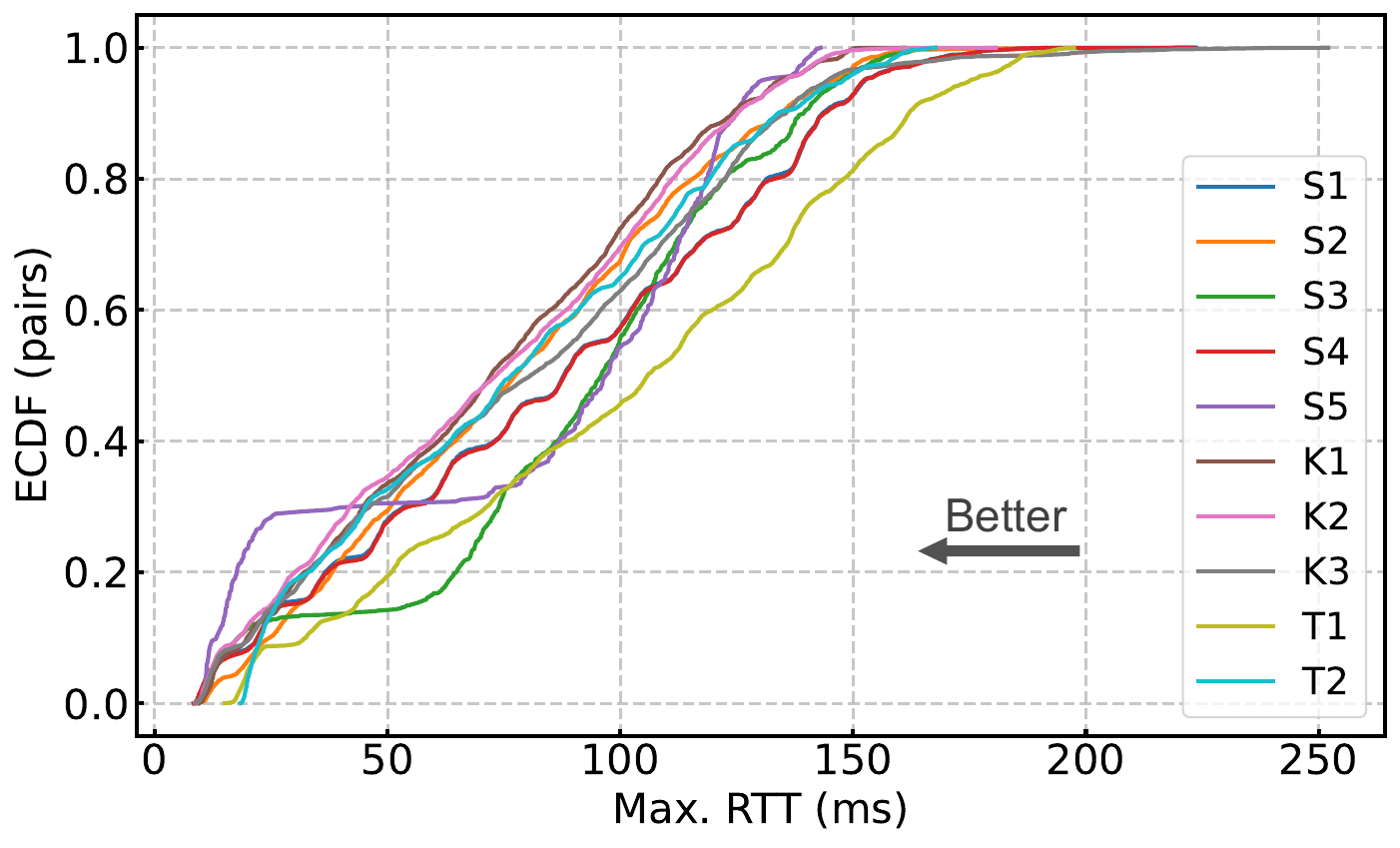}
                \label{all Commercial 1}
		\end{minipage}%
	}%
	\subfigure[]{
	\vspace{-0.5cm}
		\begin{minipage}[t]{0.49\linewidth}
			\centering
			\includegraphics[scale=.28]{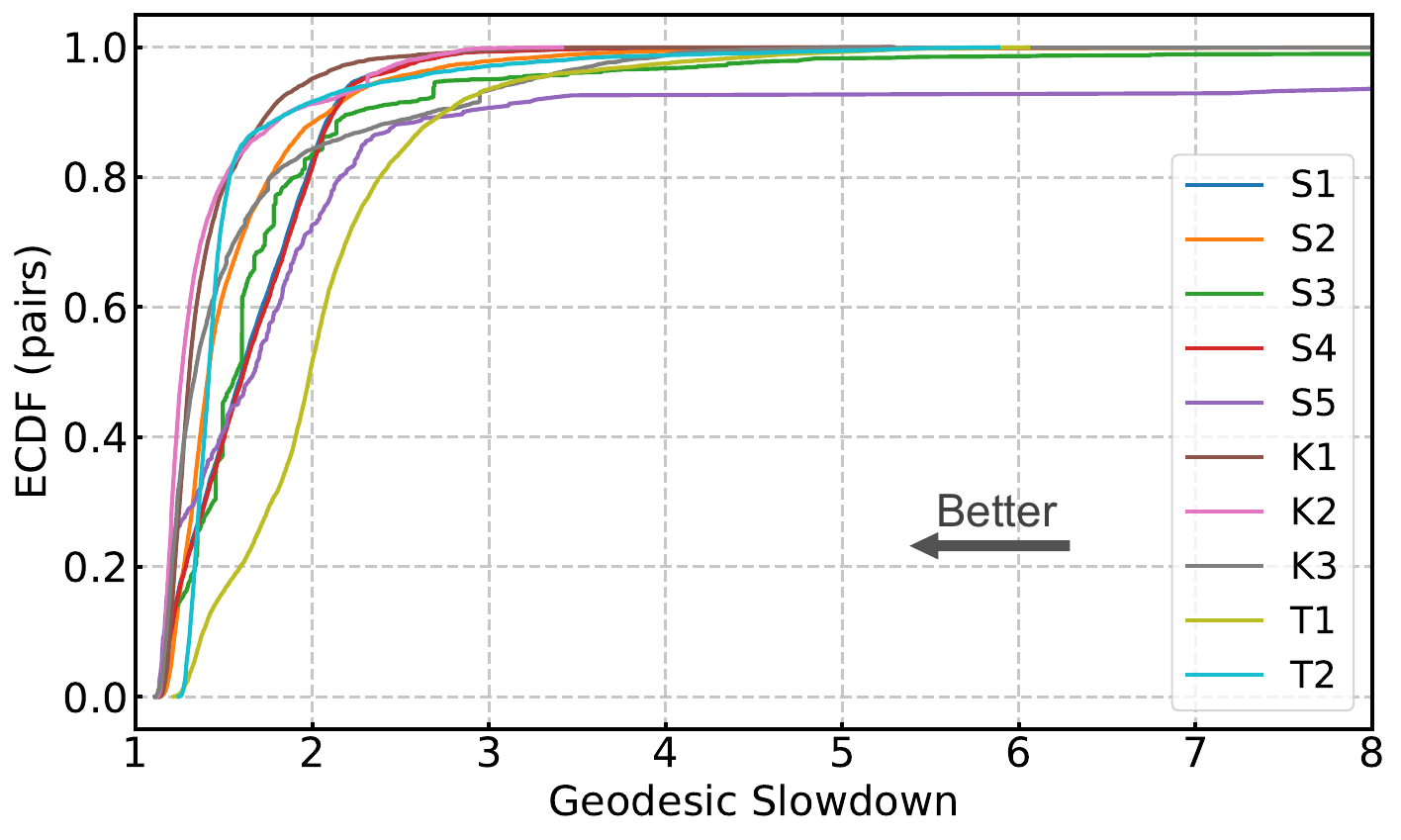}
                \label{all Commercial 2}
		\end{minipage}%
	}%
 
	\subfigure[]{
	\vspace{-0.5cm}
		\begin{minipage}[t]{0.49\linewidth}
			\centering
			\includegraphics[scale=.28]{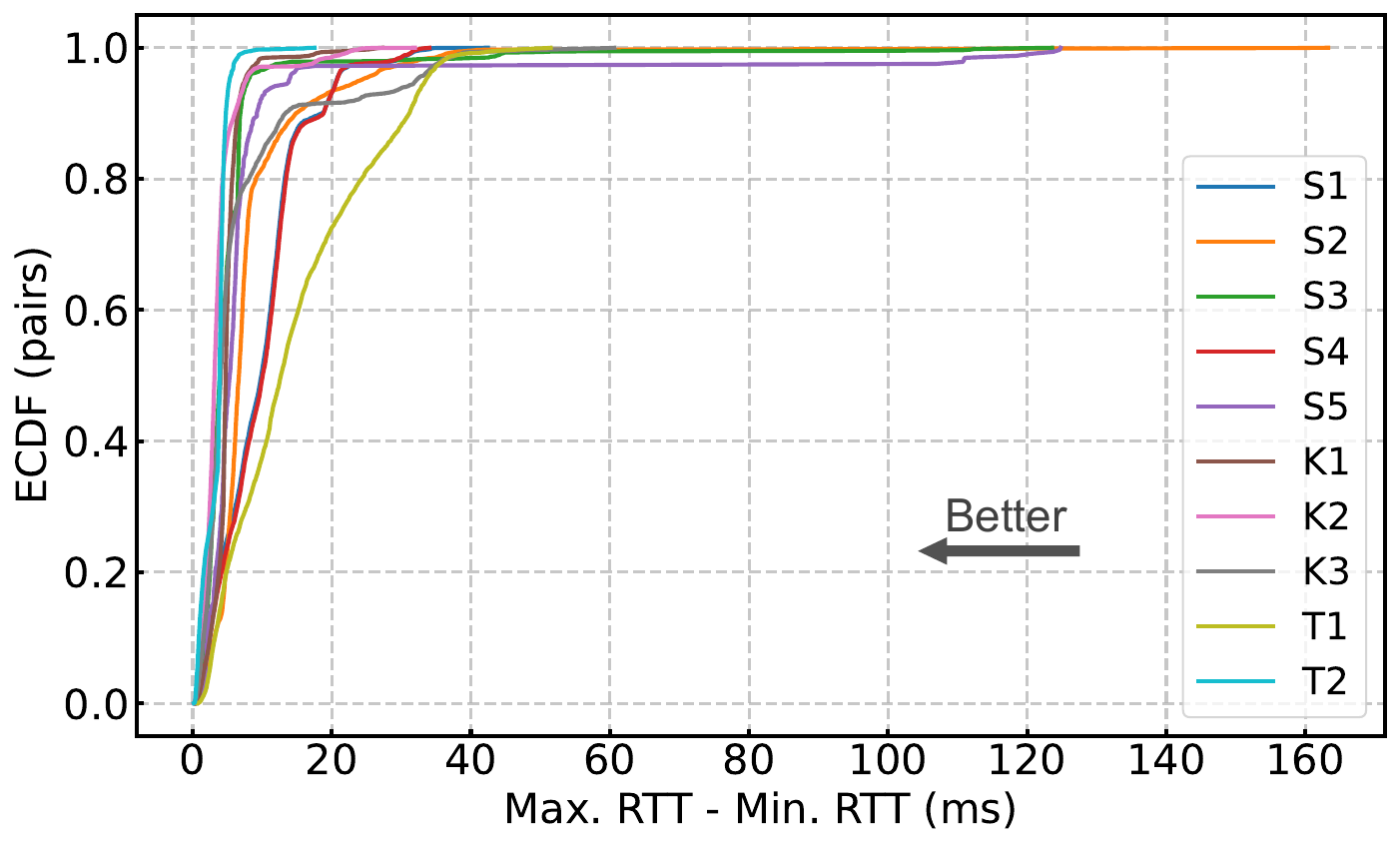}
                \label{all Commercial 3}
		\end{minipage}%
	}%
	\subfigure[]{
	\vspace{-0.5cm}
		\begin{minipage}[t]{0.49\linewidth}
			\centering
			\includegraphics[scale=.28]{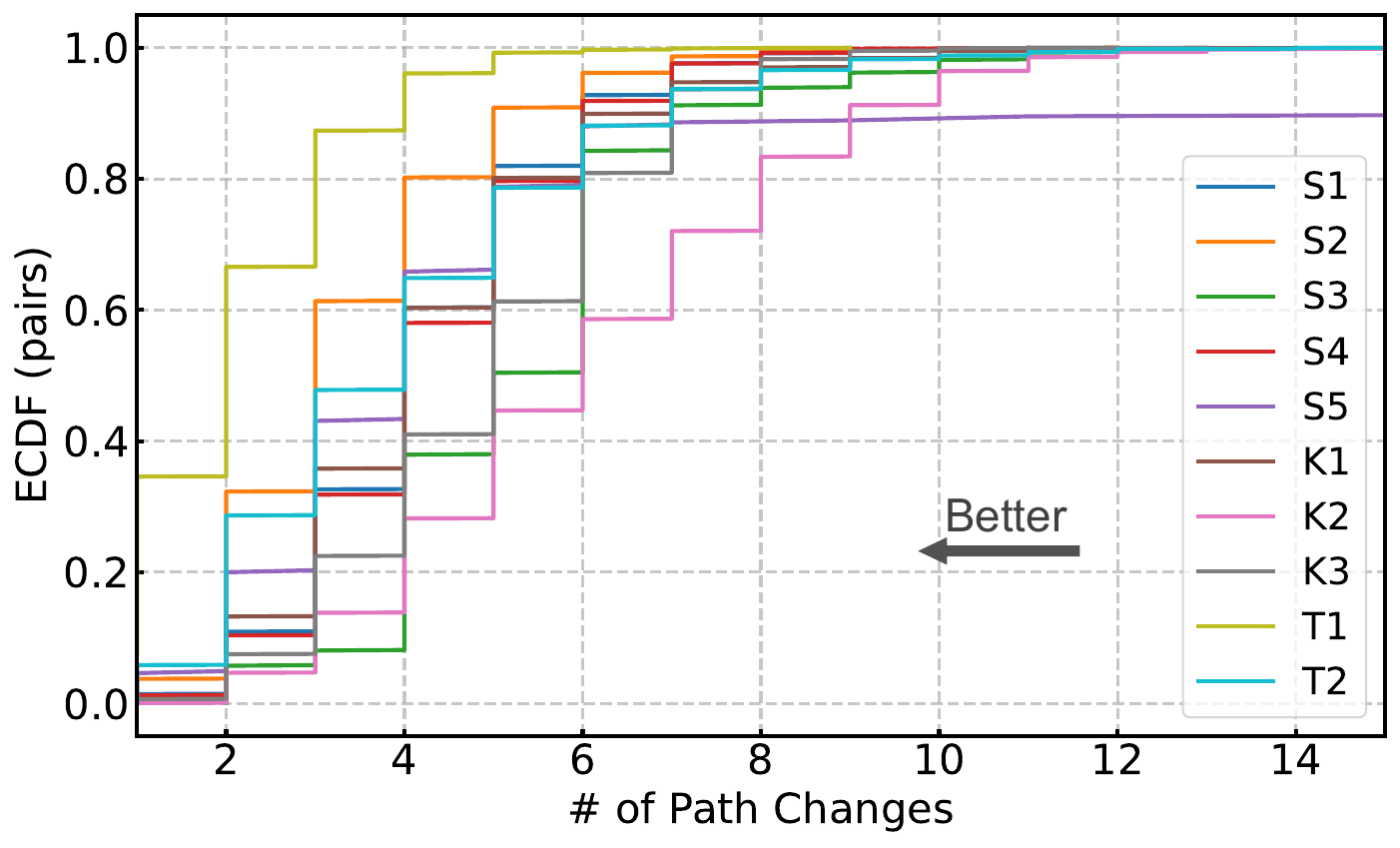}
                \label{all Commercial 4}
		\end{minipage}%
	}
	\caption{The network performance of all shells of Starlink, Kuiper, and Telesat.}
	\label{all Commercial}
\end{figure*}

As mentioned in~\S~\ref{Comparing Different Shells}, we estimate the network performance of different shells of existing commercial LEO mega-constellations. Shown in Fig.~\ref{all Commercial}, we plot all the shells' curves in one figure to compare their performance intuitively. Noted that we cut the long tail in Fig.~\ref{all Commercial 2} and Fig.~\ref{all Commercial 4}. We can see that Kuiper Shell 1, Shell 2, Starlink Shell 2, and Telesat shell 2 all perform well in Max. RTT, Geodesic Slowdown, and RTT fluctuation. For the \# of Path Changes, Telesat shell 2 performs the best because of its fewer satellite number.  

\clearpage

\begin{figure*}[ht]
	\centering
	\subfigure[20 Orbits]{
		\begin{minipage}[t]{0.24\linewidth}
			\centering
			\includegraphics[scale=.25]{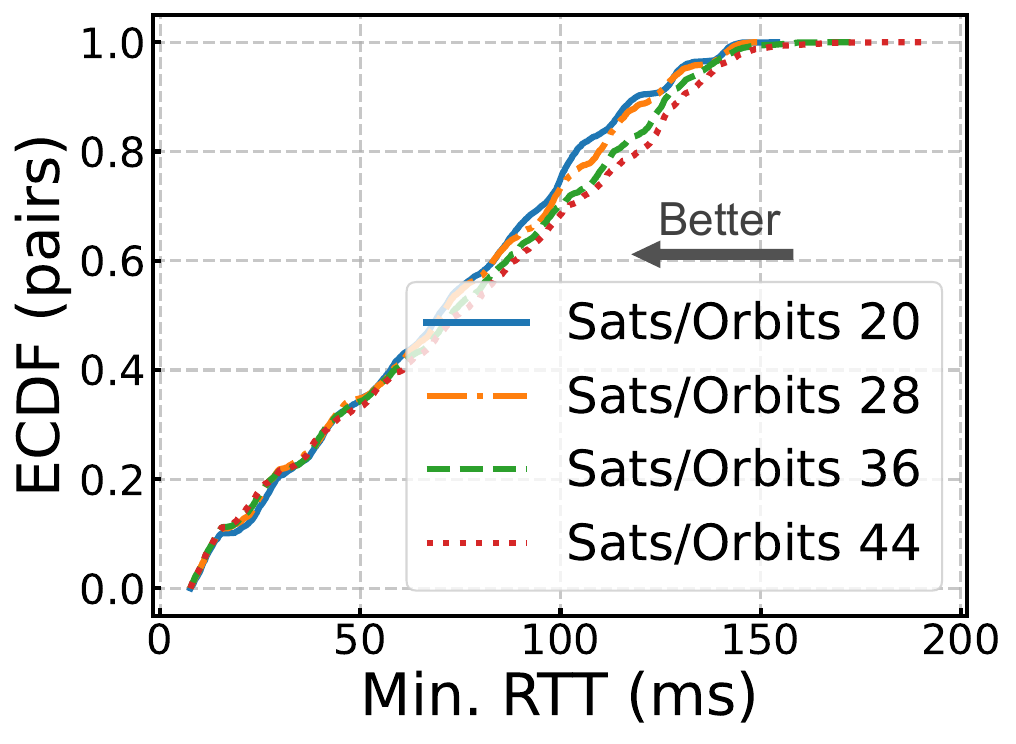}
			\vspace{-0.5cm}
      	\label{Min RTT Sats/Orbits 1}
		\end{minipage}%
	}%
	\subfigure[33 Orbits]{
		\begin{minipage}[t]{0.24\linewidth}
			\centering
			\includegraphics[scale=.25]{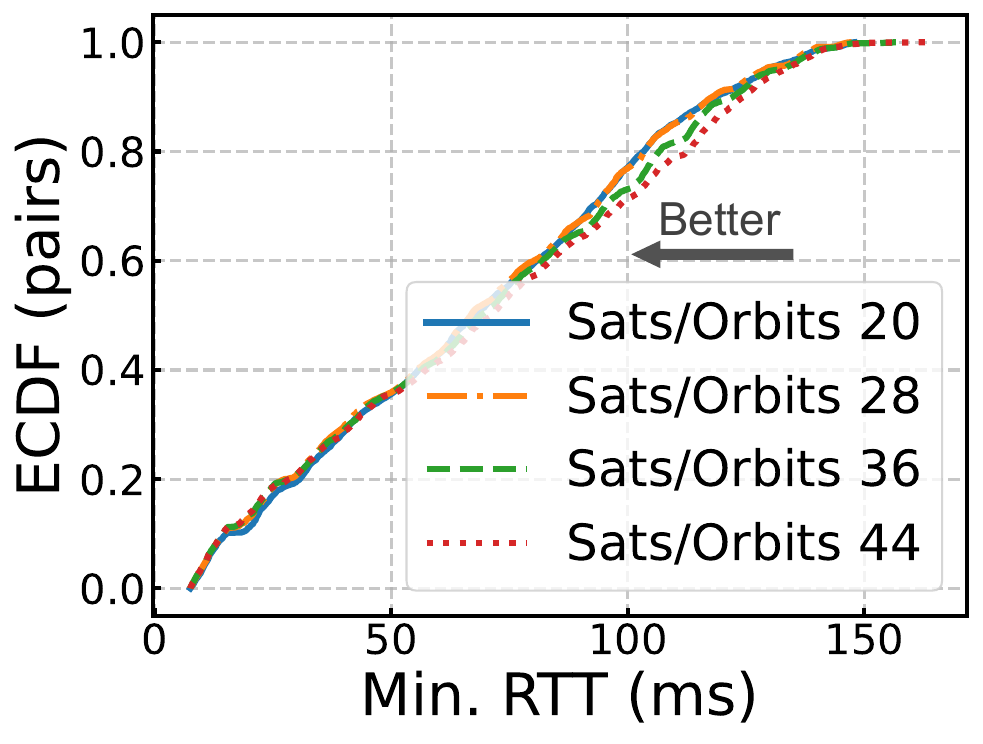}
			\vspace{-0.5cm}
      	\label{Min RTT Sats/Orbits 2}
		\end{minipage}%
	}%
	\subfigure[46 Orbits]{
		\begin{minipage}[t]{0.24\linewidth}
			\centering
			\includegraphics[scale=.25]{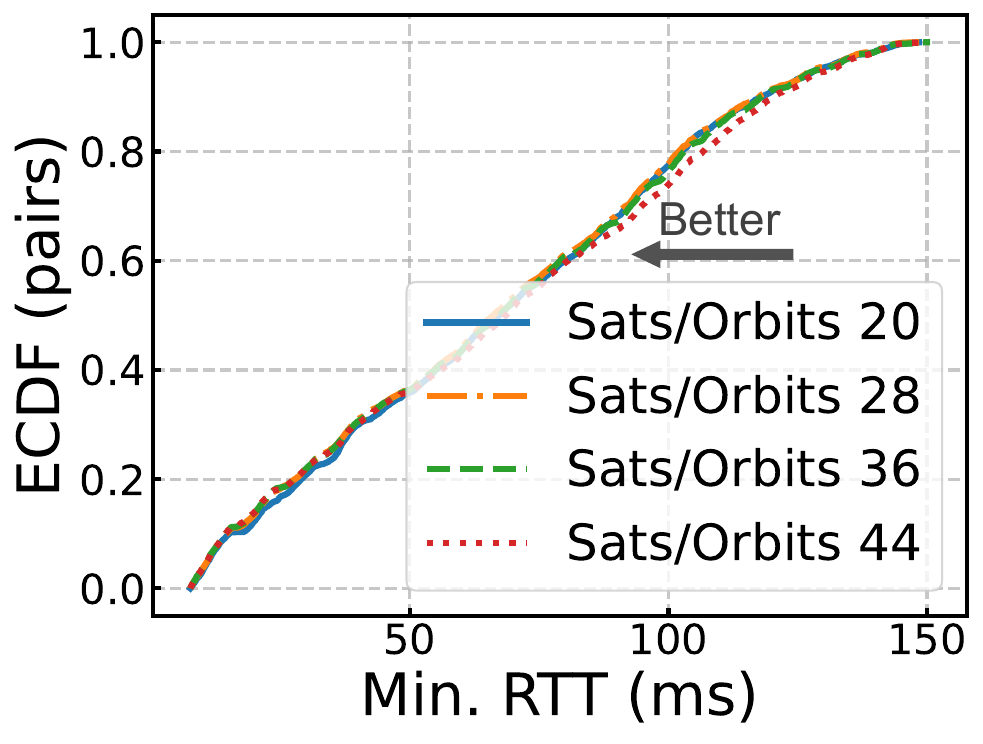}
			\vspace{-0.75cm}
      	\label{Min RTT Sats/Orbits 3}
		\end{minipage}%
	}%
	\subfigure[59 Orbits]{
		\begin{minipage}[t]{0.24\linewidth}
			\centering
			\includegraphics[scale=.25]{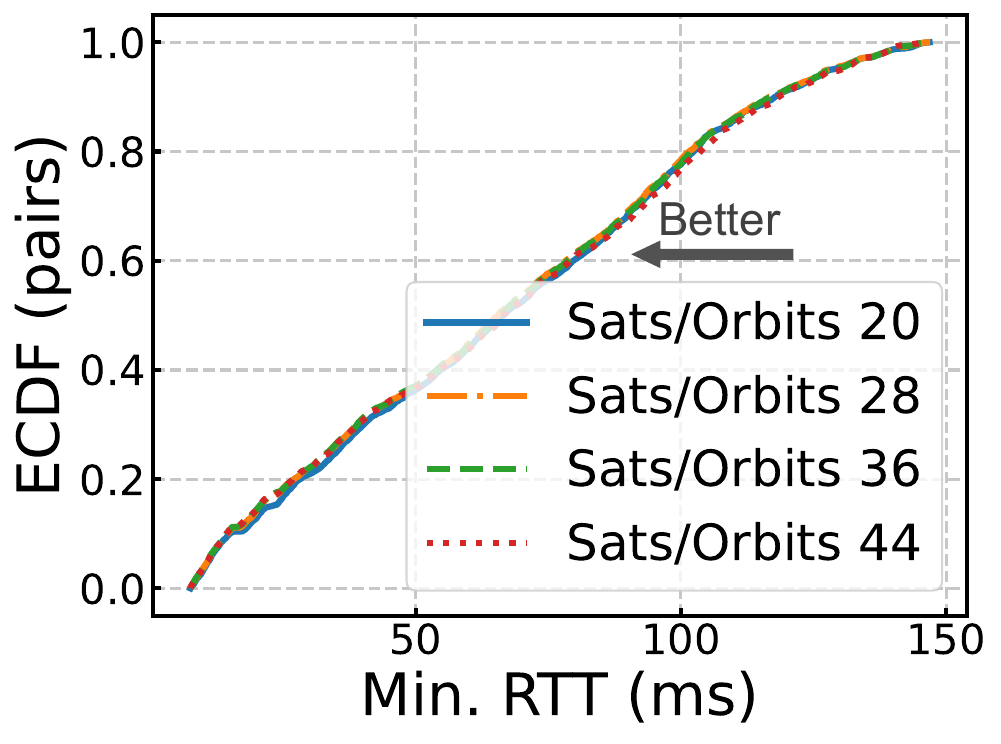}
			\vspace{-0.75cm}
   	\label{Min RTT Sats/Orbits 4}
		\end{minipage}%
	}%
    \vspace{-0.5cm}
	\caption{The Min. RTT (ms) ECDFs of changing Orbits number with different Sats/Orbits.}
     \vspace{-0.5cm}
	\label{Min RTT Sats/Orbits}
\end{figure*}

\begin{figure*}[ht]
	\centering
	\subfigure[20 Orbits]{
	\vspace{-0.75cm}
		\begin{minipage}[t]{0.24\linewidth}
			\centering
			\includegraphics[scale=.25]{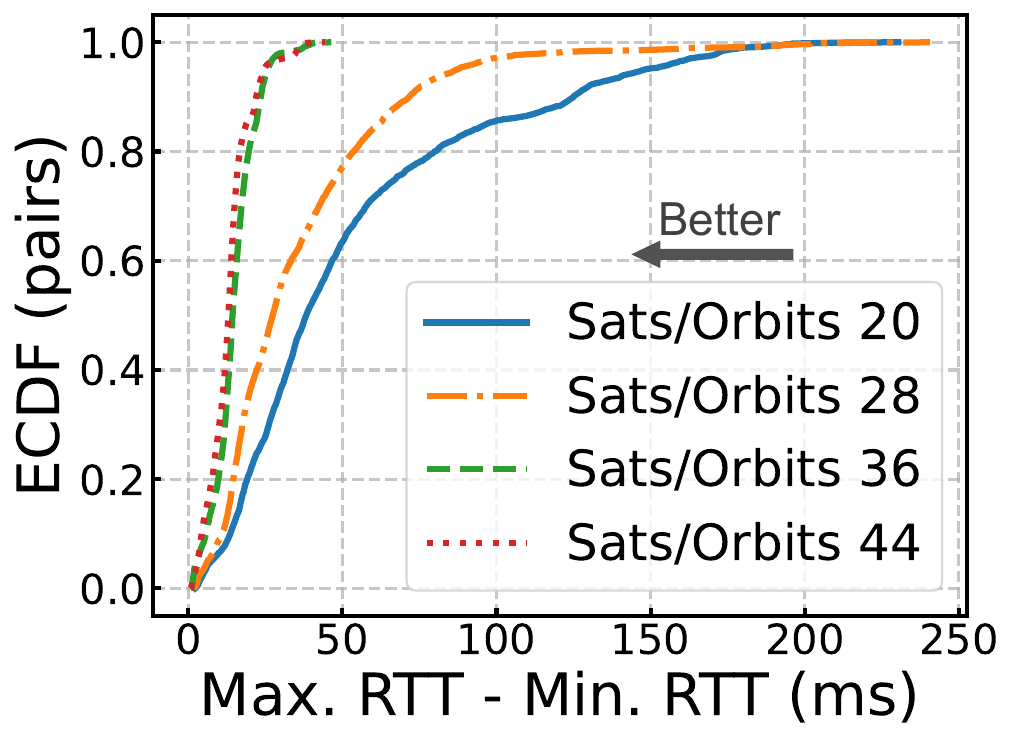}
			\vspace{-0.5cm}
      	\label{minus RTT Sats/Orbits 1}
		\end{minipage}%
	}%
	\subfigure[33 Orbits]{
	\vspace{-0.75cm}
		\begin{minipage}[t]{0.24\linewidth}
			\centering
			\includegraphics[scale=.25]{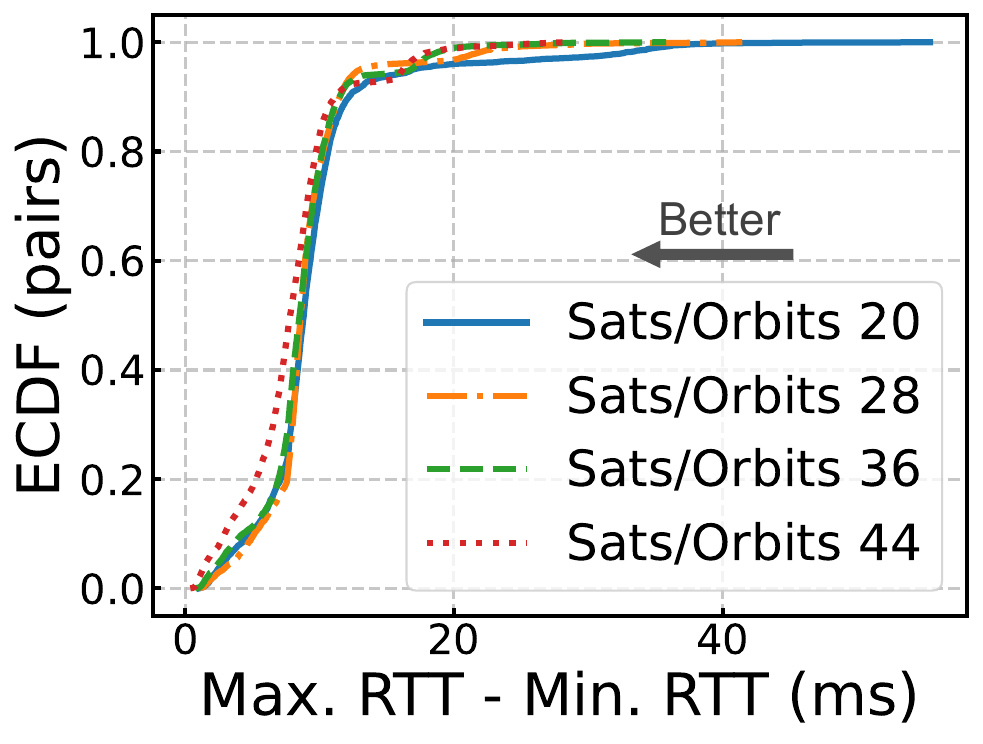}
			\vspace{-0.5cm}
      	\label{minus RTT Sats/Orbits 2}
		\end{minipage}%
	}%
	\subfigure[46 Orbits]{
	\vspace{-0.75cm}
		\begin{minipage}[t]{0.24\linewidth}
			\centering
			\includegraphics[scale=.25]{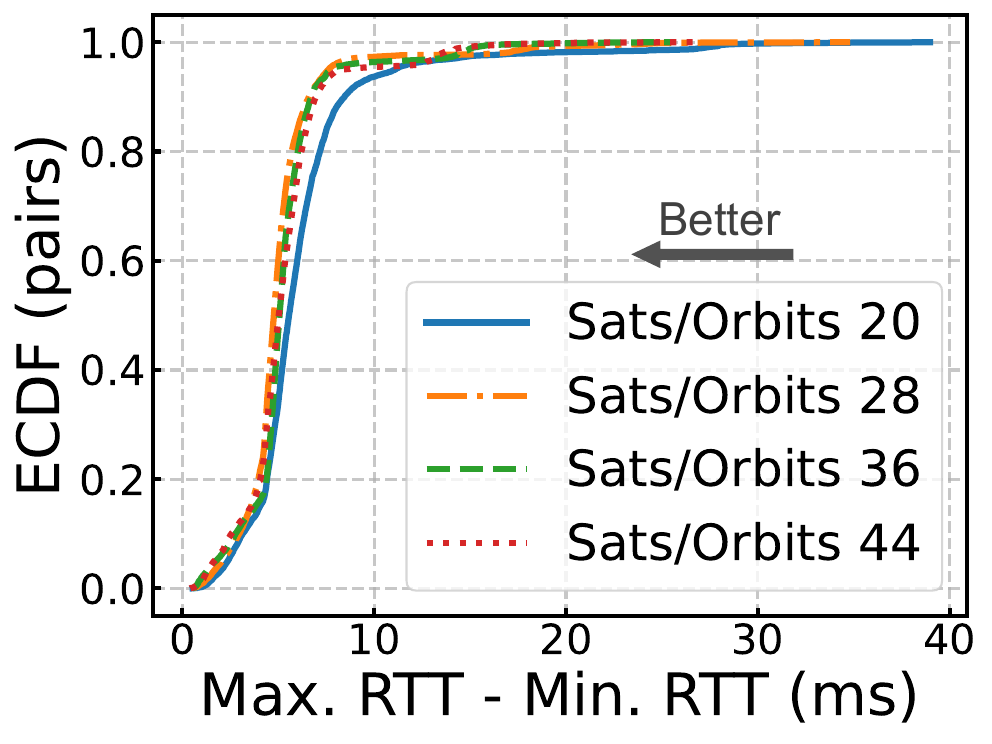}
			\vspace{-0.5cm}
      	\label{minus RTT Sats/Orbits 3}
		\end{minipage}%
	}%
	\subfigure[59 Orbits]{
	\vspace{-0.75cm}
		\begin{minipage}[t]{0.24\linewidth}
			\centering
			\includegraphics[scale=.25]{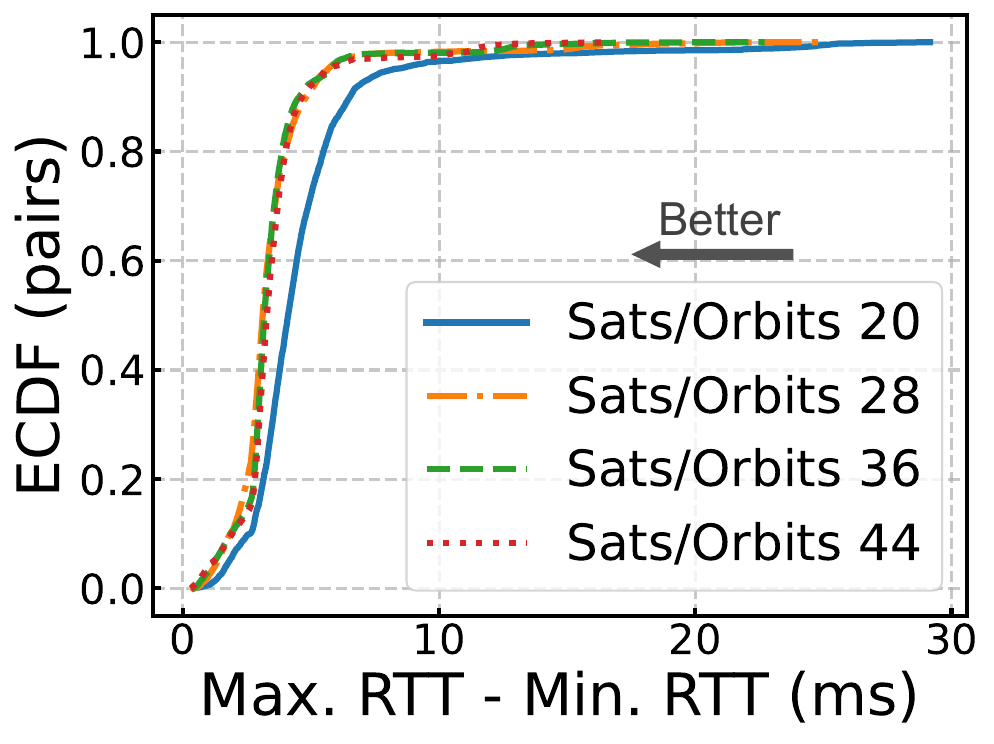}
			\vspace{-0.5cm}
   	\label{minus RTT Sats/Orbits 4}
		\end{minipage}%
	}%
    \vspace{-0.5cm}
	\caption{The Max. RTT - Min. RTT (ms) ECDFs of changing Orbits number with different Sats/Orbits.}
     \vspace{-0.5cm}
	\label{minus RTT Sats/Orbits}
\end{figure*}

\begin{figure*}[ht]
	\centering
	\subfigure[20 Orbits]{
	\vspace{-0.75cm}
		\begin{minipage}[t]{0.24\linewidth}
			\centering
			\includegraphics[scale=.25]{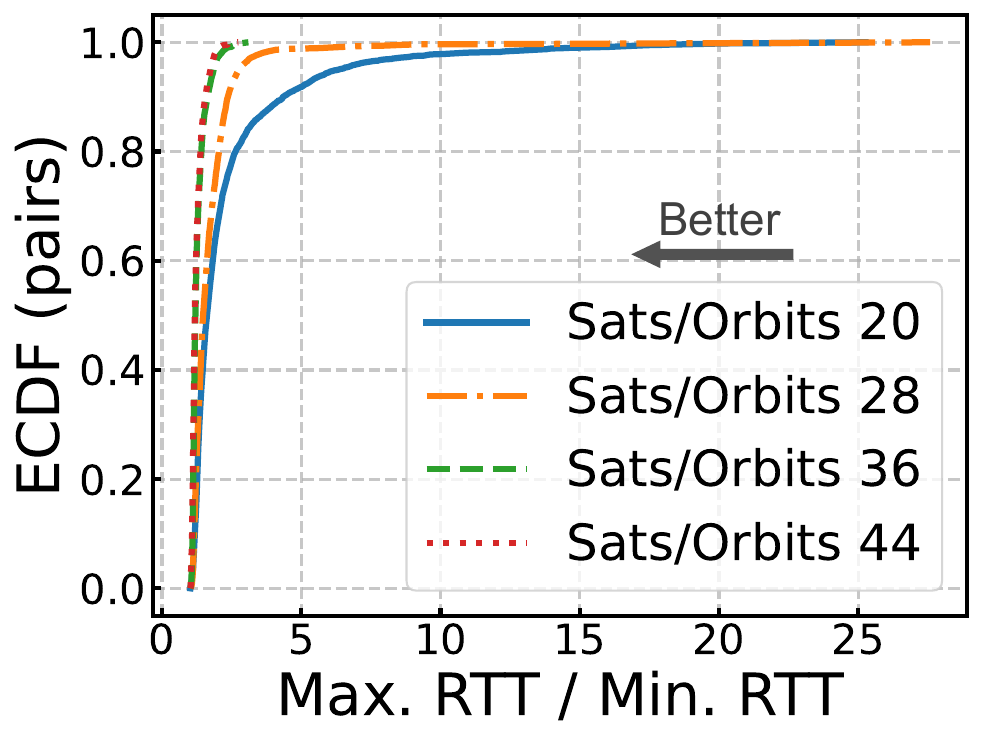}
			\vspace{-0.5cm}
      	\label{To RTT Sats/Orbits 1}
		\end{minipage}%
	}%
	\subfigure[33 Orbits]{
	\vspace{-0.75cm}
		\begin{minipage}[t]{0.24\linewidth}
			\centering
			\includegraphics[scale=.25]{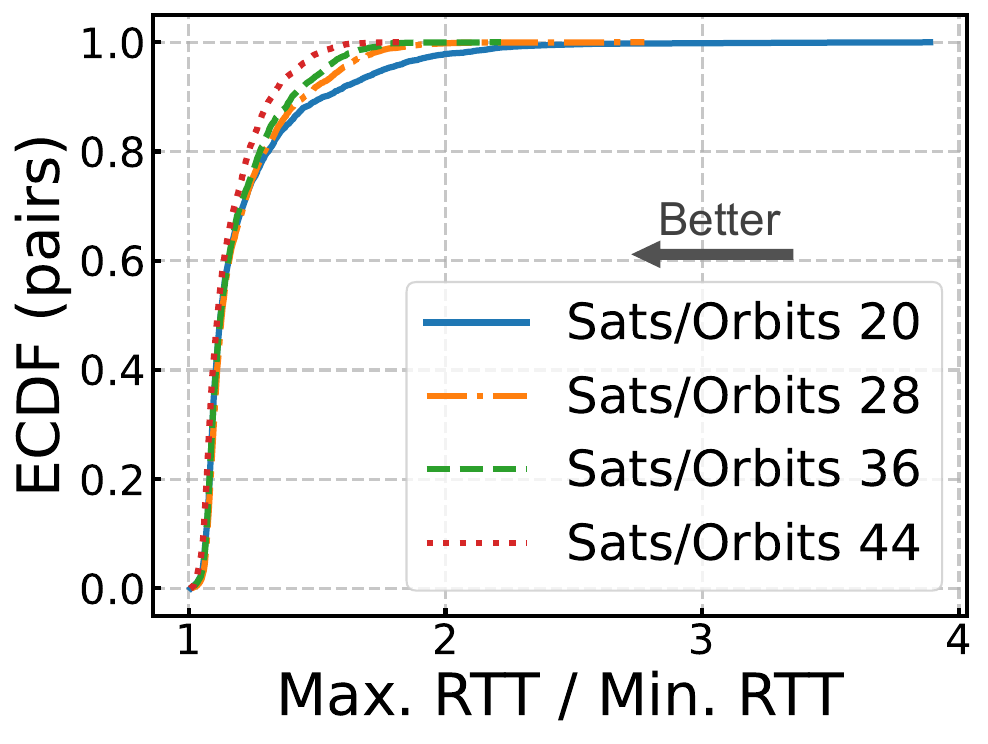}
			\vspace{-0.5cm}
      	\label{To RTT Sats/Orbits 2}
		\end{minipage}%
	}%
	\subfigure[46 Orbits]{
	\vspace{-0.75cm}
		\begin{minipage}[t]{0.24\linewidth}
			\centering
			\includegraphics[scale=.25]{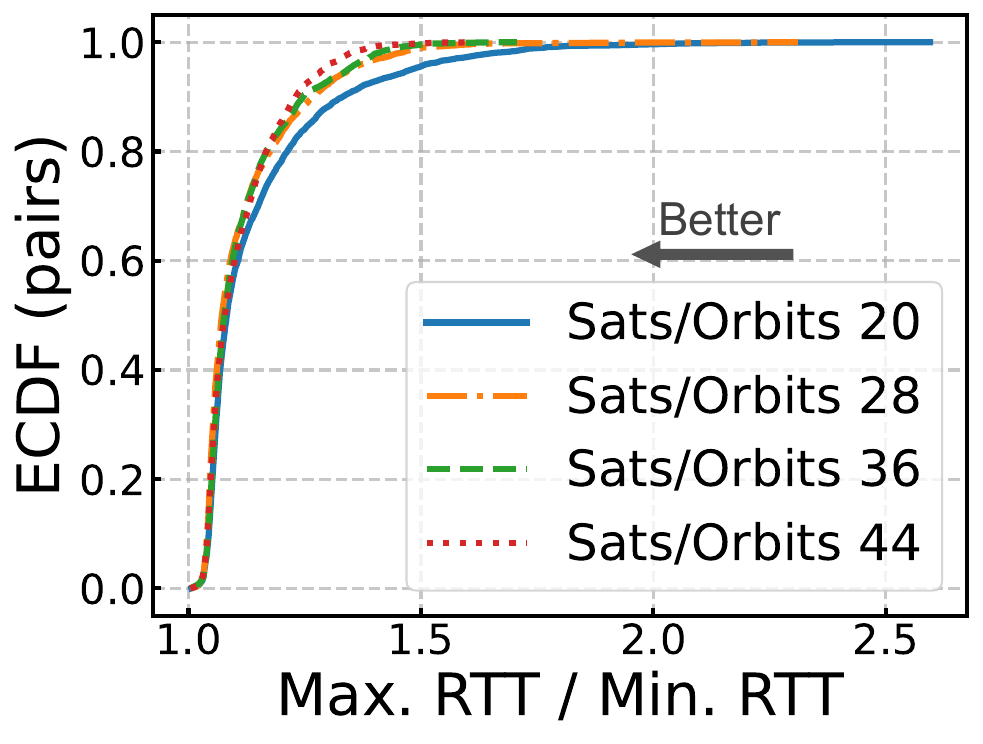}
			\vspace{-0.5cm}
      	\label{To RTT Sats/Orbits 3}
		\end{minipage}%
	}%
	\subfigure[59 Orbits]{
	\vspace{-0.75cm}
		\begin{minipage}[t]{0.24\linewidth}
			\centering
			\includegraphics[scale=.25]{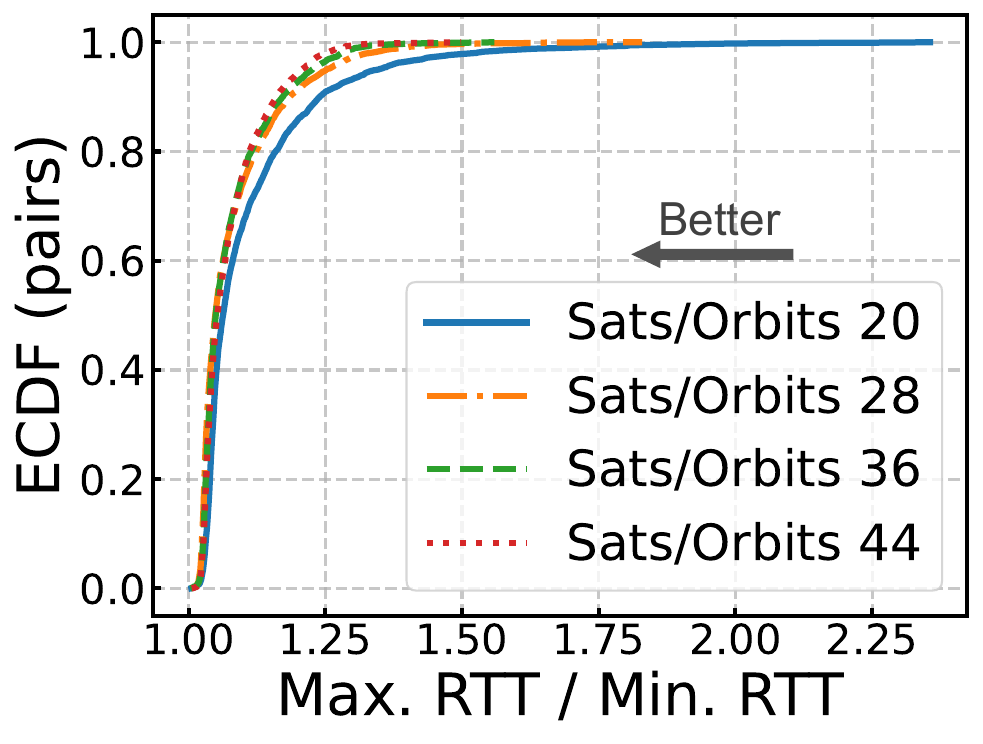}
			\vspace{-0.5cm}
   	\label{To RTT Sats/Orbits 4}
		\end{minipage}%
	}%
    \vspace{-0.5cm}
	\caption{The Max. RTT to Min. RTT ECDFs of changing Orbits number with different Sats/Orbits.}
     \vspace{-0.5cm}
	\label{To RTT Sats/Orbits}
\end{figure*}

\begin{figure*}[ht]
	\centering
	\subfigure[20 Orbits]{
	\vspace{-0.75cm}
		\begin{minipage}[t]{0.24\linewidth}
			\centering
			\includegraphics[scale=.25]{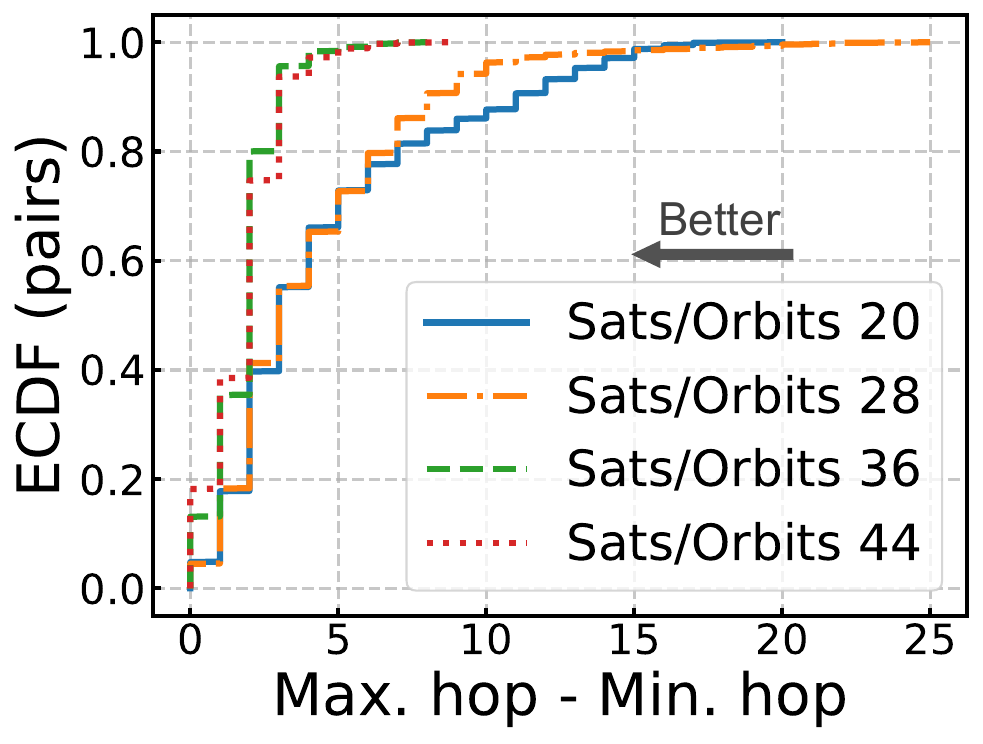}
			\vspace{-0.5cm}
      	\label{minus hop Sats/Orbits 1}
		\end{minipage}%
	}%
	\subfigure[33 Orbits]{
	\vspace{-0.75cm}
		\begin{minipage}[t]{0.24\linewidth}
			\centering
			\includegraphics[scale=.25]{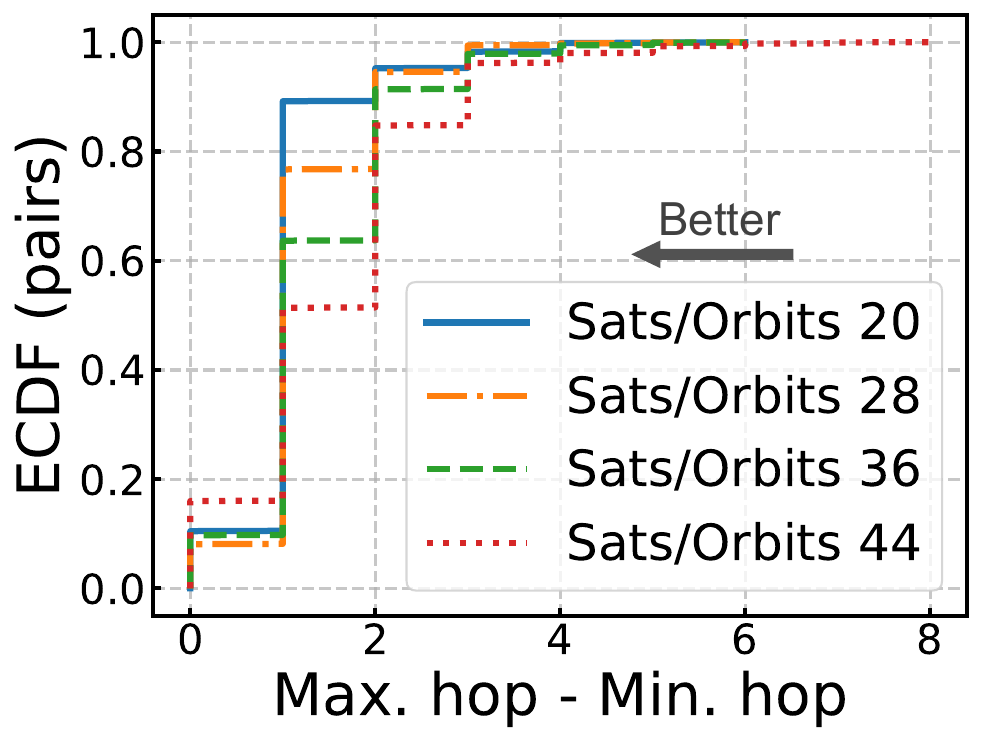}
			\vspace{-0.5cm}
      	\label{minus hop Sats/Orbits 2}
		\end{minipage}%
	}%
	\subfigure[46 Orbits]{
	\vspace{-0.75cm}
		\begin{minipage}[t]{0.24\linewidth}
			\centering
			\includegraphics[scale=.25]{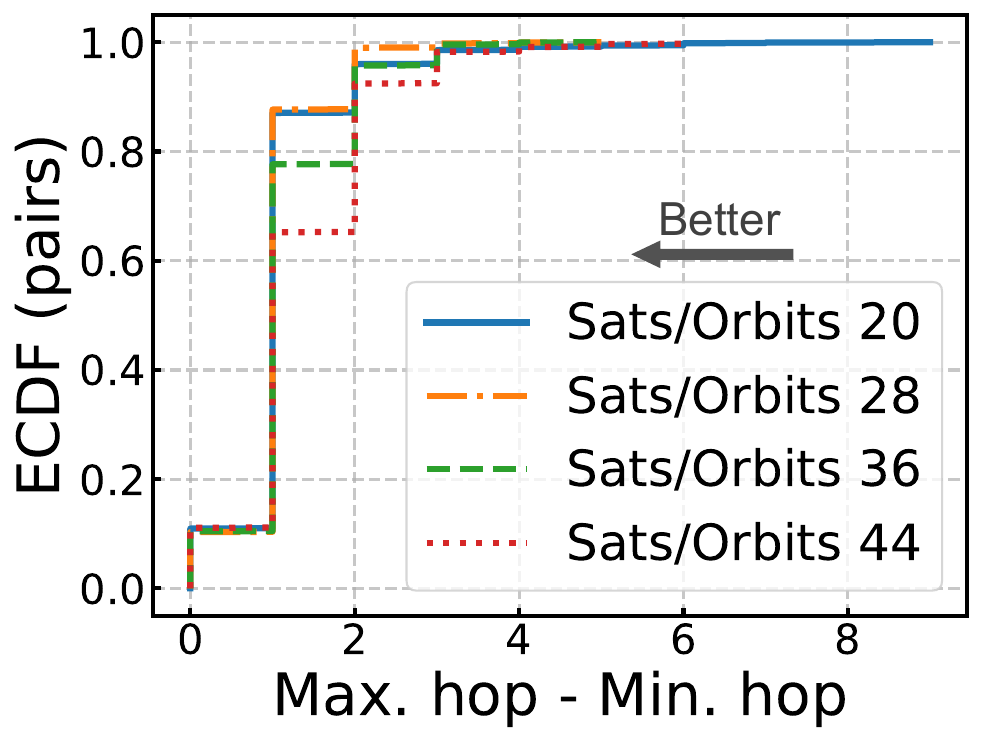}
			\vspace{-0.5cm}
      	\label{minus hop Sats/Orbits 3}
		\end{minipage}%
	}%
	\subfigure[59 Orbits]{
		\begin{minipage}[t]{0.24\linewidth}
			\centering
			\includegraphics[scale=.25]{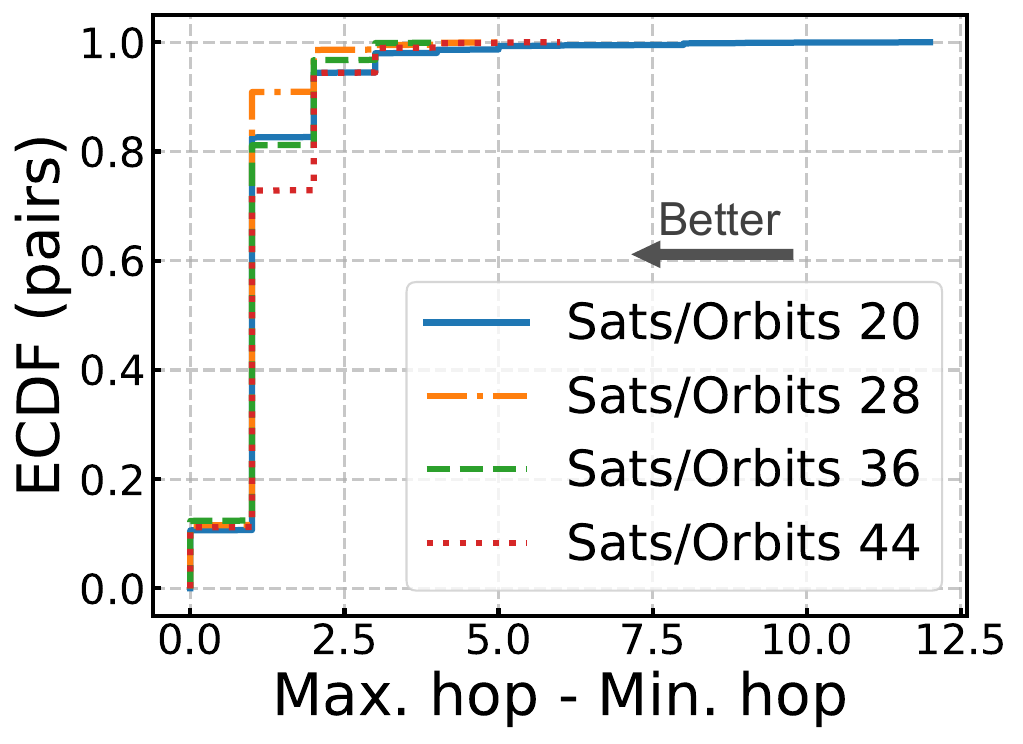}
			\vspace{-0.5cm}
   	\label{minus hop Sats/Orbits 4}
		\end{minipage}%
	}%
    \vspace{-0.5cm}
	\caption{The Max. hop - Min. hop ECDFs of changing Orbits number with different Sats/Orbits.}
     \vspace{-0.5cm}
	\label{minus hop Sats/Orbits}
\end{figure*}

\begin{figure*}[ht]
	\centering
	\subfigure[20 Orbits]{
	\vspace{-0.75cm}
		\begin{minipage}[t]{0.24\linewidth}
			\centering
			\includegraphics[scale=.25]{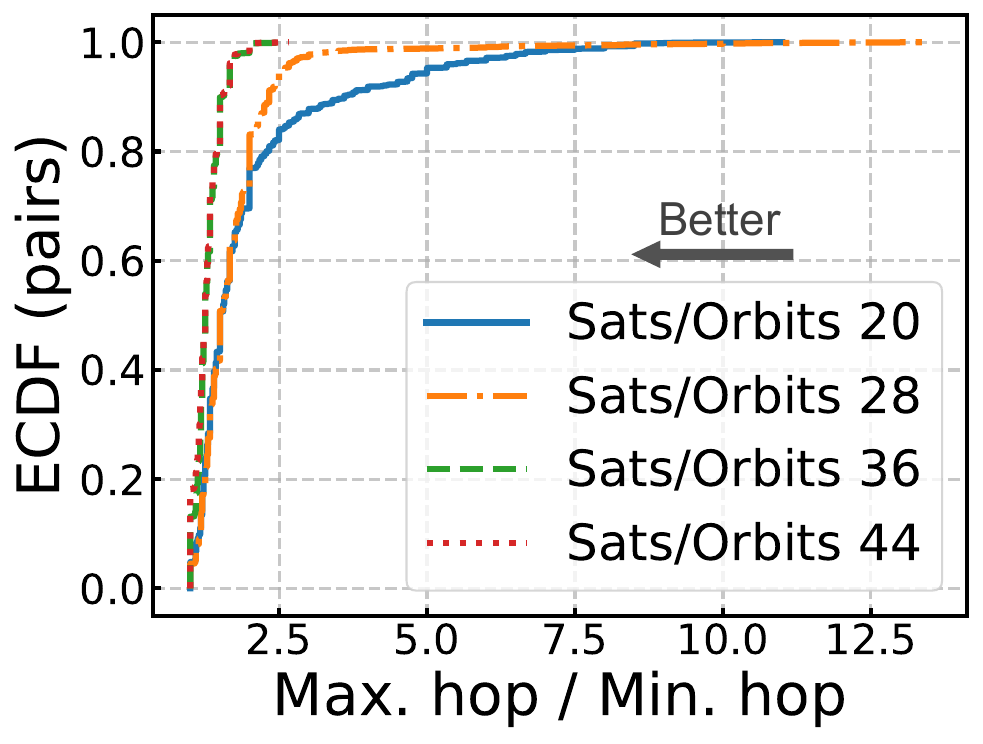}
			\vspace{-0.5cm}
      	\label{to hop Sats/Orbits 1}
		\end{minipage}%
	}%
	\subfigure[33 Orbits]{
	\vspace{-0.75cm}
		\begin{minipage}[t]{0.24\linewidth}
			\centering
			\includegraphics[scale=.25]{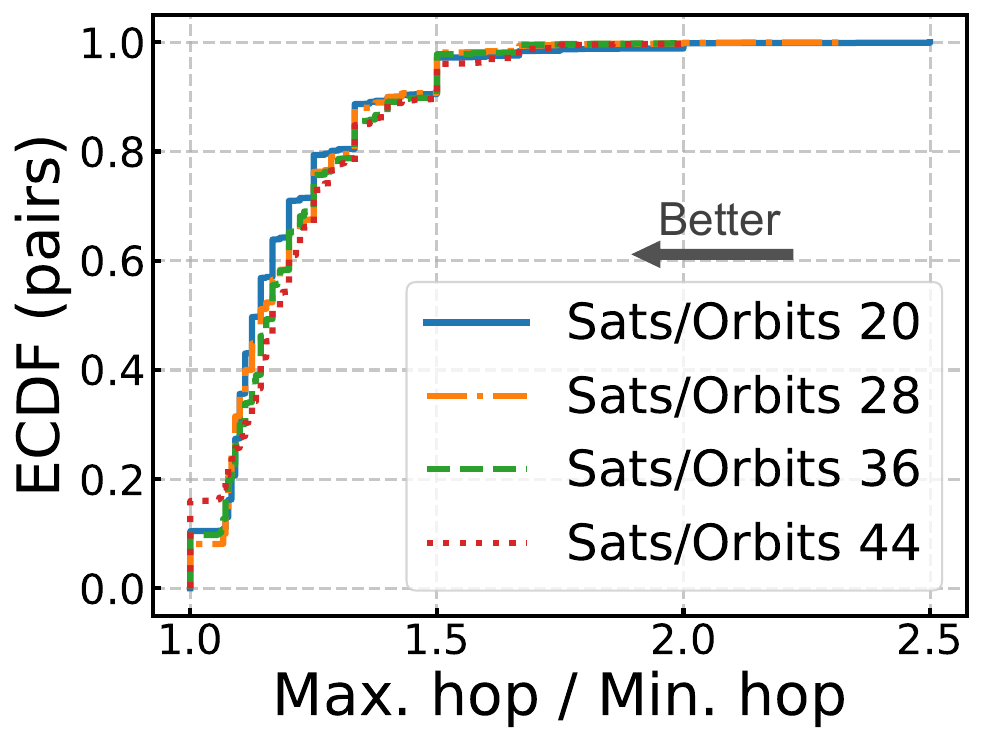}
			\vspace{-0.5cm}
      	\label{to hop Sats/Orbits 2}
		\end{minipage}%
	}%
	\subfigure[46 Orbits]{
	\vspace{-0.75cm}
		\begin{minipage}[t]{0.24\linewidth}
			\centering
			\includegraphics[scale=.25]{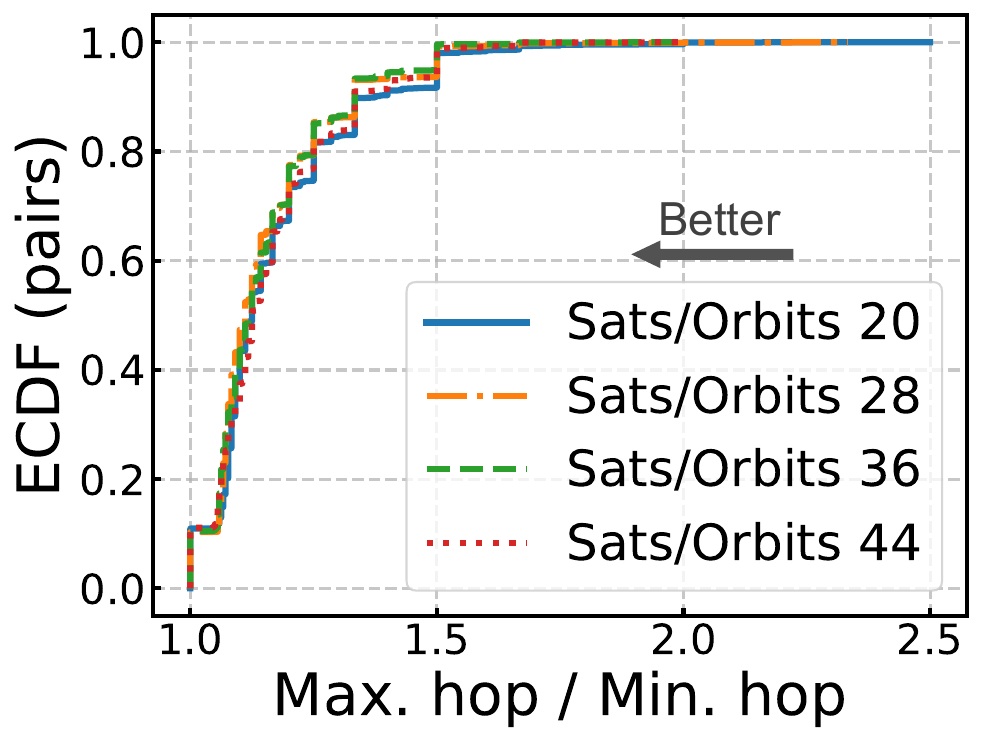}
			\vspace{-0.75cm}
      	\label{to hop Sats/Orbits 3}
		\end{minipage}%
	}%
	\subfigure[59 Orbits]{
	\vspace{-0.75cm}
		\begin{minipage}[t]{0.24\linewidth}
			\centering
			\includegraphics[scale=.25]{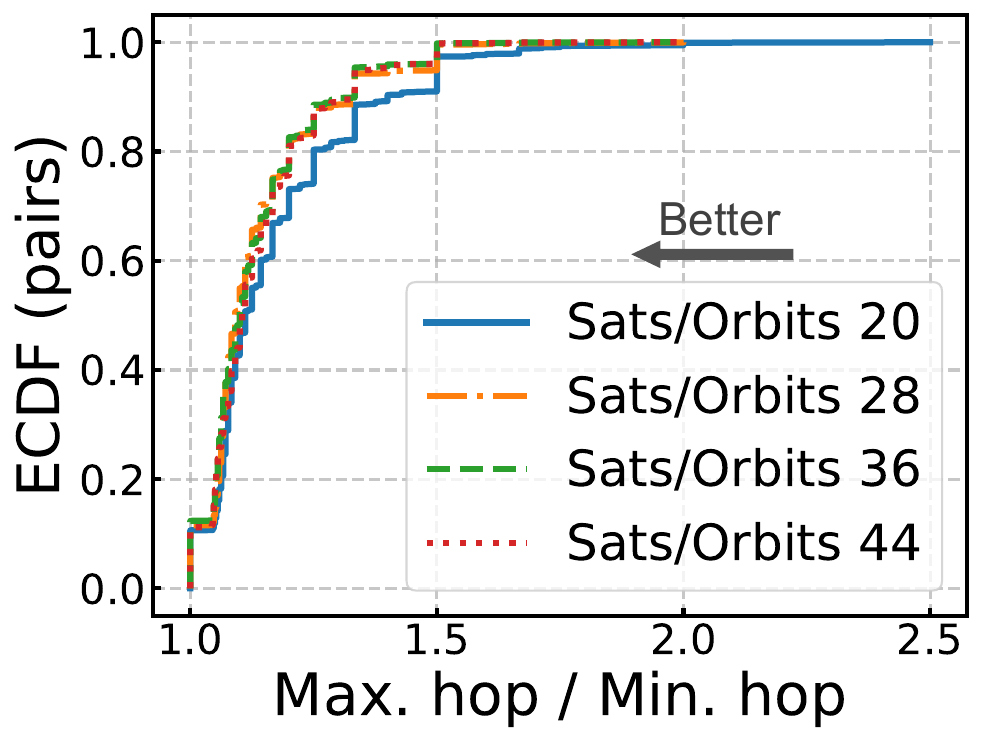}
			\vspace{-0.75cm}
   	\label{to hop Sats/Orbits 4}
		\end{minipage}%
	}%
    \vspace{-0.5cm}
	\caption{The Max. hop to Min. hop ECDFs of changing Orbits number with different Sats/Orbits.}
     \vspace{-0.5cm}
	\label{to hop Sats/Orbits}
\end{figure*}

\section{SYNTHETIC LEO
MEGA-CONSTELLATIONS}
In this section, we show more experiment result for the synthetic LEO mega-constellations. 

\subsection{Satellites Per Orbit}
\label{appendix: Satellites Per Orbit}

At an inclination angle of 53°, we measured the synthetic LEO mega-constellations' network performance with the Satellite Number per Orbit = 20, 28, 36, 44 under the same Orbit Number = 20, 33, 46, 59, respectively. The extension measurement results of Min. RTT, Max. RTT - Min. RTT, Max. RTT to Min. RTT, Max. hop - Min. hop, Max. hop to Min. hop are show in Fig.~\ref{Min RTT Sats/Orbits}, Fig.~\ref{minus RTT Sats/Orbits}, Fig.~\ref{To RTT Sats/Orbits}, Fig.~\ref{minus hop Sats/Orbits}, Fig.~\ref{to hop Sats/Orbits}, respectively. Because the long-tail effect of some parameter configurations is too significant, we are forced to select only part of the ECDF curves and ignore the excessively long tails to facilitate the presentation. From Fig.~\ref{Min RTT Sats/Orbits}, we can see that the difference in changing the Sats/Orbits number is not that much, and it also follows the rule that when Sats/Orbits number increases, the Min. RTT is fewer than before. For Fig.~\ref{minus RTT Sats/Orbits} and Fig.~\ref{To RTT Sats/Orbits}, we can find that with the Sats/Orbits increases, the RTT fluctuation is less. Also, when the Orbits number achieves 33+, the decrease of RTT fluctuation becomes saturation, although Sats/Orbits increases. The same rules apply to hop fluctuation, shown in Fig.~\ref{minus hop Sats/Orbits} and Fig.~\ref{to hop Sats/Orbits}. Moreover, we can see that at least 20 Sats/Orbits is far from the ideal satellite network performance, and 28+ Sats/Orbits are needed to achieve relatively low delay, low network churn, and stable network connection.

\clearpage

\begin{figure*}[htp]
	\centering
	\subfigure[33 Orbits, 28 Sats/Orbits]{
	\vspace{-0.75cm}
		\begin{minipage}[t]{0.24\linewidth}
			\centering
			\includegraphics[scale=.25]{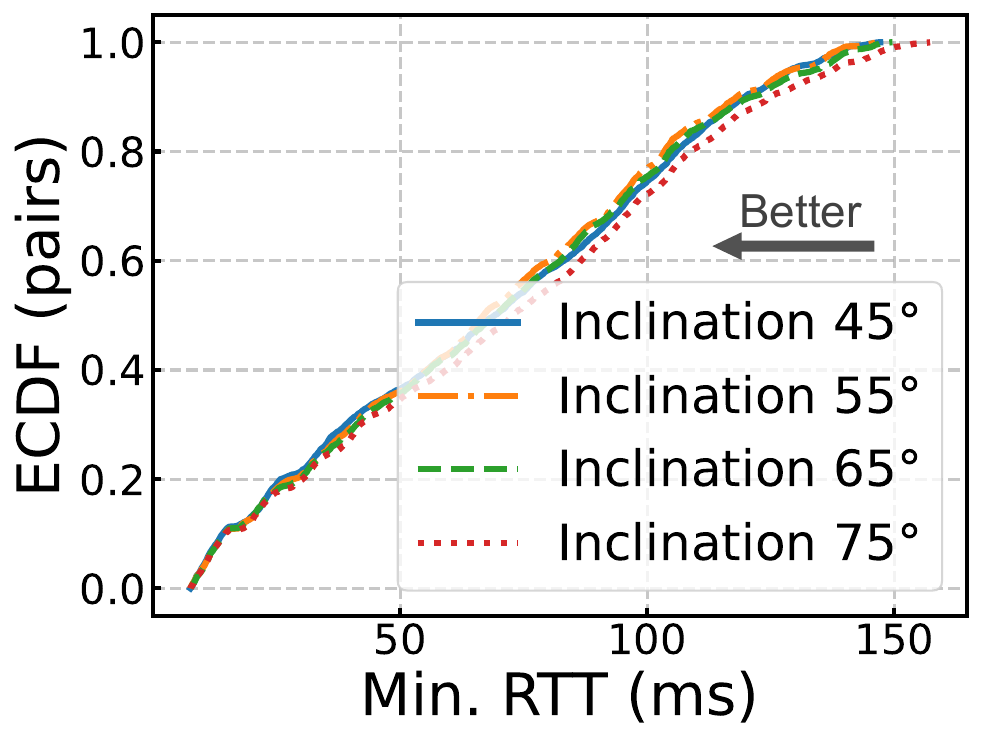}
			\vspace{-0.5cm}
   \label{A Inclination 1}
		\end{minipage}%
	}%
	\subfigure[33 Orbits, 28 Sats/Orbits]{
	\vspace{-0.75cm}
		\begin{minipage}[t]{0.24\linewidth}
			\centering
			\includegraphics[scale=.25]{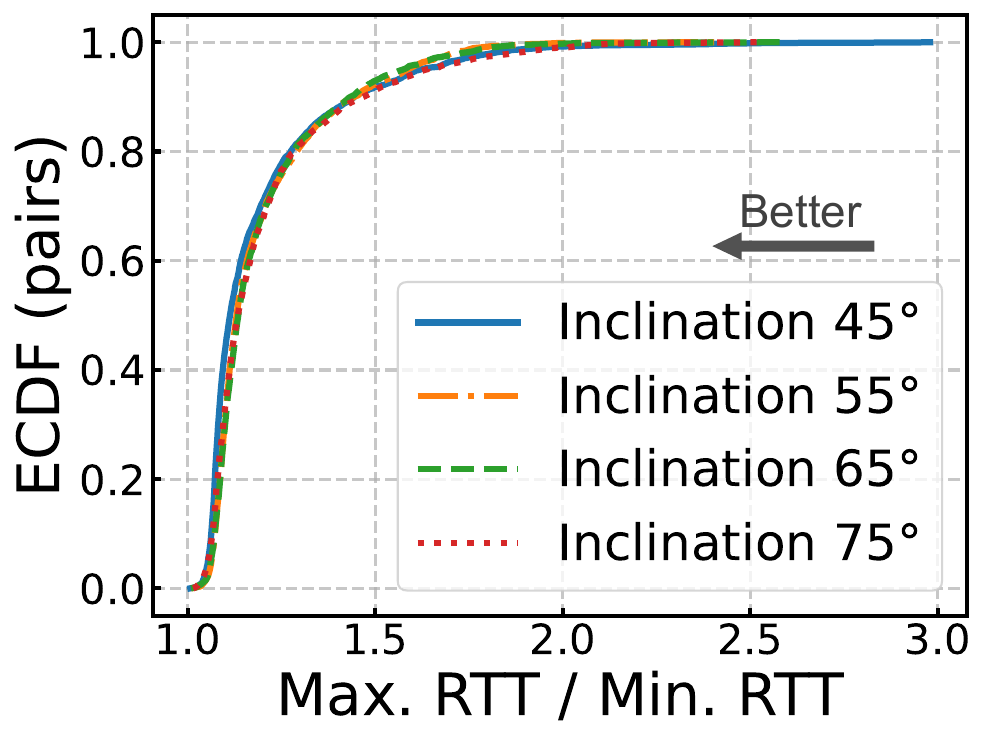}
			\vspace{-0.5cm}
   \label{A Inclination 2}
		\end{minipage}%
	}%
	\subfigure[33 Orbits, 28 Sats/Orbits]{
	\vspace{-0.75cm}
		\begin{minipage}[t]{0.24\linewidth}
			\centering
			\includegraphics[scale=.25]{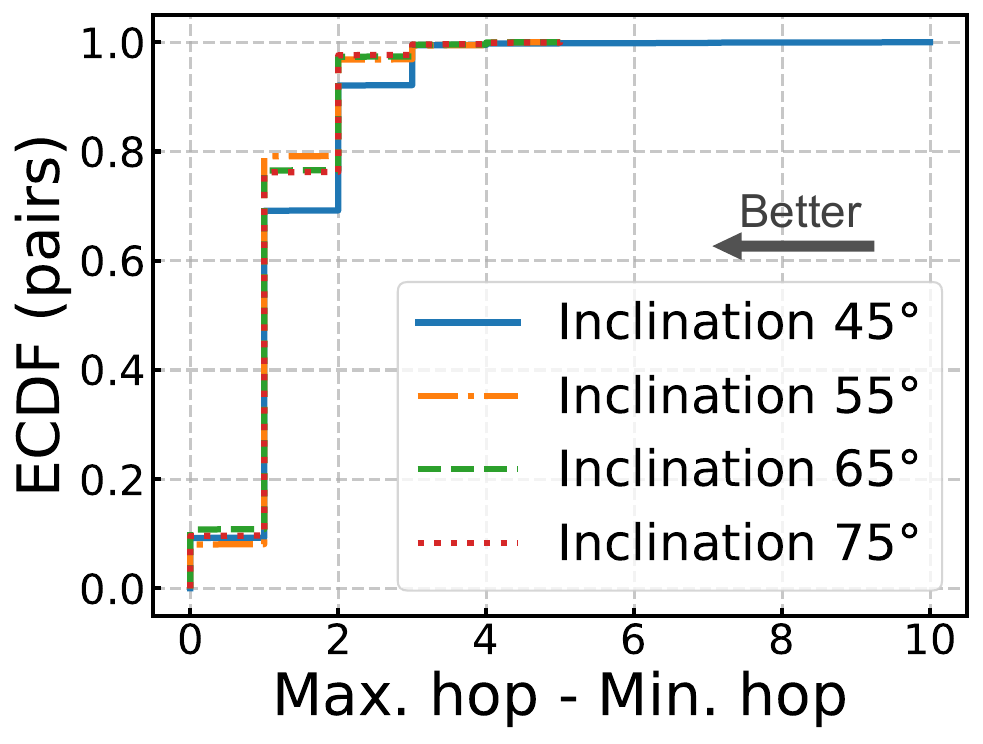}
			\vspace{-0.5cm}
   \label{A Inclination 3}
		\end{minipage}%
	}%
	\subfigure[33 Orbits, 28 Sats/Orbits]{
	\vspace{-0.75cm}
		\begin{minipage}[t]{0.24\linewidth}
			\centering
			\includegraphics[scale=.25]{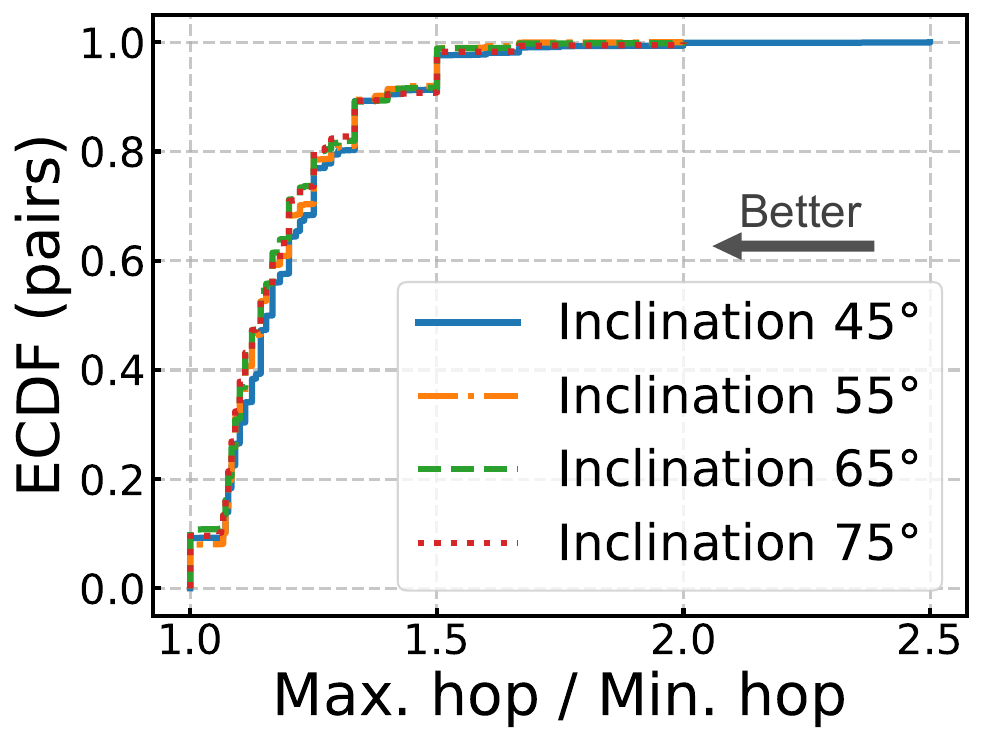}
			\vspace{-0.5cm}
   \label{A Inclination 4}
		\end{minipage}%
	}%
\vspace{-0.5cm}
	\caption{The performance of changing LEO mega-constellations' Inclination.}
	\vspace{-0.75cm}
	\label{A Inclination}
\end{figure*}
\subsection{Inclination}
\label{appendix: Inclination}

The results of Min. RTT, Max. RTT to Min. RTT, Max. hop - Min. hop, Max. hop to Min. hop are shown in Fig.~\ref{A Inclination}. From the result, we can see that the change in Inclination is not apparent. For the Min. RTT, the Inclination 75° one has the largest RTT, while the Inclination 55° one has the best performance of the median value of RTT. However, the difference between the best and the worst is still negligible. This is especially true for the maximum to minimum ratio of RTT and hop. The only point that can be found is that 30\% Endpoints pairs of constellation Inclination 45° have more than two hop fluctuations, the group with the worst performance among the four parameters. From the results of these images, it is still difficult to see the impact of Inclination change on LEO mega-constellation's network performance, so further experiments are necessary.

\begin{figure*}[htp]
	\centering
	\subfigure[20 Sats/Orbits]{
	\vspace{-0.75cm}
		\begin{minipage}[t]{0.24\linewidth}
			\centering
			\includegraphics[scale=.25]{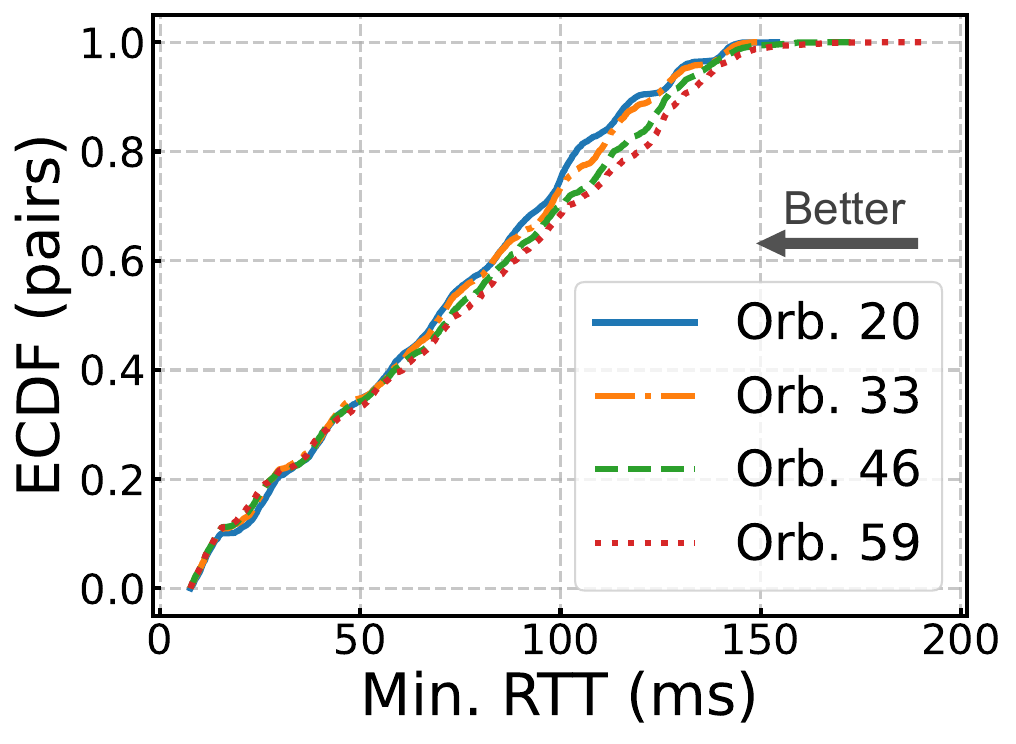}
			\vspace{-0.5cm}
      	\label{Min RTT Orbits 1}
		\end{minipage}%
	}%
	\subfigure[28 Sats/Orbits]{
	\vspace{-0.75cm}
		\begin{minipage}[t]{0.24\linewidth}
			\centering
			\includegraphics[scale=.25]{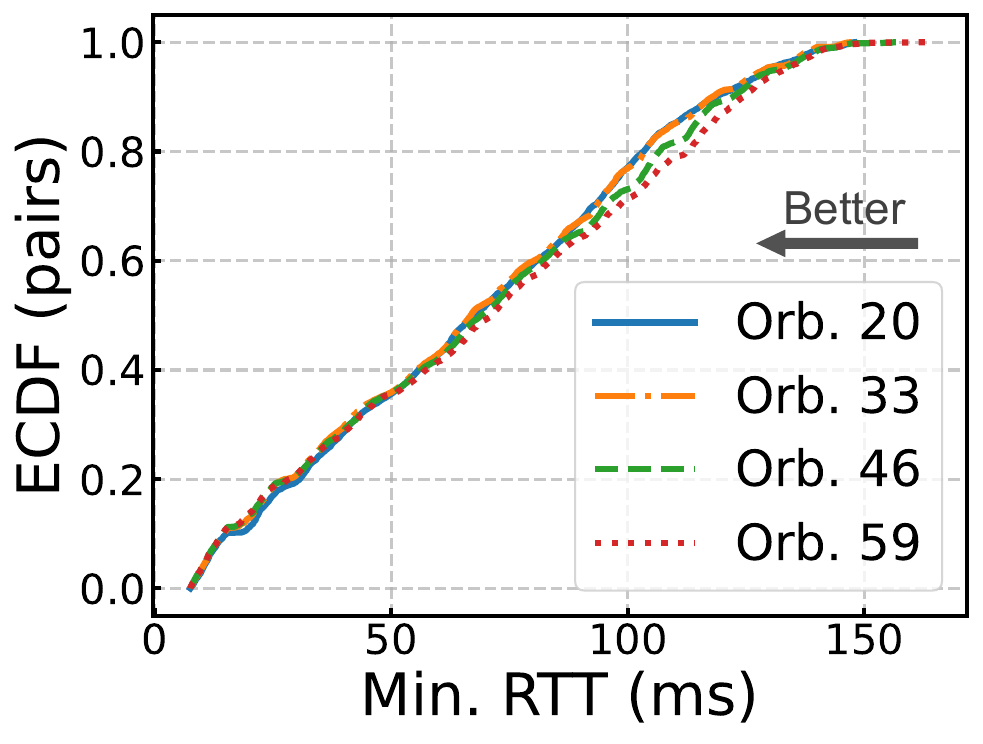}
			\vspace{-0.5cm}
      	\label{Min RTT Orbits 2}
		\end{minipage}%
	}%
	\subfigure[36 Sats/Orbits]{
	\vspace{-0.75cm}
		\begin{minipage}[t]{0.24\linewidth}
			\centering
			\includegraphics[scale=.25]{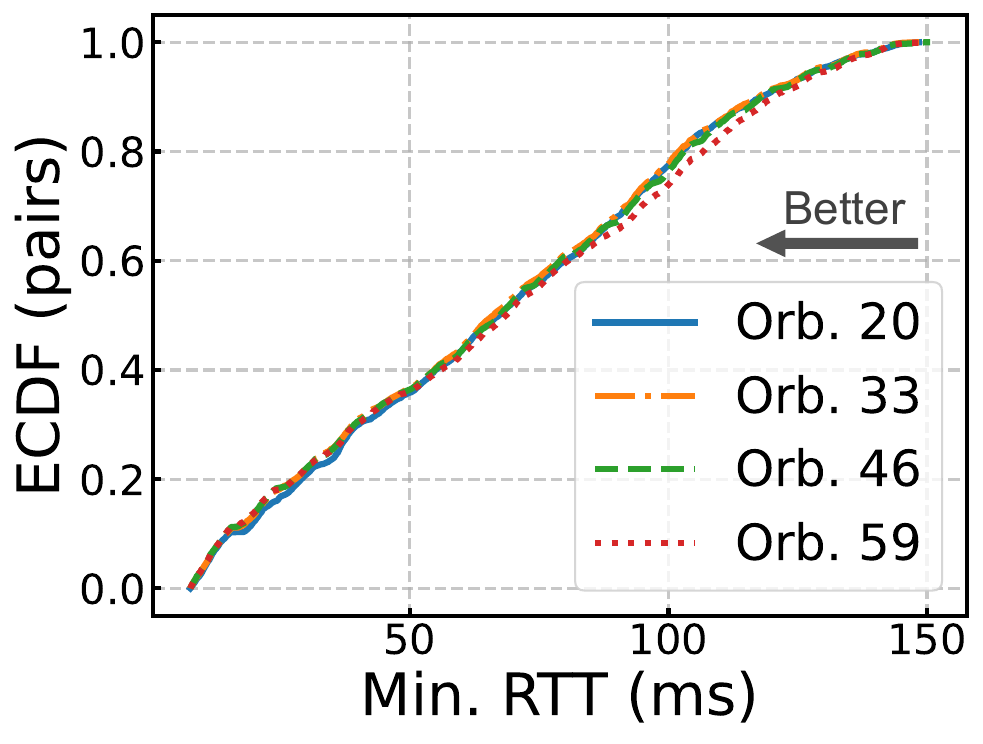}
			\vspace{-0.5cm}
      	\label{Min RTT Orbits 3}
		\end{minipage}%
	}%
	\subfigure[44 Sats/Orbits]{
	\vspace{-0.75cm}
		\begin{minipage}[t]{0.24\linewidth}
			\centering
			\includegraphics[scale=.25]{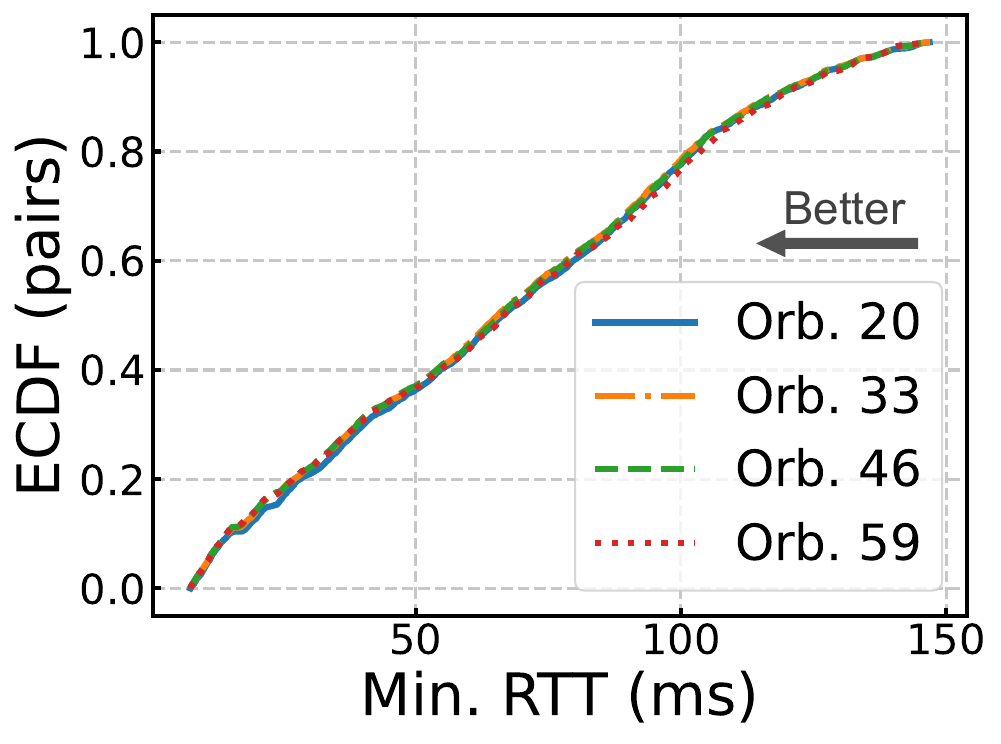}
			\vspace{-0.5cm}
   	\label{Min RTT Orbits 4}
		\end{minipage}%
	}%
    \vspace{-0.5cm}
	\caption{The Min. RTT (ms) ECDFs of changing Sats/Orbits number with different Orbits.}
     \vspace{-0.5cm}
	\label{Min RTT Orbits}
\end{figure*}

\begin{figure*}[ht]
	\centering
	\subfigure[20 Sats/Orbits]{
	\vspace{-0.75cm}
		\begin{minipage}[t]{0.24\linewidth}
			\centering
			\includegraphics[scale=.25]{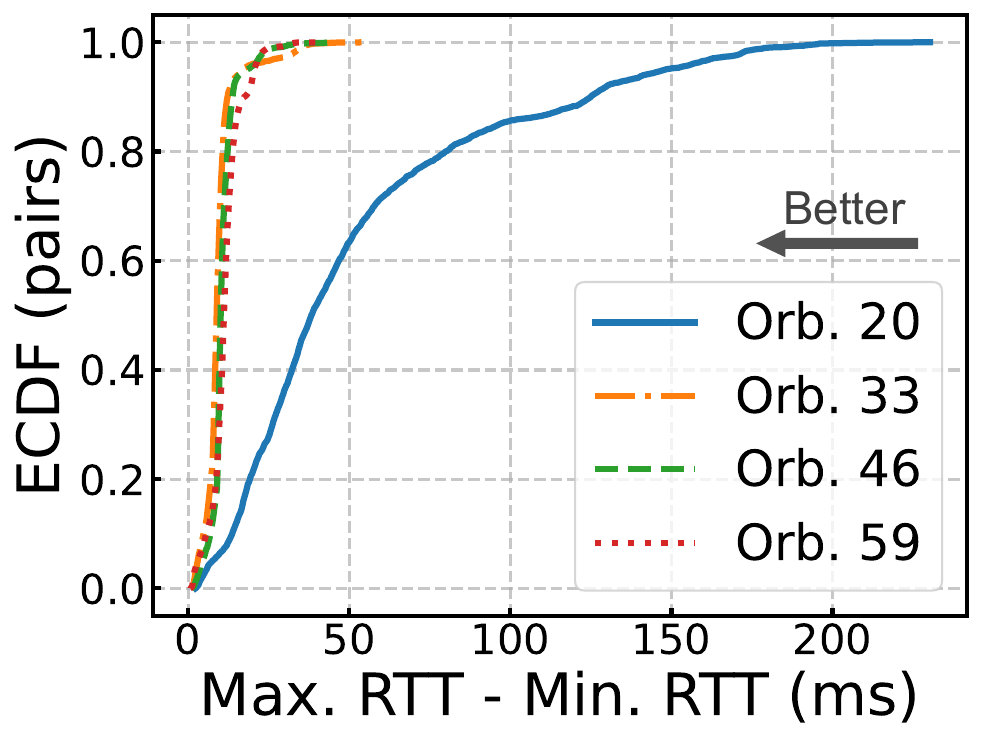}
			\vspace{-0.5cm}
      	\label{minus RTT Orbits 1}
		\end{minipage}%
	}%
	\subfigure[28 Sats/Orbits]{
	\vspace{-0.75cm}
		\begin{minipage}[t]{0.24\linewidth}
			\centering
			\includegraphics[scale=.25]{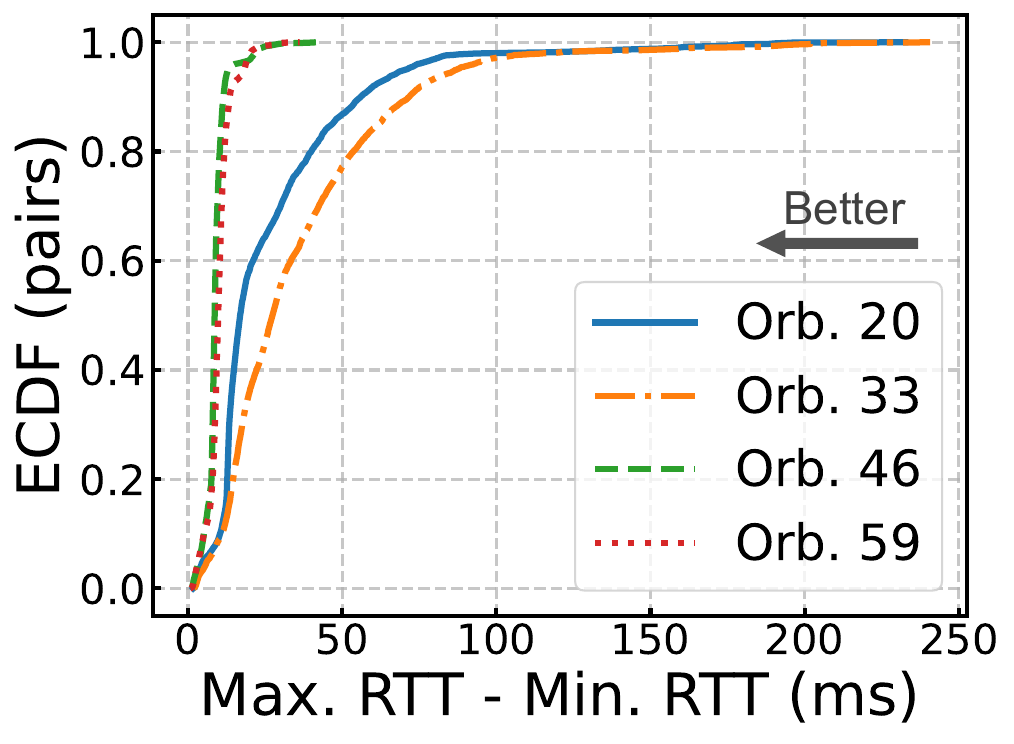}
			\vspace{-0.5cm}
      	\label{minus RTT Orbits 2}
		\end{minipage}%
	}%
	\subfigure[36 Sats/Orbits]{
	\vspace{-0.75cm}
		\begin{minipage}[t]{0.24\linewidth}
			\centering
			\includegraphics[scale=.25]{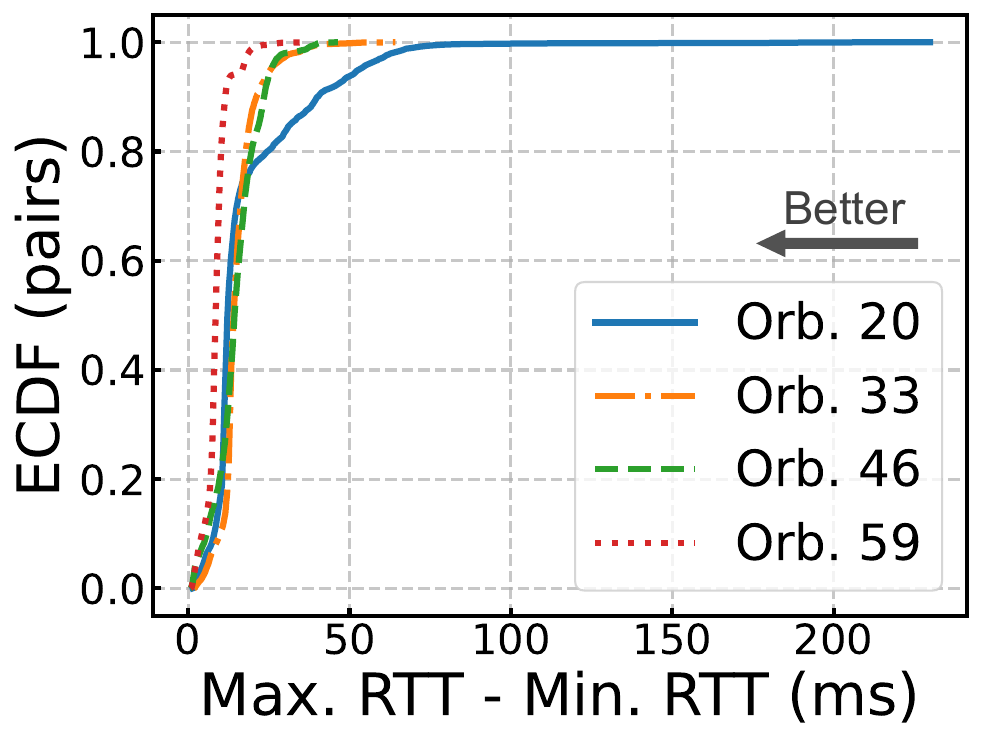}
			\vspace{-0.5cm}
      	\label{minus RTT Orbits 3}
		\end{minipage}%
	}%
	\subfigure[44 Sats/Orbits]{
	\vspace{-0.75cm}
		\begin{minipage}[t]{0.24\linewidth}
			\centering
			\includegraphics[scale=.25]{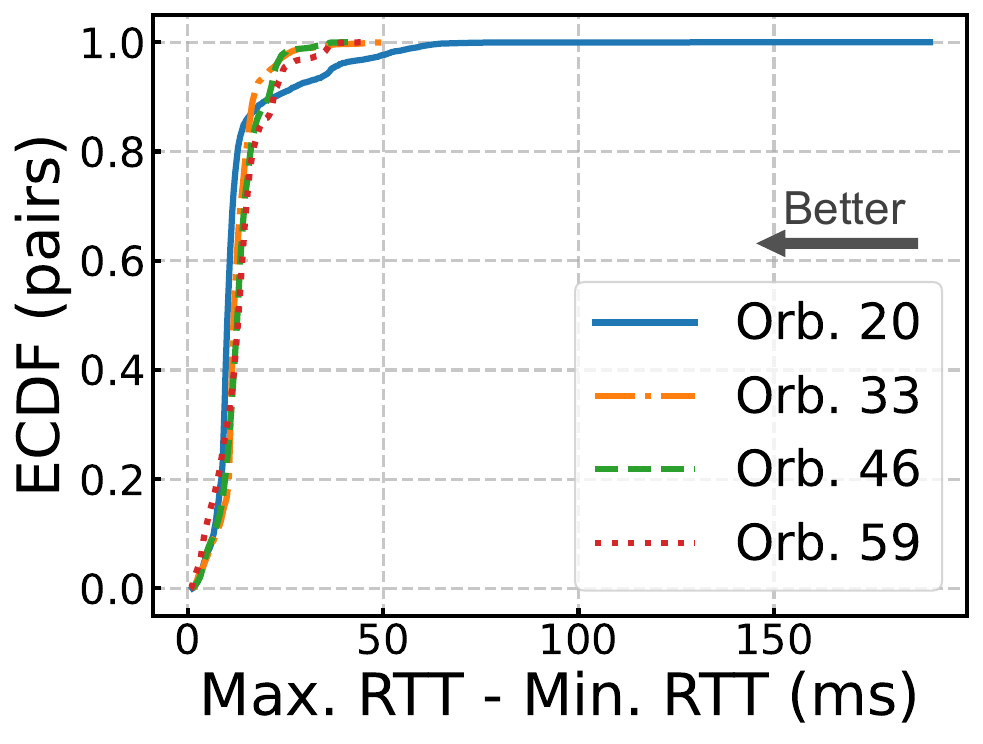}
			\vspace{-0.5cm}
   	\label{minus RTT Orbits 4}
		\end{minipage}%
	}%
    \vspace{-0.5cm}
	\caption{The Max. RTT - Min. RTT (ms) ECDFs of changing Sats/Orbits number with different Orbits.}
     \vspace{-0.5cm}
	\label{minus RTT Orbits}
\end{figure*}

\begin{figure*}[ht]
	\centering
	\subfigure[20 Sats/Orbits]{
	\vspace{-0.75cm}
		\begin{minipage}[t]{0.24\linewidth}
			\centering
			\includegraphics[scale=.25]{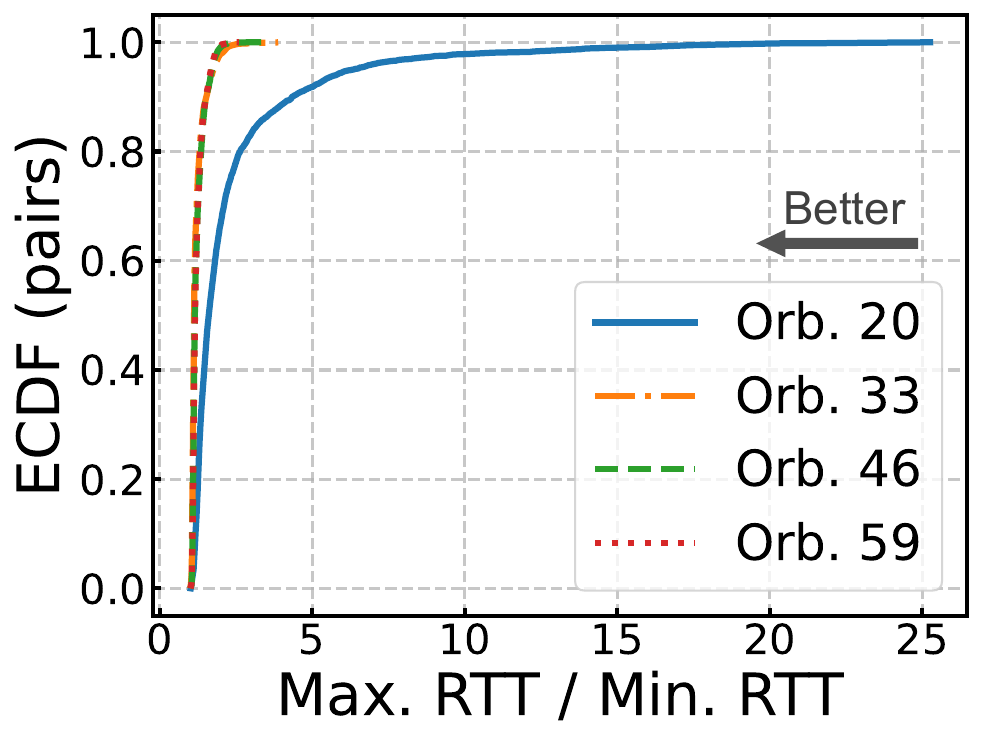}
			\vspace{-0.5cm}
      	\label{To RTT Orbits 1}
		\end{minipage}%
	}%
	\subfigure[28 Sats/Orbits]{
	\vspace{-0.75cm}
		\begin{minipage}[t]{0.24\linewidth}
			\centering
			\includegraphics[scale=.25]{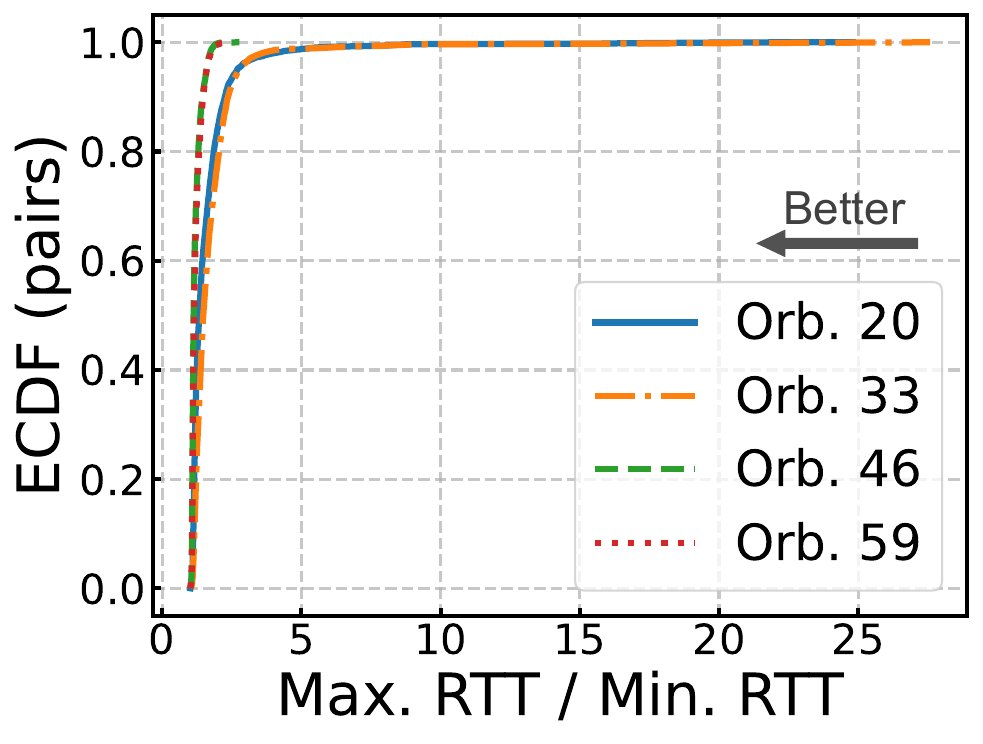}
			\vspace{-0.5cm}
      	\label{To RTT Orbits 2}
		\end{minipage}%
	}%
	\subfigure[36 Sats/Orbits]{
	\vspace{-0.75cm}
		\begin{minipage}[t]{0.24\linewidth}
			\centering
			\includegraphics[scale=.25]{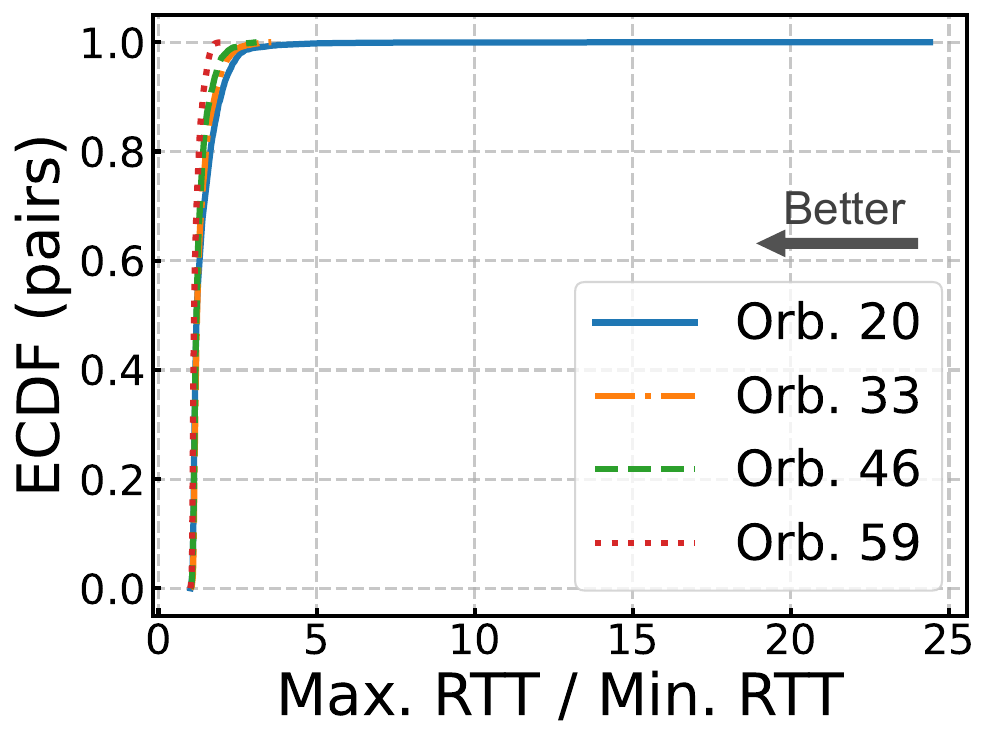}
			\vspace{-0.75cm}
      	\label{To RTT Orbits 3}
		\end{minipage}%
	}%
	\subfigure[44 Sats/Orbits]{
	\vspace{-0.75cm}
		\begin{minipage}[t]{0.24\linewidth}
			\centering
			\includegraphics[scale=.25]{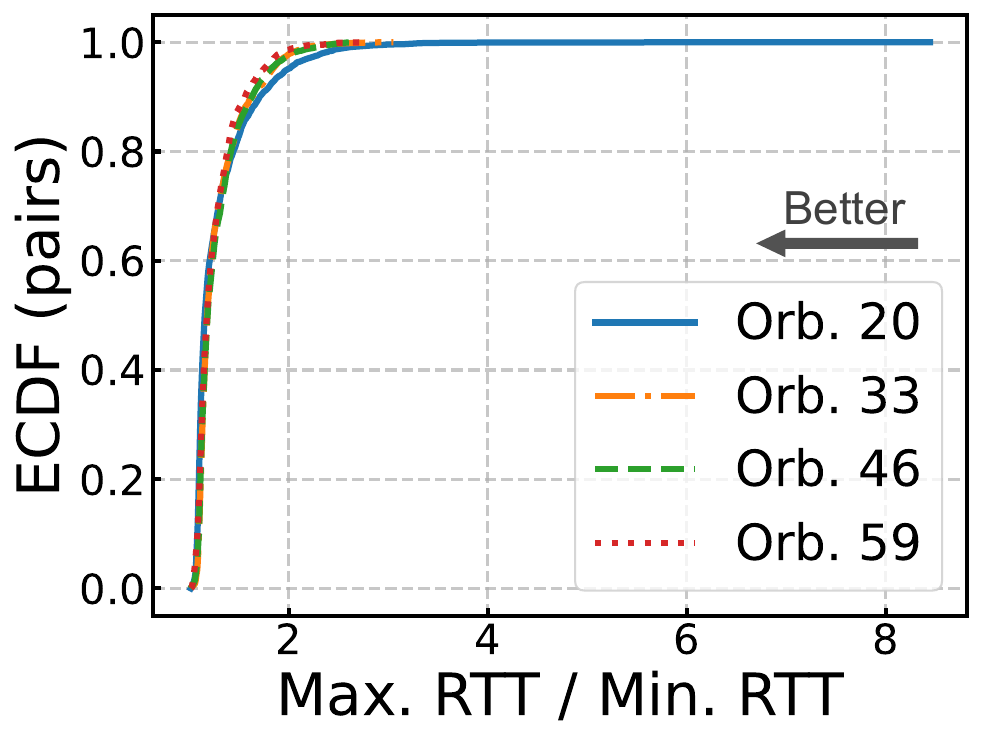}
			\vspace{-0.75cm}
   	\label{To RTT Orbits 4}
		\end{minipage}%
	}%
    \vspace{-0.5cm}
	\caption{The Max. RTT to Min. RTT ECDFs of changing Sats/Orbits number with different Orbits.}
     \vspace{-0.45cm}
	\label{To RTT Orbits}
\end{figure*}

\begin{figure*}[ht]
	\centering
	\subfigure[20 Sats/Orbits]{
	\vspace{-0.75cm}
		\begin{minipage}[t]{0.24\linewidth}
			\centering
			\includegraphics[scale=.25]{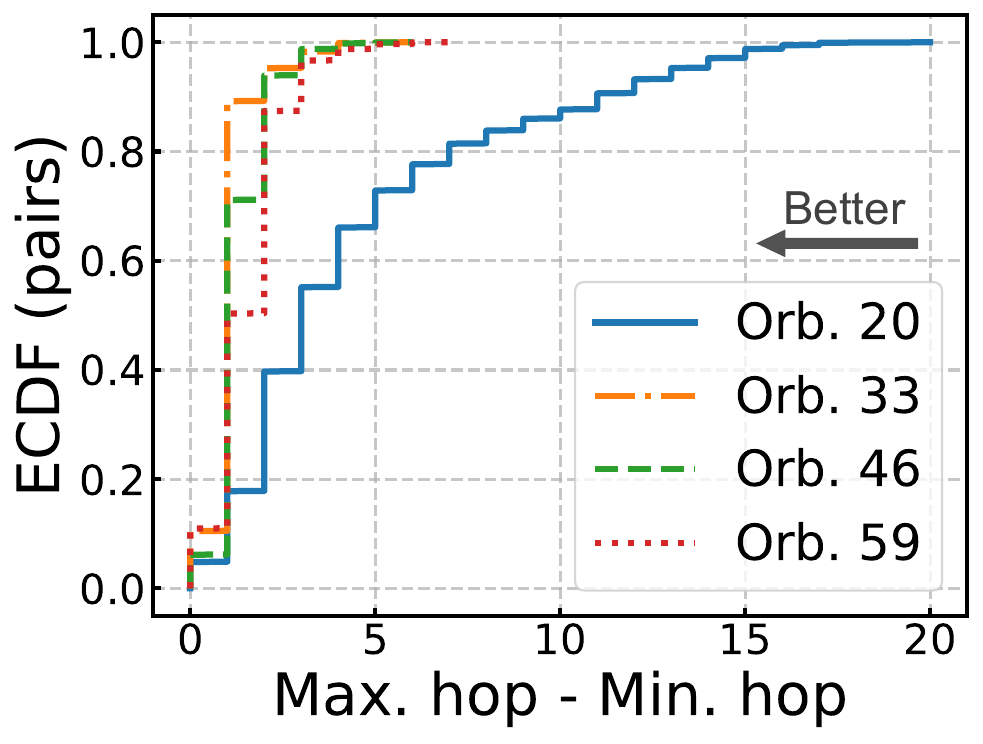}
			\vspace{-0.5cm}
      	\label{minus hop Orbits 1}
		\end{minipage}%
	}%
	\subfigure[28 Sats/Orbits]{
	\vspace{-0.75cm}
		\begin{minipage}[t]{0.24\linewidth}
			\centering
			\includegraphics[scale=.25]{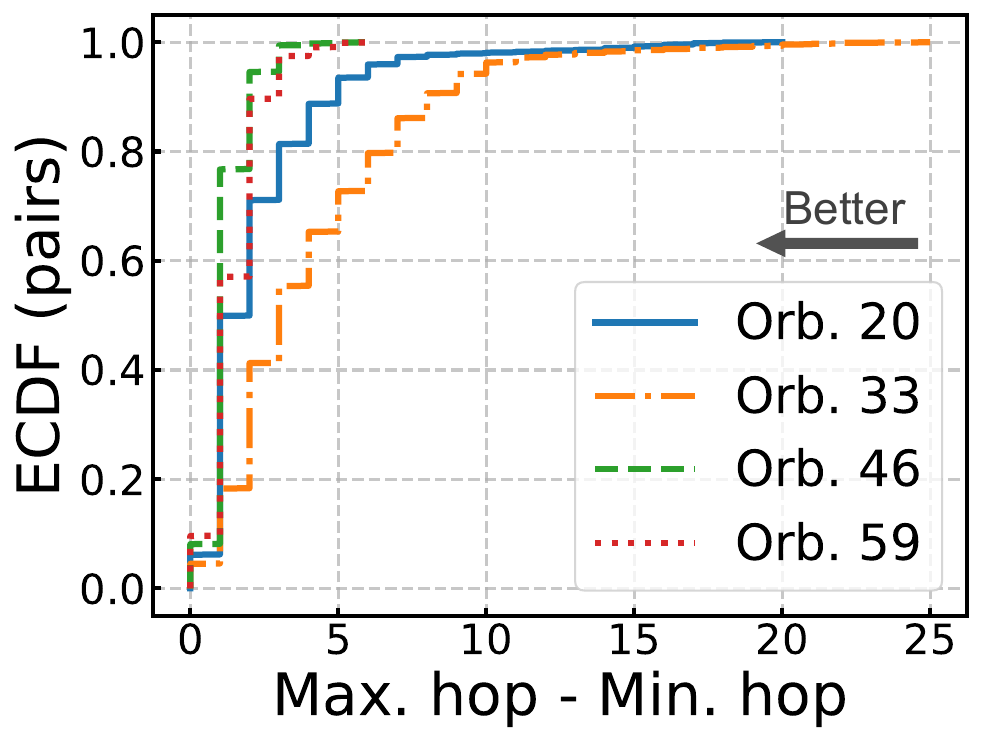}
			\vspace{-0.5cm}
      	\label{minus hop Orbits 2}
		\end{minipage}%
	}%
	\subfigure[36 Sats/Orbits]{
	\vspace{-0.75cm}
		\begin{minipage}[t]{0.24\linewidth}
			\centering
			\includegraphics[scale=.25]{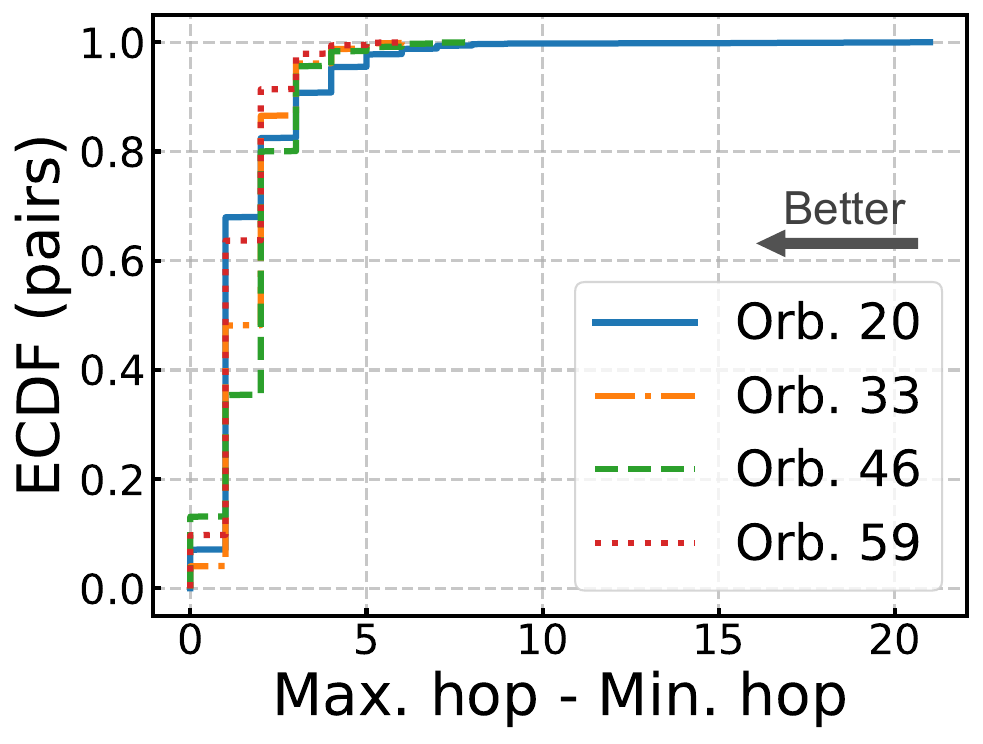}
			\vspace{-0.75cm}
      	\label{minus hop Orbits 3}
		\end{minipage}%
	}%
	\subfigure[44 Sats/Orbits]{
	\vspace{-0.75cm}
		\begin{minipage}[t]{0.24\linewidth}
			\centering
			\includegraphics[scale=.25]{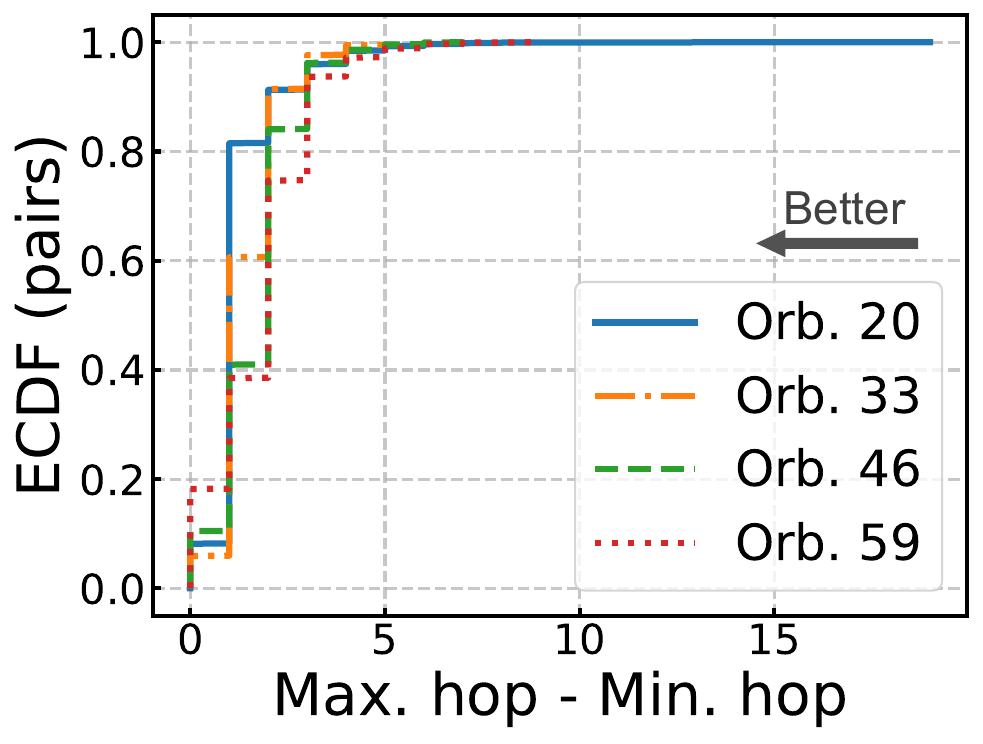}
			\vspace{-0.75cm}
   	\label{minus hop Orbits 4}
		\end{minipage}%
	}%
    \vspace{-0.5cm}
	\caption{The Max. hop - Min. hop ECDFs of changing Sats/Orbits number with different Orbits.}
     \vspace{-0.45cm}
	\label{minus hop Orbits}
\end{figure*}

\begin{figure*}[ht]
	\centering
	\subfigure[20 Sats/Orbits]{
	\vspace{-0.75cm}
		\begin{minipage}[t]{0.24\linewidth}
			\centering
			\includegraphics[scale=.25]{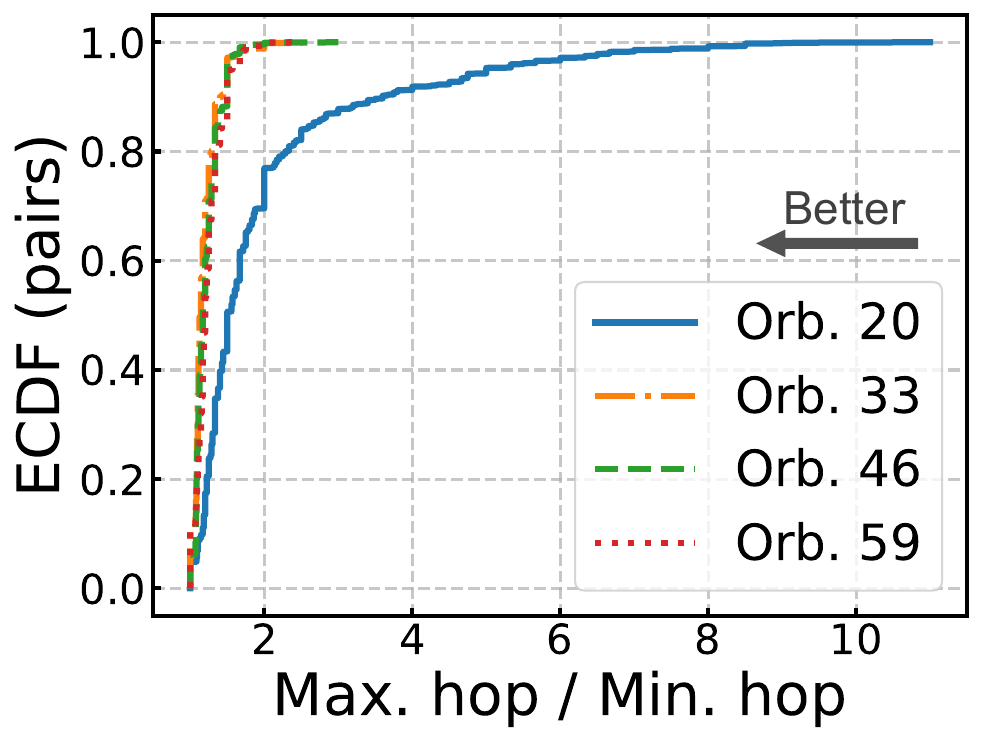}
			\vspace{-0.5cm}
      	\label{to hop Orbits 1}
		\end{minipage}%
	}%
	\subfigure[28 Sats/Orbits]{
	\vspace{-0.75cm}
		\begin{minipage}[t]{0.24\linewidth}
			\centering
			\includegraphics[scale=.25]{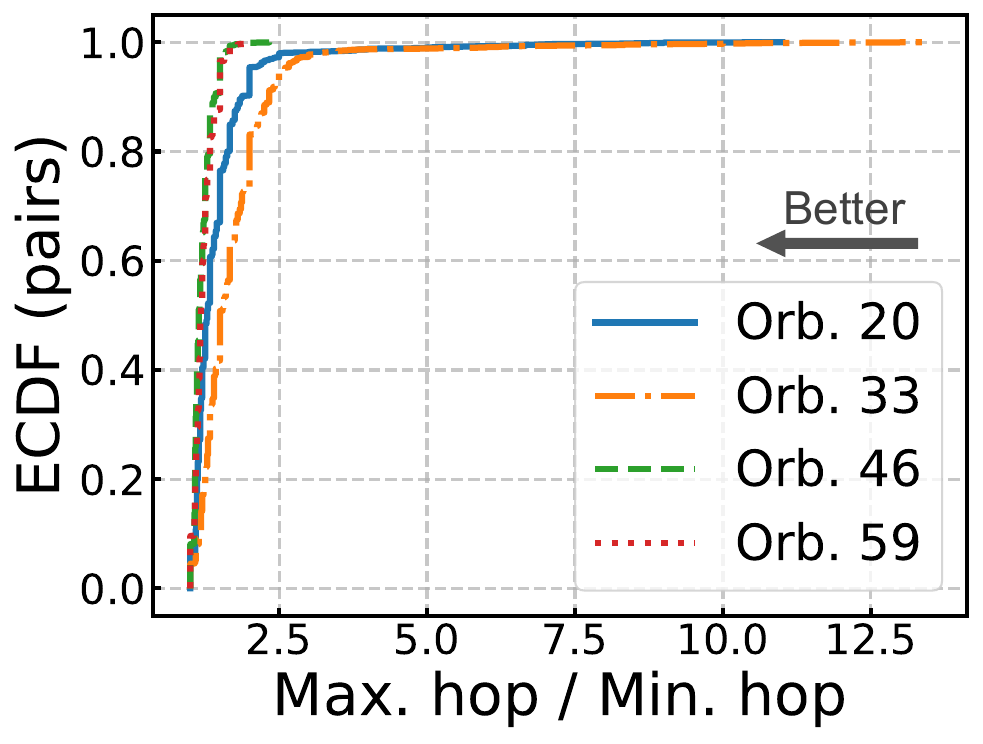}
			\vspace{-0.5cm}
      	\label{to hop Orbits 2}
		\end{minipage}%
	}%
	\subfigure[36 Sats/Orbits]{
	\vspace{-0.75cm}
		\begin{minipage}[t]{0.24\linewidth}
			\centering
			\includegraphics[scale=.25]{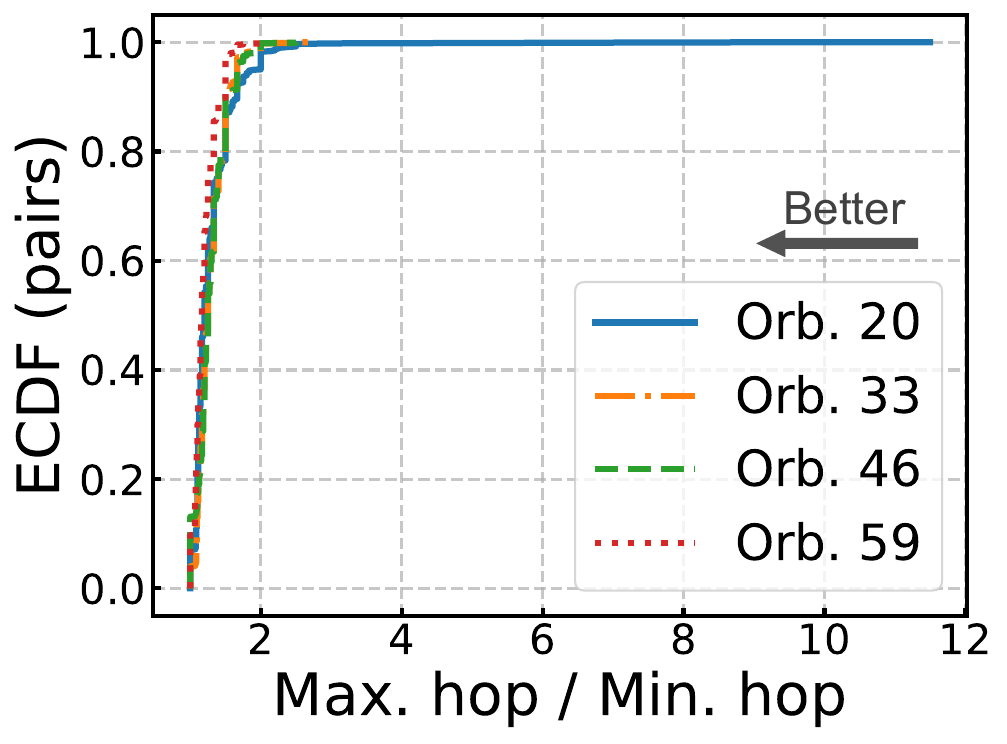}
			\vspace{-0.75cm}
      	\label{to hop Orbits 3}
		\end{minipage}%
	}%
	\subfigure[44 Sats/Orbits]{
	\vspace{-0.75cm}
		\begin{minipage}[t]{0.24\linewidth}
			\centering
			\includegraphics[scale=.25]{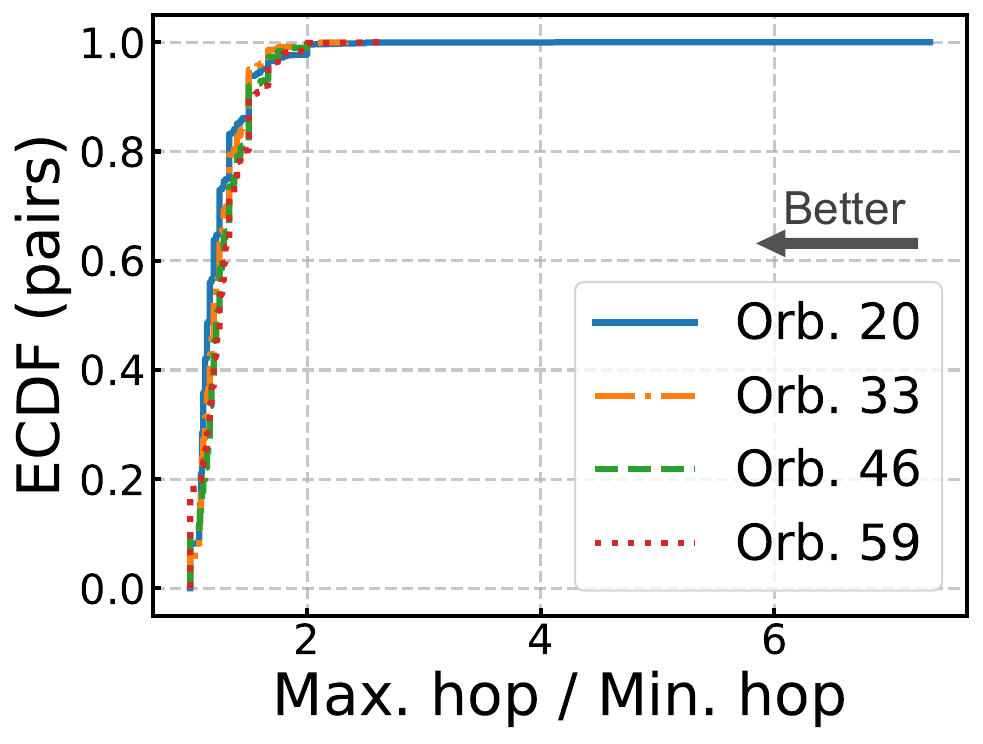}
			\vspace{-0.75cm}
   	\label{to hop Orbits 4}
		\end{minipage}%
	}%
    \vspace{-0.5cm}
	\caption{The Max. hop to Min. hop ECDFs of changing Sats/Orbits number with different Orbits.}
     \vspace{-0.5cm}
	\label{to hop Orbits}
\end{figure*}

\subsection{Orbit Number}
\label{appendix: Orbit Number}
With the inclination of 53°, the satellite network performance is measured when the Orbits number = 20, 33, 46, 59 under the same Sats/Orbits = 20, 28, 36, 44. The extension result of Min. RTT, Max. RTT - Min. RTT, Max. RTT to Min. RTT, Max. hop - Min. hop, Max. hop to Min. hop are show in Fig.~\ref{Min RTT Orbits}, Fig.~\ref{minus RTT Orbits}, Fig.~\ref{To RTT Orbits}, Fig.~\ref{minus hop Orbits}, Fig.~\ref{to hop Orbits}, respectively. Because the long-tail effect of some parameter configurations is too significant, we are forced to select only part of the ECDF curves and ignore the excessively long tails to facilitate the presentation. From Fig.~\ref{Min RTT Orbits}, we can see that the Min. RTT does not differ much between different orbits numbers for 28+ Sats/Orbits. This phenomenon means that LEO mega-constellation with 20+ Sats/Orbits and 20+ orbits are close enough to the upper limit performance of Min. RTT. Moreover, by observing the tail of the ECDF curve in Fig.~\ref{Min RTT Orbits}, we can find that with the increase of the number of Sats/Orbits, the maximum value of Min. RTT gradually decreases from nearly 200ms to less than 150ms. From Fig.~\ref{minus RTT Orbits} and Fig.~\ref{To RTT Orbits}, we can see the fluctuation amplitude of RTT. When Sats/Orbits = 20, the 33- orbits constellation has a large RTT fluctuation. However, for Sats/Orbits > 28 LEO mega-constellations, when the orbits number increases, the influence of the number of Orbits on the network RTT fluctuation reaches saturation and can no longer significantly improve the stability of RTT fluctuation.As shown in Fig.~\ref{minus hop Orbits} and Fig.~\ref{to hop Orbits}, hop fluctuations follow the same pattern. Combining the above experiments result, we can conclude that when the orbits number increases, both the latency and hop performance will improve. We also notice a relative saturation state value for Sat/Orbits between 20 and 28. To understand this phenomenon better, we further generated different satellite constellations with smaller intervals of Sats/Orbits between 20 and 28, hoping to find one Sats/Orbits number that could help the LEO mega-constellation reach the saturation state no matter the change in the orbits number, the reults are shown in~\S~\ref{subsec: Further Exploration}.

\subsection{Further Exploration}
\label{subsec: Further Exploration}
\begin{figure*}[ht]
	\centering
	\subfigure[20 Sats/Orbits]{
	\vspace{-0.75cm}
		\begin{minipage}[t]{0.24\linewidth}
			\centering
			\includegraphics[scale=.25]{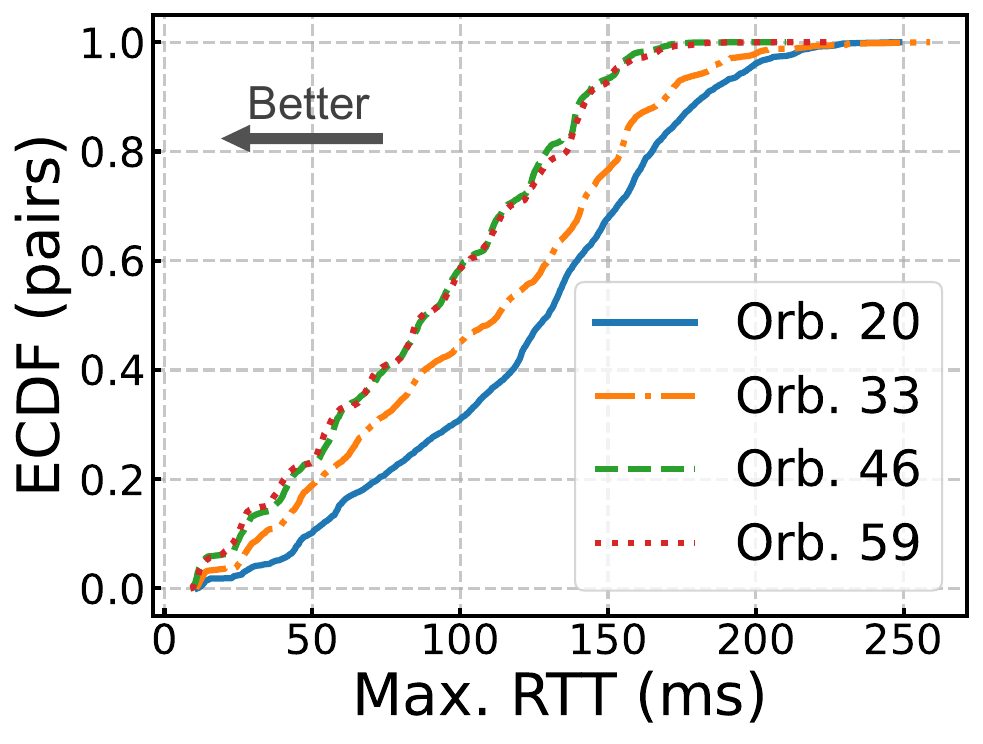}
			\vspace{-0.5cm}
      	\label{RTT Orbits 5}
		\end{minipage}%
	}%
	\subfigure[22 Sats/Orbits]{
	\vspace{-0.75cm}
		\begin{minipage}[t]{0.24\linewidth}
			\centering
			\includegraphics[scale=.25]{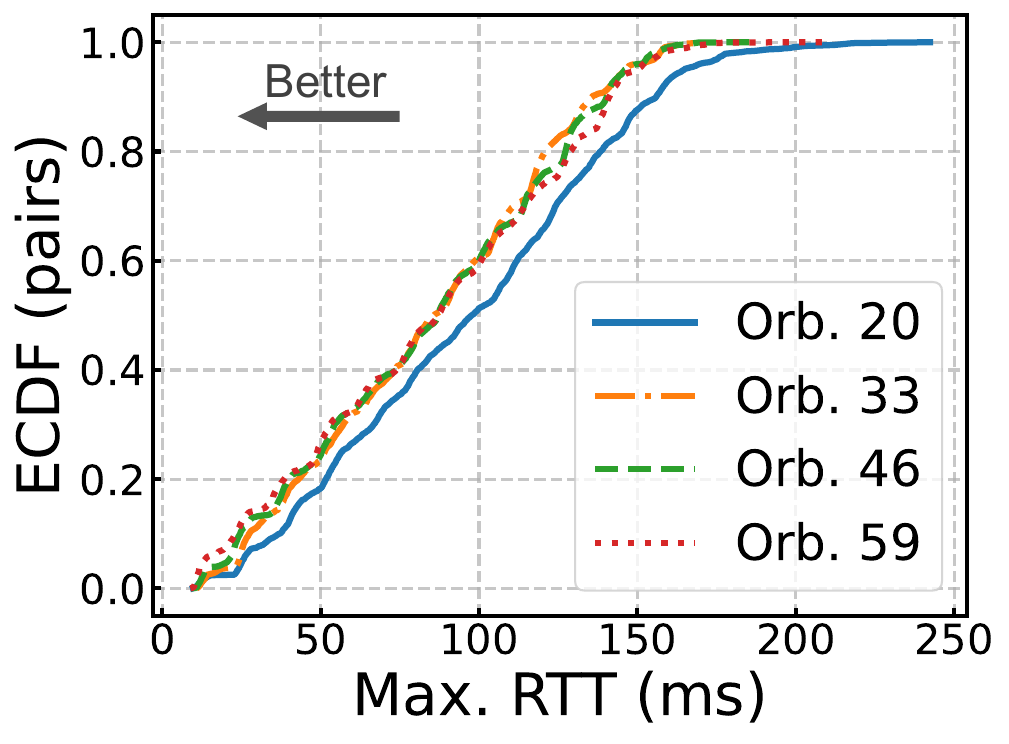}
			\vspace{-0.5cm}
      	\label{RTT Orbits 6}
		\end{minipage}%
	}%
	\subfigure[24 Sats/Orbits]{
	\vspace{-0.75cm}
		\begin{minipage}[t]{0.24\linewidth}
			\centering
			\includegraphics[scale=.25]{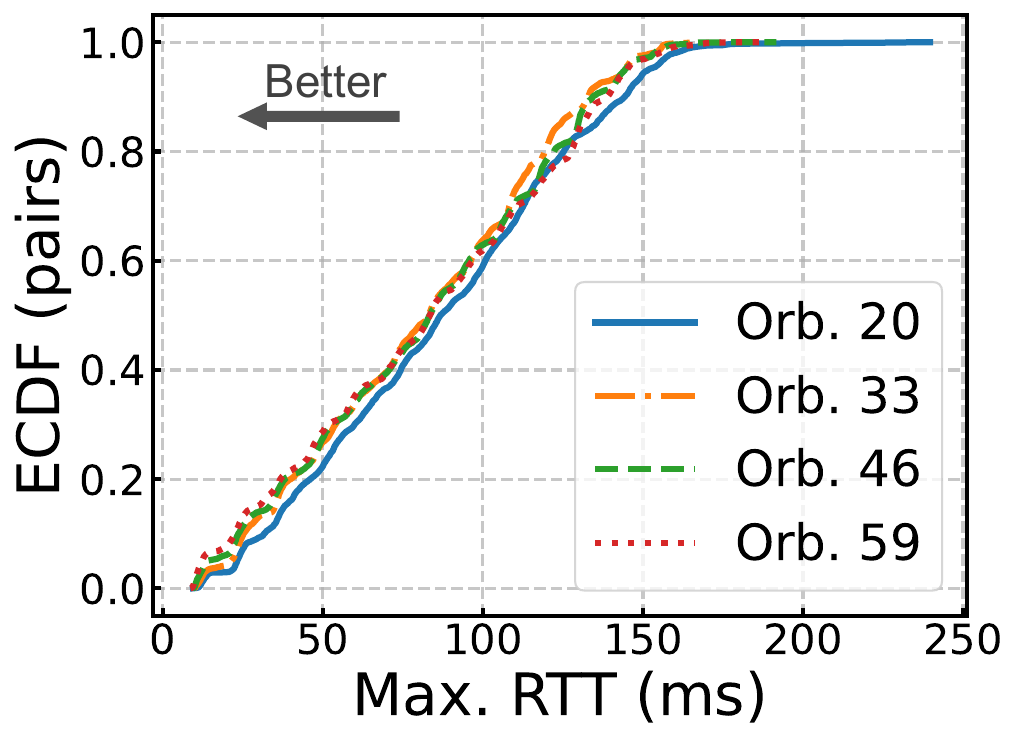}
			\vspace{-0.5cm}
      	\label{RTT Orbits 7}
		\end{minipage}%
	}%
	\subfigure[26 Sats/Orbits]{
	\vspace{-0.75cm}
		\begin{minipage}[t]{0.24\linewidth}
			\centering
			\includegraphics[scale=.25]{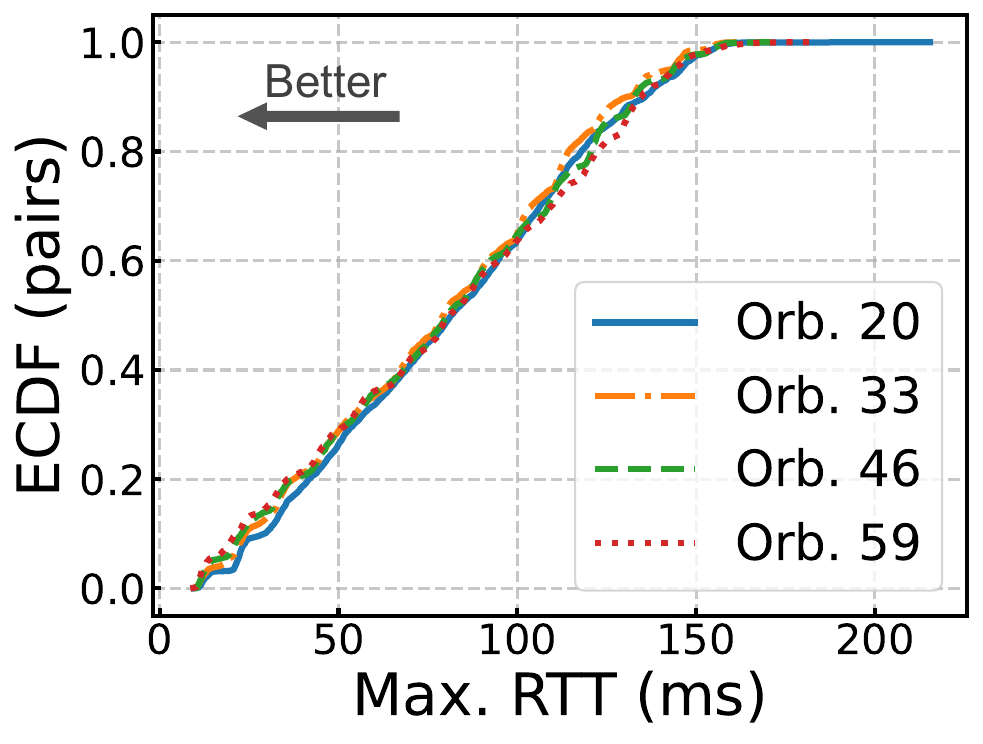}
			\vspace{-0.5cm}
      	\label{RTT Orbits 8}
		\end{minipage}%
	}%
    \vspace{-0.5cm}
	\caption{The Max. RTT(ms) ECDFs of changing orbits number with different Sats/Orbits.}
	\vspace{-0.5cm}
   	\label{RTT Orbits (2)}
\end{figure*}

\begin{figure*}[ht]
	\centering
	\subfigure[20 Sats/Orbits]{
	\vspace{-0.75cm}
		\begin{minipage}[t]{0.24\linewidth}
			\centering
			\includegraphics[scale=.25]{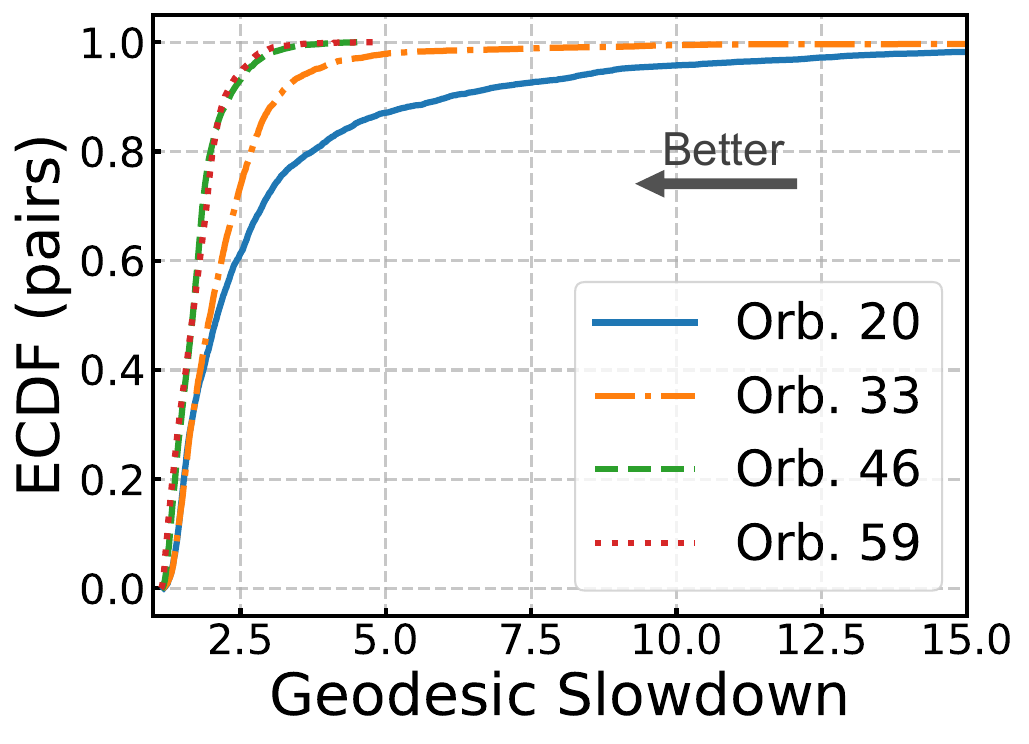}
			\vspace{-0.75cm}
      	\label{slowdown Orbits 5}
		\end{minipage}%
	}%
	\subfigure[22 Sats/Orbits]{
	\vspace{-0.75cm}
		\begin{minipage}[t]{0.24\linewidth}
			\centering
			\includegraphics[scale=.25]{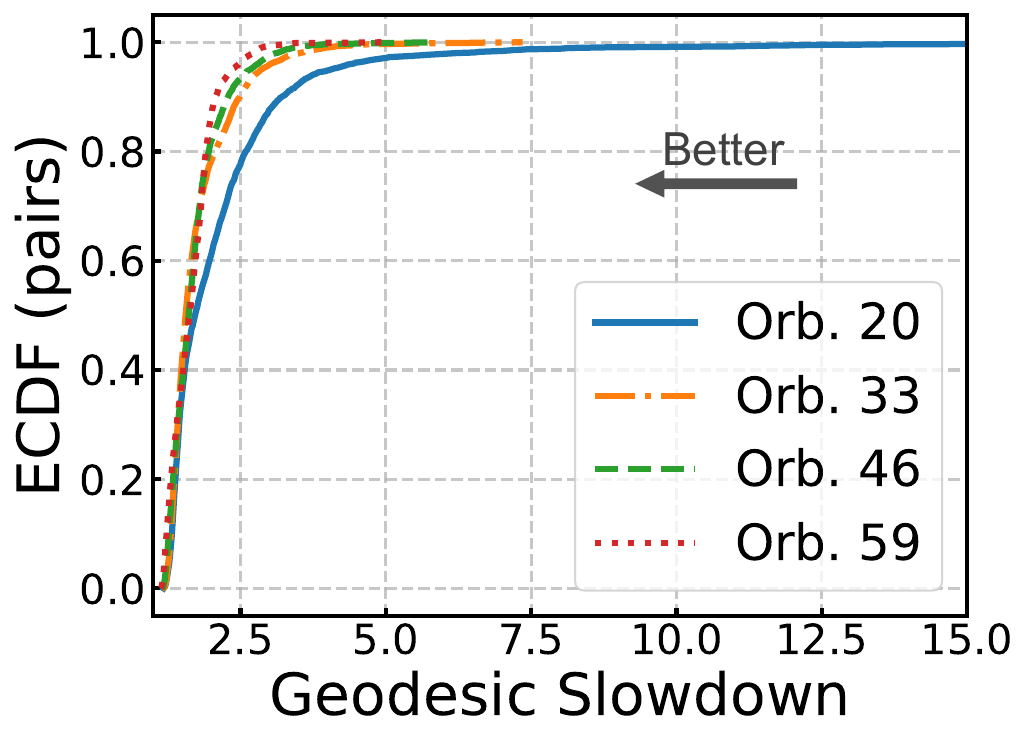}
			\vspace{-0.75cm}
      	\label{slowdown RTT Orbits 6}
		\end{minipage}%
	}%
	\subfigure[24 Sats/Orbits]{
	\vspace{-0.75cm}
		\begin{minipage}[t]{0.24\linewidth}
			\centering
			\includegraphics[scale=.25]{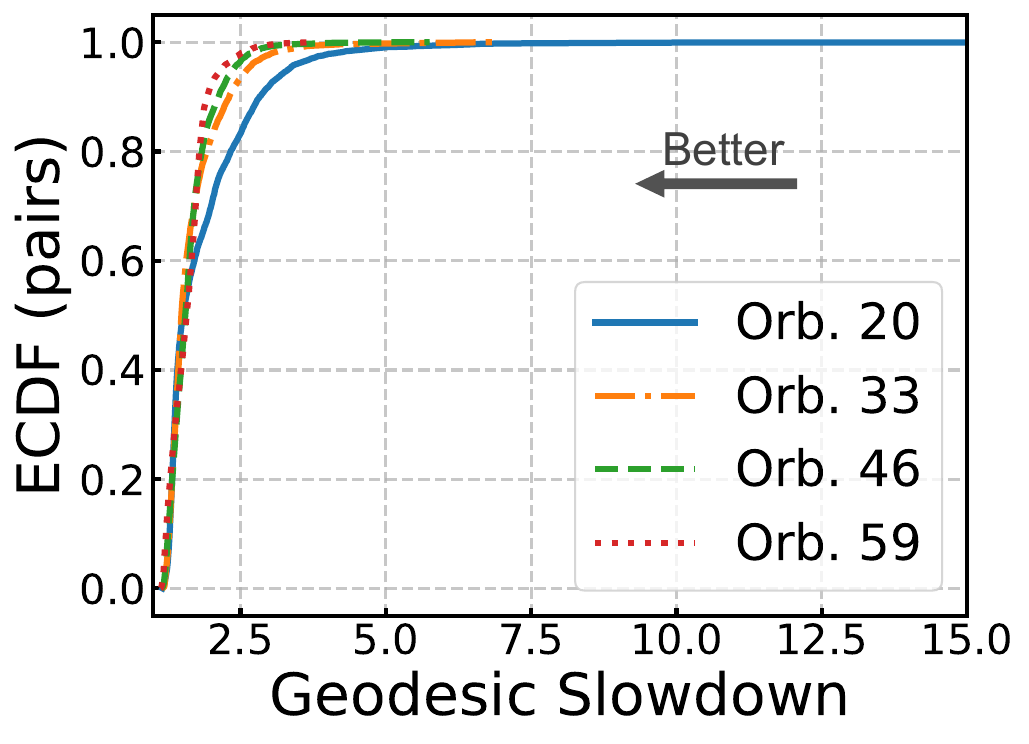}
			\vspace{-0.75cm}
      	\label{slowdown RTT Orbits 7}
		\end{minipage}%
	}%
	\subfigure[26 Sats/Orbits]{
	\vspace{-0.75cm}
		\begin{minipage}[t]{0.24\linewidth}
			\centering
			\includegraphics[scale=.25]{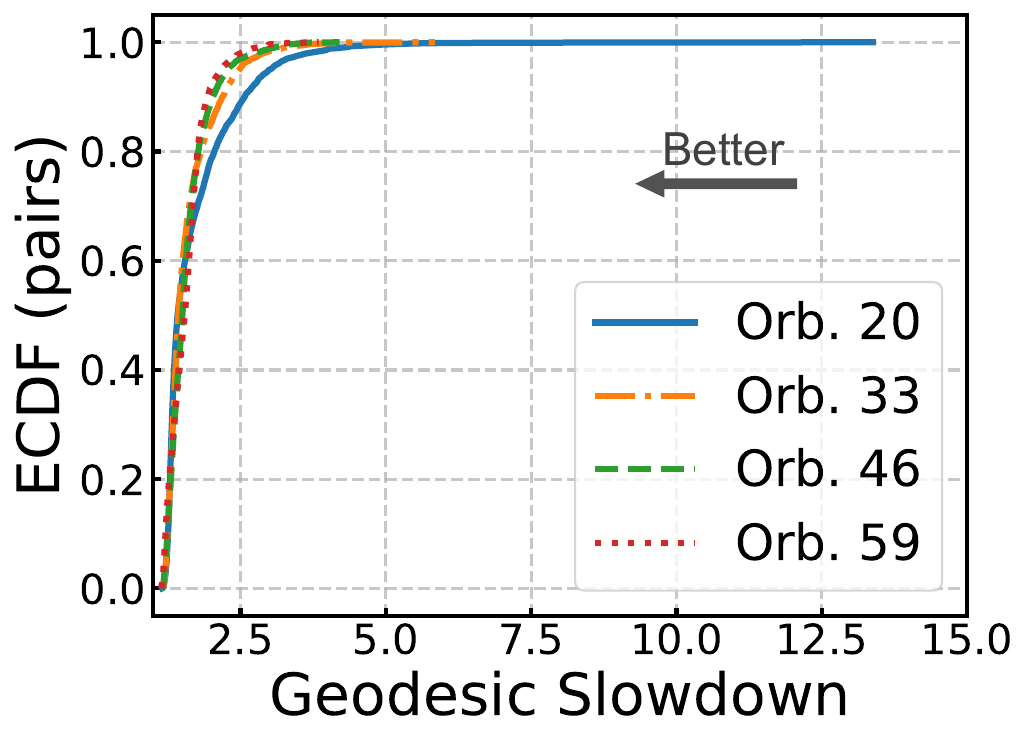}
			\vspace{-0.75cm}
      	\label{slowdown RTT Orbits 8}
		\end{minipage}%
	}%
    \vspace{-0.5cm}
	\caption{The RTT Geodesic Slowdown ECDFs of changing orbits number with different Sats/Orbits.}
	\vspace{-0.5cm}
   	\label{slowdown RTT Orbits (2)}
\end{figure*}

\begin{figure*}[ht]
	\centering
	\subfigure[20 Sats/Orbits]{
	\vspace{-0.75cm}
		\begin{minipage}[t]{0.24\linewidth}
			\centering
			\includegraphics[scale=.25]{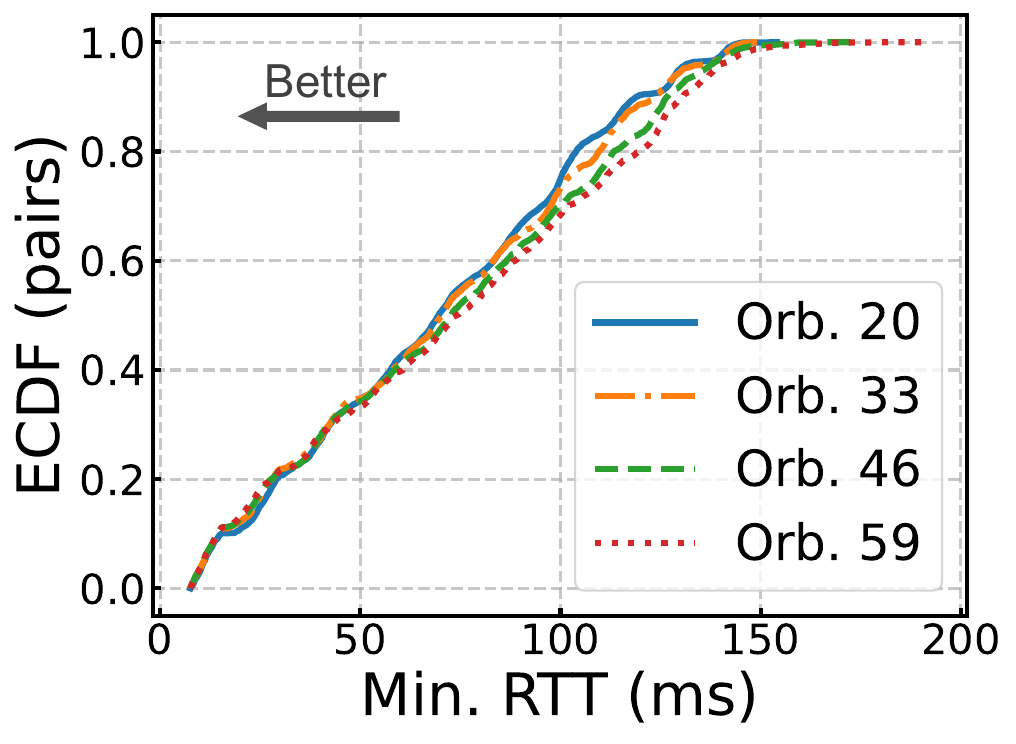}
			\vspace{-0.5cm}
      	\label{Min RTT Orbits 5}
		\end{minipage}%
	}%
	\subfigure[22 Sats/Orbits]{
	\vspace{-0.75cm}
		\begin{minipage}[t]{0.24\linewidth}
			\centering
			\includegraphics[scale=.25]{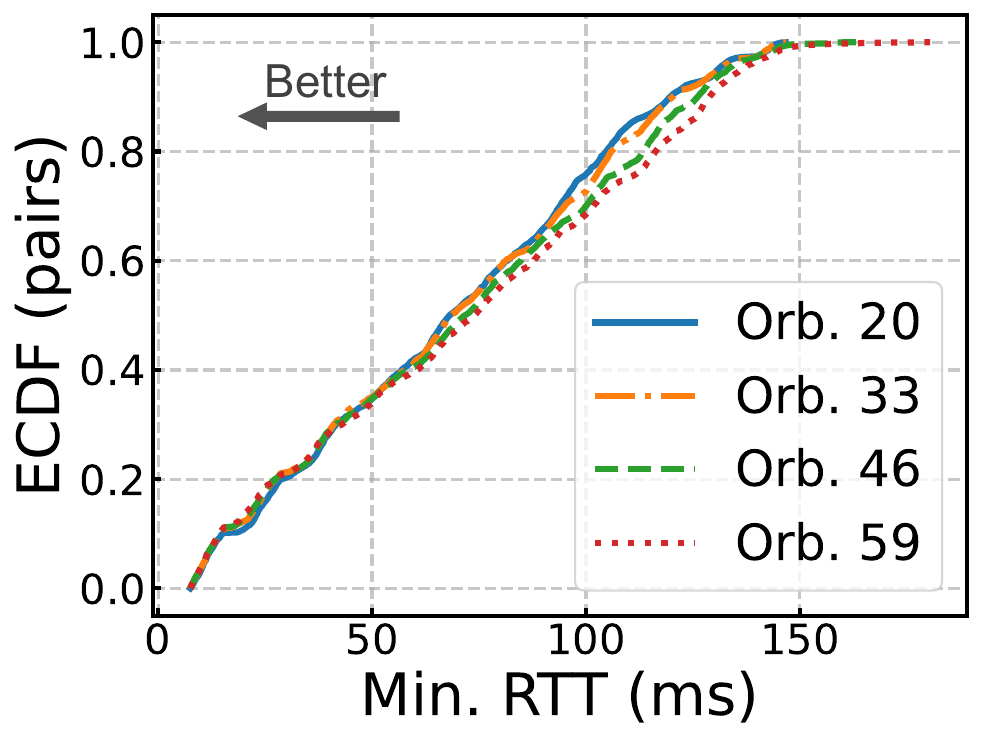}
			\vspace{-0.5cm}
      	\label{Min RTT Orbits 6}
		\end{minipage}%
	}%
	\subfigure[24 Sats/Orbits]{
	\vspace{-0.75cm}
		\begin{minipage}[t]{0.24\linewidth}
			\centering
			\includegraphics[scale=.25]{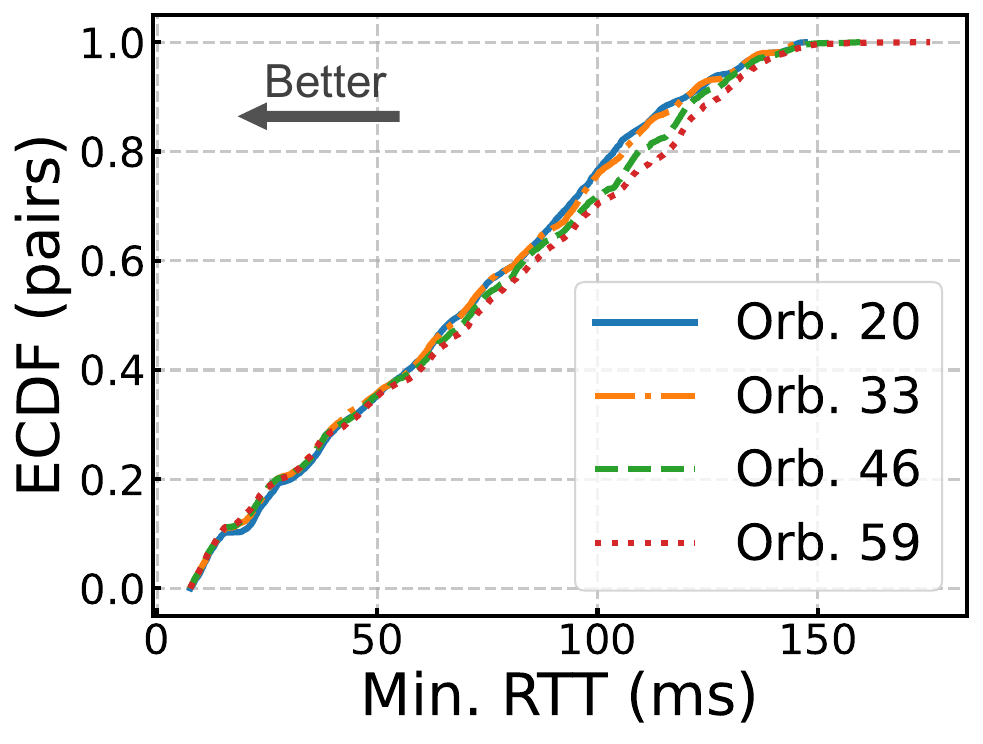}
			\vspace{-0.5cm}
      	\label{Min RTT Orbits 7}
		\end{minipage}%
	}%
	\subfigure[26 Sats/Orbits]{
	\vspace{-0.75cm}
		\begin{minipage}[t]{0.24\linewidth}
			\centering
			\includegraphics[scale=.25]{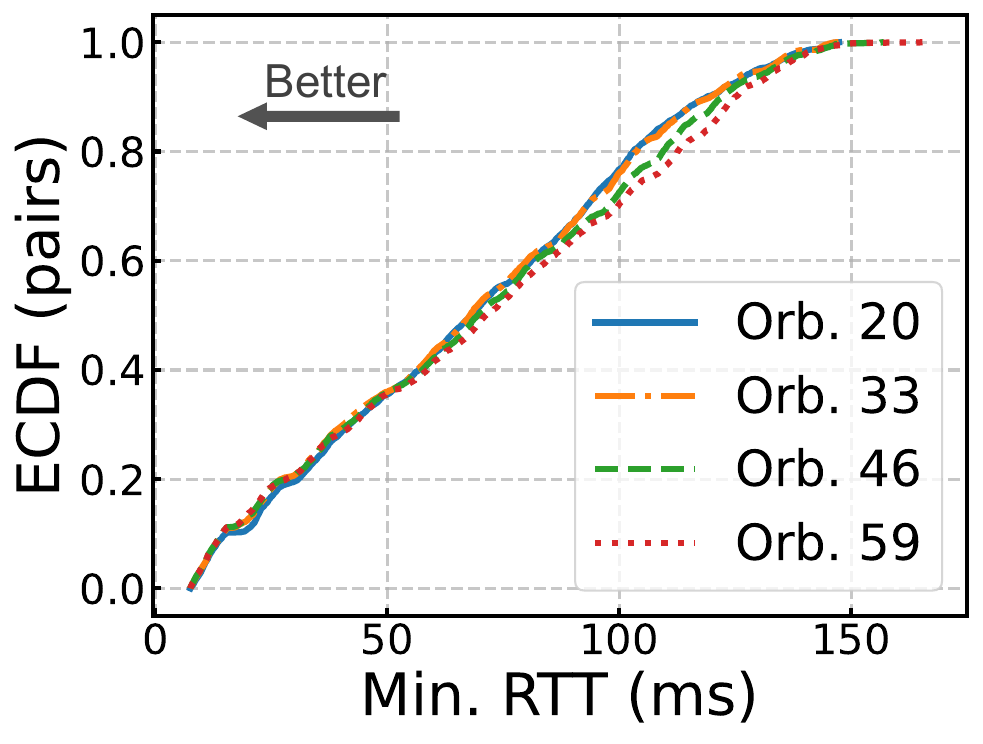}
			\vspace{-0.5cm}
      	\label{Min RTT Orbits 8}
		\end{minipage}%
	}%
    \vspace{-0.5cm}
	\caption{The Min. RTT(ms) ECDFs of changing orbits number with different Sats/Orbits.}
	\vspace{-0.5cm}
   	\label{Min RTT Orbits (2)}
\end{figure*}

\begin{figure*}[ht]
	\centering
	\subfigure[20 Sats/Orbits]{
	\vspace{-0.75cm}
		\begin{minipage}[t]{0.24\linewidth}
			\centering
			\includegraphics[scale=.25]{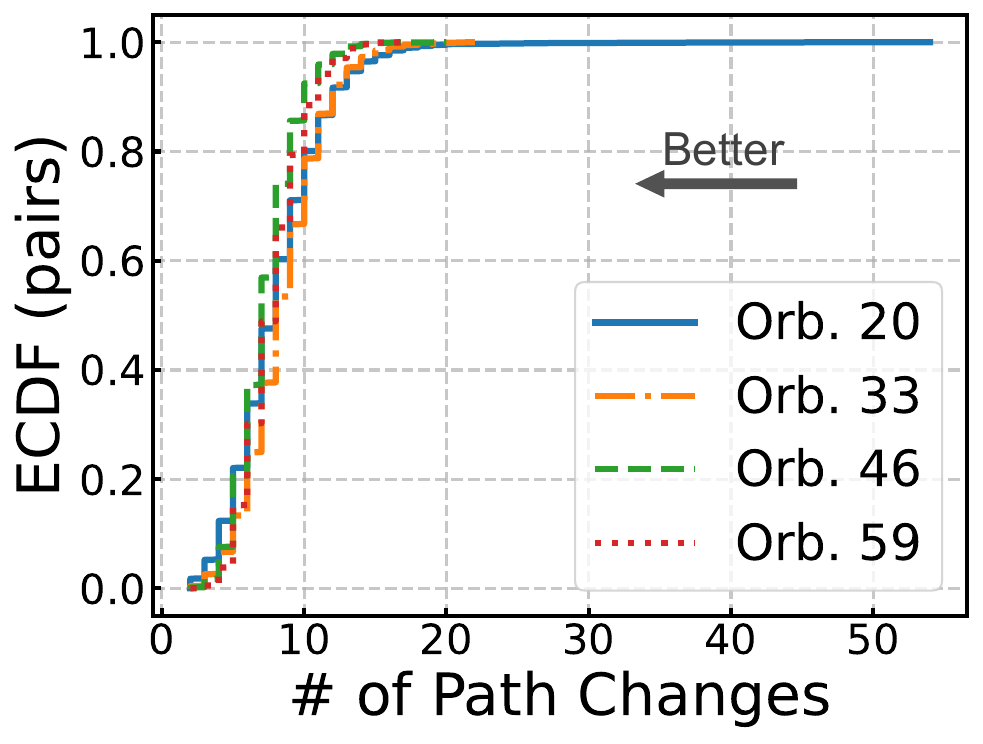}
			\vspace{-0.75cm}
      	\label{path Orbits 5}
		\end{minipage}%
	}%
	\subfigure[22 Sats/Orbits]{
	\vspace{-0.75cm}
		\begin{minipage}[t]{0.24\linewidth}
			\centering
			\includegraphics[scale=.25]{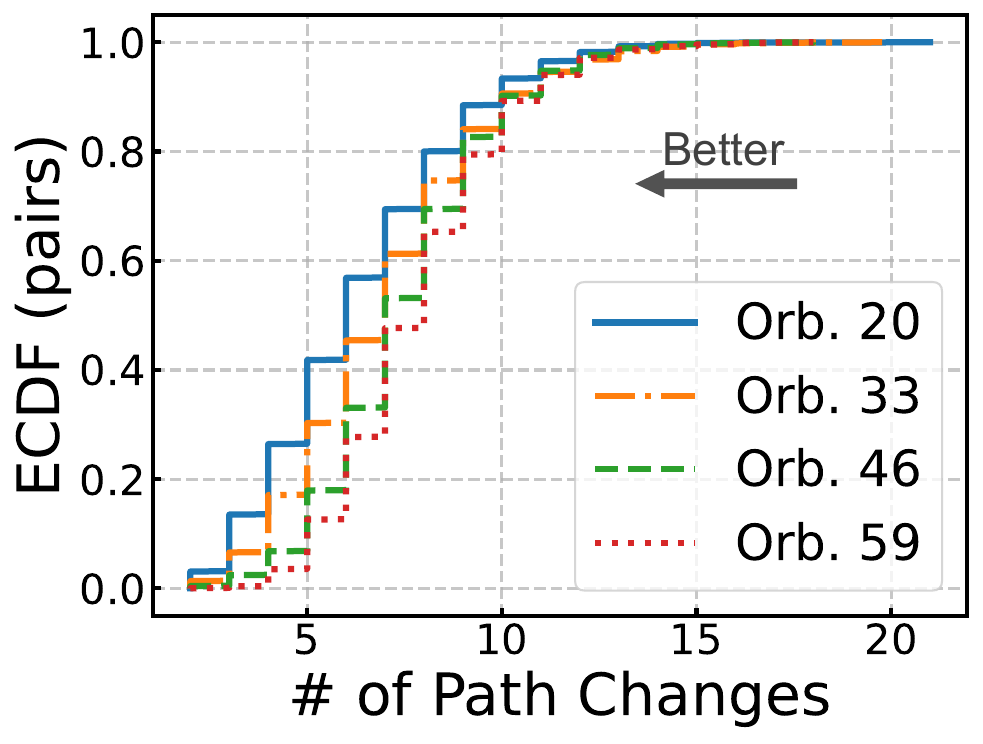}
			\vspace{-0.75cm}
      	\label{path Orbits 6}
		\end{minipage}%
	}%
	\subfigure[24 Sats/Orbits]{
	\vspace{-0.75cm}
		\begin{minipage}[t]{0.24\linewidth}
			\centering
			\includegraphics[scale=.25]{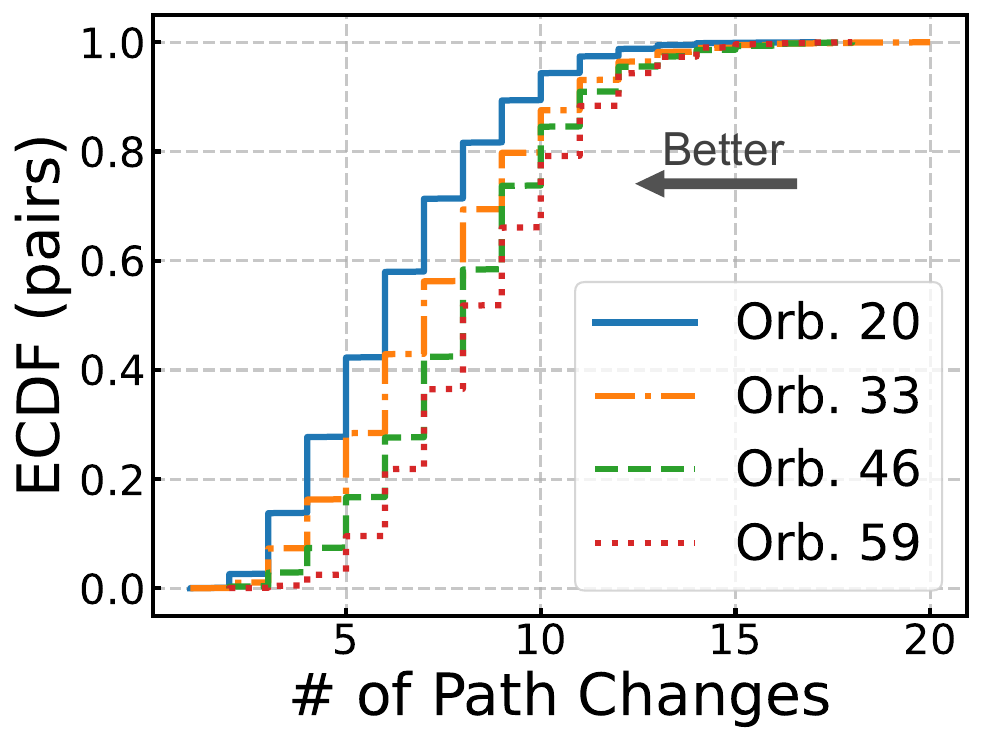}
			\vspace{-0.75cm}
      	\label{path Orbits 7}
		\end{minipage}%
	}%
	\subfigure[26 Sats/Orbits]{
	\vspace{-0.75cm}
		\begin{minipage}[t]{0.24\linewidth}
			\centering
			\includegraphics[scale=.25]{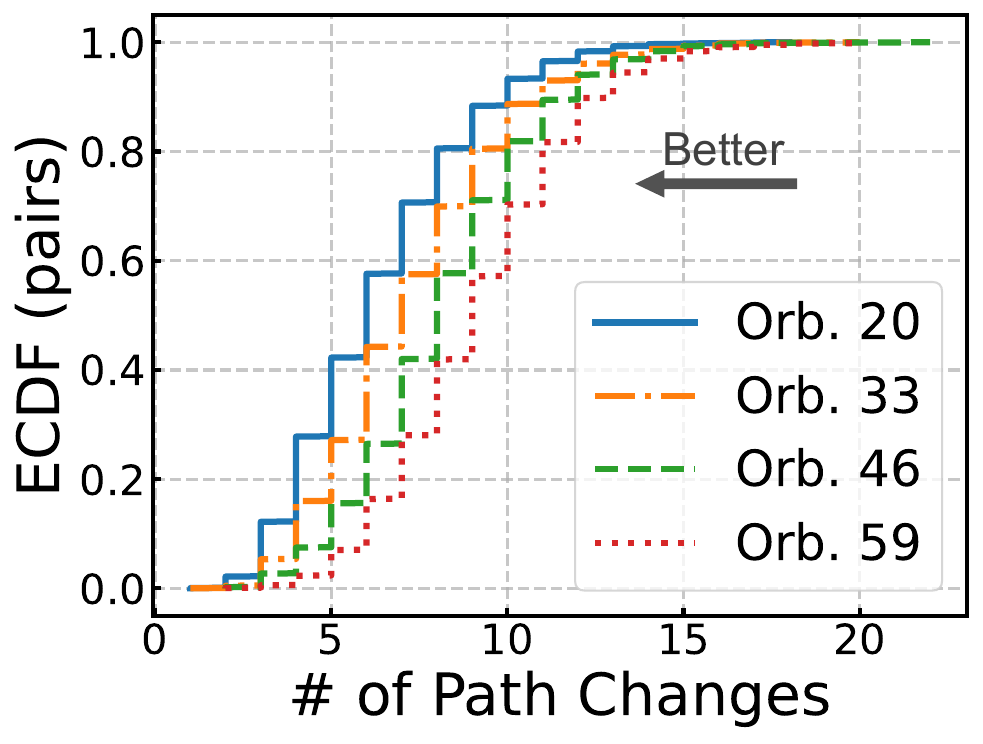}
			\vspace{-0.75cm}
      	\label{path Orbits 8}
		\end{minipage}%
	}%
    \vspace{-0.5cm}
	\caption{The Path Changes ECDFs of changing orbits number with different Sats/Orbits.}
	\vspace{-0.45cm}
   	\label{path Orbits (2)}
\end{figure*}

We further explore performance threshold of number of satellites per orbit. According to Fig.~\ref{RTT Orbits (2)}, Fig.~\ref{slowdown RTT Orbits (2)}, Fig.~\ref{Min RTT Orbits (2)}, Fig.~\ref{path Orbits (2)}, the improvement of increasing number of satellites per orbit converges at 22 Satellites per orbit. When exceeding 22 satellites, the performance does not greatly improve.

\newpage

\section{Endpoints pairs and Inclination}
\label{Appendix:geographic angle}

\begin{figure}[!htp] %
  \centering 
  \includegraphics[width=0.8\linewidth]{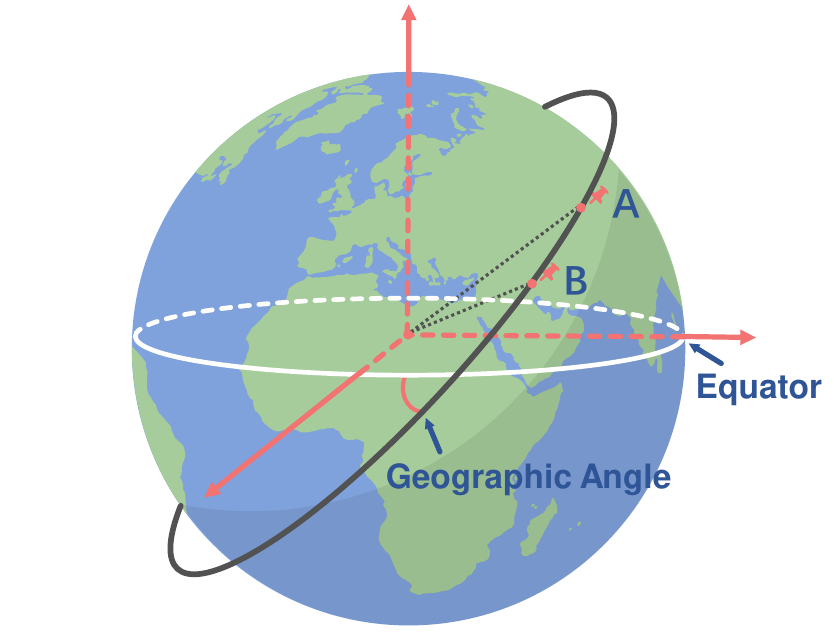}  %
  \caption{\small Geographic Angle between City A and B.} 
  \label{Geographic Angle}                 
\end{figure}

\subsection{Geographic Angle (Bearing Angle)}
\label{appendix: Geographic Angle (Bearing Angle)}

Illustrated by Fig.~\ref{Geographic Angle}, the Geographic Angle, also known as the Bearing angle, is a measure of the angle between the plane that contains two geographic locations (such as cities A and B) and the center of the Earth, and the plane that represents the Earth's equator. It is a fundamental concept for understanding and describing the relative positions of locations on the Earth's surface, which provides information about the orientation or direction from one point to another on the Earth's surface, with a range of 0 to 90 degrees.

\subsection{Visualization of User Endpoints}
\label{appendix: Visualization of User Endpoints}

As mentioned in~\S~\ref{subsec:Different Geographic Angle Ground Station Implementation}, we divided the endpoints pairs into nine groups based on their geographic angle. The visualization of these nine groups and the positions of the top 100 GDP cities on the earth is shown in Fig.~\ref{Vis City pairs}.

\subsection{Geodesic Slowdown}
\label{appendix: Geodesic Slowdown}
\begin{figure*}[tp]
	\subfigure[0-10° Endpoints pairs.]{
		\begin{minipage}[t]{0.33\linewidth}
			\centering
			\includegraphics[scale=.26]{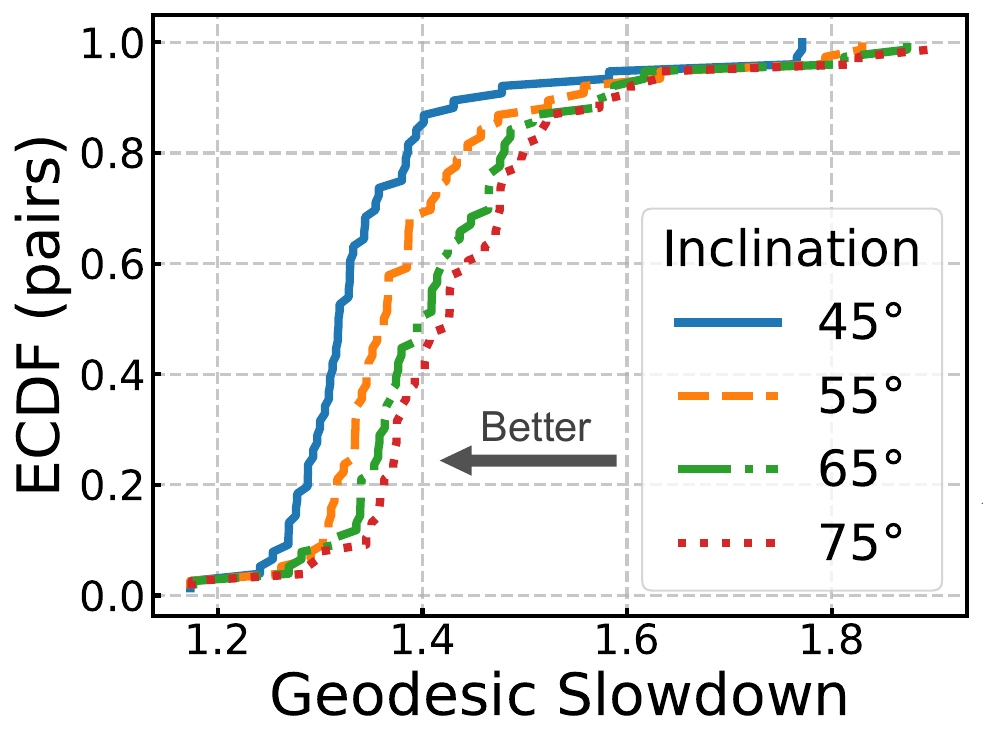}
                \label{appendix:inclination geodesic slowdown 1}
		\end{minipage}%
	}%
	\subfigure[10-20° Endpoints pairs.]{
		\begin{minipage}[t]{0.33\linewidth}
			\centering
			\includegraphics[scale=.26]{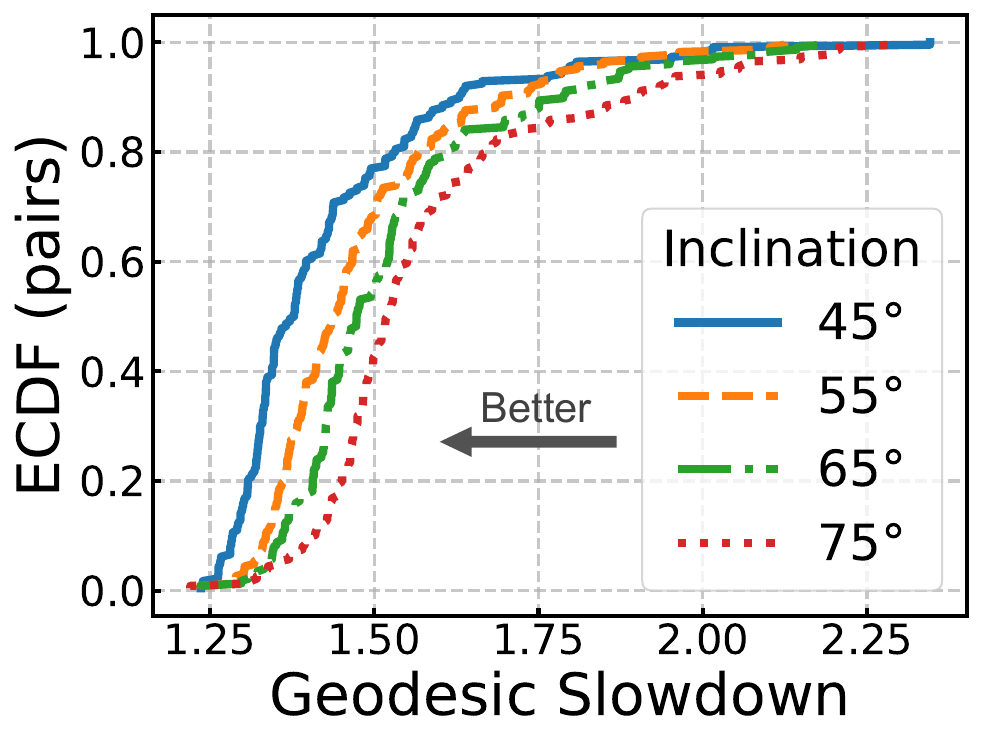}
                \label{appendix:inclination geodesic slowdown 2}
		\end{minipage}%
	}%
	\subfigure[20-30° Endpoints pairs.]{
		\begin{minipage}[t]{0.33\linewidth}
			\centering
			\includegraphics[scale=.26]{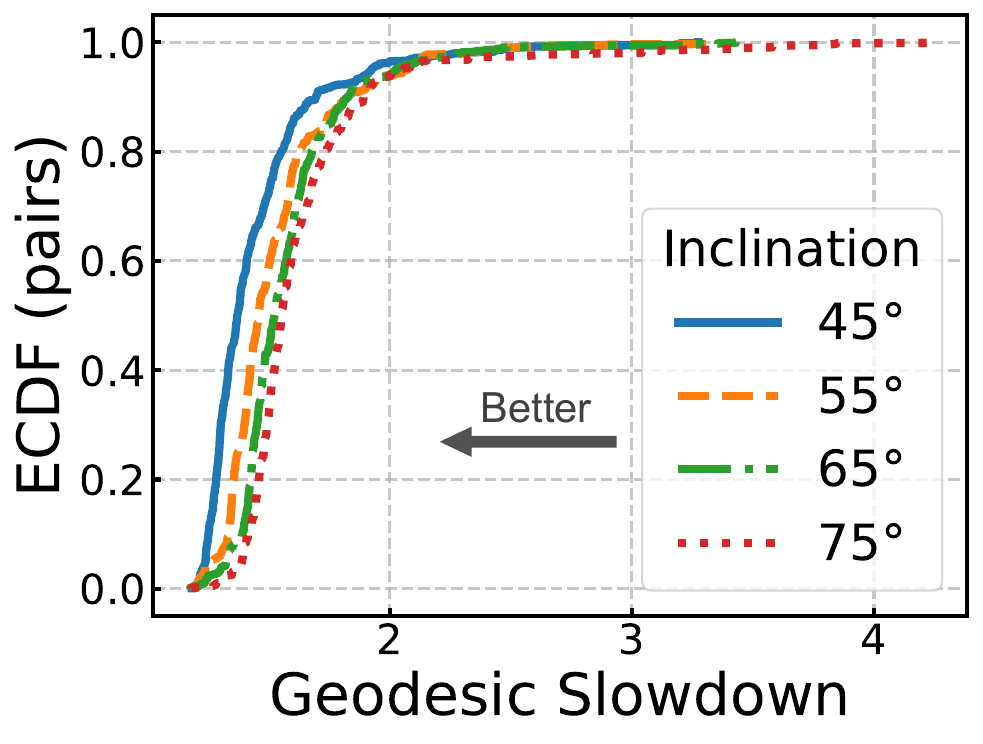}
                \label{appendix:inclination geodesic slowdown 3}
		\end{minipage}%
	}%

	\subfigure[30-40° Endpoints pairs.]{
		\begin{minipage}[t]{0.33\linewidth}
			\centering
			\includegraphics[scale=.26]{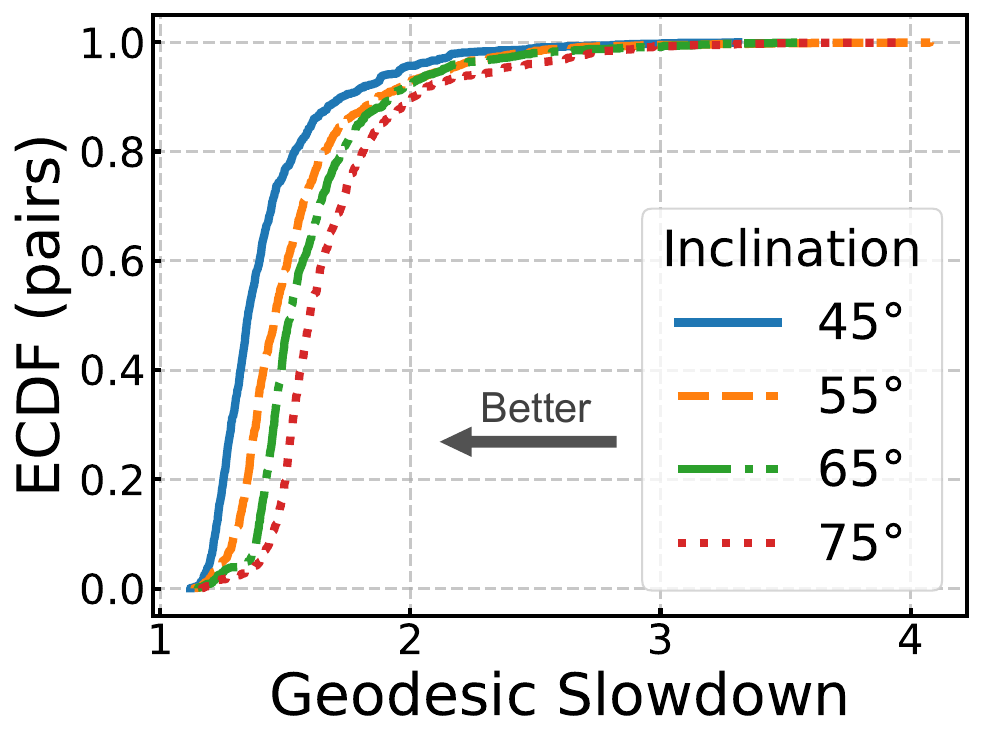}
                \label{appendix:inclination geodesic slowdown 4}
		\end{minipage}%
	}%
	\subfigure[40-50° Endpoints pairs.]{
		\begin{minipage}[t]{0.33\linewidth}
			\centering
			\includegraphics[scale=.26]{figures/City_pairs_and_Inclination/40_50_city_pairs_rtt_max_to_geodesic_slowdown.pdf}
                \label{appendix:inclination geodesic slowdown 5}
		\end{minipage}%
	}%
	\subfigure[50-60° Endpoints pairs.]{
		\begin{minipage}[t]{0.33\linewidth}
			\centering
			\includegraphics[scale=.26]{figures/City_pairs_and_Inclination/50_60_city_pairs_rtt_max_to_geodesic_slowdown.pdf}
                \label{appendix:inclination geodesic slowdown 6}
		\end{minipage}%
	}%

	\subfigure[60-70° Endpoints pairs.]{
		\begin{minipage}[t]{0.33\linewidth}
			\centering
			\includegraphics[scale=.26]{figures/City_pairs_and_Inclination/60_70_city_pairs_rtt_max_to_geodesic_slowdown.pdf}
                \label{appendix:inclination geodesic slowdown 7}
		\end{minipage}%
	}%
	\subfigure[70-80° Endpoints pairs.]{
		\begin{minipage}[t]{0.33\linewidth}
			\centering
			\includegraphics[scale=.26]{figures/City_pairs_and_Inclination/70_80_city_pairs_rtt_max_to_geodesic_slowdown.pdf}
                \label{appendix:inclination geodesic slowdown 8}
		\end{minipage}%
	}%
	\subfigure[80-90° Endpoints pairs.]{
		\begin{minipage}[t]{0.33\linewidth}
			\centering
			\includegraphics[scale=.26]{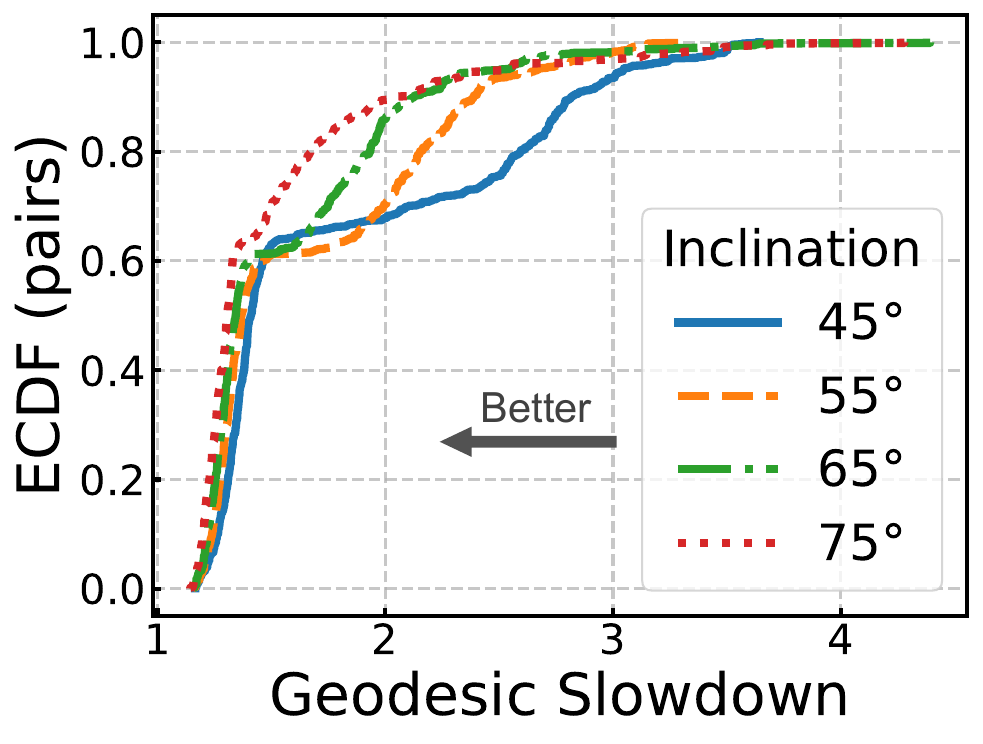}
		\end{minipage}%
	}%
	\caption{Evaluation of Geodesic Slowdown of nine groups of endpoint pairs with varying Inclination of orbits (1).  For all curves, lower values indicate better performance. Long tails indicate outliers with poor performance.}
	\vspace{-0.25cm}
	\label{appendix:inclination geodesic slowdown}
\end{figure*}

Fig.~\ref{appendix:inclination geodesic slowdown} shows the geodesic slowdown of nine endpoints pairs groups from 0° to 90° under inclinations = 45°, 55°, 65°, and 75° LEO mega-constellations shells.

We can observe that the LEO mega-constellations' Inclination aligned with the geographic angle of the endpoints pairs performs better. Shell with a 45° inclination has the lowest slowdown over endpoints pairs with a 0-50° geographic angle. The rest of the shells also perform the best in the endpoints pairs with the closest geographic angle to Inclination.

\begin{figure*}
	\subfigure[Inclination 45°.]{
		\begin{minipage}[t]{0.49\linewidth}
			\centering
			\includegraphics[scale=.25]{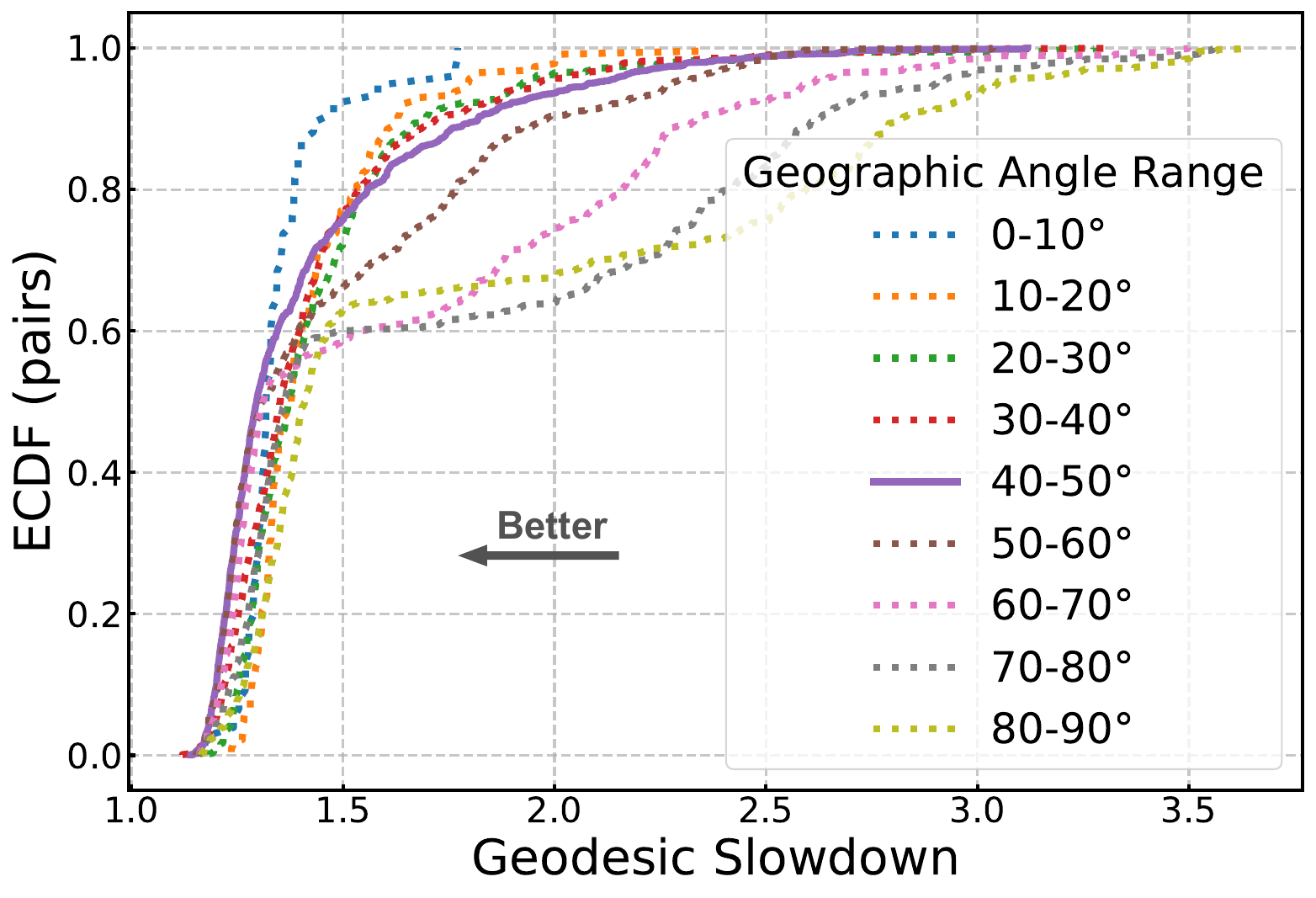}
                \label{9 inclination 45}
		\end{minipage}%
	}%
	\subfigure[Inclination 55°.]{
		\begin{minipage}[t]{0.49\linewidth}
			\centering
			\includegraphics[scale=.25]{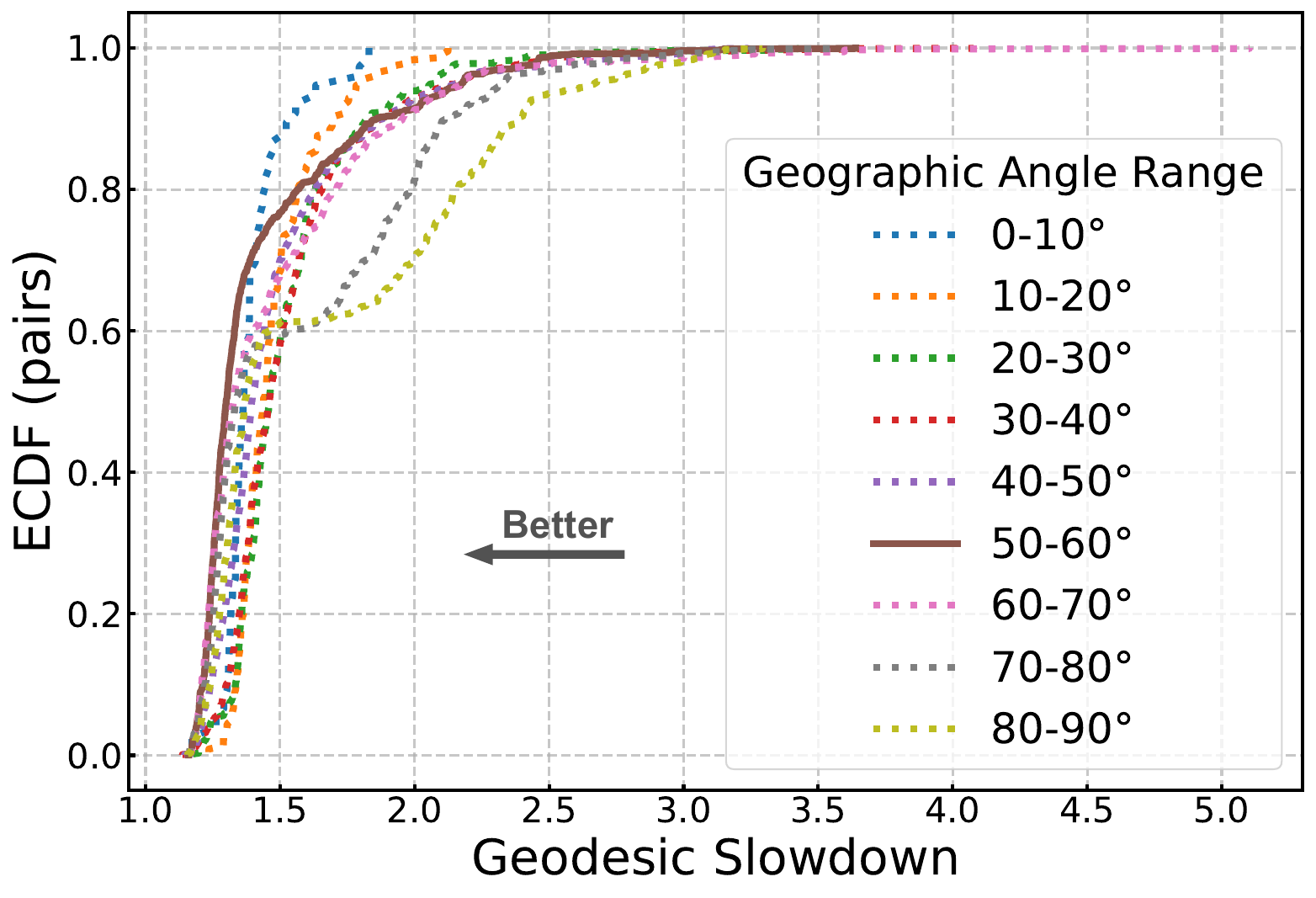}
                \label{9 inclination 55}
		\end{minipage}%
	}%
 
	\subfigure[Inclination 65°.]{
		\begin{minipage}[t]{0.49\linewidth}
			\centering
			\includegraphics[scale=.25]{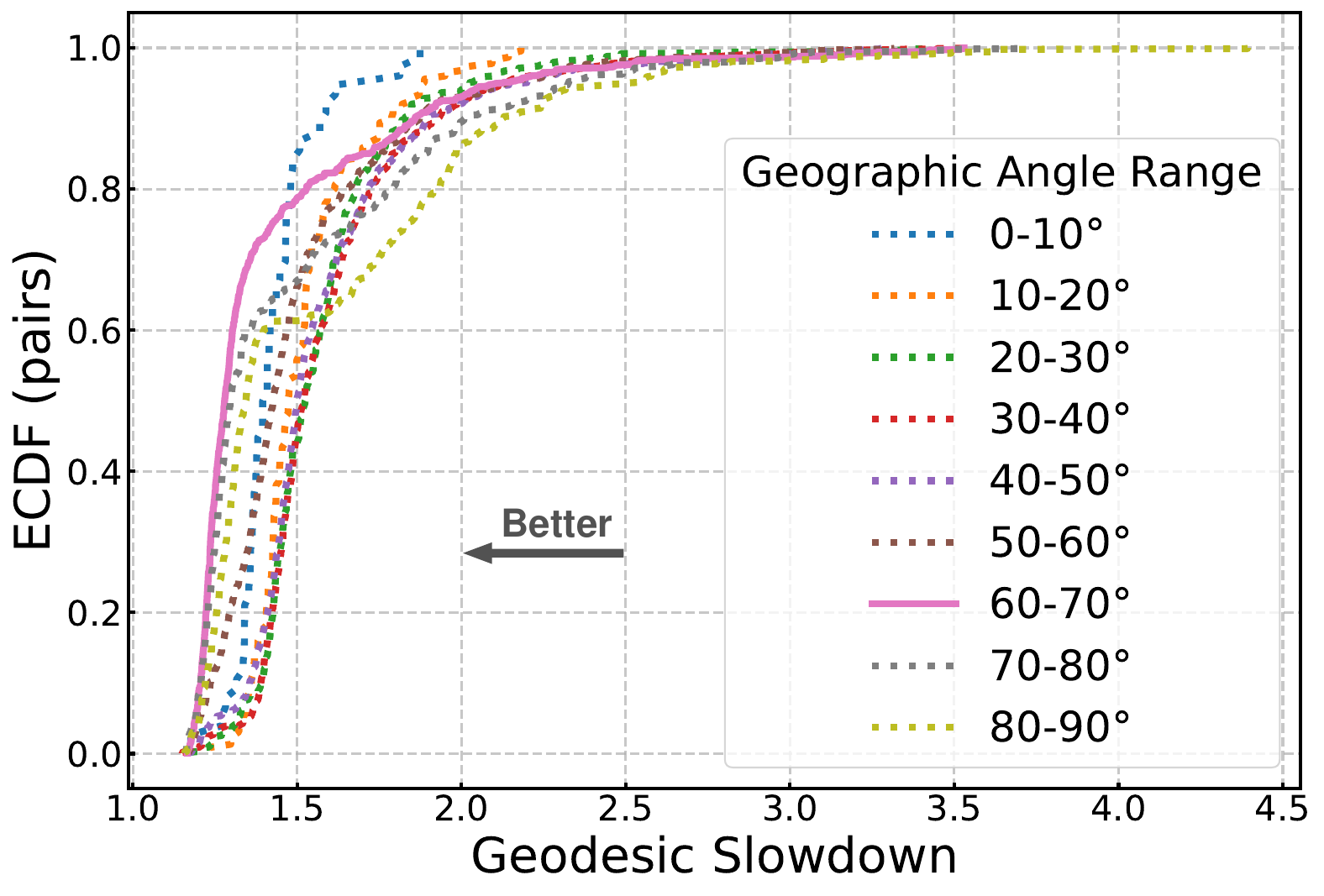}
                \label{9 inclination 65}
		\end{minipage}%
	}%
	\subfigure[Inclination 75°.]{
		\begin{minipage}[t]{0.49\linewidth}
			\centering
			\includegraphics[scale=.25]{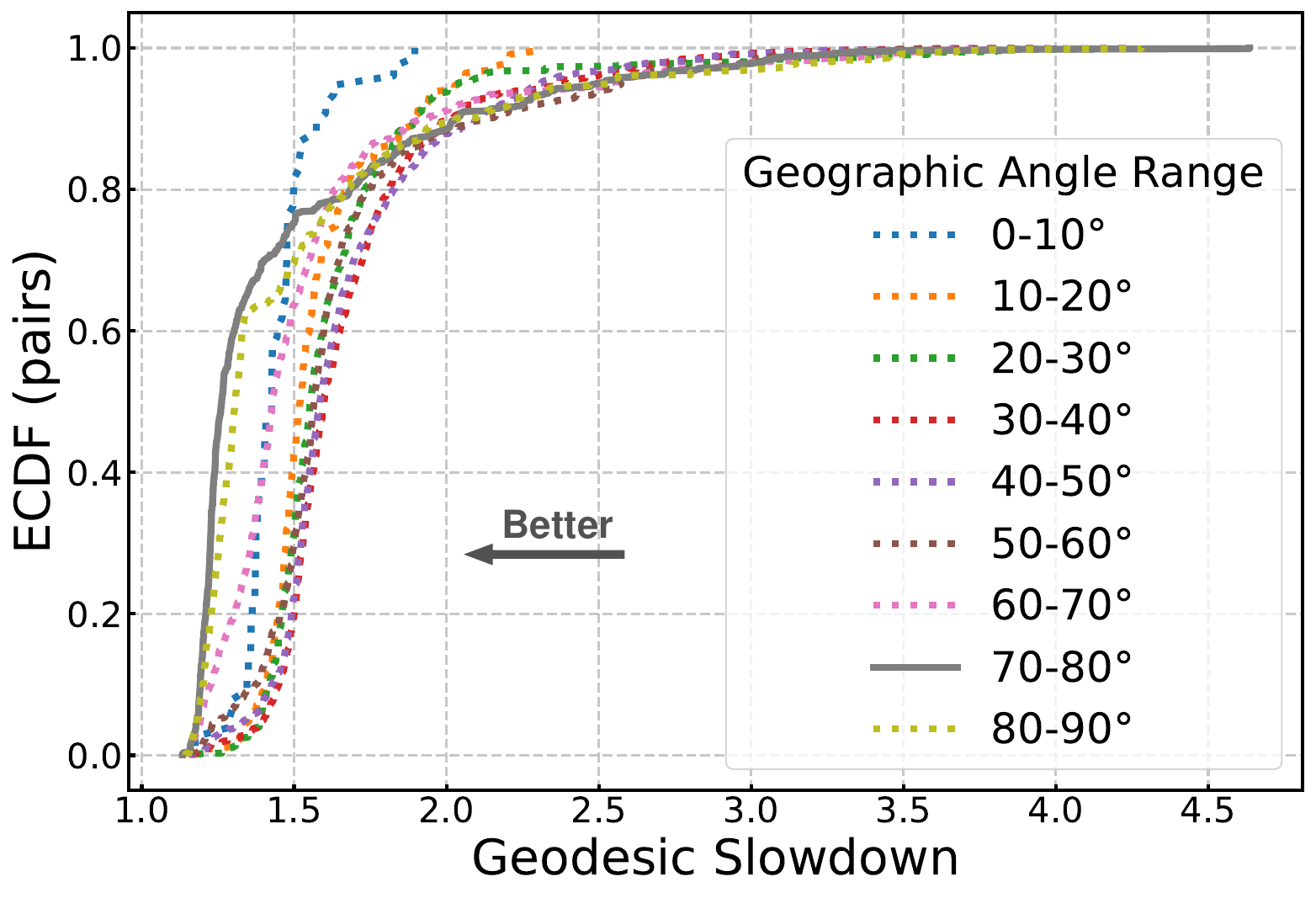}
                \label{9 inclination 75}
		\end{minipage}%
	}%

	\caption{Evaluation of Geodesic Slowdown of nine groups of Endpoints pairs under LEO mega-constellations with Inclination 45-75° (2).  For all curves, lower values indicate better performance. Long tails indicate outliers with poor performance.}
	\label{appendix:inclination all pairs}
\end{figure*}

Fig.~\ref{appendix:inclination all pairs} shows the geodesic slowdown of nine groups of user points pairs in constellations with different Inclinations. We can see that solid line in each graph and conclude that for the median value of Geodesic Slowdown, the endpoints pairs aligned most with the inclination angle of LEO mega-constellations have the least value. 

Therefore, the optimal range of the LEO mega-constellation should be more aligned with the endpoints pairs' geographic angle. 

\subsection{Average Hop}
\label{appendix: Average Hop}

\begin{figure*}[ht]
	\centering
	\subfigure[Inclination 65°]{
	\vspace{-0.75cm}
		\begin{minipage}[t]{\linewidth}
			\centering
			\includegraphics[scale=.30]{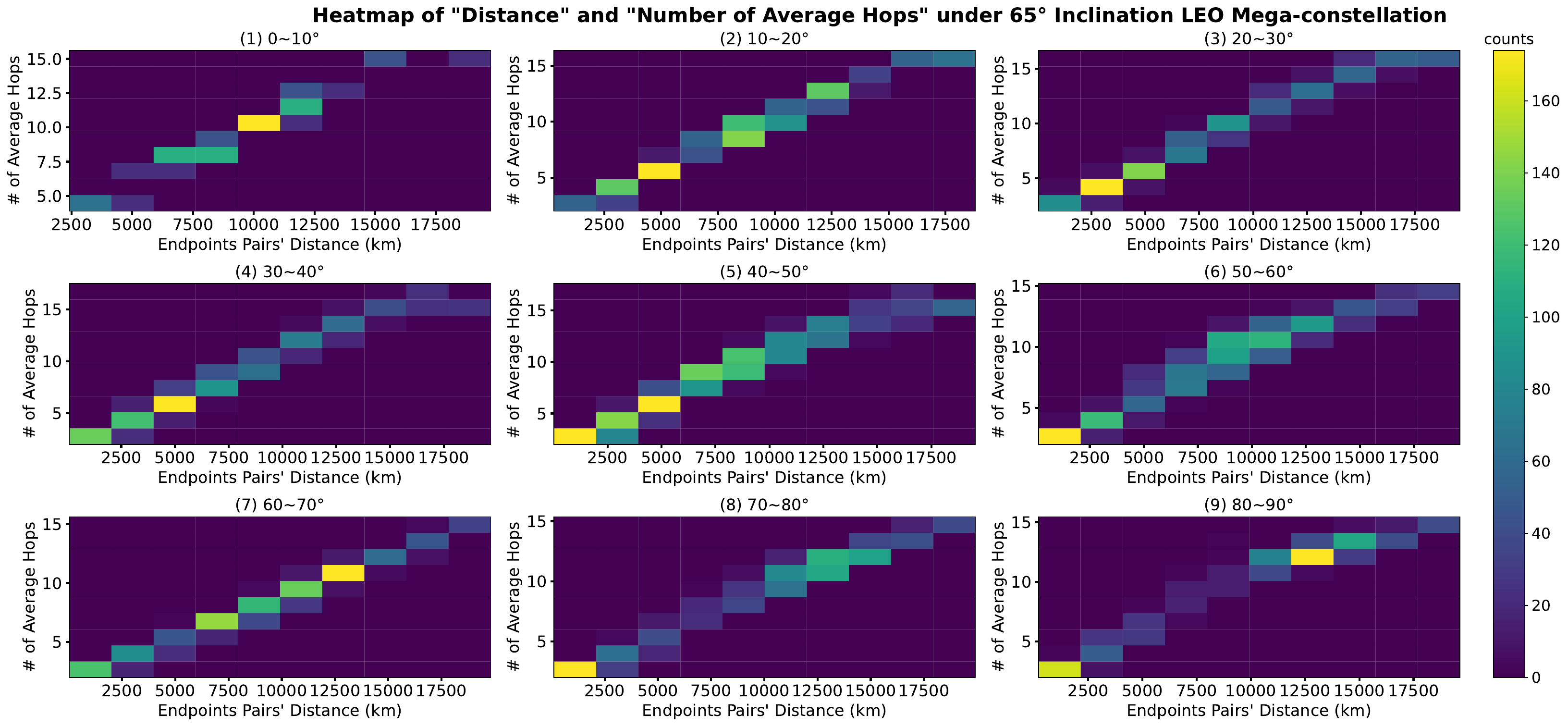}
			\vspace{-0.75cm}
   \label{65 75 heatmap 1}
		\end{minipage}%
	}%
 
	\subfigure[Inclination 75°]{
	\vspace{-0.75cm}
		\begin{minipage}[t]{\linewidth}
			\centering
			\includegraphics[scale=.30]{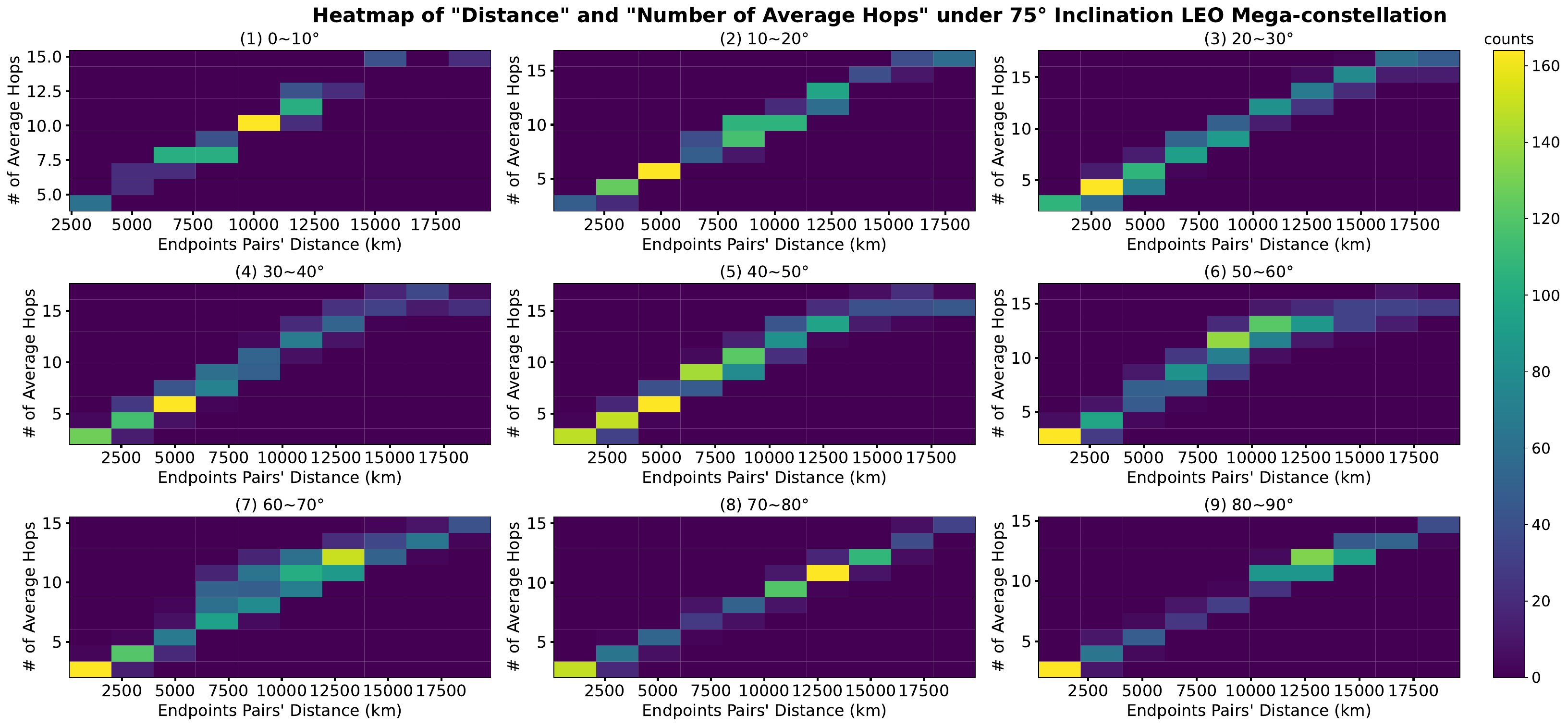}
			\vspace{-0.75cm}
   \label{65 75 heatmap 2}
		\end{minipage}%
	}%
    \vspace{-0.5cm}
	\caption{Heatmap of "Distance" and "Number of Average Hops" under 65° and 75° Inclination LEO Mega-constellation.}
	\vspace{-0.45cm}
	\label{65 75 heatmap}
\end{figure*}

As mentioned in ~\S~\ref{subsubsec:IURD}, we plot the heatmaps according to the distance of the endpoints in each Geographic Angles group and the corresponding number of average hops, showing inclination changes Intuitively. We only put the result of Inclination 45° and Inclination 55° in the main text. Here we also plot the heatmaps of Inclination 45°, 55°, 65° and 75°, shown in Fig.~\ref{heatmap} and Fig.~\ref{65 75 heatmap}.

Let us call endpoints within the distance interval of 6000-8000km and a geographic angle of 40-50° (i.e., the fourth column in the heatmap (5) at the center of each figure) as bin A and endpoints within the distance interval of 8000-10000km and a geographic angle of 50-60° (i.e., the fifth column in the heatmap (6) of each figure) as  bin B. Under the inclination of 45° (Fig.~\ref{heatmap 1}), more than 160 pairs of bin A have an average hop count in the range 6-7, with only a tiny fraction with an average hop count of 5-6 or 7-8. However, under the inclination of 55° (Fig.~\ref{heatmap 2}), only about 60 pairs in bin A have an average hop count in the range of 6-7. The remaining have an average hop count in the range of 7-8, thereby increasing the overall average hop count. In contrast, for bin B, as we increase the orbital angle (Fig.~\ref{heatmap 1} vs. Fig.~\ref{heatmap 2}), the average hop count decreases. This is because the orbital angle of 55° in Fig.~\ref{heatmap 2} is closer to the geographic angle for bin B, while the orbital angle of 45° is closer to the geographic angle for bin A. The same pattern holds for all other groups of endpoint pairs. 

As shown in Fig.~\ref{65 75 heatmap 1} and Fig.~\ref{65 75 heatmap 2}, we can observe the changes in average hops of endpoints pairs with different Geographic Angles and different distances. For user points pairs between 60-70°, under the inclination of 65°, more than 160 pairs of endpoints of 12000-14000 km have an average hop of 10-11. Only a tiny percentage‘s average hop is 11-12 or 10. However, under the inclination of 75°, the same endpoints pair with 12000-14000 km and Geographic Angles of 60-70° only has an average hop of 12-13, close to 140 pairs. The rest are distributed in the higher average hops 13-14, and the overall number of average hops is improved.
In contrast, the opposite result is shown for endpoints pairs with Geographic Angles 70-80°. Under the 45° inclination LEO mega-constellation, 12000-14000 km endpoints pairs have more than 120 pairs with an average hop of 12. However, at Inclination 55° one, the average hop of endpoints within the same range decreases significantly, and most of the inclination is concentrated at eleven average hops. The same pattern holds for the other groups of endpoints pairs.

Combined these two experimental results with the result of 
Inclination 45° and 55°, shown in Fig.~\ref{heatmap 1} and Fig.~\ref{heatmap 2}, we can confirm that the congruency between the inclination and distribution of endpoints pairs Geographic Angles is positively correlated with the performance of the average hop number of the LEO mega-constellation. When the LEO mega-constellation’s orbits are more aligned to the lines between endpoints pairs, the route between source and destination will favor the use of in-orbit satellites, avoiding more inter-track zig-zag and bringing less hop count. Therefore, when we design the LEO mega-constellation, adjusting the inclination of LEO mega-constellations as far as possible to make it more consistent with the distribution of endpoints can bring less hop and network delay.

\begin{figure*}[htp]
	\centering  %
	\subfigure[0-10°.]{
		\includegraphics[width=0.16\textwidth]{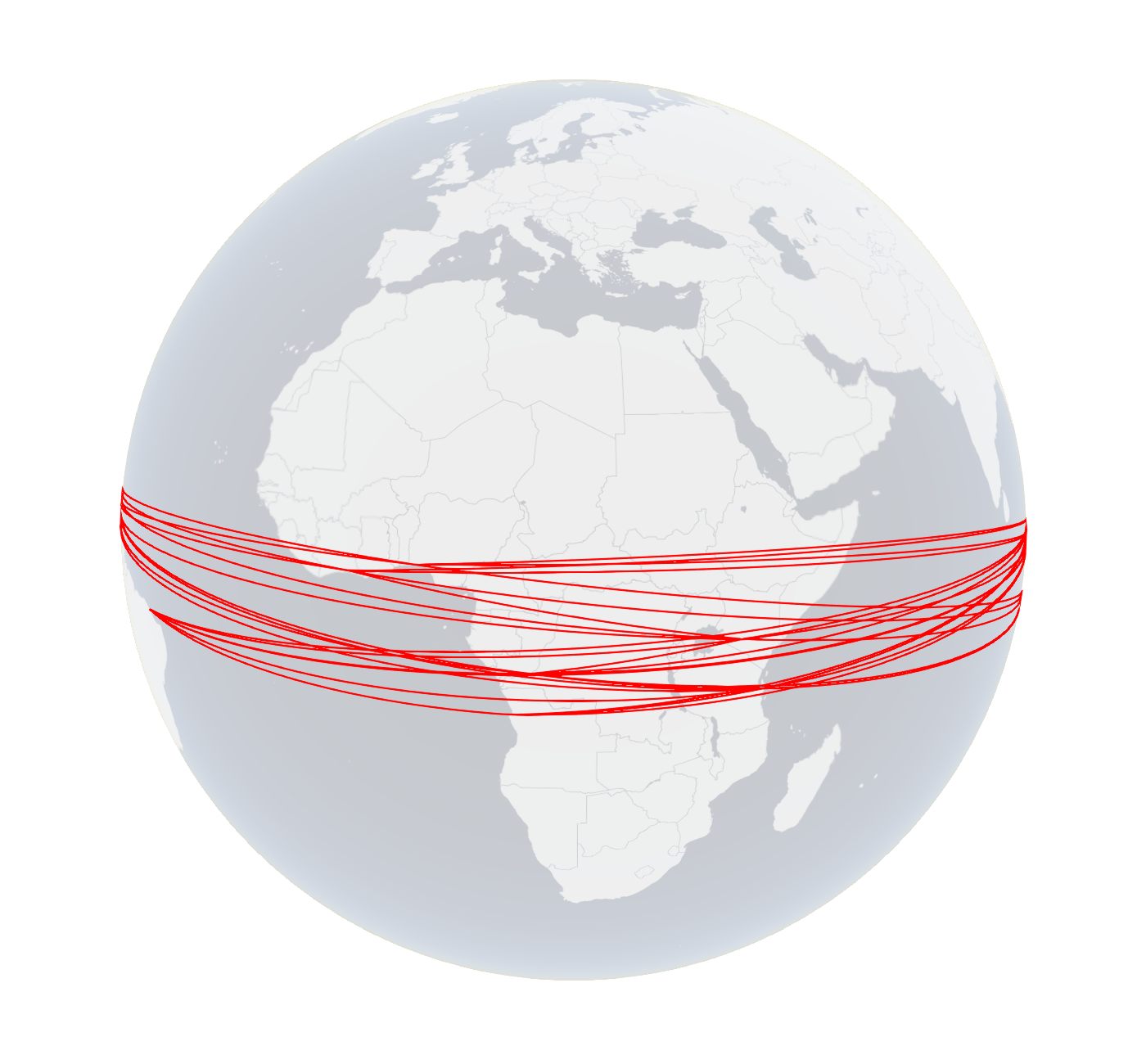}
    }
	\subfigure[10-20°.]{
		\includegraphics[width=0.16\textwidth]{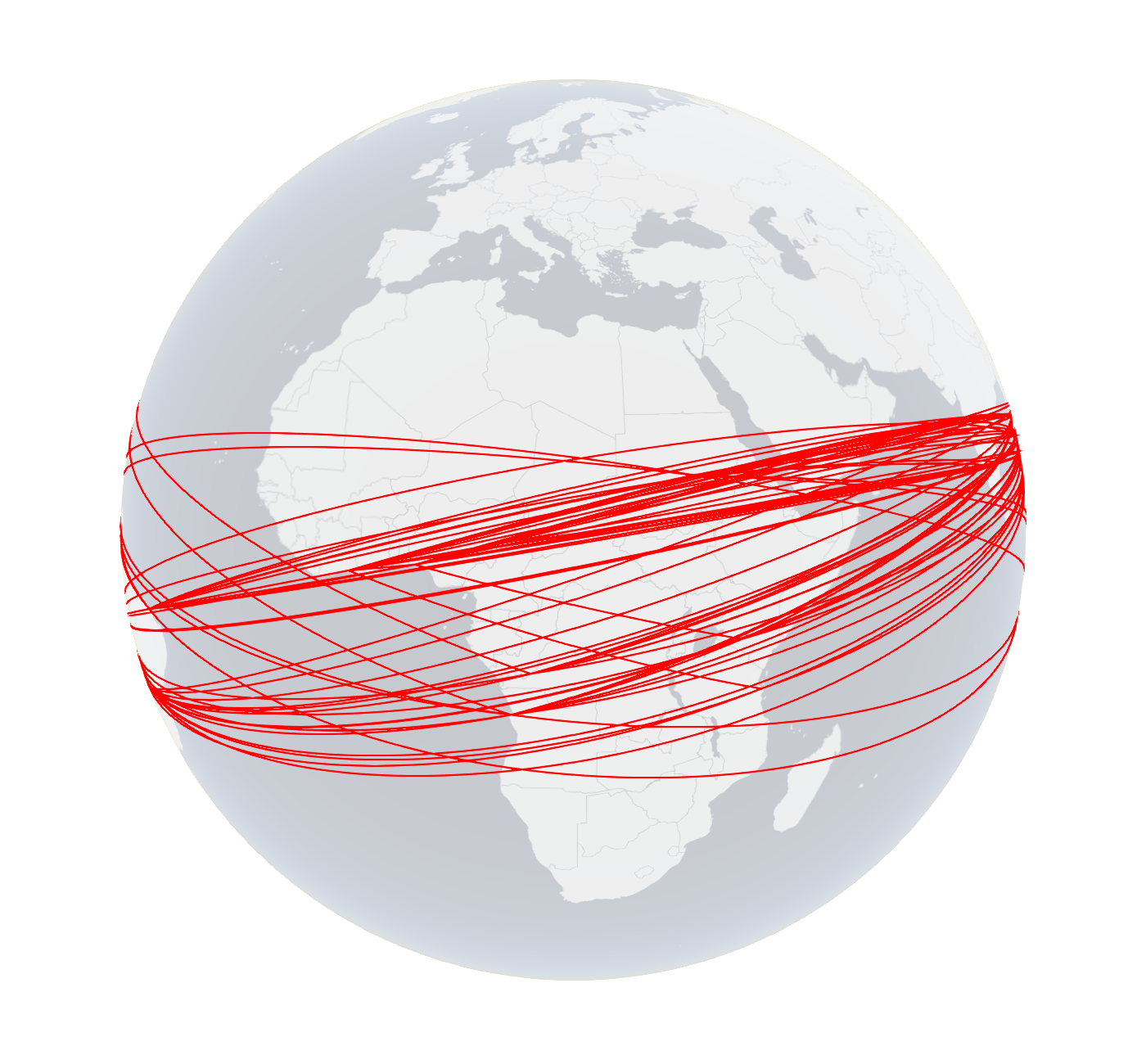}}
	\subfigure[20-30°.]{
		\includegraphics[width=0.16\textwidth]{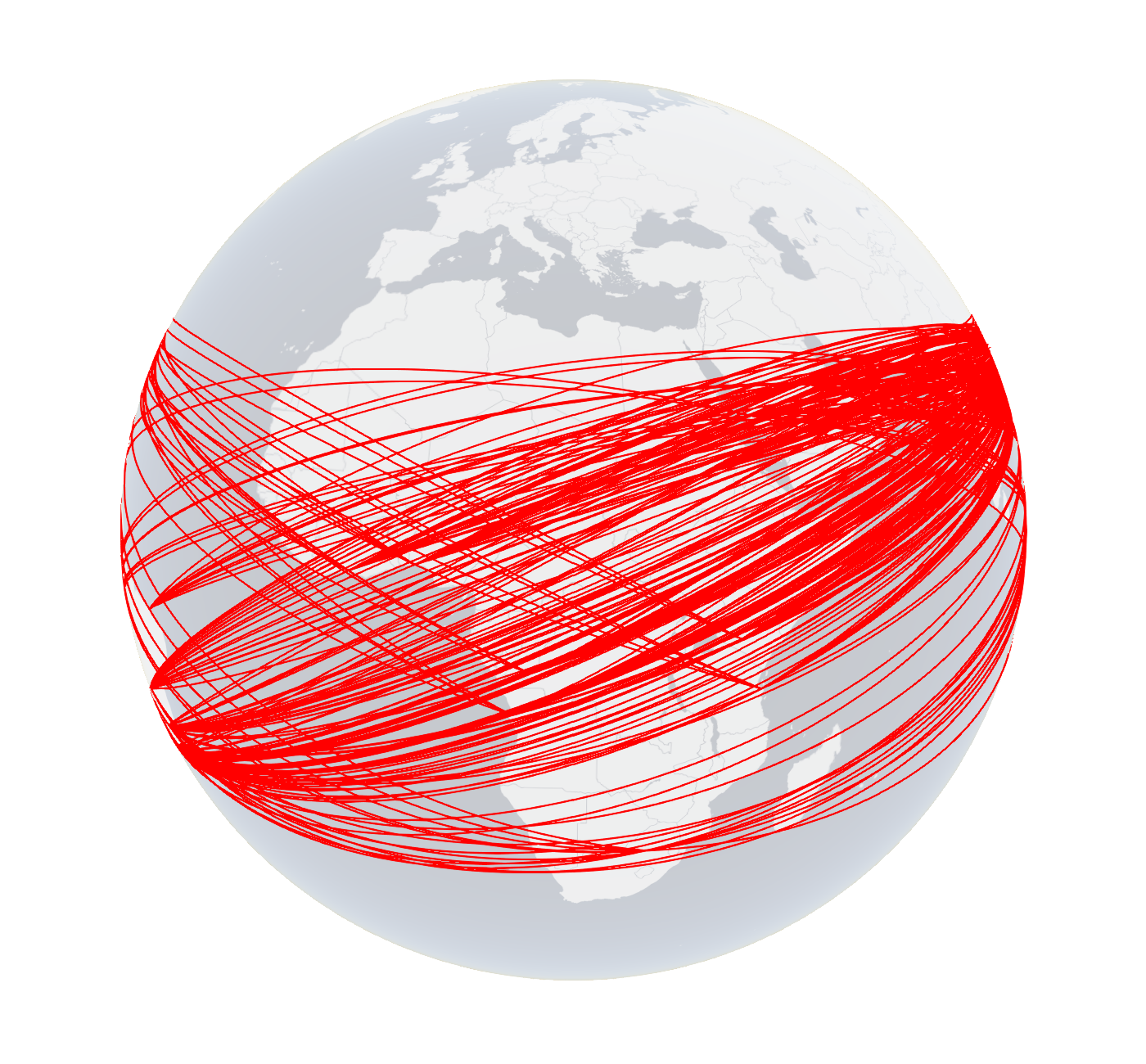}}
	\subfigure[30-40°.]{
		\includegraphics[width=0.16\textwidth]{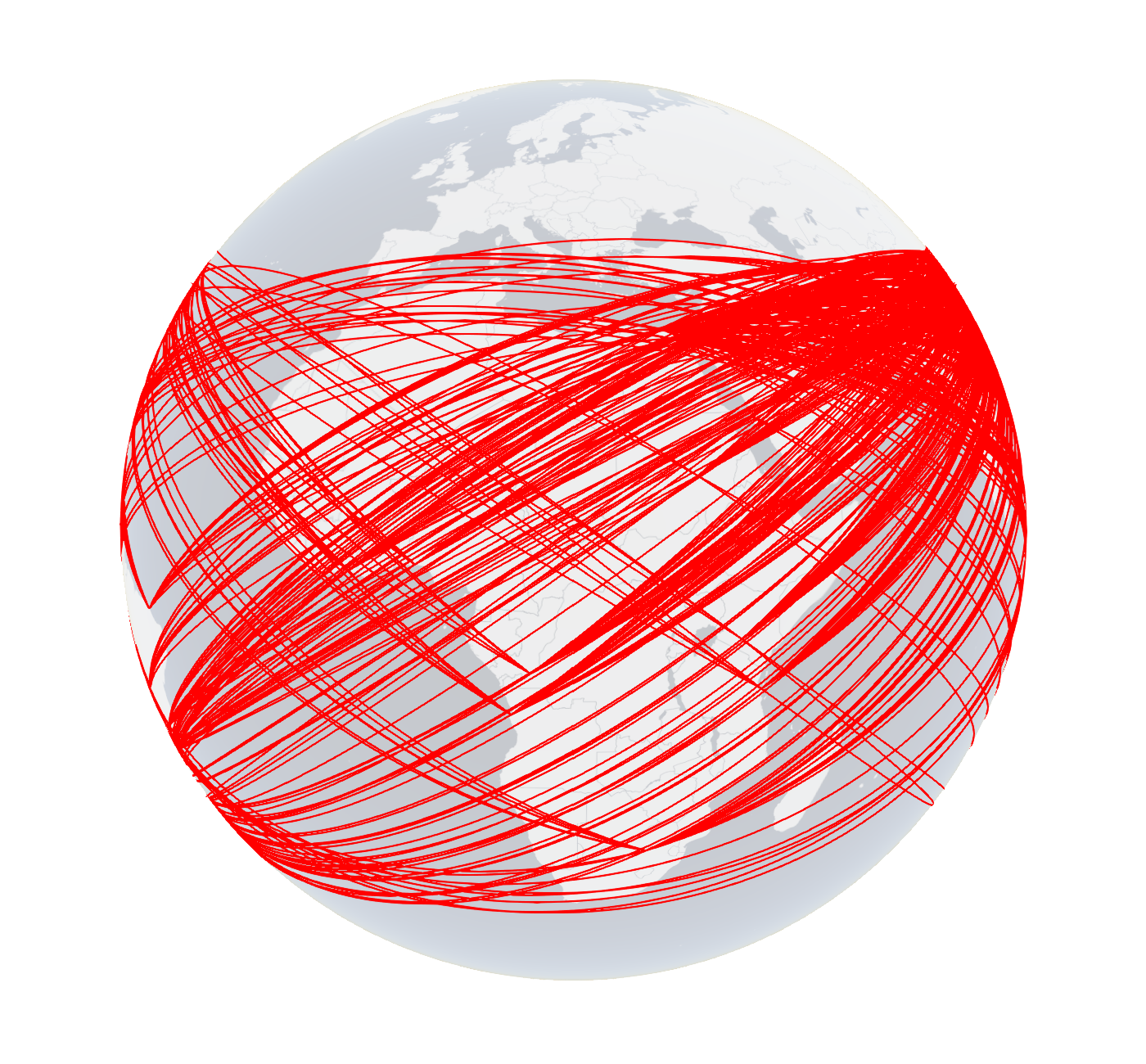}}
    \subfigure[40-50°.]{
		\includegraphics[width=0.16\textwidth]{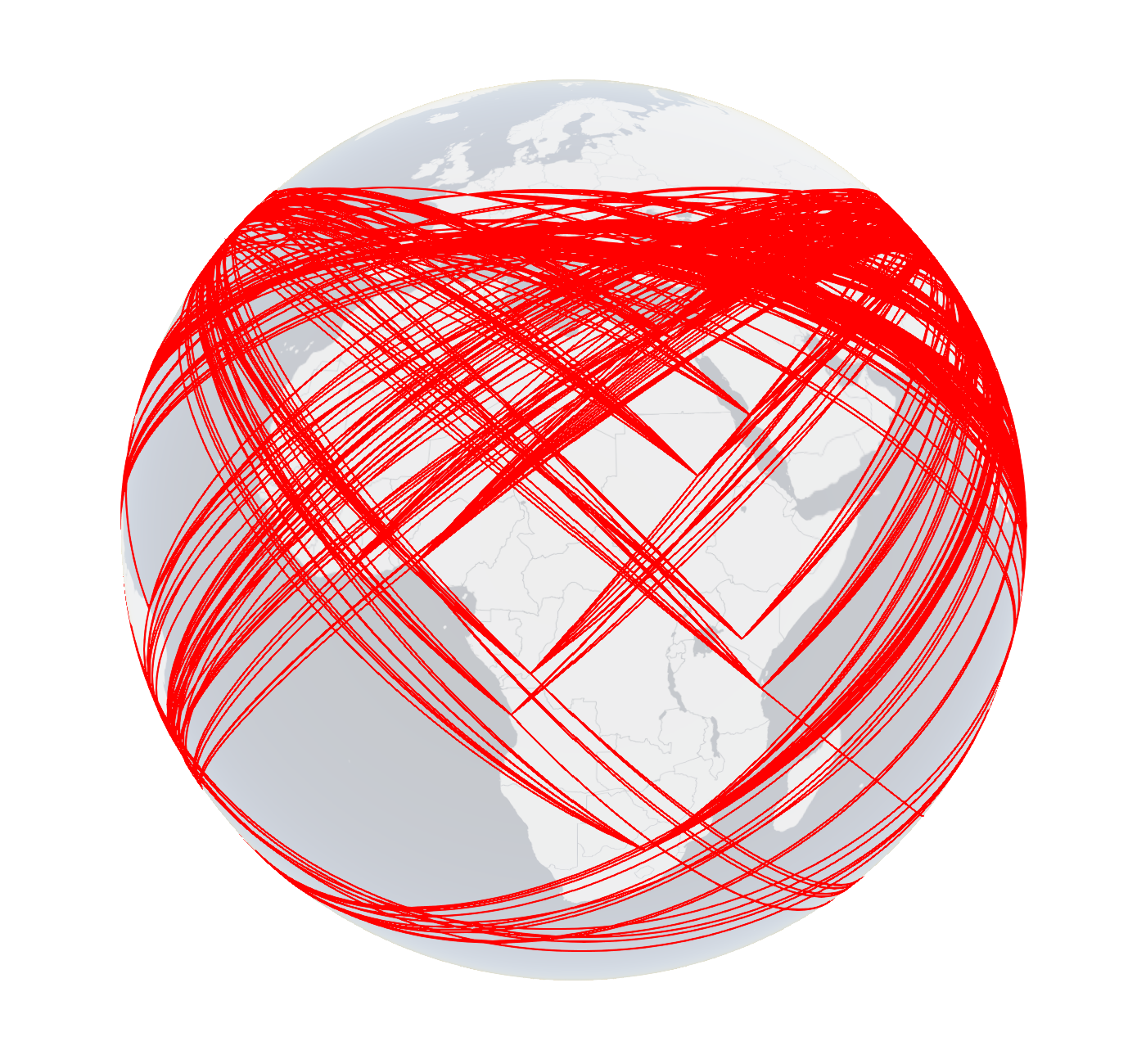}}
    \subfigure[50-60°.]{
		\includegraphics[width=0.16\textwidth]{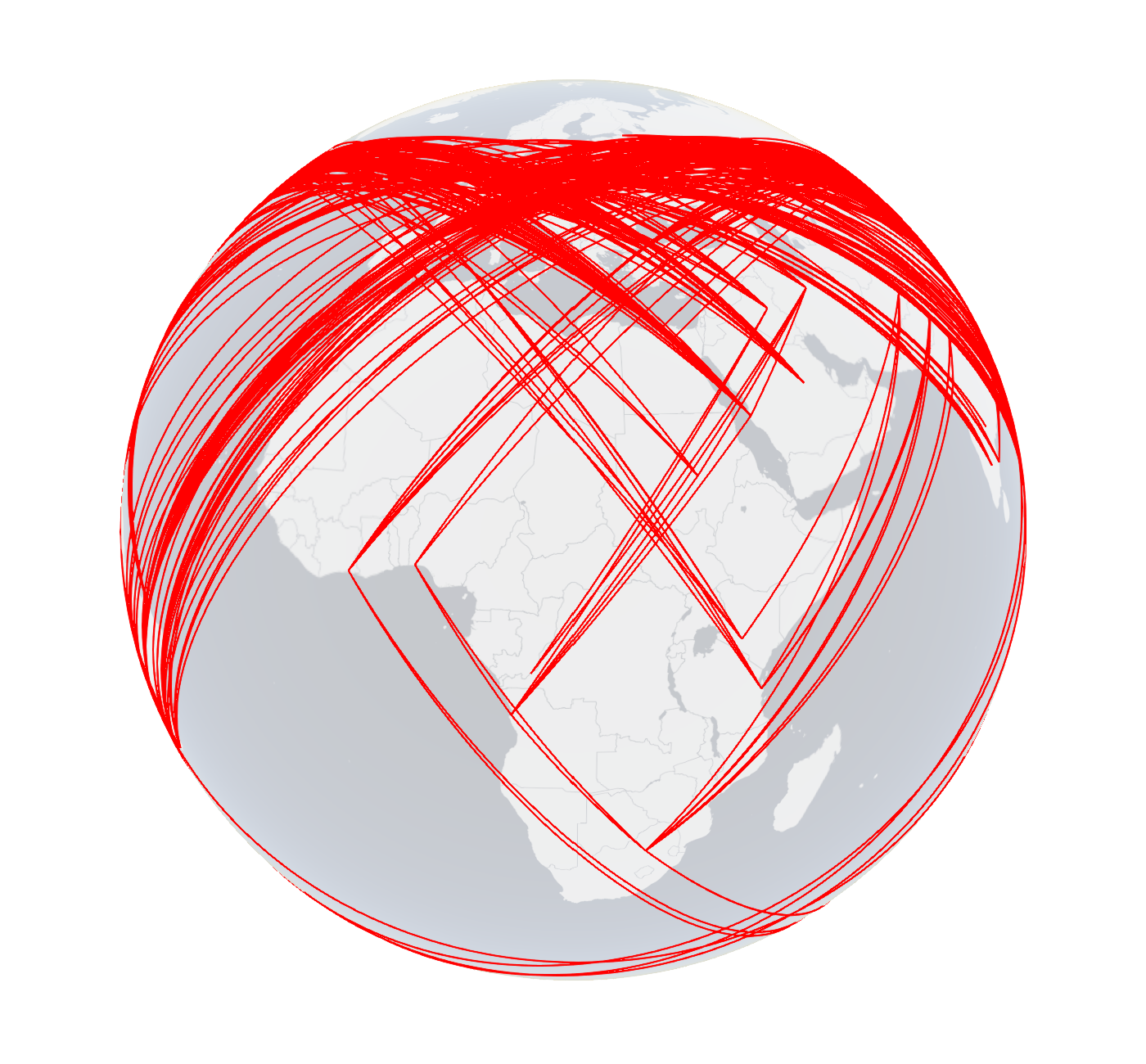}}
	\subfigure[60-70°.]{
		\includegraphics[width=0.16\textwidth]{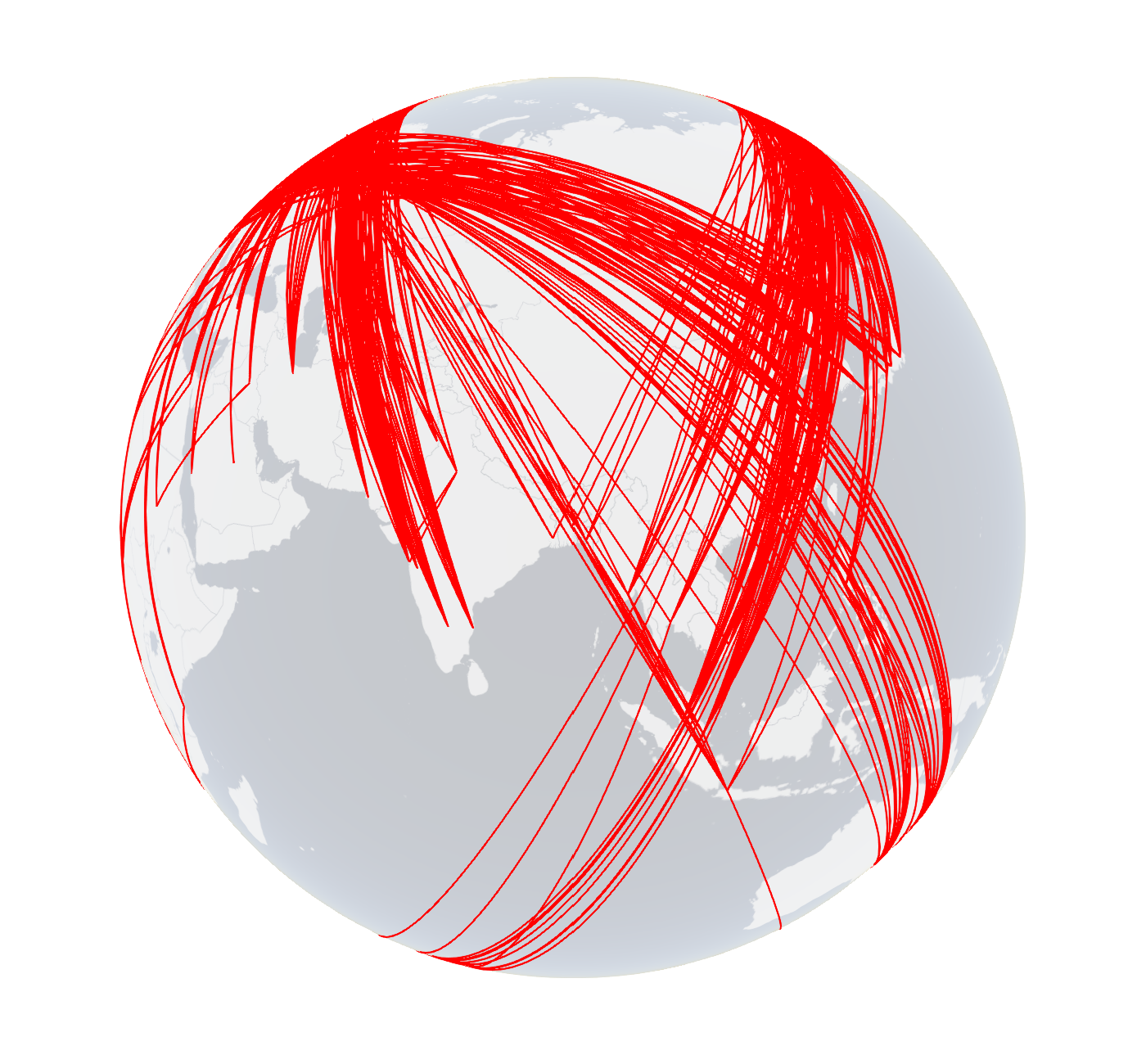}}
	\subfigure[70-80°.]{
		\includegraphics[width=0.16\textwidth]{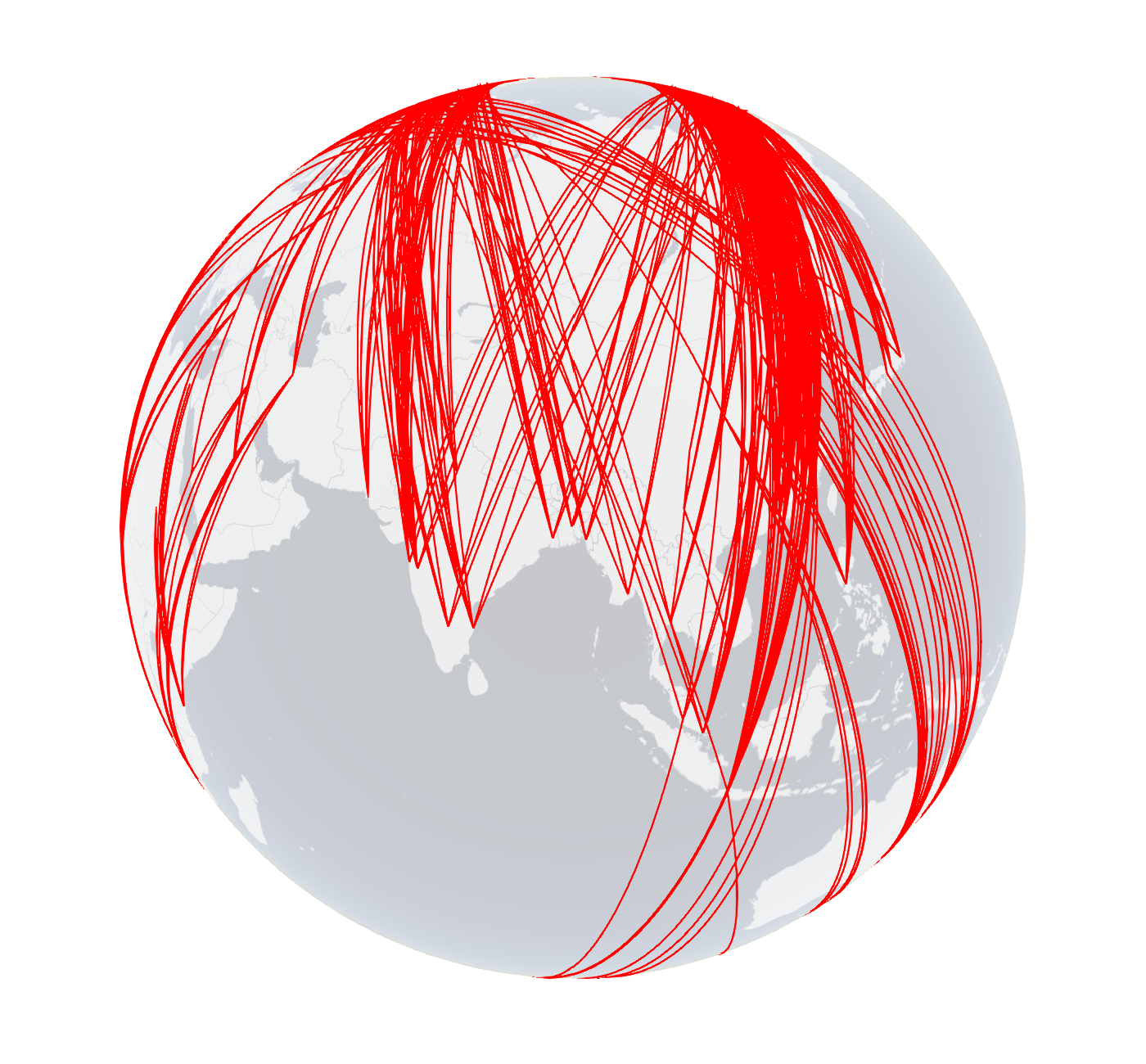}}
	\subfigure[80-90°.]{
		\includegraphics[width=0.16\textwidth]{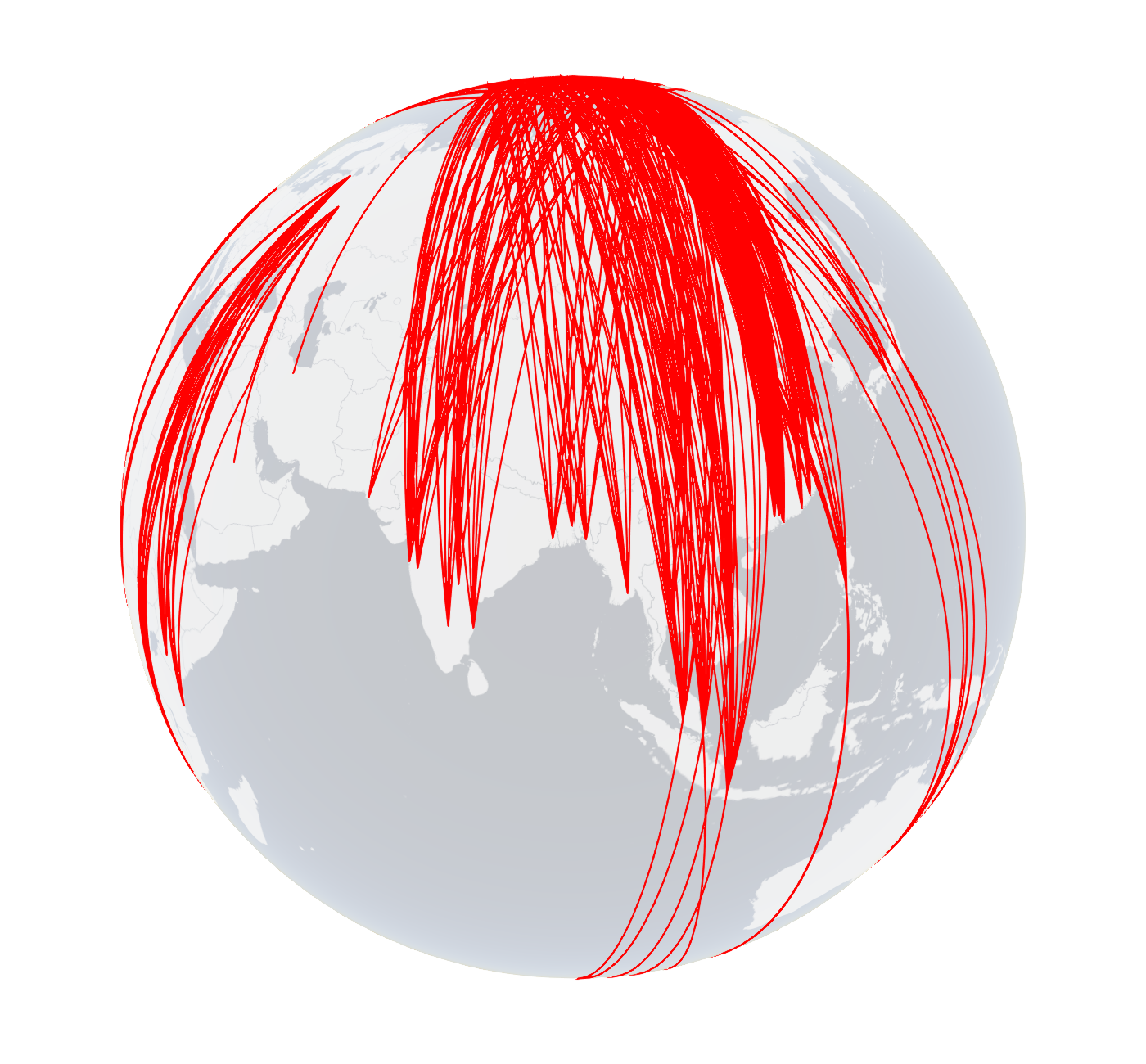}}
    \subfigure[Top 100 GDP Cities.]{
		\includegraphics[width=0.16\textwidth]{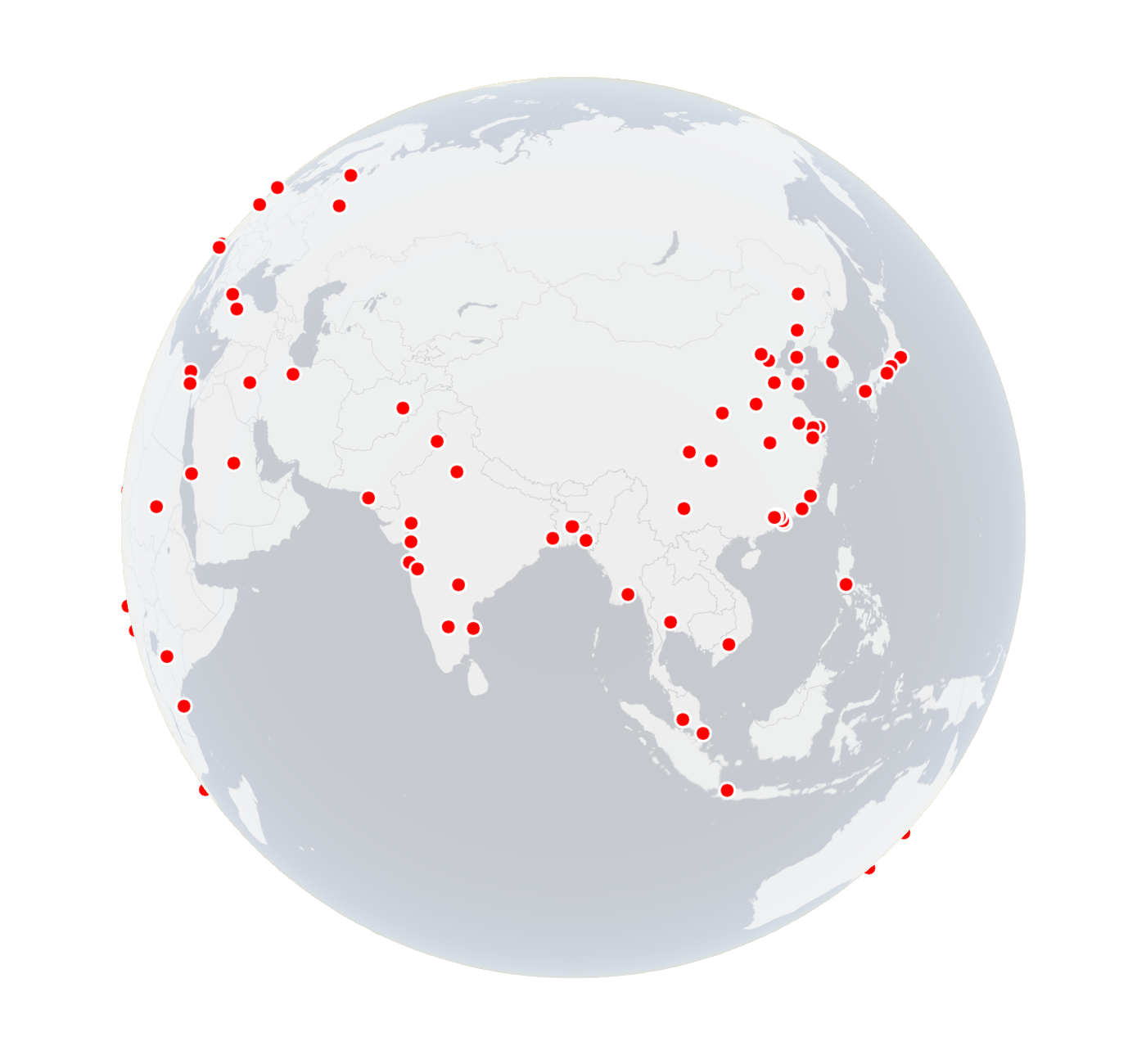}}
	\caption{Visualization of nine groups of endpoints pairs with different Geographic Angle (a)-(i), and the positions of all 100 endpoints (j).}
        \label{Vis City pairs}
\end{figure*}

\begin{figure*}[ht]
	\centering
	\subfigure{
        \ContinuedFloat
		\begin{minipage}[t]{\linewidth}
			\centering
			\includegraphics[scale=.35]{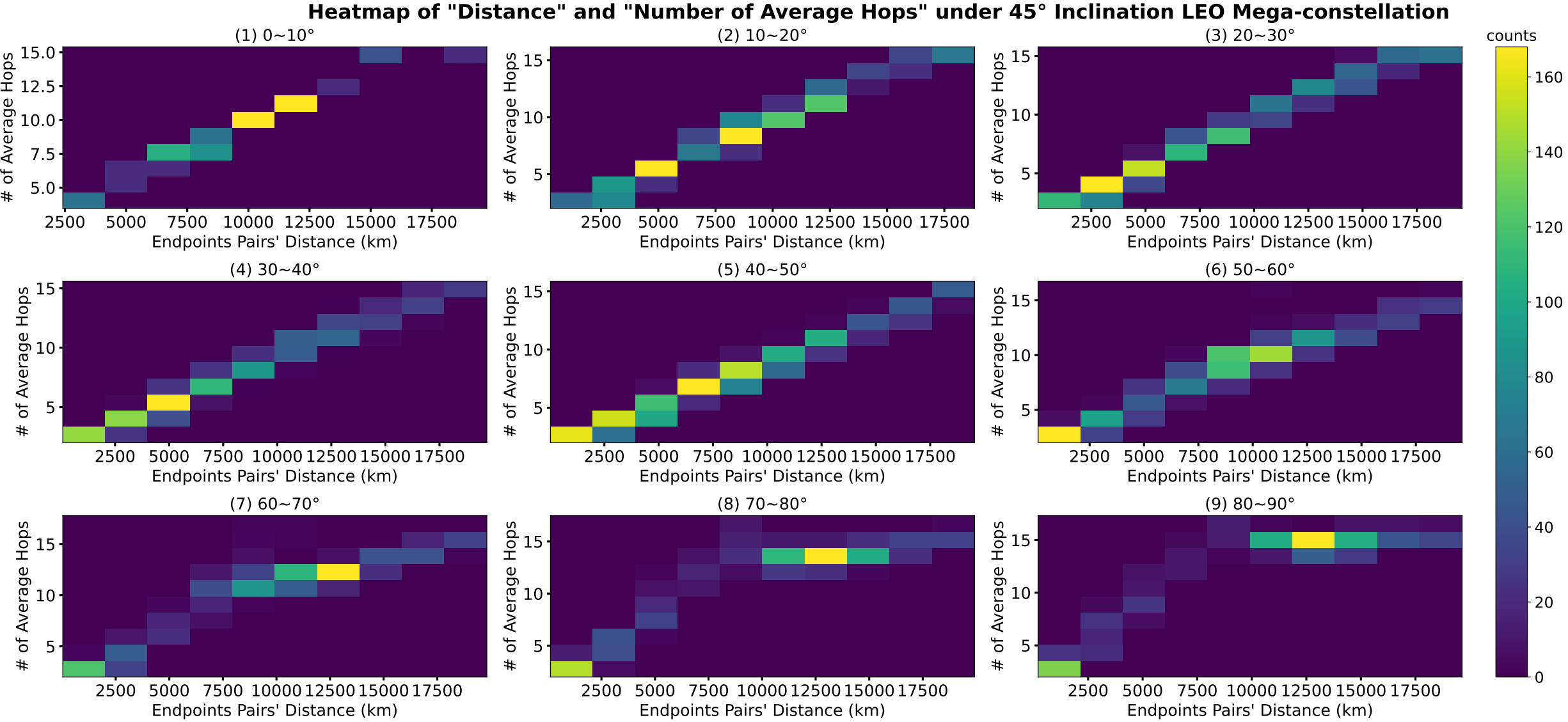}
			\vspace{-0.25cm}
  \label{heatmap 1} 
            \caption{(a) Satellite Orbit Inclination 45° }

		\end{minipage}%
	}%
 
	\subfigure{
    \ContinuedFloat
	\vspace{-0.75cm}
		\begin{minipage}[t]{\linewidth}
			\centering
			\includegraphics[scale=.35]{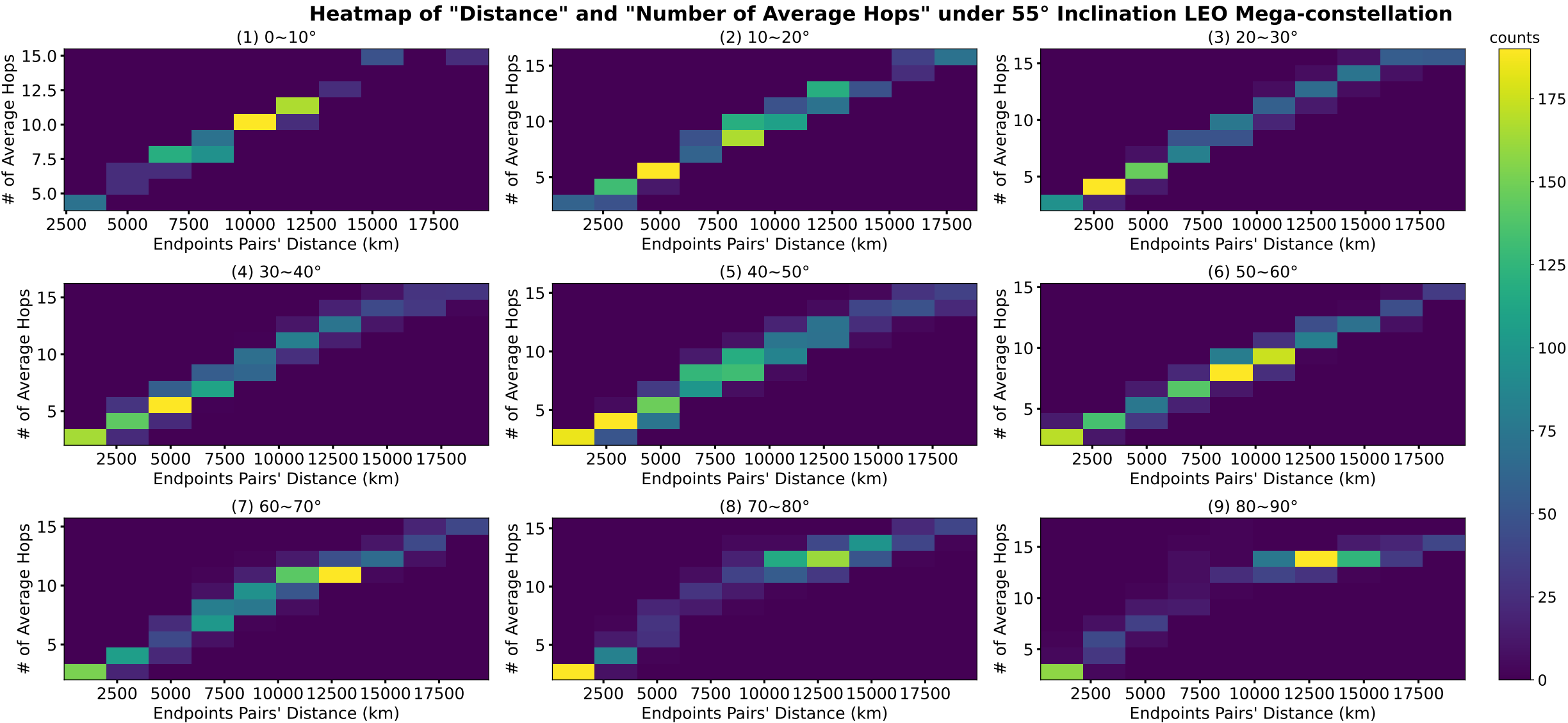}   \label{heatmap 2}
   \caption{(b) Satellite Orbit Inclination 55° }
			\vspace{-0.25cm}

		\end{minipage}%
	}
	\caption{\fontsize{10}{12}\selectfont Heatmap of "Distance" vs. "Number of Average Hops" for constellations under satellite orbit inclination of (a) 45° and (b) 55°. The city pairs are divided into nine categories based on their geographic angle, each spanning a geographic angle interval of 10° (into one of the nine sub-plots). Within each sub-plot (i.e., city pairs with similar inclination), the city pairs are further divided into bins based on their geodesic distance (along the x-axis).  Comparing the same bin across (a) and (b) helps us understand the change in the number of average hops for the same city pairs under different inclinations. For example, city pairs in the bin 50°-60° have a lower average hop count in (b) when the inclination angle of the orbit is more closely aligned with their geographic angle.}
	\vspace{-0.45cm}
	\label{heatmap}
\end{figure*}

\FloatBarrier

\section{Visualization of routing path}
\label{apendix:visulization}

\begin{figure*}[htp]
	\centering  %
	\subfigure[Beijing-Jakarta]{
		\includegraphics[width=0.17\textwidth]{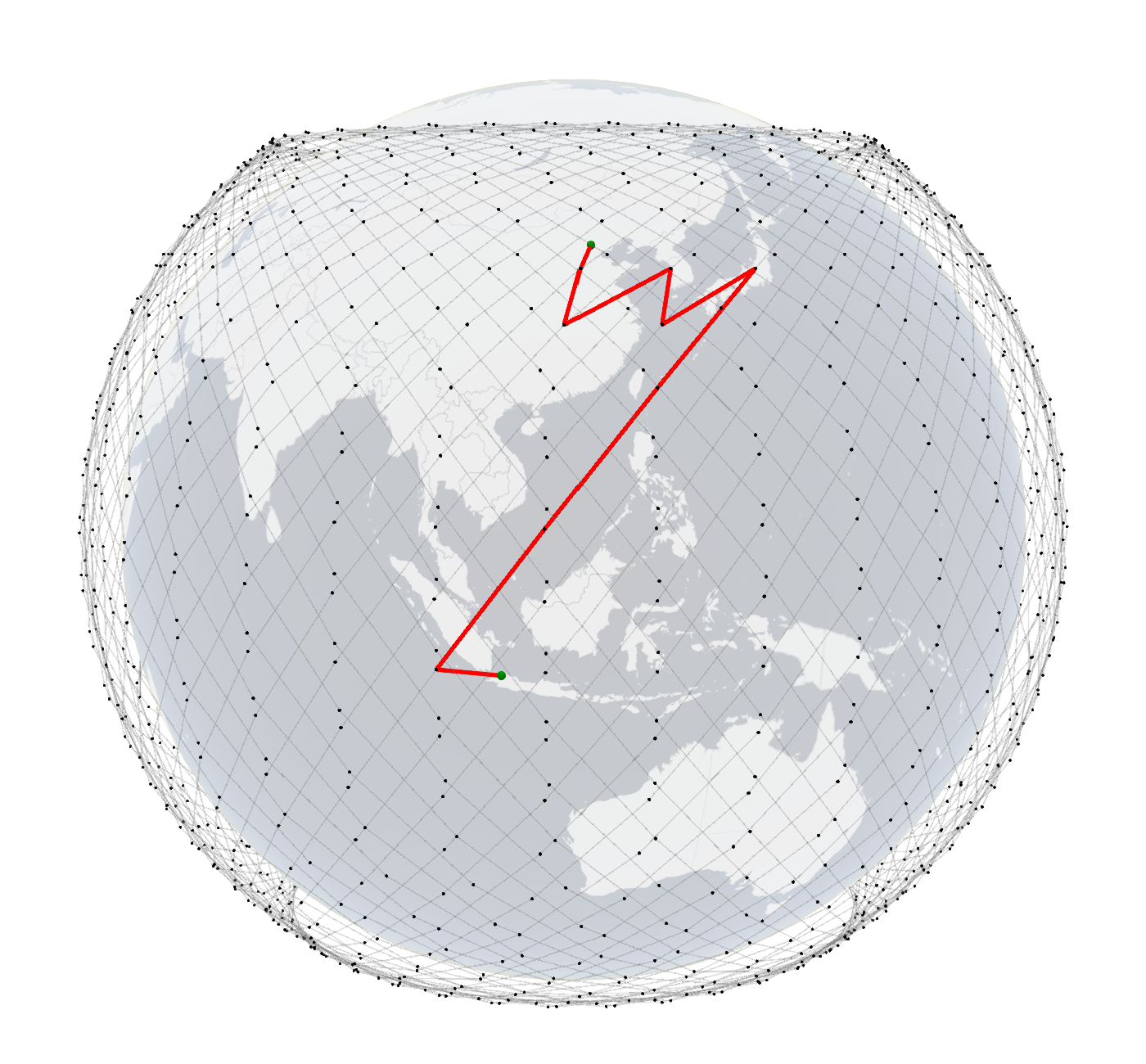}
    }
	\subfigure[London-Dalian]{
		\includegraphics[width=0.17\textwidth]{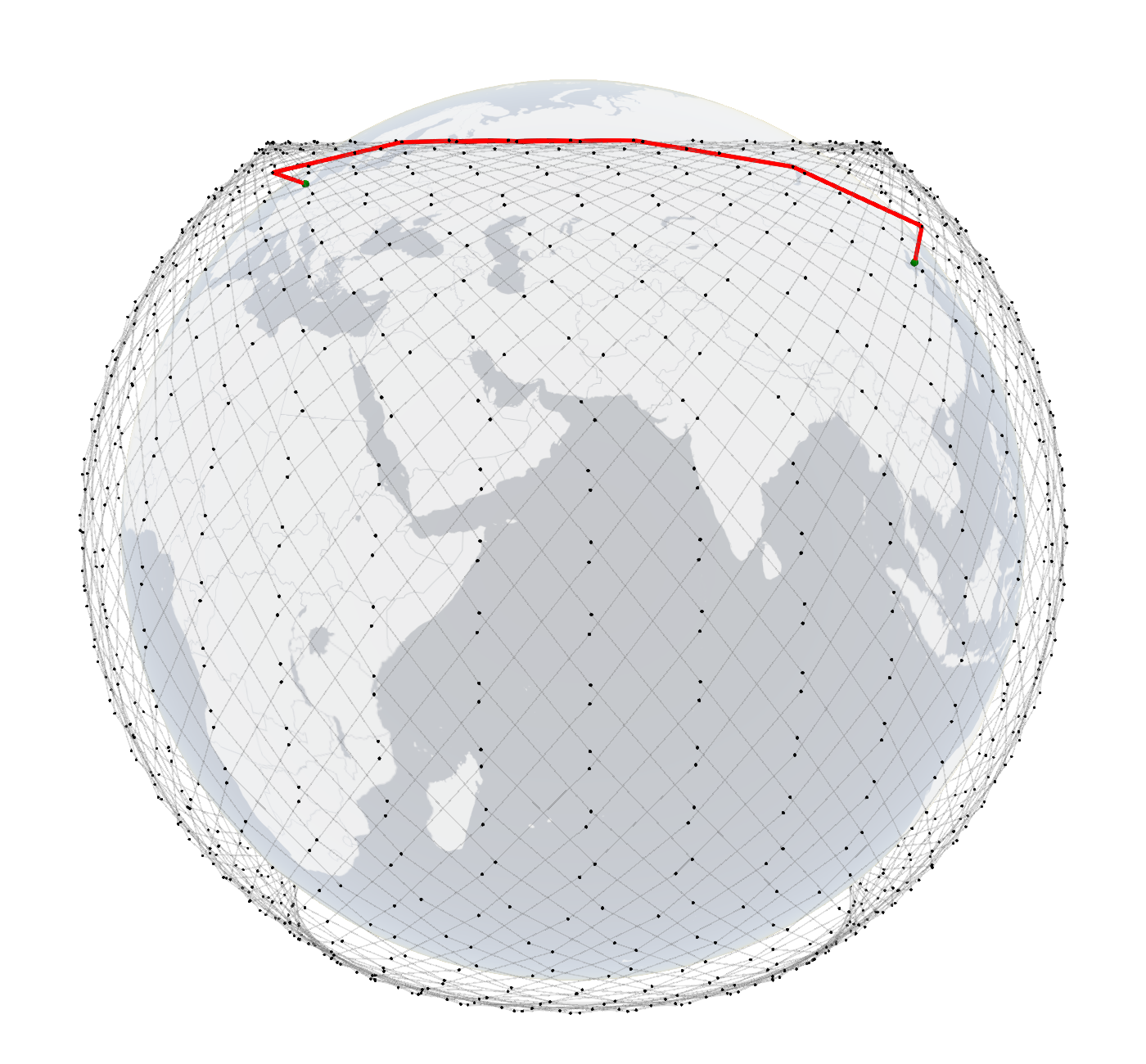}}
	\subfigure[Madrid-Tehran]{
		\includegraphics[width=0.17\textwidth]{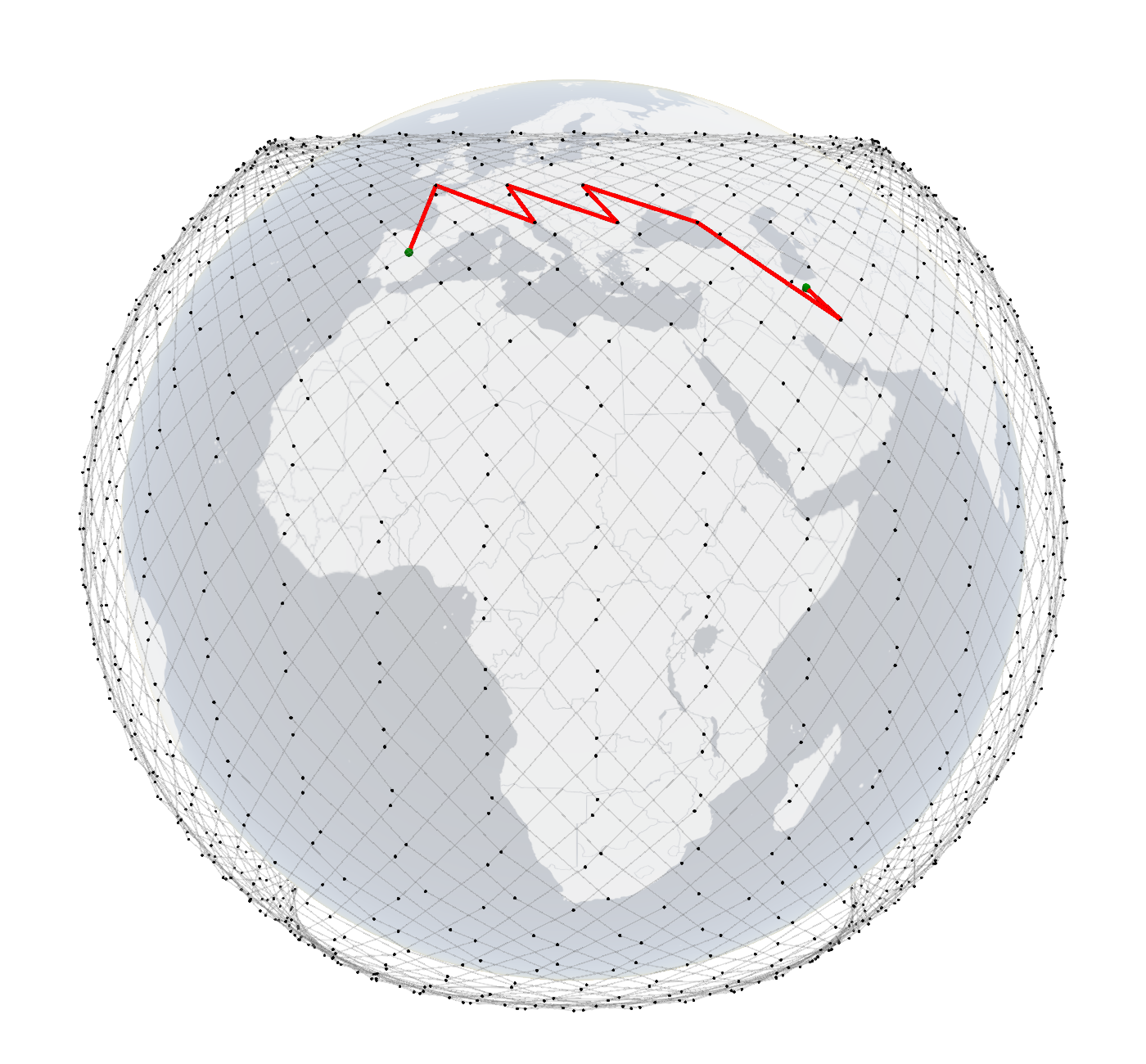}}
	\subfigure[Shanghai-Buenos Aires]{
		\includegraphics[width=0.17\textwidth]{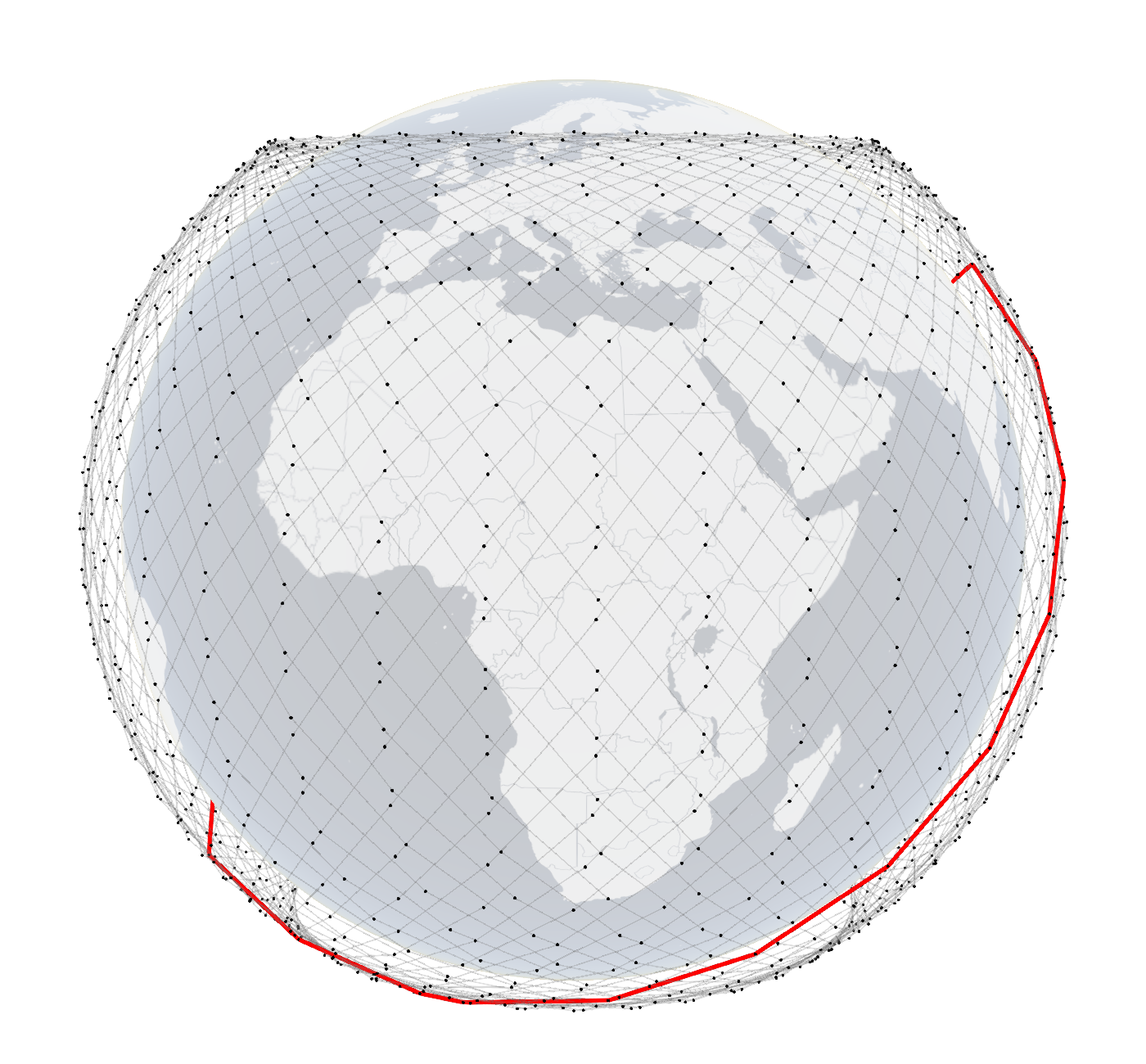}}
    \subfigure[Singapore-Nairobi]{
		\includegraphics[width=0.17\textwidth]{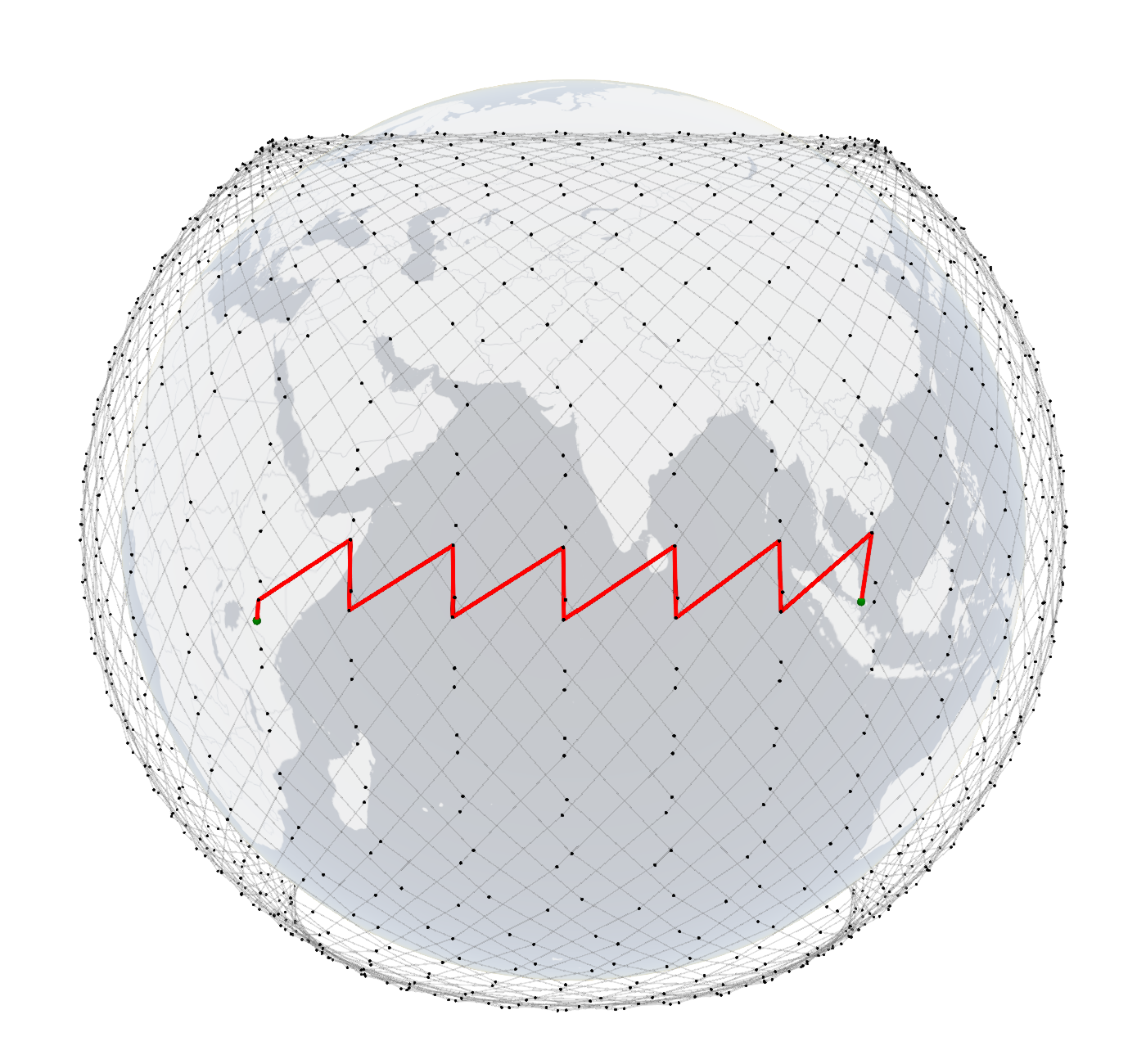}}
    \subfigure[Delhi-Brasília (1)]{
		\includegraphics[width=0.17\textwidth]{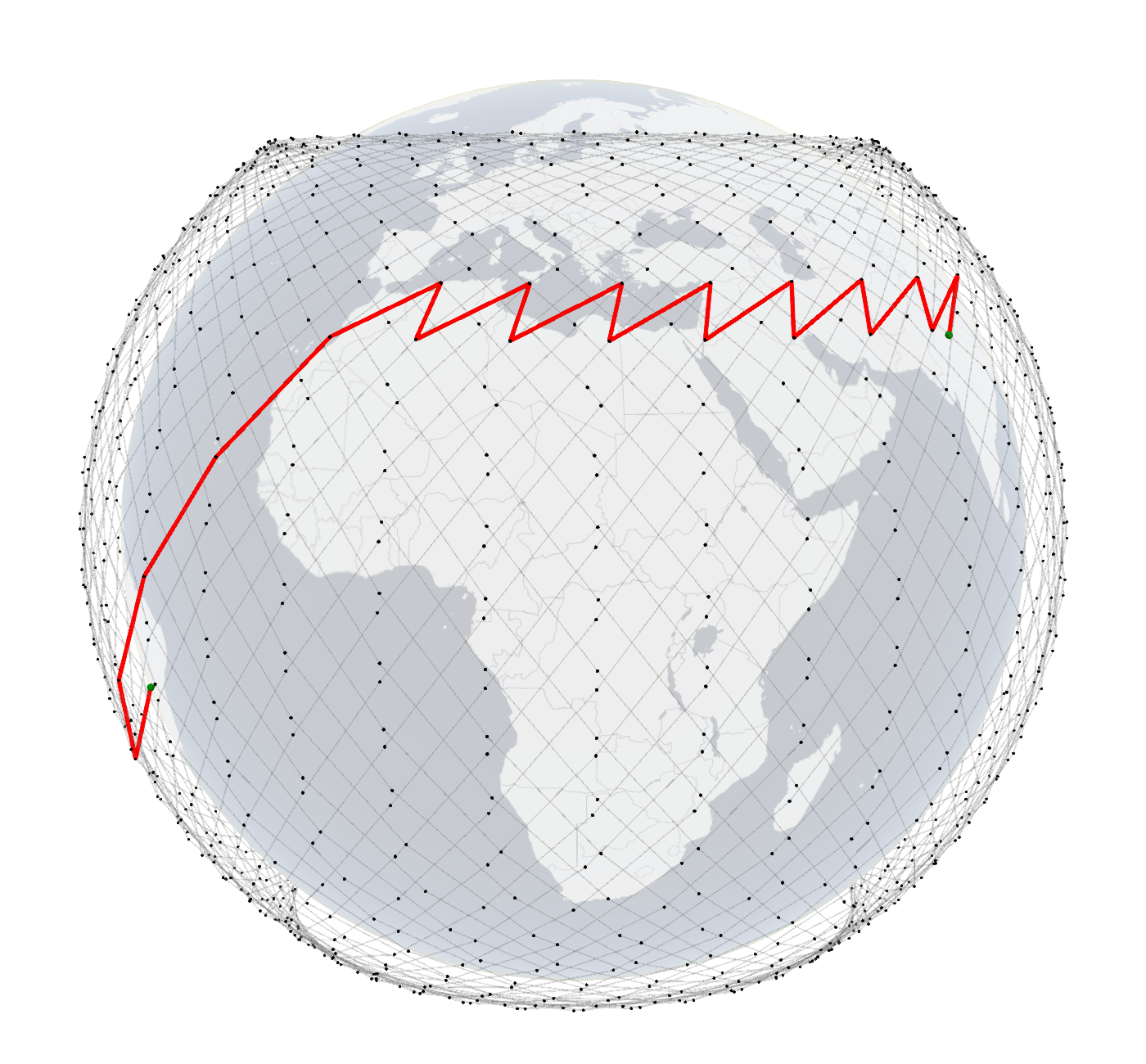}}
	\subfigure[Delhi-Brasília (2)]{
		\includegraphics[width=0.17\textwidth]{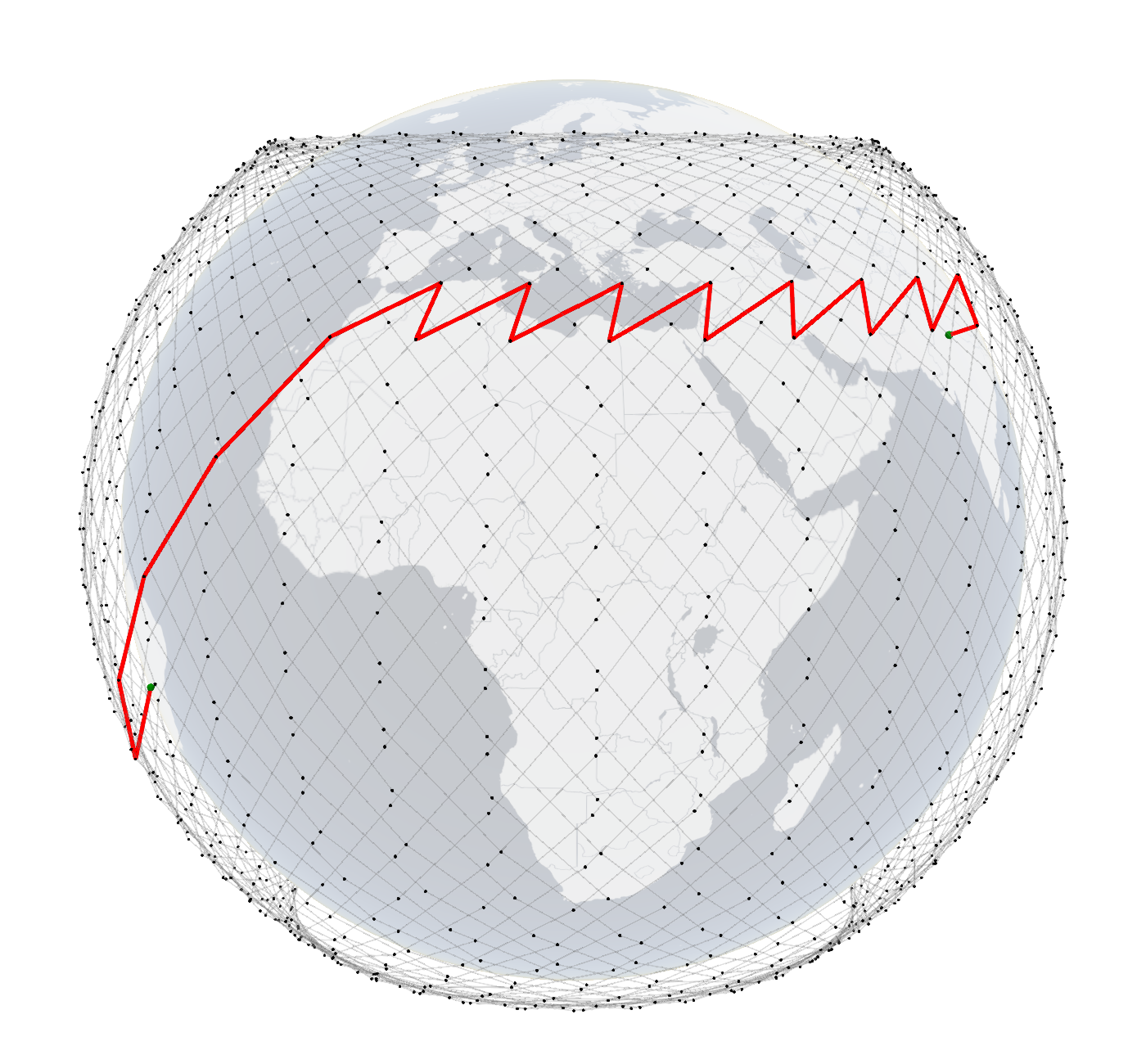}}
	\subfigure[Delhi-Brasília (3)]{
		\includegraphics[width=0.17\textwidth]{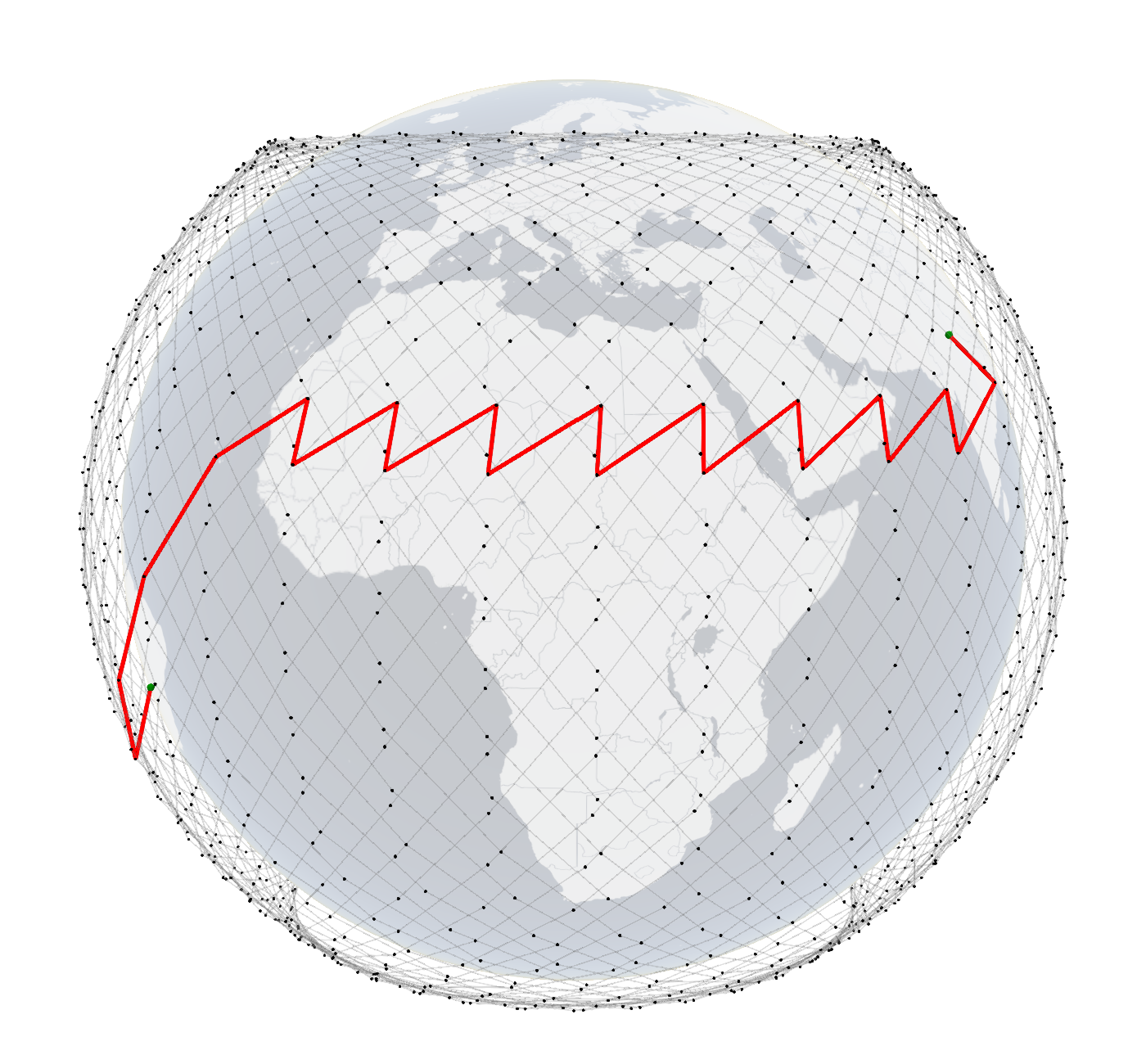}}
	\subfigure[Delhi-Brasília (4)]{
		\includegraphics[width=0.17\textwidth]{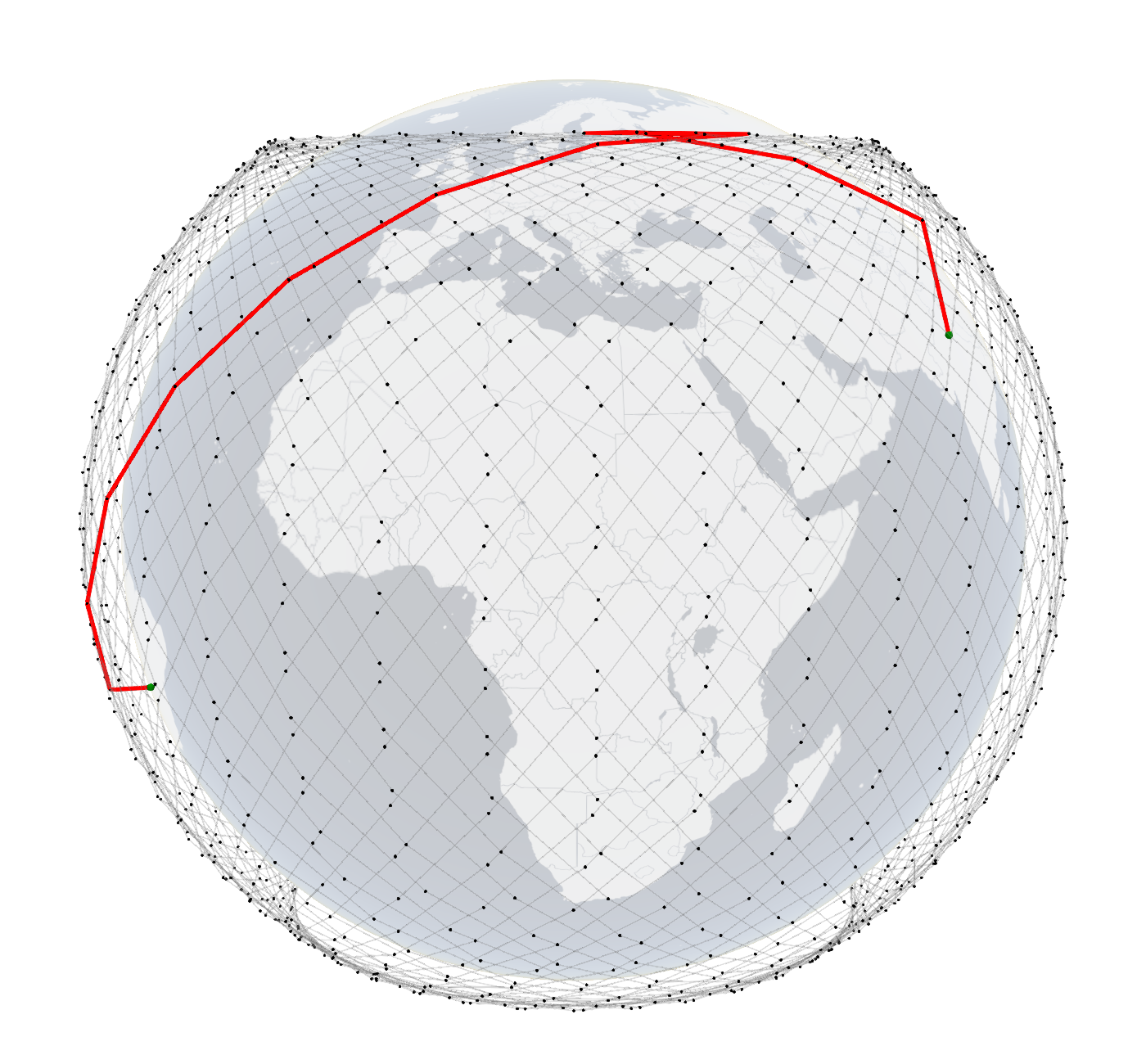}}
    \subfigure[Delhi-Brasília (5)]{
		\includegraphics[width=0.17\textwidth]{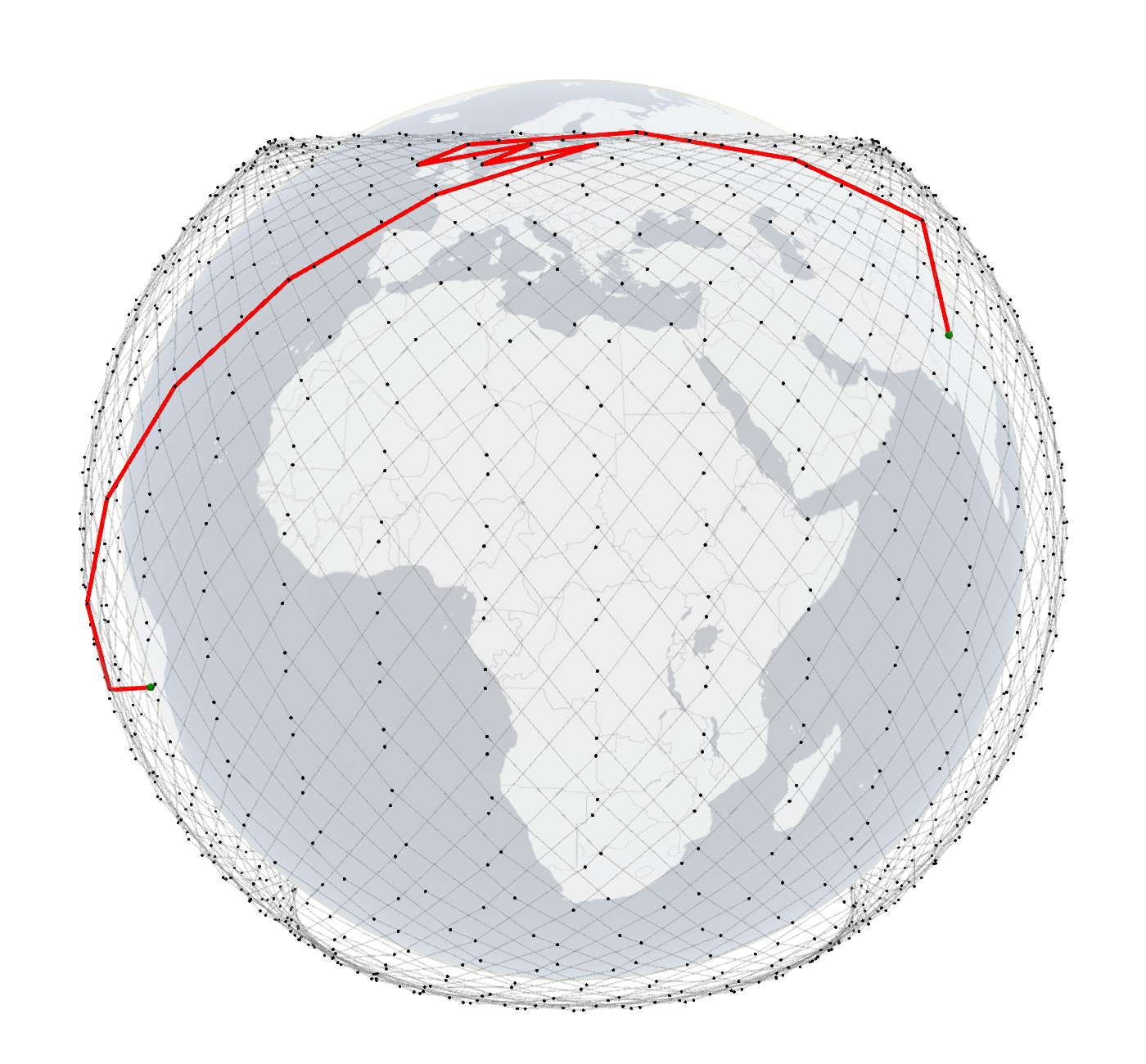}}
	\caption{Visualization of routing path of different Endpoints pairs (a)-(e), and changes of routing path overtime (f)-(j).}
        \label{vis path}
\end{figure*}

Fig.~\ref{vis path} shows the routing path of different Endpoints pairs, and shows the changes of routing path of one particular Endpoints pairs. We can observe that when the data is transmitted to north or south direction, the routing path tends to go through intra-orbit hops. When transmitting to east or west direction, the routing path tends to go through inter-orbit hops, resulting a zig-zag pattern.

\end{document}